\documentclass[classica,12pt,twoside,cucitura]{toptesi} 

\pdfoutput=1

\usepackage[applemac]{inputenc} 
\usepackage[english,italian]{babel}

\usepackage[bookmarks=false,hyperfootnotes=false]{hyperref}
\hypersetup{ 
pdfpagemode={UseOutlines}, 
bookmarksopen, 
pdfstartview={FitH}, 
colorlinks, 
linkcolor={black}, 
citecolor={black}, 
urlcolor={blue} 
pdftitle={Tesi Magistrale - Fraternale Federico},
pdfauthor={Federico Fraternale},
}
\usepackage{gfsbodoni}
\usepackage[T1]{fontenc}
\usepackage{placeins}
\usepackage{graphicx}
\usepackage{multirow}
\usepackage{tabularx}
\usepackage{epspdfconversion}

\usepackage[font=footnotesize]{caption} 
\addto\captionsenglish{%
}

\usepackage[margin=0pt,skip=-10pt,font=footnotesize]{subcaption}
\usepackage[a4paper]{geometry}

\usepackage{amssymb,amsmath,amsbsy}       
\usepackage{latexsym} 
\usepackage{amsthm} 
\usepackage{eucal}
\usepackage{bm}
\usepackage{bbm}
\usepackage{layaureo}
\usepackage{fancyhdr}
\usepackage{titlesec}
\titleformat{\chapter}[display]
{\normalfont\sc\Large\centering}
{{Chapter\ }{\thechapter}}{30pt}{\rm\huge}
\titlespacing{\chapter}{0pt}{120pt}{54.5pt}
\titleformat{\section}
{\normalfont\rm\Large}
{\thesection}{1em}{}
\titlespacing{\section}{0pt}{30pt}{18.6pt}
\titleformat{\subsection}
{\normalfont\it\large}
{\thesubsection}{1em}{\itshape}
\titlespacing{\subsection}{0pt}{18pt}{10pt}
\titleformat{\subsubsection}
{\normalfont\rm\centering}
{\thesubsubsection}{1em}{\itshape}
\titlespacing{\subsubsection}{0pt}{15pt}{10pt}

{\setlength{\rightskip}{-\marginparsep}\addtolength{\rightskip}{-\marginparwidth
}}{\par}

\usepackage[titletoc]{appendix}
\usepackage{url}
\usepackage{natbib}

\newcommand{\hv}{\hat v}
\newcommand{\hu}{\hat u}
\newcommand{\hw}{\hat w}
\newcommand{\he}{\hat \eta}
\newcommand{\de}[2]{\frac{\partial #1 }{\partial #2}}
\newcommand{\dde}[2]{\frac{\partial^2 #1}{\partial #2^2}}
\newcommand{\dddde}[2]{\frac{\partial^4 #1}{\partial #2^4}}
\newcommand{\dtot}[2]{\frac{\mathrm{d} #1}{\mathrm{d} #2}}
\newcommand{\BU}[1]{\mathop{}\!\bm{\mathrm{ #1}}}
\newcommand{\ud}{\mathop{}\!\mathrm{d}}  
\newcommand{\up}[1]{\mathrm{#1}}  
\newcommand{\figref}[1]{Fig.~\ref{#1}}
\newcommand{\tabref}[1]{Tab.~\ref{#1}}
\newcommand{\secref}[1]{\S \ref{#1}}
\newcommand{\idop}{\mathbbm I}
\newcommand{\matlab}{Matlab{\textsuperscript{\rm{\textregistered}}} }
\newcommand{\eqdef}{\overset{def}{=}}

\usepackage{listings}
\usepackage[svgnames]{xcolor}
\usepackage{tikz}
\usetikzlibrary{decorations.markings,shapes,snakes}
\pgfdeclarelayer{edgelayer}
\pgfdeclarelayer{nodelayer}
\pgfsetlayers{edgelayer,nodelayer,main}

\tikzstyle{none}=[inner sep=0pt]

\tikzstyle{circ}=[circle,fill=White,draw=Black]
\tikzstyle{rett}=[rectangle,fill=White,draw=Black,rounded corners]
\tikzstyle{rombo}=[diamond,fill=White,draw=Black]

\tikzstyle{link}=[->,draw,thin]
\tikzstyle{link_plain}=[-,draw,thin]
\tikzstyle{arrow}=[->,draw,thin]

\begin{document}
\begin{titlepage}

\begin{center}
{\fontsize{20}{1.2}\textrm{POLITECNICO DI TORINO}}\\[0.5cm]
{\fontsize{16}{1.2}\textrm{I Facoltà di Ingegneria}}\\[0.2cm]
{\fontsize{13}{1.2}\textrm{Corso di Laurea Magistrale in Ingegneria Aerospaziale}}\\[2.78cm]
{\fontsize{18}{1.2}\textrm{Tesi di Laurea Magistrale}}\\[1.5cm]
\begin{minipage}[t]{1.0\textwidth}
{ \fontsize{18}{1.2}\bfseries\textit{Frequency Transient of Three-Dimensional Perturbations in Shear Flows.
Similarity Properties and Wave Packets Linear Formation.}}
\end{minipage}\\[2cm]

\includegraphics[width=4cm]{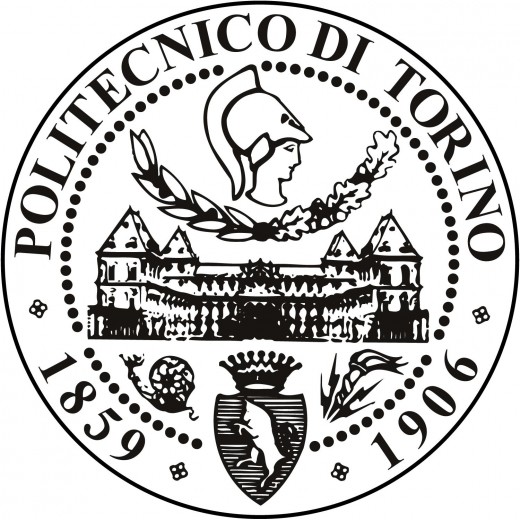}\\[1.5cm]


\begin{minipage}[t]{0.4\textwidth}
\begin{flushleft} \large
\emph{Relatore:}\\
Prof. Daniela Tordella\\ 
\ \\ \emph{Correlatore:}\\
Prof. Gigliola Staffilani\\
(Massachusetts Institute of Technology)
\end{flushleft}
\end{minipage}
\hfill
\begin{minipage}[t]{0.35\textwidth}
\begin{flushleft} \large
\emph{Candidato:} \\

Federico Fraternale
\end{flushleft}
\end{minipage}

\vfill

{\large Marzo 2013}

\end{center}
\end{titlepage}

\
\thispagestyle{empty}
\newpage
\pagenumbering{Roman}
\sommario
La Stabilit\`a Idrodinamica e la transizione alla turbolenza sono stati oggetto di studio sin dalla fine del XIX
secolo. In particolare, la discrepanza tra teoria e osservazioni sperimentali nel caso di transizioni
\textit{subcritiche}, ha costituito un problema complicato che ha promosso la ricerca di altri meccanismi che potessero
generare la transizione, differenti da quello classico che prevede la crescita esponenziale asintotica delle onde di
Tollmien-Schlichting. Nonostante il ruolo delle nonlinearit\`a sia universalmente riconosciuto, un rinnovato interesse
verso l'analisi lineare a partire dalla fine del XIX secolo \`e derivato dai risultati dell'analisi non modale.
La possibilit\`a di una crescita di tipo algebrico, pur significativa e anche per perturbazioni asintoticamente stabili
come nel caso del flusso piano di Couette, ha aperto un nuovo scenario nello studio sulla transizione
laminare-turbolento. Si osserva infatti che alcuni meccanismi, come il \textit{vortex tilting} oppure il ruolo
delle perturbazioni ortogonali al flusso medio, vengono riscontrati gi\`a dall'analisi lineare e tridimensionale.
Inoltre si \`e mostrato che la crescita in energia cinetica della perturbazione \`e soltanto attribuibile al un
meccanismo lineare.\par
Scopo del presente lavoro \`e quello di contribuire alle conoscenze attuali sull'evoluzione temporale di piccole
perturbazioni tridimensinali in flussi confinati, in particolare il flusso di Couette piano, tramite l'analisi delle
velocit\`a di fase
e delle frequenze. La loro evoluzione
temporale \`e stata, ed \`e tuttora, poco analizzata ma contiene in realt\`a preziose informazioni sulla vita delle
perturbazioni. I risultati ottenuti per flussi di Couette e Poiseuille mostrano la possibilit\`a di velocit\`a di fase
diverse per le tre componenti di velocit\`a, e soprattutto la presenza di brusche variazioni o salti nell'evoluzione
temporale delle frequenze. Tali variazioni permettono di distinguere tre periodi distinti della vita della singola
onda, l'\textit{Early transient}, l'\textit{Intermediate transient} e il \textit{Far transient}, e sembrano essere
correlate con l'instaurarsi di certe condizioni di self-similarit\`a nei profili di velocit\`a o vorticit\`a.
Tali analisi non sarebbero state possibili senza lo sviluppo di un codice di calcolo in ambiente \matlab basato
su una soluzione semi analitica (per flussi confinati) del problema ai valori iniziali di Orr-Sommerfeld e Squire. Tale
soluzione \`e espressa come serie di funzioni ortogonali, e la soluzione approssimata viene ottenuta applicando il
metodo variazionale di Galerkin. Il codice risultante risulta decisamente vantaggioso i termini di tempi di calcolo e
accuratezza. A concludere il lavoro, viene mostrata l'evoluzione lineare di disturbi localizzati in forma di pacchetti
d'onda per flusso di Couette e di Strato Limite. Si evidenziano le analogie con uno scenario di transizione in presenza
di \textit{spot} turbolenti, le quali portano a supporre che alcune propriet\`a di tali strutture risiedano gi\`a
nelle equazioni di governo linearizzate.\par \ \vspace{1cm} \\
Il presente lavoro di tesi \`e stato in parte svolto al dipartimento di Matematica del \textit{Massachusetts Institute
of Technology} di Cambridge (USA), sotto la supervisione della Prof.ssa Gigliola Staffilani e tramite il progetto di
mobilit\`a extra-UE FP (Final Froject). Tale opportunit\`a \`e il risultato della collaborazione tra la Prof.ssa
Tordella e la Prof.ssa Staffilani.
\selectlanguage{english}

\tableofcontents
\cleardoublepage

\clearpage
\pagenumbering{arabic}
\chapter{Introduction}
\section{Linear Stability and transition}
The reasons for the breakdown of a laminar flow to turbulence has been one of the central issues in fluid mechanics for
over a hundred years, for the many applications in the engineering, meteorology, oceanography and astrophysics. The
theoretical work on transition is mainly based on the linear stability studies, which were firstly initiated  in the
nineteenth century by \citet{Helmholts1868}, Rayleigh and Kelvin. \citet{Reynolds1883} dedicated to  experiments on
the instability of the pipe flow, and was the first to find the existence of a \textit{critical velocity} (actually,
the non-dimensional parameter that now brings his name, the \textit{Reynolds number}) above which the transition to
turbulence occurs. He
observed the intermittent character of this phase as well, naming \textit{flashes} the objects that we now call
\textit{turbulent spots}.\\
The formulation for the viscous stability problem is due to \citet{Orr1907} and \citet{Sommerfeld1908}, who dedicated
respectively to the Plane Couette flow and to the Plane Poiseuille flow. The Orr-Sommerfeld equation has become the
basis of the modal theory of hydrodynamic stability. Many years later \citet{Tollmien1929} calculated the first neutral
eigenvalues for Plane Poiseuille flow, and Schlichting continued his work, leading to the definition of the TS-waves,
whose role in the transition process is salient. \\
Only in the second half of the twentieth century the three-dimensional initial value problem was considered. The
transient dynamics of perturbations revealed aspects that made the non-modal problem even more of interest than than
the past analysis on the asymptotic states. The most important result is the presence of an algebraic behavior in the
early and intermediate stages of a perturbation's life; three main reasons for the transient growth were found: the
non-orthogonality of the eigenfunctions, the possible resonance between the Orr-Sommerfeld and the Squire solutions
and, for unbounded or semi-bounded flows, the presence of a continuous spectrum (see e.g. the works by Criminale and
Gustavsson). The role of these mechanisms, though linear, in a transitional scenario is evident, and it is easy
to understand why many efforts were made in the last two decades to investigate the conditions for ``optimal growth''.
Only in the recent years the role of the linear mechanisms in the \textit{subcritical} transition to turbulence has
been pointed out by many authors (see, among others, Henningson). 

 \section{Thesis motivations and layout}
The aim of the present thesis is to contribute to the actual knowledge about the transient behavior of small
perturbations in channel flows. The focus will be  on a quantity whose temporal evolution had not been
considered in detail before: the phase velocity or, equivalently, the frequency of the components of
velocity and vorticity of a perturbation. Throughout the present work, it will be shown as from the analysis of the
wave frequency, three terms of a disturbance's life can clearly be discerned. Some properties of similarity of the
velocity and vorticity profiles will also be highlighted.\par
In Chapter 2 the mathematical background is given, and the principal equations and definitions are introduced. In
Chapter 3 an analytical method to solve the Orr-Sommerfeld and Squire initial value problem is
presented,
together with the implementation of a \matlab code to obtain approximate solutions. The suggested method is verified
and used for the further analysis. The focus of Chapter 4 is on the perturbation frequency and phase velocity. Numerical
results are shown in terms of both the vorticity and velocity components, and similarity properties of the profiles are
investigated. The last Chapter concerns the evolution of wave packets and linear spots.
\begin{figure}[h!]
        \centering
        \advance\leftskip0cm
        \begin{subfigure}{1\textwidth}
        \centering
\includegraphics[width=14.0cm]{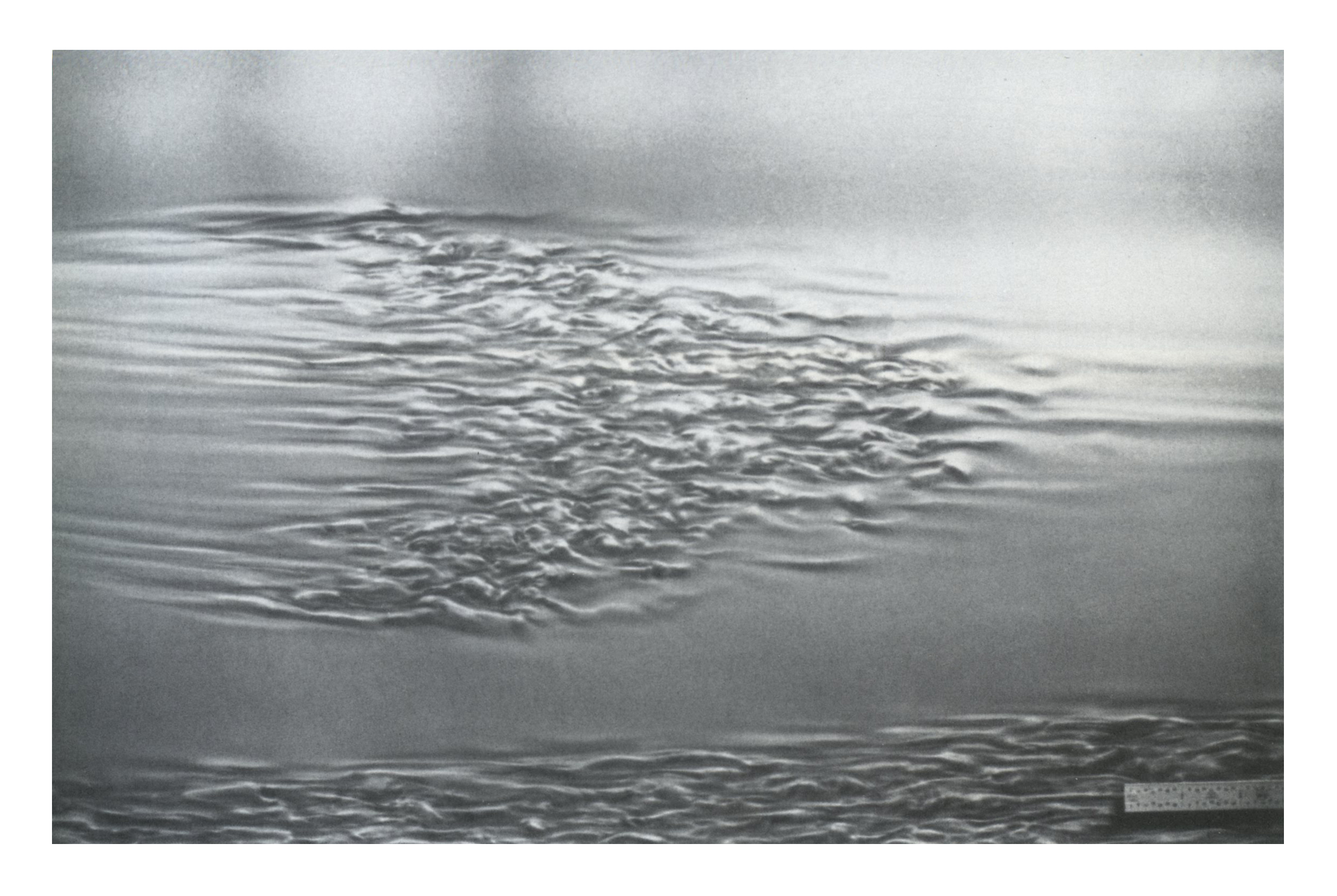}
	\vspace{-0.6cm}
	 \end{subfigure}
        \begin{subfigure}{1\textwidth}
        \centering 
\includegraphics[width=14cm]{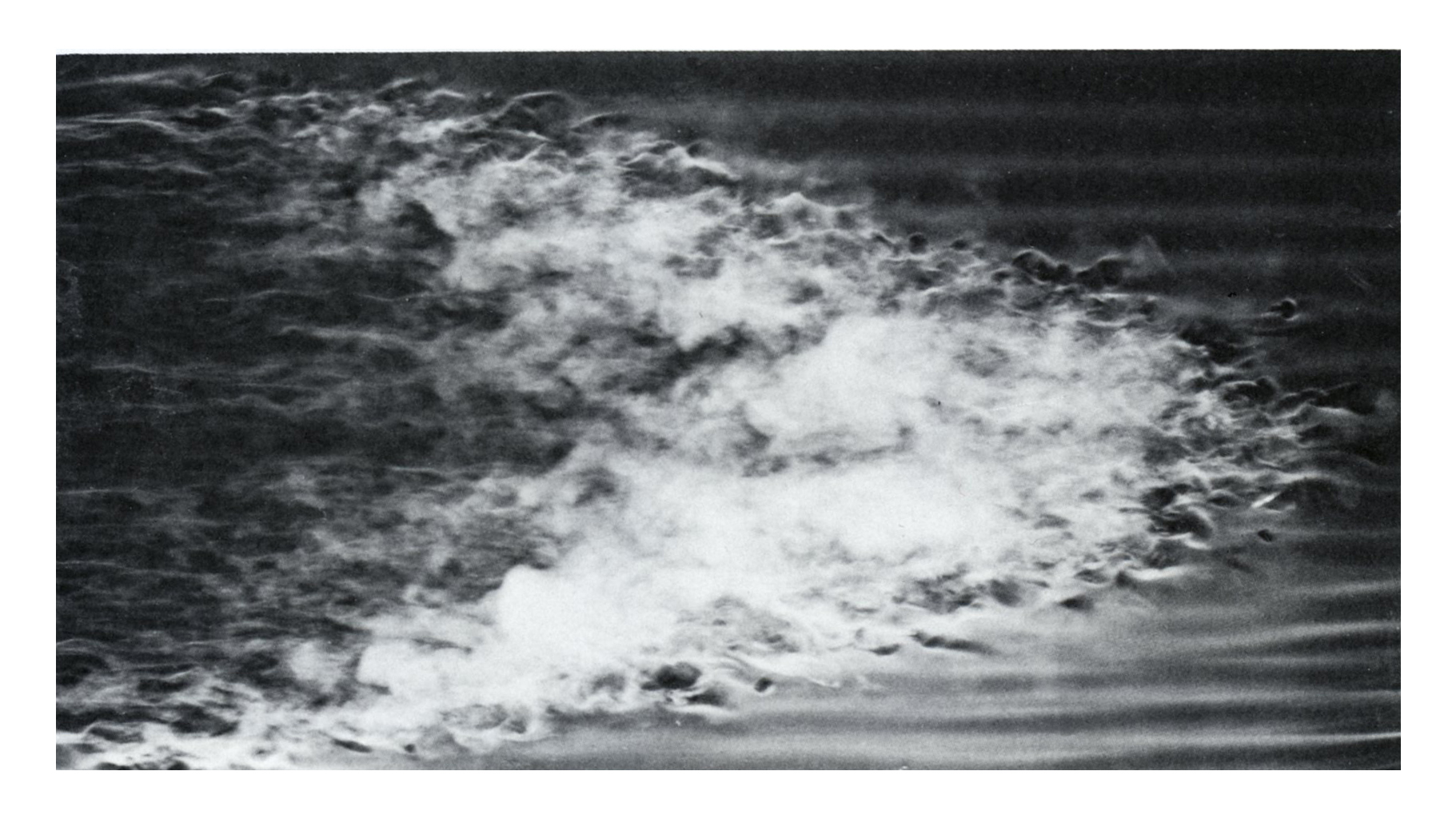}
	\vspace{-0.0cm}
	 \end{subfigure}
	\caption{Turbulent spot on a flat plate.\ \textit{Top}: $Re=200000$, the
sublayer of the spot is shown from the glass wall of the water channel, by a suspension of aluminium
flakes \citep{Cantwell1978}. \textit{Bottom}: $Re=400000$, the typical arrowshape angle becomes steeper; visualization
by smoke in air with flood lighting, photograph by R. E. Falco, taken from the book of \citet{VanDyke_book} }
\label{fig:Spot_real}
\end{figure}
\FloatBarrier
\ \newpage \ \thispagestyle{plain}
\chapter{Mathematical background}
\section{Initial value problem for shear flows: viscous linear analysis}
 \subsection{Base governing equations}
\begin{figure}
        \centering
         \advance\leftskip-2cm
         \advance\rightskip-2cm
        \begin{subfigure}{0.39\textwidth}
	\includegraphics[width=1.05\textwidth]{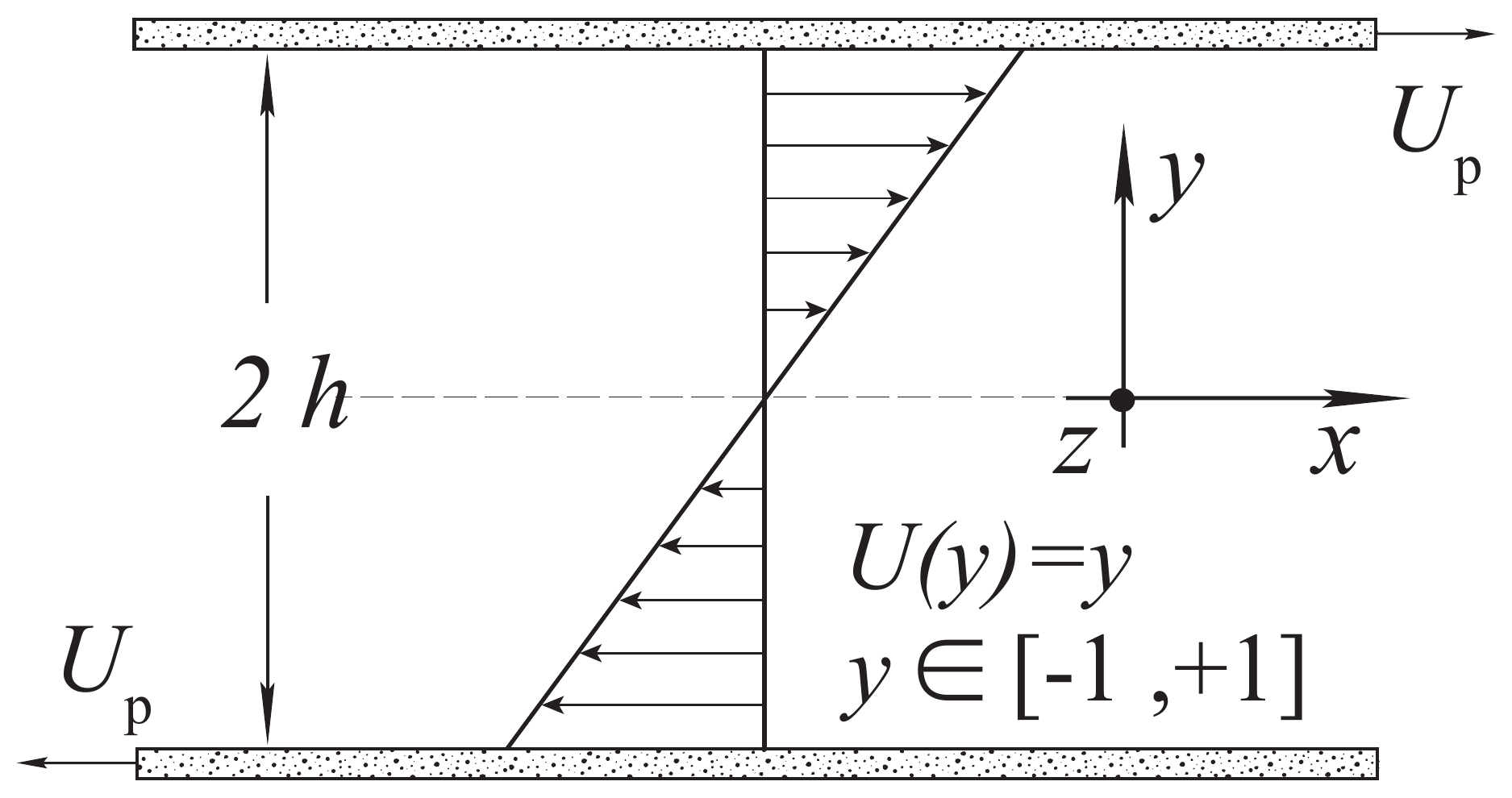}
	\vspace{5pt}
	\subcaption{PCf: $L_{ref}=h$, $U_{ref}=U_p$}
	 \end{subfigure}\hspace{10pt}
        \begin{subfigure}{0.39\textwidth}
\includegraphics[width=0.95\textwidth]{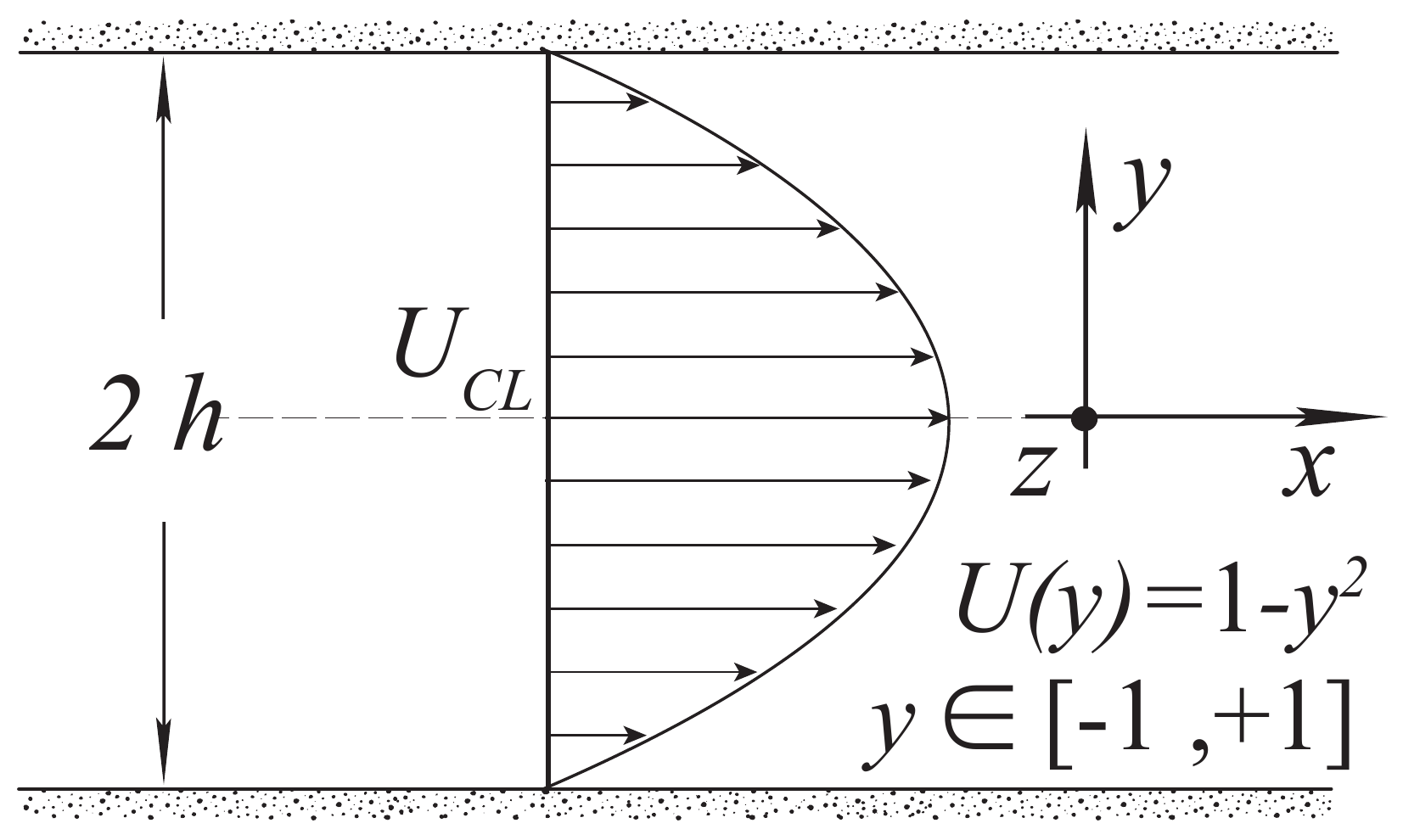}
	\vspace{19pt}
	\subcaption{PPf: $L_{ref}=h$, $U_{ref}=U_{CL}$}
	 \end{subfigure}
	  \begin{subfigure}{0.39\textwidth}
	\includegraphics[width=1\textwidth]{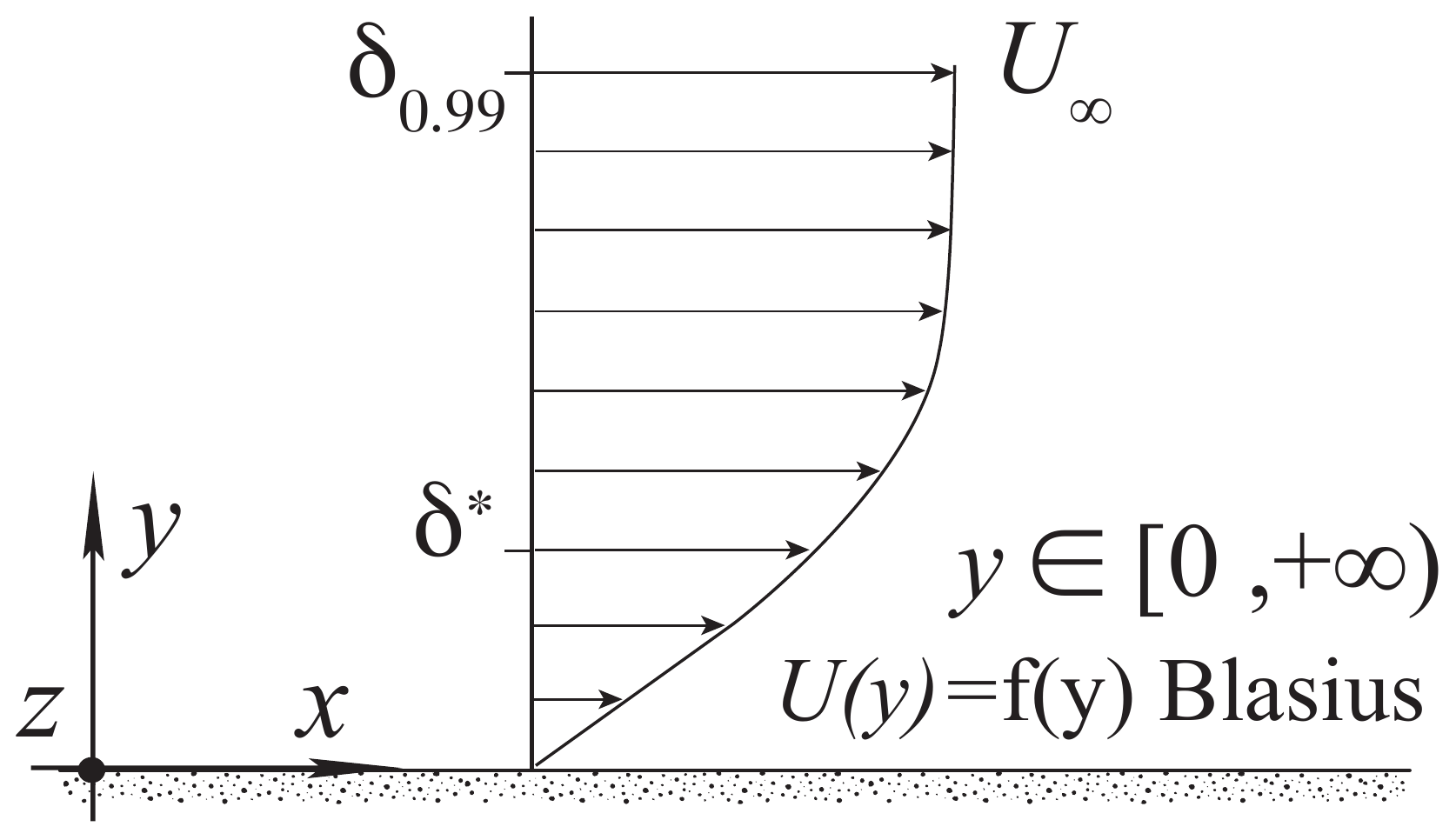}
	\vspace{5pt}
	\subcaption{Bbl: $L_{ref}=\delta^*$, $U_{ref}=U_\infty$}
	 \end{subfigure}
	\caption{Flow schemes: reference quantities, base flow and reference systems.}
\label{fig:base_flow_schemes}
\end{figure}

  In the present analysis, the flow is taken to be incompressible and the
governing equations for infinitesimal disturbances in parallel flows are
considered. The base flow general expression is $U_i=U(y)\delta_{1i}$, i.e. the
streamwise direction is $x$, and it only depends on the wall-normal
direction $y$. The origin of the reference system is set on the channel
symmetry plane $xz$ for Plane Couette flow and Plane Poiseuille flow (PCf and PPf, in the following), and
on the wall for Blasius boundary layer flow (Bbl), i.e. the flow along a flat
plate with zero pressure gradient(\figref{fig:base_flow_schemes}). The equations governing the
general evolution of fluid flow are the Navier-Stokes equations, that using
Cartesian tensor notation read

\begin{align}
\label{NS}
\de{ u_i}{t} &=
-u_j\de{ u_i}{\tilde x_j}-\de{ p}{x_i}+\frac{1}{Re}\nabla^2
u_i \\
\de{ u_i}{ x_i}&=0
\end{align}

supported with the typical initial and boundary conditions of the form

\begin{equation}
\centering
\begin{aligned}
u_i(\BU{x},0)&=u^0_i(\BU{x})\\
u_i(\BU{x},t)&=0\ \ \ \textrm{on walls}
\end{aligned}
\end{equation}
The physical quantities \textit{u},\textit{v} and \textit{w} represent the
velocity components, \textit{p} represents the flow static pressure, and they
appear in the system \eqref{NS} in nondimensional form. For PCf and PPf the reference
length is the channel semi-height \textit{h}, the reference velocity is
assumed to be the medium wall velocity $U_p=(U^+-U^-)/2$ for PCf, and the centeline velocity $U_{CL}$ for PPf. For Bbl,
the
velocity scale is the freestream velocity $U_\infty$ and the length scale is the
boundary layer displacement thickness $\delta^*$, which takes the following
expression, as exact solution of the Blasius equation \citep[p.
141]{Schlichting_book}:

 \begin{align}
  \delta^* = 1.7208\sqrt{\frac{\nu x}{U_{\infty}}}\ \ \ (displacement\
thickness) 
 \end{align}
 The approximate expression for the geometric thickness, defined as the
distance for which $u=0.99U_\infty$, is found to be
 \begin{align}
  \delta_{0.99} = 4.91\sqrt{\frac{\nu x}{U_{\infty}}}\ \ \ (geometric\
thickness) 
 \end{align}
 So the following definitions for the Reynolds number will be considered
\begin{gather}
Re=\frac{U_ph}{\nu}\ \ \textrm{Plane Couette flow}\hspace{8mm} Re=\frac{U_{CL}h}{\nu}\ \ \textrm{Plane Poiseuille
flow}\\
Re=\frac{U_\infty\delta^*}{\nu}\ \ \textrm{Blasius boundary layer} 
\end{gather}

where $\nu$ is the kinematic viscosity.
The evolution equation for the disturbances can be obtained by splitting the
flow
in two components, the Base flow $(U_i(y),P(y))$ and the perturbed state
$(\tilde u_i(\mathbf{x},t),\tilde p(\mathbf{x},t))$ so that the complete fluid
field
can be written as $u_i=U_i+\tilde u_i$ and $p=P+\tilde p$. The nonlinear
disturbance equations read
\begin{align}
\label{NS_pert_nonlin}
\de{\tilde u_i}{t} &= -U_j\de{\tilde
u_i}{x_j}-\tilde u_j\de{U_i}{x_j}-\de{\tilde
p}{x_i} + \frac{1}{Re}\nabla^2 \tilde u_i -\tilde u_j\de{\tilde u_i}{x_j} \\
\de{\tilde u_i}{x_i}&=0
\end{align}
toghether with the appropriate initial and boundary conditions.\vfill
\newpage
 \subsection{Linearized perturbative equations}

Considering the $x$ reference axis oriented as the base flow streamwise
direction, so that it assumes the general expression  $U_i=U(y)\delta_{1i}$, the
complete velocity field becomes $\mathbf{u}=(U+\tilde u,
\tilde v, \tilde w)$. In particular, for PCf $U(y)=y$, and for Bbl the base
velocity profile is tabulated in self-similar coordinates \citep[Chap.
V]{Rosenhead}.
Introducing the mean velocity profile and assuming small perturbations, the
following linear equations can be written, as shown by \citet{Schmid_book} and
\citet{Criminale_book}:
\begin{align}
\label{NS_lin_cont}
&\de{\tilde u}{x}+\de{\tilde v}{y}+\de{\tilde w}{z}=0\\
\label{NS_lin_mom_u}
&\de{ \tilde u}{t} +U\de{ \tilde u}{x}+\tilde v\frac{dU}{dy} =-\de{ \tilde
p}{x}+ \frac{1}{Re}\nabla^2 \tilde u\\
\label{NS_lin_mom_v}
&\de{\tilde v}{t}+U\de{\tilde v}{x} =-\de{\tilde p}{y}+ \frac{1}{Re}\nabla^2
\tilde v\\
\label{NS_lin_mom_w}
&\de{\tilde w}{t} +U\de{\tilde w}{x} =-\de{\tilde p}{z}+ \frac{1}{Re}\nabla^2
\tilde w
\end{align}
Taking the divergence of the linearized momentum equations \eqref{NS_lin_mom_u},
\eqref{NS_lin_mom_v}, \eqref{NS_lin_mom_w}, and using the continuity equation
\eqref{NS_lin_cont}, an equation for the fluctuating pressure can be obtained
and used to eliminate the pressure terms, in combination with
\eqref{NS_lin_mom_v}, leading to the following equation for the
wall-normal velocity:

\begin{equation}
\label{NS_lin_v}
\bigg[(\de{}{t}+U\de{}{x})\nabla^2-\frac{d^2U}{dy^2}\de{}{x}-\frac{1}{Re}
\nabla^4\bigg]\tilde v=0
\end{equation}
To completely describe the three-dimensional flow field, a second equation
is necessary, and it is convenient to write an equation for wall-normal
vorticity, defined as $\tilde \eta=\de{\tilde u}{z}-\de{\tilde w}{x}$
\begin{equation}
\label{NS_lin_omega_y}
\bigg[\de{}{t}+U\de{}{x}-\frac{1}{Re}\nabla^2\bigg]\tilde
\eta=-\frac{dU}{dy}\de{ \tilde v}{z}
\end{equation}

The quantity $\tilde \Gamma$ is then defined as $\tilde \Gamma=\nabla^2 \tilde
v$, so that the system becomes
\begin{align}
\tilde \Gamma&=\nabla^2 \tilde
v\\
\label{NS_lin_vv}
\bigg[\de{}{t}+U\de{}{x}-\frac{1}{Re}\nabla^2\bigg]\tilde
\Gamma&=\frac{d^2U}{dy^2}\de{ \tilde v}{x}\\
\bigg[\de{}{t}+U\de{}{x}-\frac{1}{Re}\nabla^2\bigg]\tilde
\eta&=-\frac{dU}{dy}\de{ \tilde v}{z}
\end{align}
The perturbations are Fourier transformed in x and z directions: two real
wavenumbers, $\alpha$ and $\beta$ are introduced along the $x$ and $z$
coordinates, respectively. The generic quantity $\hat f$ is hence expressed as
\begin{equation}
\hat f(y,t;\alpha,\beta)=\int_{-\infty}^{+\infty}\int_{-\infty}^{+\infty}\tilde
f(x,y,z,t)e^{-i\alpha x-i\beta z}dxdz
\end{equation}
The system can now be written in the following form
\begin{align}
&\dde{\hv}{y}-k^2\hv=\tilde \Gamma\\
&\de{\hat \Gamma}{t}=-ikcos(\phi)U\hat
\Gamma+ikcos(\phi)\frac{d^2U}{dy^2}\hv+\frac{1}{Re}\bigg(\dde{\hat
\Gamma}{y}-k^2\hat \Gamma \bigg)\\
&\de{\he}{t}=-ikcos(\phi)U\he-iksin(\phi)\frac{d^2U}{dy^2}\hv+\frac{1}{Re}
\bigg(\dde{\he}{y}-k^2\he \bigg)
\end{align}

where $\phi=tan^{-1}(\beta/\alpha)$ is the perturbation obliquity angle, and
$k=\sqrt{\alpha^2+\beta^2}$ is the polar wavenumber. The following boundary
conditons applies in the wavenumber space, respectively for Bbl and PCf:
\begin{gather}
 \hv(y=\pm 1,t)= \de{\hv}{y} (y=\pm 1,t)= \he(y=\pm 1,t)=0\ \ \ \ \\
 \hv(y \to +\infty,t)= \de{\hv}{y} (y \to +\infty,t)= \he(y \to +\infty,t)=0
\end{gather}
The streamwise velocity $\hu$ and the spanvise velocity $\hw$ can be recovered from
the following expressions
\begin{gather}
\label{eq:sol_u}
 \hu=\frac{i}{k^2}(\alpha\de{\hv}{y}-\beta\he)\\
 \label{eq:sol_w}
 \hw=\frac{i}{k^2}(\beta\de{\hv}{y}+\alpha\he)
\end{gather}

 \subsection{Energy amplification factor}\label{sec:G}
 In order to quantify the growth of the perturbations, a natural choice is the
\textit{kinetic energy density}, defined as

\begin{align}
e(t;\alpha,\beta)&=\frac{1}{2}\int_{y_a}^{y_b}
\big(|\hu|^2+|\hv|^2+|\hw|^2\big)dy \\
&=\frac{1}{2k^2}\int_{y_a}^{y_b}\bigg(\bigg|\de{\hv}{y}
\bigg|^2+k^2|\hv|^2+|\he|^2\bigg)dy
\end{align}
where $y_a$ and $y_b$ are the limits of the domain.
As a disturbance measure,
the proper quantity is
the \textit{energy amplification factor}, $G$, defined as the kinetic energy
density normalized with respect to its initial value
\citep{Criminale1997,Lasseigne1999}
\begin{gather}
G(t;\alpha,\beta)=\frac{e(t;\alpha,\beta)}{e(t=0;\alpha,\beta)}
\end{gather}
the \textit{temporal growth rate} of the kinetic energy $r$ is then introduced to
evaluate the beginning of the exponential asymptotic period, when $dr/dt\to0$
\begin{equation}
 r(t;\alpha,\beta)=\frac{log|e(t;\alpha,\beta)|}{2t}, t>0
\end{equation}

\chapter{Wave transient analysis: an eigenfunction expansion solution method}
\section{Introduction}
In the present chapter an analytical solution to the Orr-Sommerfeld and
Squire initial value problem (eq.~\ref{Orr-Somm} and \ref{Squire}) is
researched for channel flows, aiming to a better understanding of
the early and intermediate terms of a perturbation's life. 
As a starting point, the IVP in the normal-velocity and
normal-vorticity form is considered
\begin{align}
\label{Orr-Somm}
&\de{}{t}\dde{\hv}{y}-k^2\de{\hv}{t}+i\alpha U(y)\dde{\hv}{y}-i\alpha k^2
U(y)\hv-i\alpha
U''(y)\hv-\frac{1}{Re}\bigg(\dddde{\hv}{y}-2k^2\dde{\hv}{y}+k^4\hv\bigg)=0\\
\label{Squire}
&\de{\he}{t}+i\alpha U(y)\he-\frac{1}{Re}\bigg(\dde{\he}{y}-k^2\he\bigg)=-i\beta
U(y)^{\prime}\hv
\end{align}
\begin{gather}
\hv(y=\pm 1,t)= \de{\hv}{y} (y=\pm 1,t)= \he(y=\pm 1,t)=0\\
\hv(y,t=0)=\hv_0(y)\ \ \ \ \he(y,t=0)=\he_0(y)
\end{gather}

where the prime symbol indicates a total derivative along $y$. The evolution of
the wall-normal velocity $\hv$ is described by the Orr-Sommerfeld PDE
\eqref{Orr-Somm}, which is of fourth order in the spatial coordinate $y$ and homogeneous, with homogeneous
boundary conditions. The Squire equation \eqref{Squire} is inhomogeneous and
the forcing term $-i\beta U(y)^{\prime}\hv$ is known as \textit{vortex
tilting}, being the product of the main vorticity in the spanwise direction
($\Omega_z=-U^{\prime}$) and the perturbation velocity $\hv$. This term is
responsible of the increase of the normal vorticity, for three-dimensional
perturbations \citep[see][]{Criminale1997}.\\
About the initial conditions, the following will be used in the present work
\begin{align}
 &\hv_0(y)=(1-y^2)^2\ \ \ \he_0(y)=0\hspace{1cm}Symmetrical\\
 &\hv_0(y)=y(1-y^2)^2\ \ \ \he_0(y)=0\hspace{1cm}Antisymmetrical
\end{align}

It is known  that for bounded flows all eigenvalues of the
Orr-Sommerfeld and Squire ODE are
discrete and infinite in number and
that the eigensolutions of the problem form a complete set as proved by \citet{Schensted1960} and \citet{DiPrima1969}.
For unbounded or semi-bounded flows (as the Wake  or the Boundary layer flows) \citet{Miklavcic1982} and
\citet{Miklavcic1983} proved that if the base flow decays in an exponential way, then only a finite number of
eigenvalues exists and a continuum is present, while if the decay is algebraic there exists a infinite discrete set
(without the continuum).\par
The focus of this chapter is on channel flows. Most of the studies in the past century deal with the modal analysis.
About the Orr-Sommerfeld ODE, it is possible to express the solution as a
generalized Fourier series once a base of orthogonal functions is found, and
variational or Galerkin methods can be applied to provide very accurate approximations when a finite number of trial
functions are used.\\ 
The Orr-Sommerfeld and Squire modes can be used to express the solution \citep[see][]{Schmid_book}, however it was shown
that some sets of normal functions can give better results in
terms of accuracy and computational  cost.
\citet{Orszag1971} solved the Orr-Sommerfeld ODE  numerically using expansions
in Chebyshev polynomials and used the Lanczos's tau method to determine the series coefficients. He showed that this
series gives the highest convergence rate, since the error after $N$ terms is smaller than any power of $N^{-1}$. 
Before him, \citet{Dolph1958} were the first applying a Galerkin method to obtain the  coefficients (reduction to a
system of $N$ algebraic equations), together with the $QR$ algorithm. They used normal functions that guarantee a $N^4$
rate. \citet{Gallagher1962} used the Chandrasekhar functions, adopted in the present work as well, which
provide a rate of convergence of $N^5$.\par
About the solution of the initial value problem, there is no conceptual
difficulty in using the eigensolutions of the Orr-Sommerfeld and Squire ODE
system but, as outlined in \citet{Drazin_book}, this requires the solution of
the adjoint differential equation. \\
In the first part of this  chapter an eigenfunction expansion method for the initial value problem
\eqref{Orr-Somm}-\eqref{Squire} is proposed; the method does not
involve the eigensolutions to the Orr-Sommerfeld and Squire ODE system, and the approximate time-dependent coefficients
are obtained with the
variational minimization principle. The two PDEs are then reduced to a system of $N$ ODEs. A \matlab code is
implemented
and verified and afterwards used for the analysis of Chapter \ref{chap:wave_transient}, whose focus will be on the wave
frequency and velocity profiles.\newpage
\section{Solution to $\hv$ equation}\label{sec:v_solution}
\subsection{Choice of a base of orthogonal functions}\label{sec:base_v}
\begin{figure}
        \centering
         \advance\leftskip-2.5cm
         \advance\rightskip-2cm
        \begin{subfigure}{0.6\textwidth}
        \centering 
	\includegraphics[width=9.5cm]{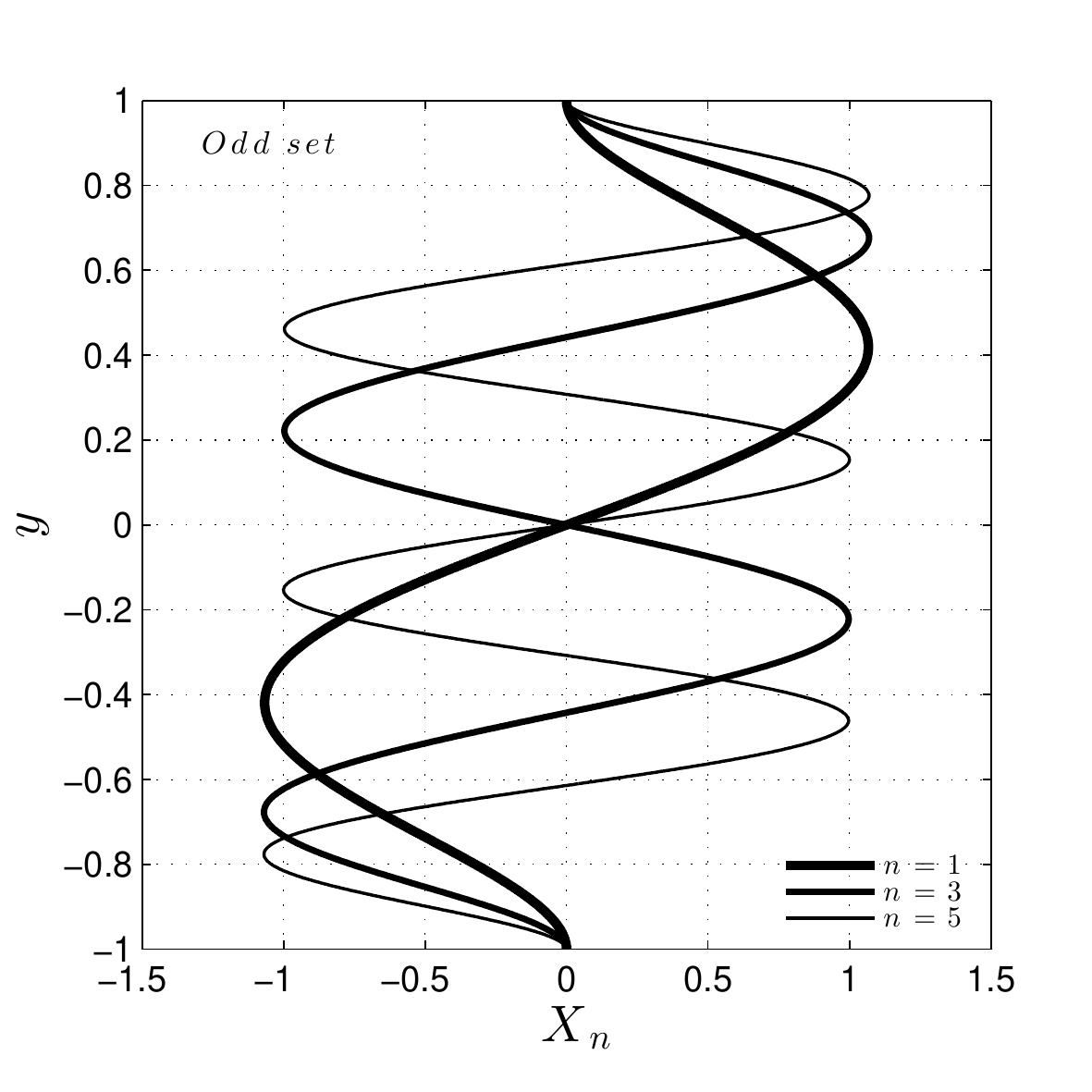}
	 \end{subfigure}
        \begin{subfigure}{0.6\textwidth}
        \centering 
	\includegraphics[width=9.5cm]{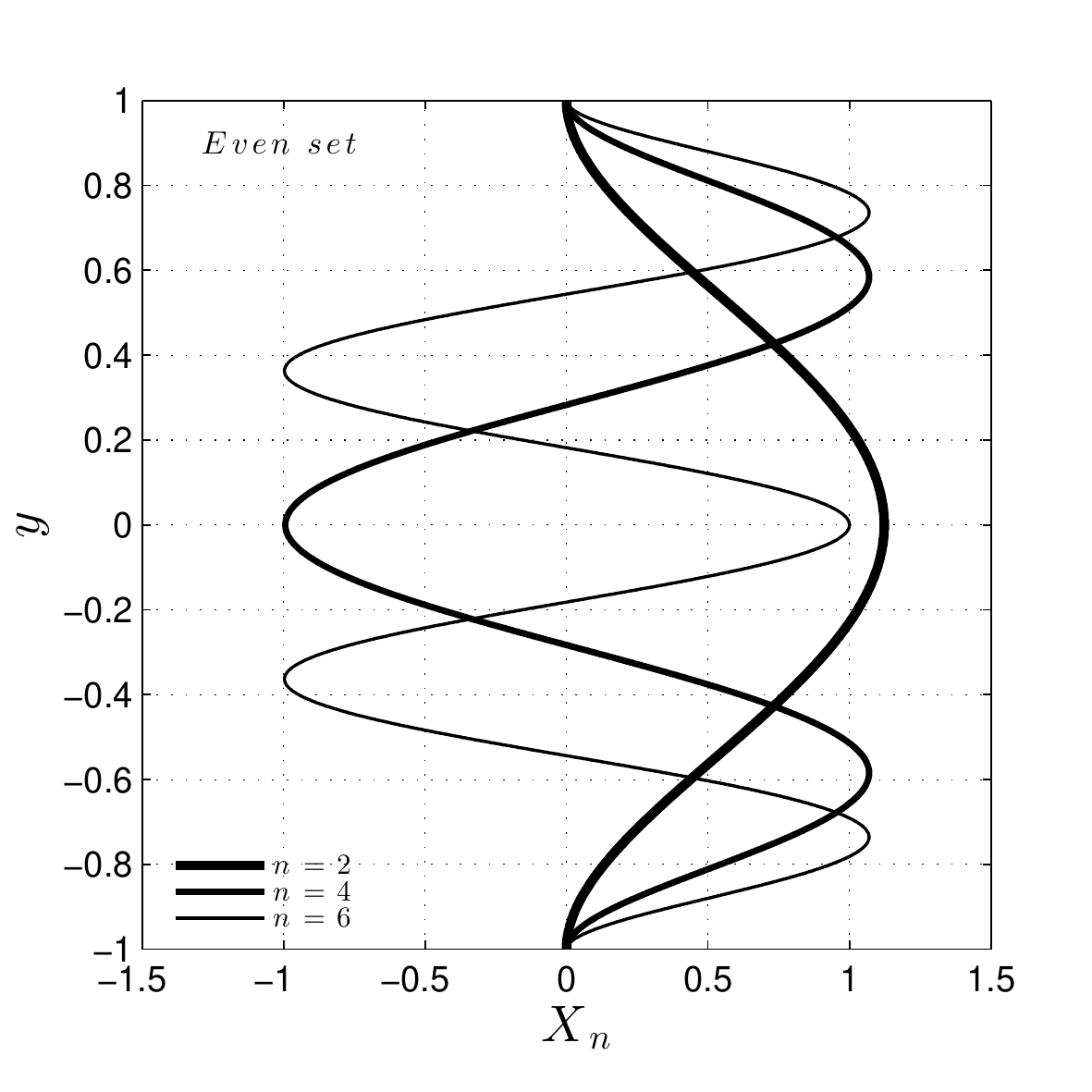}
	 \end{subfigure}
	\caption{The basis eigenfunctions}
\label{fig:modes_v}
\end{figure}

The solution of \eqref{Orr-Somm} can be expressed as a generalized Fourier
expansion, with time-dependent coeffcients:
\begin{gather}
\label{v_expansion}
\hv(y,t)=\sum_{n=1}^{\infty}c_n(t)X_n(y)\ \ \ \ y \in[-1,1]
\end{gather}
where $X_n(y)$ are orthogonal functions, and the following inverse
transform applies \citep[see][]{PDE_book}:
\begin{gather}
\label{inverse_trasf}
c_n(t)=\frac{\int_{-1}^1 \hv(y,t)X_n(y)\ud y}{\int_{-1}^1 X_n(y)X_n(y)\ud y}
\end{gather}
Since in the initial value problem both the initial condition
and the boundary conditions need to be imposed,  
it is worthwhile to consider functions that satisfy the boundary conditions of the problem considered.
Moreover, note that the coefficients $c_n$ of the series are in general
complex, since $\hv$ is complex-valued and the spatial modes are
considered as real. 
The particular orthogonal functions which we use are those defined by
the following fourth order eigenvalue problem satisfying the same boundary
conditions of the original equation. This choice for the simplified problem is not the only possible
but revealed to be appropriate; note that this model equation is contained in the diffusive part of the PDE
\eqref{Orr-Somm} 
\begin{gather}
\label{problem4}
\dddde{X(y)}{y}=\lambda^4 X(y)\ \ \ \ \ y \in[-1,1]\\
\label{bc_problem4}
X(y=\pm 1)=0\ \ \ \ \de{X}{y}(y=\pm 1)=0
\end{gather}
A solution to this problem is obtained considering sines, cosines,
hyperbolic sines and hyperbolic cosines (see appendix \ref{sec:Appendix_A1} for
the complete solution). Two different sets of eigenvalues and the corresponding
eigenfunctions are found, respectively odd and even, by numerically solving 
the following transcendental equations
\begin{gather}
\label{trasc_odd}
tan(\lambda_n)-tanh(\lambda_n)=0\ \ \ (odd\
set)\\
\label{trasc_even}
tan(\lambda_n)+tanh(\lambda_n)=0\ \ \ (even\ set)
\end{gather}
The corresponding normalized eigenfunctions
(\figref{fig:modes_v}) are
\begin{align}
&X_n=\frac{1}{\sqrt{2}}\bigg[\frac{sinh(\lambda_n
y)}{sinh(\lambda_n)}-\frac{sin(\lambda_n
y)}{sin(\lambda_n)}\bigg]\hspace{1cm}{n=1,3,5..,N-1}\ \ \ (odd\
set)\\[10pt]
&X_n=\frac{1}{\sqrt{2}}\bigg[\frac{cosh(\lambda_n
y)}{cosh(\lambda_n)}-\frac{cos(\lambda_n
y)}{cos(\lambda_n)}\bigg]\hspace{1cm}{n=2,4,6..,N}\ \ \ \ \ (even\
set)
\end{align}
Similar functions, in a different domain, have been used by
\citet[][app. V]{Chandrasekhar_book}, in the study of the circular Couette flow
between coaxial cylinders, and by \citet{Gallagher1962} to solve the
Orr-Sommerfeld ODE.

Since the imaginary and the real part of the solution $\hv$ usually have
opposite parity, independently on the initial condition, both the odd
and the
even set are necessary to completely describe the problem and obtain the
correct result. \\
In the following paragraphs a compact notation for the space derivatives is
introduced. In order to simplify the reading, the $y$-derivatives will be
indicated with a subscript. The temporal derivatives will be indicated
explicitly or with a dot. 
\newpage

\subsection{Weak formulation and approximate solution to $\hv$ equation by
Galerkin method}\label{sec:gal_v}
Substituting the expansion \eqref{v_expansion} in $\hv$ equation
\eqref{Orr-Somm} yields
\begin{equation}
\begin{split}
0 &=\sum_{n=1}^{\infty}\dtot{}{t}
c_n(t)X_{n_{yy}}-k^2\sum_{n=1}^{\infty}\dtot{}{t} c_n(t)X_n+i\alpha
U(y)\sum_{n=1}^{\infty}c_n(t)X_{n_{yy}}\\
&-i\alpha k^2 U(y)\sum_{n=1}^{\infty}c_n(t)X_n
-i\alpha 
\frac{\ud^2
U(y)}{\ud y^2}\sum_{n=1}^{\infty}c_n(t)X_n-\frac{1}{Re}\sum_{n=1}^{\infty}
c_n(t)X_ {
n_{yyyy}}\\
&+\frac{2k^2}{Re}\sum_{n=1}^{\infty}c_n(t)X_{
n_{yy}}-\frac{k^4}{Re}\sum_{n=1}^{\infty}c_n(t)X_n
\end{split}
\end{equation}
The above expression represents an exact form. If only a finite number of modes
is considered, the equation is not satisfied exactly, so a residual
$\epsilon$ (dependent on the choice of the functions $X_n$) appears at the left hand
side
\begin{equation}
\begin{split}
\label{residual}
\epsilon(y,t;\ \alpha,\ \beta) &=\sum_{n=1}^{N}\dtot{}{t}
c_n(t)X_{n_{yy}}-k^2\sum_{n=1}^{N}\dtot{}{t} c_n(t)X_n+i\alpha
U(y)\sum_{n=1}^{N}c_n(t)X_{n_{yy}}\\
&-i\alpha k^2 U(y)\sum_{n=1}^{N}c_n(t)X_n
-i\alpha 
\frac{\ud^2
U(y)}{\ud y^2}\sum_{n=1}^{N}c_n(t)X_n-\frac{1}{Re}\sum_{n=1}^{N}
c_n(t)X_ {
n_{yyyy}}\\
&+\frac{2k^2}{Re}\sum_{n=1}^{N}c_n(t)X_{
n_{yy}}-\frac{k^4}{Re}\sum_{n=1}^{N}c_n(t)X_n
\end{split}
\end{equation}
Galerkin (1915) focused on the problem of minimizing the
functional $\epsilon$, so his method consists of a variational approach \citep[see also][p. 27-32]{Chandrasekhar_book}.
He
showed that the best approximation of the solution is obtained when the
error is orthogonal to the space of the linearly independent trial functions
$X_n$ with $n={1,2,..N}$. In this context, given two functions $u(y)$ and
$v(y)$ with $y\in\ \Omega=[-1,\ 1]$, the following definition of scalar product
applies
\begin{equation}
 \langle u,v\rangle\eqdef\int_\Omega u\cdot v\ud y
\end{equation}
so, using the above notation, the Galerkin orthogonality condition can be
expressed as
\begin{equation}
 \langle \epsilon,X_m\rangle=0\ \hspace{1.5cm} m=1,2,...,N
\end{equation}
 Substituting $\epsilon$ with its expression \eqref{residual}, inverting
the integral and the sum signs, and taking the time dependent coefficients out
of the integral sign, leads to the following system of equations
\begin{equation}
\begin{split}
\label{ODE_system}
0 &=\sum_{n=1}^{N}\dtot{}{t}
c_n(t)\langle X_{n_{yy}},X_m\rangle-k^2\sum_{n=1}^{N}\dtot{}{t}
c_n(t)\langle X_n, X_m\rangle+i\alpha
\sum_{n=1}^{N}c_n(t)\langle U(y)X_{n_{yy}},X_m\rangle\\
&-i\alpha k^2\sum_{n=1}^{N}c_n(t)\langle
U(y)X_n,
X_m\rangle-i\alpha \sum_{n=1}^{N}c_n(t)\langle\frac{\ud^2
U(y)}{\ud y^2}X_n X_m\rangle-\frac{1}{Re}\sum_{n=1}^{N}
c_n(t)\langle X_{
n_{yyyy}},X_m\rangle\\
&+\frac{2k^2}{Re}\sum_{n=1}^{N}c_n(t)\langle X_{
n_{yy}},X_m\rangle-\frac{k^4}{Re}\sum_{n=1}^{N}c_n(t)\langle X_n, X_m\rangle
\hspace{1.5cm} n,\ m=1,2,3,...,N
\end{split}
\end{equation}
The original partial differential equation is now reduced to a system of $N$
ordinary differential equations of the first order, where the time dependent coefficients $c_n(t)$ are the only unknown.
The scalar products can be
evaluated analytically or computed by numerical integration
and take the following expressions\\
\begin{gather}
D_{m,n}=\langle X_n, X_m\rangle=\delta_{m,n}\\[15pt]
S_{m,n}=\langle X_{n_{yy}}, X_m\rangle=\\[8pt] \nonumber
=
\begin{cases}
+4\frac{\lambda_n^2\lambda_m^2}{ \lambda_n^4-\lambda_m^4}
(\lambda_n\gamma_n-\lambda_m\gamma_m) & \mbox{if } (n+m) \mbox{ is even, }n\ne
m \\ 
0  & \mbox{if } (n+m) \mbox{ is odd} \\  
-\lambda_n^2\gamma_n^2+\lambda_m\gamma_m & \mbox{if }n=m 
\end{cases}\\[15pt]
F_{m,n}=\langle X_{n_{yyyy}}, X_m\rangle=\lambda_n^4\delta_{m,n}
\end{gather}
where
\begin{equation}
\gamma_n=\frac{cosh(2\lambda_n)-cos(2\lambda_n)}{
sinh(2\lambda_n)-sin(2\lambda_n)}\hspace{1cm} \lim_{n\to\infty}\gamma_n=1
\end{equation}
For Plane Couette flow, in all the present work the following expression of the
base flow will be considered
\begin{equation}
 U(y)=y
\end{equation}
so that the other integrals take the expressions\\
\begin{gather}
U^{(1)}_{m,n}=\langle U(y)X_{n_{yy}}, X_m\rangle=\\[8pt] \nonumber
=
\begin{cases}
0 & \mbox{if } (n+m) \mbox{ is even, }n\ne
m \\ 
4\frac{\lambda_n^2\lambda_m^2}{ \lambda_n^4-\lambda_m^4}
(\lambda_n\gamma_n-\lambda_m\gamma_m-1)-8\frac{\gamma_n^4+\gamma_m^4}{
(\lambda_n^4-\lambda_m^4)^2}\lambda_n^2\lambda_m^2  & \mbox{if } (n+m) \mbox{ is
odd}  \\
0 & \mbox{if }n=m 
\end{cases}\\[15pt]
\end{gather}

\begin{gather}
U^{(2)}_{m,n}=\langle U(y)X_n, X_m\rangle=\\[8pt]\nonumber
=
\begin{cases}
0 & \mbox{if } (n+m) \mbox{ is even, }n\ne
m \\ 
16\frac{\lambda_n^3\lambda_m^3\gamma_n\gamma_m}{( \lambda_n^4-\lambda_m^4)^2}  &
\mbox{if } (n+m) \mbox{ is
odd}  \\
0 & \mbox{if }n=m 
\end{cases}\\[15pt]
U^{(3)}_{m,n}=\langle \frac{\ud^2U(y)}{\ud y^2}X_n, X_m\rangle=0\ \ \
\ \forall\ n,m
\end{gather}
It is convenient to express the ODEs system \eqref{ODE_system} in a more compact
notation: in the following, vectors will be indicated either explicitly using
braces or with bold lower case letters; matrices will be indicated with bold
capital letters; constants with roman capital
letters and physical parameters in italic. 
The system can be written as
\begin{gather}
\underbrace{\big(\BU{S}-k^2\BU{D}\big)}_{\BU{H}}\BU{\dot{c}}-\underbrace{\big(\
-i\alpha\BU{U^{(1)}}+i\alpha k^2\BU{U^{(2)}}+i\alpha
\BU{U^{(3)}}+\frac{1}{Re}\BU{F}-\frac{2k^2}{Re}\BU{S}+\frac{k^4}{Re}\BU{D}
\big)}_{\BU{G}}\BU { c }=0\\
\BU{H}\BU{\dot{c}}-\BU{G}\BU{c}=0
\end{gather}
where $\BU{D}=[D_{m,n}]$ etc., i.e. the element $D_{m,n}$ is placed at the
$n^{th}$ column and at the $m^{th}$ row of the matrix. $\BU{H}$ is invertible,
so denoting $\BU A=\BU{H}^{-1}\BU{G}$ yields
\begin{gather}
\label{ODE_system2}
 \BU{\dot{c}}-\BU{A}\BU{c}=0
\end{gather}
The general solution to the ODEs system \eqref{ODE_system2} in the case of
matrix $\BU{A}$ having $N$ distinct eigenvalues $\mu_i$ \citep[either real or
complex, see][]{ODE_book}, reads
\begin{gather}
\label{ODE_solution2}
\BU{c}(t)=\up{K_1}\BU{l}_1 e^{\mu_1 t}+\up{K_2}\BU{l}_2 e^{\mu_2
t}+\hdots+\up{K_N}\BU{l}_N e^{\mu_N t}
\end{gather}
where $\BU{l}_i$ are the eigenvectors corresponding to $\mu_i$ and $\up{K_i}$
are constants to be determined by imposing the initial condition, or
alternatively using the Matrix Exponential notation
\begin{gather}
\BU{c}(t)=e^{\BU{A}t}\{K_i\}
\end{gather}
The coefficients at the initial time, $\BU{c_0}$, can be obtained from the inverse transformation
\eqref{inverse_trasf} since the initial condition $\hv(t=0)$ is known, so finally the solution is get by
solving the algebraic system
\begin{gather}
\BU{c_0}=\up{K_1}\BU{l}_1 +\up{K_2}\BU{l}_2 +........+\up{K_N}\BU{l}_N\\
\BU{h_0}=\{K_i\}=\BU{L}^{-1}\BU{c_0}
\end{gather}
where $\BU{h}(t)=\BU{L}^{-1}\BU{c}(t)$, and $\BU{L}$ is the matrix whose columns
are the eigenvectors $\BU{l}_i$.
Their linear independence  ensures that  $\BU{L}$ is invertible.
\section{Solution to the forced $\he$ equation}\label{sec:eta_solution}
\subsection{Choice of a base of orthogonal functions}\label{sec:base_eta}
Following the same procedure of \secref{sec:base_v}, we now focus on the
normal-vorticity equation \eqref{Squire}, which is forced by the solution $\hv$
of the Orr-Sommerfeld PDE equation \eqref{Orr-Somm}. Together with the
non-orthogonality of the
Orr-Sommerfeld differential operator, a resonance phenomenon has been pointed
out as one of the
reasons for large energy transient growths, if there is sufficient wave obliquity
\citep[see][]{Gustavsson1991}. In order to solve the $\he$ equation a set of normal functions different from the one
adopted in
\secref{sec:base_v} is needed, since the second order PDE only requires $\he$
to vanish at the boundaries, but not its first derivative . The simplest
choice for the basis functions, here adopted, is the following
\begin{align}
&Y_n=sin(\xi_n y)\hspace{1.5cm}n=1,3,5,...N-1\ \ &(odd\
set)\\
&Y_n=cos(\xi_n y)\hspace{1.5cm}n=2,4,6,...N\ \ \ &(even\
set)
\end{align}
where
\begin{align}
&\xi_n=\frac{(n+1)\pi}{2}\hspace{1.0cm}n=1,3,5,...N-1\ \ &(odd\
set)\\
&\xi_n=\frac{(n-1)\pi}{2}\hspace{1.0cm}n=2,4,6,...N\ \ \ &(even\
set)
\end{align}
Also in this case, note that two sets of eigenfunctions are put together to
form a unique set, since both are necessary to completely describe the
complex-valued normal vorticity.
The general solution is then obtained as the sum of a particular solution
$\he_p$ and the solution to the corresponding homogeneous equation $\he_h$
\begin{equation}
 \he(y,t)=\he_h(y,t)+\he_p(y,t)
\end{equation}

\subsection{Weak formulation and approximate  solution to $\he$ equation by
Galerkin method}\label{sec:gal_eta}
Considering the complete equation \eqref{Squire}, we proceed as done for the
normal-velocity and expand the solution as follows
\begin{equation}
 \he(y,t)=\sum_{n=1}^{\infty}b_n(t)Y_n(y)
\end{equation}
Substituting, the equation reads
\begin{equation}
\begin{split}
&\sum_{n=1}^{\infty}\dtot{}{t}b_n(t)Y_n+i\alpha
U(y)\sum_{n=1}^{\infty}b_n(t)Y_n-\frac{1}{Re}\sum_{n=1}^{\infty}
b_n(t)Y_{n_{yy}}\\ 
&+\frac{k^2}{Re}\sum_{n=1}^{\infty}b_n(t)Y_n=-i\beta\dtot{U(y)}{y}\sum_{n=1}^{
\infty}c_n(t)X_n(y)
\end{split}
\end{equation}
Considering a finite number $N$ of terms of the expansion and applying the
Galerkin method yields 
\begin{equation}
\begin{split}
\label{system_eta_discr}
&\sum_{n=1}^{N}\dtot{}{t}b_n(t)\langle Y_n,Y_m\rangle+i\alpha
\sum_{n=1}^{N}b_n(t)\langle U(y)
Y_n,Y_m\rangle-\frac{1}{Re}\sum_{n=1}^{N}
b_n(t)\langle Y_{n_{yy}},Y_m\rangle\\ 
&+\frac{k^2}{Re}\sum_{n=1}^{N}b_n(t)\langle
Y_n,Y_m\rangle=-i\beta\sum_{n=1 } ^ {
N}c_n(t)\langle \dtot{U(y)}{y}X_n,Y_m\rangle\ \ \ \ n,m=1,2,3,...,N
\end{split}
\end{equation}
The scalar products can be evaluated analytically and take the following
expressions
\begin{gather}
 D^*_{m,n}=\langle Y_n,Y_m\rangle=\delta_{m,n}\\[15pt]
 S^*_{m,n}=\langle Y_{n_{yy}},Y_m\rangle=-\xi_n^2\delta_{m,n}
\end{gather}
For Plane Couette flow:
\begin{gather}
 U^*_{m,n}=\langle U(y)Y_n,Y_m\rangle=\\[8pt]\nonumber
 =
\begin{cases}
0 & \mbox{if } (n+m) \mbox{ is even, or }n=
m \\ 
\frac{(-1)^{\frac{n+m+1}{2}}4\xi_n\xi_m}{(\xi_n^2-\xi_m^2)^2}  & \mbox{if }
(n+m) \mbox{ is
odd}  
\end{cases}\\[15pt]
F^*_{m,n}=\langle\dtot{U(y)}{y}X_n,Y_m\rangle=\\[8pt]
=
\begin{cases}
\frac{2\sqrt 2\xi_m\xi_n^2(-1)^{\frac{m+1}{2}}}{\xi_m^4-\xi_n^4} &
\mbox{if } n,m \mbox{ are odd}\\[6pt]\nonumber
\frac{2\sqrt 2 \xi_m\xi_n^2(-1)^{\frac{m}{2}}}{\xi_m^4-\xi_n^4}  & \mbox{if }
n,m \mbox{ are
even}  
\end{cases}
\end{gather}
Introducing the vector and matrix notation, the system \eqref{system_eta_discr}
reads
\begin{gather}
\BU{\dot{b}}-\underbrace{\big(\
-i\alpha\BU{U^*}+\frac{1}{Re}\BU{S^*}-\frac{2k^2}{Re}\BU{D^*}\big)}_{
\BU{G^*} } \BU { b }=\underbrace{-i\beta\BU{F^*}}_{\BU{B}}\BU{c}\\
\BU{\dot{b}}-\BU{G^*}\BU{b}=\BU{B}\BU{c}
\label{Squire_ODE_sys}
\end{gather}
which is a non-homogeneuos system of $N$ ODEs. The general solution to
\eqref{Squire_ODE_sys} consists of the superposition of the solution of the
homogeneous system and a particular one. Naming $\BU{\Phi}(t)$ the
\textit{fundamental matrix} of the system \citep[ see][]{ODE_book}, then a
formal expression for the general solution is 
\begin{equation}
 \BU{b}(t)=\BU{b}_h(t)+\BU{b}_p(t)=\BU{\Phi}(t)\BU{k}+\BU{\Phi}(t)\int_{t_0}^{t}
\BU{\Phi}^{-1}(t)\BU{B}\BU{c}(s)\ud s
\end{equation}
or, alternatively, using the Matrix Exponential form
\begin{equation}
 \BU{b}(t)=\BU{b}_h(t)+\BU{b}_p(t)=e^{\BU{G^*}t}\BU{k}+e^{\BU{G^*}t}\int_{
t_0} ^ { t }
e^{\BU{-G^*}t}\BU{B}\BU{c}(s)\ud s
\end{equation}
where $\BU{k}$ is a vector of constants. The last expressions have the
advantage that the particular integral vanishes at $t=t_0$, so it is easy to
find the constants by imposing the initial condition. Unfortunately, a
numerical evaluation of the integral can lead to non negligible errors,
especially for big times, where a product of very large and very small terms
occurs. In order to make the numerical computation possible, a different form
for the particular solution is sought.
\subsubsection{Particular solution $\BU{b}_p$}
Since the solution \eqref{ODE_solution2} in terms of the expansion coefficients
$\BU{c}(t)$ is
a combination of exponentials and represents the forcing term in
\eqref{Squire_ODE_sys}, the following particular solution $\BU{b}_p$ is sought
\begin{equation}
\label{b_p}
b_{p_n}(t)=\sum_{j=1}^N a_{nj}e^{\mu_jt}
\end{equation}
where $a_{nj}$ are constants and $\mu_j$ are the eigenvalues of $\BU{A}$,
through which the forcing term is expressed. Yields
\begin{equation}
\he_p(y,t)=\sum_{n=1}^N b_{p_n}(t)Y_n(y)
\end{equation}
Diagonalizing the system \eqref{ODE_system2}, the coefficients of the
normal-velocity result 
\begin{equation}
 \BU{c}(t)=\BU{L}\BU{h}=\BU{L}\begin{Bmatrix} 
                               h_{0_1}e^{\mu_1t}\\
               
                 h_{0_2}e^{\mu_2t}\\
                               \vdots\\
                               h_{0_N}e^{\mu_Nt}
                               \end{Bmatrix}
 \hspace{2cm} \BU{h_0}=\BU{L}^{-1}\BU{c}(t=0)
 \end{equation}
 Substituting the particular solution \eqref{b_p} in \eqref{Squire_ODE_sys}
and leads to
\begin{equation}
 \dtot{}{t}\begin{Bmatrix}
            a_{11}e^{\mu_1t}+\hdots a_{1N}e^{\mu_Nt}\\ 
            \vdots\\
            a_{N1}e^{\mu_1t}+ \hdots a_{NN}e^{\mu_Nt}
           \end{Bmatrix}
 +\BU{G^*}  \begin{Bmatrix}
            a_{11}e^{\mu_1t}+\hdots a_{1N}e^{\mu_Nt}\\ 
            \vdots\\
            a_{N1}e^{\mu_1t}+ \hdots a_{NN}e^{\mu_Nt}
           \end{Bmatrix}
 =\BU{B}\BU{L}\begin{Bmatrix} 
                               h_{0_1}e^{\mu_1t}\\
                               h_{0_2}e^{\mu_2t}\\
                               \vdots\\
                               h_{0_N}e^{\mu_Nt}
                               \end{Bmatrix}
\end{equation}
It is straightforward to find the unknown constants $a_{nj}$ by comparing terms
with the same exponential factor. This is equivalent to solve the following
set of $N$ algebraic systems
\begin{equation}
\label{aij_systems}
 (\mu_j\idop-\BU{G^*})\begin{Bmatrix} 
                               a_{1j}\\
                               a_{2j}\\
                               \vdots\\
                               a_{Nj}
                               \end{Bmatrix}
=h_{0j}\begin{Bmatrix} 
                               B^*_{1j}\\
                               B^*_{2j}\\
                               \vdots\\
                               B^*_{Nj}
                               \end{Bmatrix}
 \hspace{1cm} j=1,2,...N
\end{equation}
where $\idop$ is the identity matrix and $B^*_{ij}$ are the elements of the
matrix $\BU{B}\BU{L}$. As usual, the first subscript indicates the row and the
second one indicates the column. Finally we get the matrix of coefficients
column by column as
\begin{equation}
 \begin{Bmatrix} 
                               a_{1j}\\
                               a_{2j}\\
                               \vdots\\
                               a_{Nj}
                               \end{Bmatrix}
=(\mu_j\idop-\BU{G^*})^{-1}h_{0j}\begin{Bmatrix} 
                               B^*_{1j}\\
                               B^*_{2j}\\
                               \vdots\\
                               B^*_{Nj}
                               \end{Bmatrix}
 \hspace{1cm} j=1,2,...N
\end{equation}
\subsubsection{Homogeneous and complete solution $\BU{b}$}
The homogeneous solution of \eqref{Squire_ODE_sys}  takes the same form of the
of equation \eqref{ODE_solution2}. Indicating with $\mu^*$ and $\BU{l}^*$
respectively the eigenvalues and eigenvectors of the matrix $\BU{G^*}$, it
follows 
\begin{equation}
 \BU{b}_h(t)=\up{C_1}\BU{l}^*_1e^{\mu_1^*t}+\up{C_2}\BU{l}^*_2e^{\mu^*_2t}
+\hdots+\up{C_N}\BU{l}^*_N e^ { \mu^*_Nt }
\end{equation}
Finally the complete solution is
\begin{gather}
\BU{b}(t)=\up{C_1}\BU{l}^*_1e^{\mu_1^*t}+\up{C_2}\BU{l}^*_2e^{\mu^*_2t}
+\hdots+\up{C_N}\BU{l}^*_N e^ { \mu^*_Nt }+\BU{b}_p(t)
\end{gather}
The unknown constants $\up{C}_i$ depends on the initial condition, and can be
calculated setting $t=0$ in the above expression, leading to
\begin{equation}
 \BU{h^*_0}=\{\up{C}_i\}=\BU{L^*}^{-1}(\BU{b_0}-\BU{b_0}_p)
\end{equation}
\section{A \matlab code implementation}
\subsection{Code description}
In order to verify the proposed method and to obtain the numerical solutions, a
\matlab code has been developed and used in the further analysis. At present, it
consists of two
scrips: the main code solves the normal-velocity equation and calls
a function to
solve the normal-vorticity equation. Eventually it computes the other components
of the perturbation velocity, the frequency, the energy growth factor and other quantities of interest.
\subsubsection{The main code}
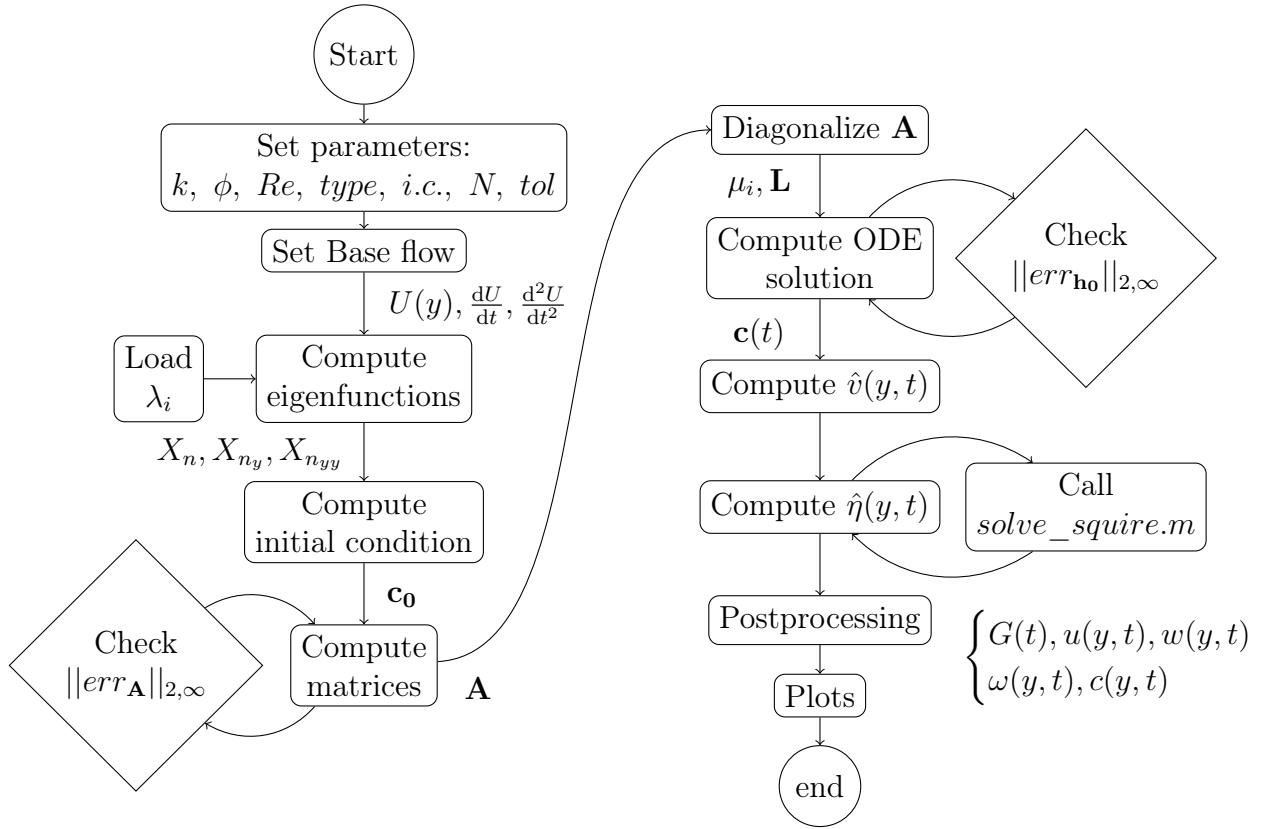
\begin{figure}
\begin{tikzpicture}[every text node part/.style={align=center},]
	\begin{pgfonlayer}{nodelayer}
		\node [style=circ] (0) at (0, 0) {Start};
		\node [style=rett] (1) at (0, -1.5) {Set\ parameters: \\$
k,\ \phi,\ Re,\ type,\ i.c.,\ N,\ tol $};
		\node [style=rett] (2) at (0, -2.6) {Set\ Base\ flow};
		\node [style=rett] (3) at (0, -4.3){Compute\\eigenfunctions};
		\node [style=rett] (31) at (-2.7,-4.3){Load\\$\lambda_i$};
		\node [style=rett] (4) at (0, -6.2) {Compute\\ initial\
condition};
		\node [style=rett] (5) at (0, -8.1) {Compute\\ matrices};
		\node [style=rombo] (6) at (-3, -8.1) {Check\\$
||err_{\BU{A}}||_{2,\infty}$};
		\node [style=rett] (7) at (6, -1.0) {Diagonalize $\BU{A}$};
		\node [style=rett] (8) at (6, -2.7) {Compute\
ODE\\solution};
		\node [style=rombo] (9) at (9.5, -2.7) {Check\\$
||{err_{\BU{h_0}}}||_{2,\infty}$};
		\node [style=rett] (10) at (6, -4.4) {Compute $\hv(y,t)$};
		\node [style=rett] (12) at (6, -6.0) {Compute $\he(y,t)$};
		\node [style=rett] (13) at (9.5, -6.0) {Call\\
$solve\_squire.m$};
		\node [style=rett] (14) at (6, -7.5) {Postprocessing};
		\node [style=rett] (15) at (6, -8.5) {Plots};
		\node [style=circ] (16) at (6, -9.7) {end};	
	\node [style=none] (11) at (1.5, -3.3)
{$U(y),\dtot{U}{t},\frac{\ud^2U}{\ud t^2}$};
\node [style=none] (22) at (-1.5, -5.3) {$X_n, X_{n_y}, X_{n_{yy}}$};
\node [style=none] (33) at (0.5, -7.2) {$\BU{c_0}$};
\node [style=none] (44) at (1.5, -8.4)
{$\BU{A}$};
\node [style=none] (55) at (5.2, -1.7){$\mu_i,\BU{L}$};
\node [style=none] (66) at (5.2, -3.7){$\BU{c}(t)$};
\node [style=none] (77) at (10, -8.0){$\begin{cases} 
                               G(t),u(y,t),w(y,t)\\
                               \omega(y,t),c(y,t)
                               \end{cases}$};
	\end{pgfonlayer}
	\begin{pgfonlayer}{edgelayer}
		\draw [style=link] (0) to (1);
		\draw [style=link] (1) to (2);
		\draw [style=link] (2) to (3);
		\draw [style=link] (3) to (4);
		\draw [style=link] (31) to (3);
		\draw [style=link] (4) to (5);
		\draw [style=arrow, bend left=40] (6) to (5);
		\draw [style=arrow, bend left=40] (5) to (6);
		\draw [style=arrow, in=180, out=3,looseness=0.75] (5) to (7);
		\draw [style=link] (7) to (8);
		\draw [style=arrow, bend left=40] (8) to (9);
		\draw [style=arrow, bend left=40] (9) to (8);
		\draw [style=link] (8) to (10);
		\draw [style=arrow, bend left=40] (12) to (13);
		\draw [style=arrow, bend left=40] (13) to (12);
		\draw [style=link] (10) to (12);
		\draw [style=link] (12) to (14);
		\draw [style=link] (14) to (15);
		\draw [style=link] (15) to (16);
	\end{pgfonlayer}
\end{tikzpicture}
\caption{Structure of the main code \textit{main\_ivp\_galerkin.m}.}
\label{fig:flowchart_v}
\end{figure}

The structure of the main program (\textit{main\_ivp\_galerkin.m}) can be represented
by the
flowchart of \figref{fig:flowchart_v}. Among the other simulation parameters,
the number of modes can be chosen. The corresponding eigenvalues $\lambda_i$
are computed by a separate script through Bisection or Newton-Raphson method,
and memorized in a \textit{.mat} file which can be loaded by the main program.
As seen in \secref{sec:gal_v}, the method requires the solution to algebraic
systems, so matrix inversions. In particular two inversions are needed to
obtain the solution $\hv$, respectively the one of the matrix $\BU{A}$ and the
one of the eigenvectors matrix $\BU{L}$. In fact, the ill-conditioning of this
matrices can influence the accuracy of the computation. In detail, the
condition number of matrix $\BU{A}$ does not represent a
problem, while the one of matrix $\BU{L}$ can reach very high values, of the
order of $10^{16}$ for some parameters settings as low obliquity angles;
moreover the condition number increases with increasing $N$. This fact has not
been investigated in details, but is quite similar to that pointed out by
\citet{Schmid_book}. The ill-conditioning of $\BU{L}$ is intrinsic, due to the non orthogonality of the Orr-Sommerfold
linear operator. In order to guarantee a certain level of accuracy, the
matrix inversion error is checked and compared to a threshold $tol$. The
absolute pointwise error in the solution of a generic system
$\BU{A}\BU{x}=\BU{b}$ is defined as follows
\begin{equation}
 err_A=\BU{A}\BU{x}-\BU{b}
\end{equation}

  The script computes by default the matrix inversion using the \matlab
\textit{backslash} command. Then, the norm (both $L_2$-norm and
$L_\infty$-norm ) of
the error is calculated and compared to the tolerance. If the threshold level is
exceeded, a different method to solve the algebraic system is performed. In
particular, the present code tries to approximately solve the system using the
\textit{GMRES} method. Both the norm of absolute and the relative error are
evaluated. Even if  a better method should be object of future
deeper analysis, it has been observed that for every parameters combination
tried, the minimum relative error is obtained using the \textit{backslash}
command, and its order of magnitude spans the range $10^{-12}-10^{-30}$.\par
The solution is obtained for every point along the space coordinate $y$ and at
all time points defined by the user. Since the time evolution is 
analytically obtained, and since the analytical expressions of the modes is
known, the accuracy of the method shouldn't depend neither on the space
 nor on the time discretization. Actually, this is not exact: the computation of
the coefficients $\BU{c_0}$ is performed through numerical integration
(trapezoidal rule), whose order of accuracy is $O(\Delta y^2)$. Moreover, in the
case of Plane Poiseuille flow, the scalar products $U^{(1)}_{m,n},\
U^{(2)}_{m,n}$ and $U^*_{m,n}$ are evaluated through numerical integration as
well, so a sufficient grid spacing is required. The computation of
the wave frequency or phase velocity can require very fine time grids, the
motivation will be given in Chapter \ref{chap:wave_transient}. In general,
fine grids
are necessary whenever a finite difference scheme for derivatives computation
needs to be applied.
In order to ensure an high accuracy of the computation of the initial
condition, and consequently of the global solution, and at the same time
allowing the user to obtain the solution only at the desired points, a
separate grid is used just for the computation of the initial condition's
coefficients $\BU{c_0}$.                       

\subsubsection{The $\he$ solving function}
\begin{figure}[h!]
        \centering
  
         \advance\leftskip-1cm
\begin{tikzpicture}[every text node part/.style={align=center},]
	\begin{pgfonlayer}{nodelayer}
		\node [style=circ] (0) at (0, 0) {Func.};
		\node [style=rett] (1) at (0, -1.5) {Inputs: \\$
\BU{L},\ \mu_i,\ \lambda_n\ X_n,\ \BU{h_0}$};
 		\node [style=rett] (2) at (0, -3.2) {Compute\\eigenfunctions};
 		\node [style=rett] (3) at (0, -5.1) {Compute\\initial\
condition};
 		\node [style=rett] (4) at (0, -7) {Compute\\ matrices};		
 		\node [style=rett] (5)  at (0, -8.9) {Compute\ $[a_{i,j}]$\\
constants\ matrix};
\node [style=rombo] (6) at (-4, -8.9) {Check\\$
||err_{\{ a_{ij}\}}||_{2,\infty}$};
 		\node [style=rett] (7) at (6, -1.5) {Compute\ ODE\\particular\
sol.};
 		\node [style=rett] (8) at (6, -3.2) {Diagonalize $\BU{G^*}$};
 		\node [style=rett] (9) at (6, -5.0) {Compute\ ODE\\complete\
sol. };
		\node [style=rombo] (10) at (9.5, -5.0) {Check\\$
||err_{\BU{h^*_0}}||_{2,\infty}$};
 		\node [style=rett] (12) at (6, -6.9) {Compute $\he(y,t)$};
 		\node [style=circ] (13) at (6, -8.7) {return};	
 \node [style=none] (11) at (1.5,-4.2){$\lambda^*,Y_n,Y_{n_{yy}}$};
 \node [style=none] (22) at (0.9, -6.1) {$\BU{b_0}$};
 \node [style=none] (33) at (-2, -10.3) {\tiny{ $for\ j=1:N$}};
 \node [style=none] (44) at (1.5, -8.0) {$\BU{G^*},\BU{B},\BU{B^*}$};
 \node [style=none] (55) at (5.2, -2.5){$\BU{b_p}(t)$};
  \node [style=none] (66) at (5.2, -4.0){$\mu^*_i,\BU{L^*}$};
\node [style=none] (77) at (5.2, -6.1){$\BU{b}(t)$};
	\end{pgfonlayer}
	\begin{pgfonlayer}{edgelayer}
		\draw [style=link] (0) to (1);
 		\draw [style=link] (1) to (2);
 		\draw [style=link] (2) to (3);
 		\draw [style=link] (3) to (4);
 		\draw [style=link] (4) to (5);
 		\draw [style=arrow, bend left=40] (6) to (5);
 		\draw [style=arrow, bend left=40] (5) to (6);
 		\draw [style=arrow, in=180, out=3,looseness=0.75] (5) to (7);
 		\draw [style=link] (7) to (8);
 		\draw [style=link] (8) to (9);
		\draw [style=arrow, bend left=40] (9) to (10);
 		\draw [style=arrow, bend left=40] (10) to (9);
 		\draw [style=link] (9) to (12);
		\draw [style=link] (12) to (13);
	\end{pgfonlayer}
\end{tikzpicture}
\caption{Structure of the function \textit{solve\_squire.m}.}
\label{fig:flowchart_eta}
\end{figure}
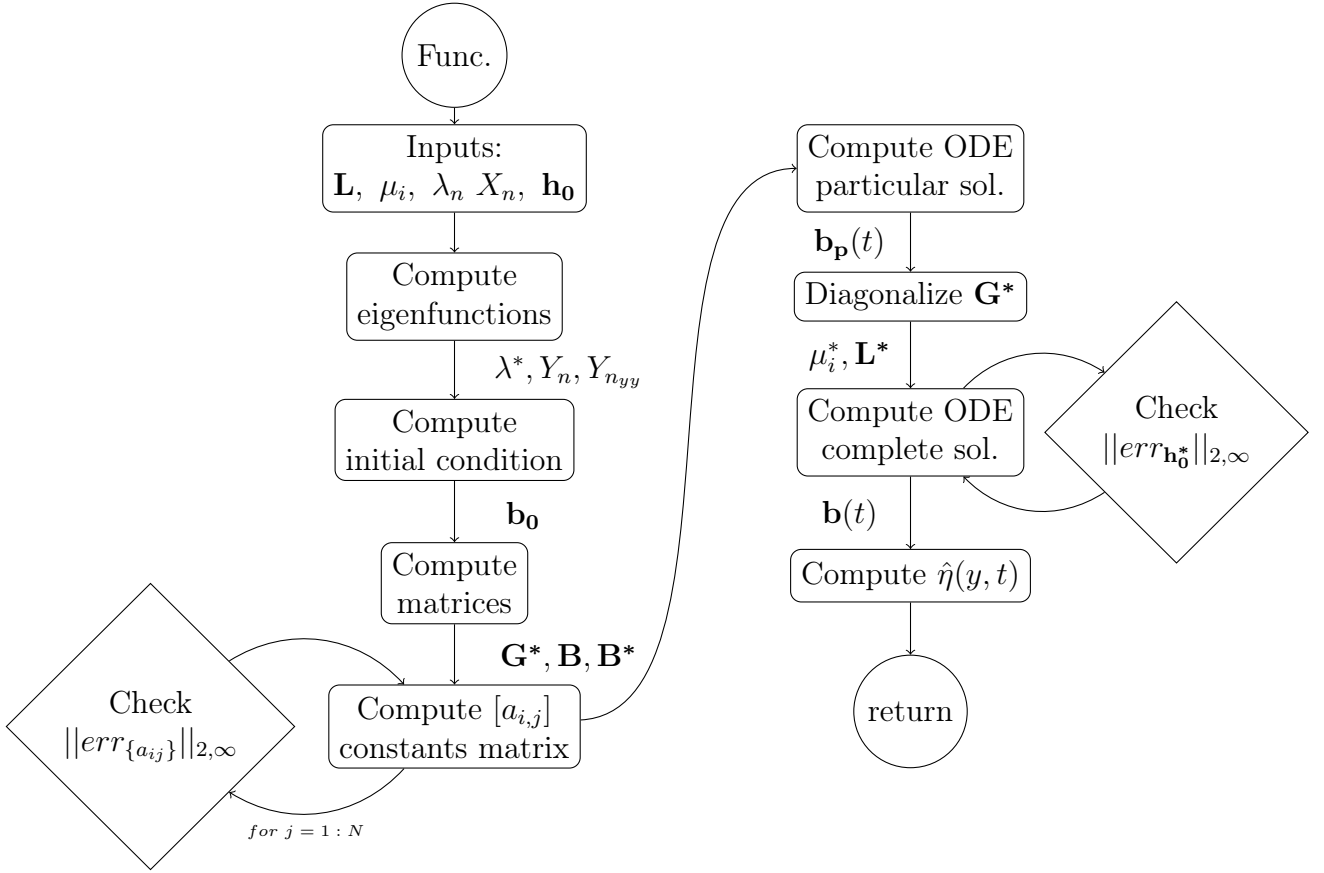

The $\he$ equation is solved by the function  \textit{solve\_squire.m}. The
script structure is represented by the flowchart of
\figref{fig:flowchart_eta}, and is quite similar to the one of the main
program. Here there is no reason to calculate the eigenvalues related to the
basis functions with another script, being known their exact expression. It can
be noticed that to obtain the normal vorticity the solution of
the $N$ algebraic systems \eqref{aij_systems} is required, as well as the inversion of the
matrix $\BU{L^*}$. Depending on the choice of the simulation parameters, these
matrices can be ill-conditioned so the same procedure described in the previous
section is implemented. During the simulation all the error norms are
displayed (about the $N$ algebraic systems \eqref{aij_systems}, only the
maximum will be shown). \\
Also in the computation of $\he$, the \textit{backslash
} command has revealed to guarantee an high level of accuracy in all simulation
performed, being all error norms of the order of $10^{-14}$ or lower.

\subsection{Rate of Convergence}
In this section, the rate of convergence of the present method to the
exact solutions of the initial value problem will be investigated. The
eigenfunction expansion method using the Chandrasekhar basis functions was
applied to the Orr-Sommerfeld modal analysis by \citet{Gallagher1962}. In that
case, the error was shown to decrease as $1/N^5$ as $N\to
\infty$, moreover the residue after $N$ terms of expansion was of order
$1/N^5$.\\
Even if the present formulation is different in the fact that the PDEs are
reduced to systems of ODEs, rather than algebraic equations, it has been
verified that the convergence ratio of order $1/N^5$
as $N\to \infty$ is kept, and it applies at all times, for what concerns $\hv$ equation.
About $\he$ equation, the method reduces to a Fourier series expansion with
time-dependent coefficients, and in this case the convergence rate is found
to be only slightly different from the one of $\hv$.\\
Both the root mean square (proportional to $L_2$-norm) and the maximum
($L_\infty$-norm) of the residual are
computed. In the following, the residuals as a function of the number of modes
used for the simulation are reported at two different times and two different
values of the obliquity angle, for Plane Couette flow and Plane
Poiseuille flow. Since the exact solution is not known, the residuals are
defined as the difference between the solution and an accurate solution computed
with 350 modes. The exact expression is only known at time zero, and the results
have been confirmed in this case. 
\begin{gather}
 \epsilon_a(y,t)=|\hv_N(y,t)-\hv_{N=350}(y,t)|\\
 rms(\epsilon_a)(t)=\frac{1}{N_y}\sqrt{\sum_{i=1}^{N_y}\epsilon_a^2(y,t)}\\
 max(\epsilon_a)(t)=\max_{y_i}(\epsilon_a(y,t))
\end{gather}

\subsubsection{Convergence of $\hv$}
\begin{figure}
        \centering
         \advance\leftskip-2.5cm
         \advance\rightskip-2cm
        \begin{subfigure}{0.6\textwidth}
        \centering 
	\includegraphics[width=9.5cm]{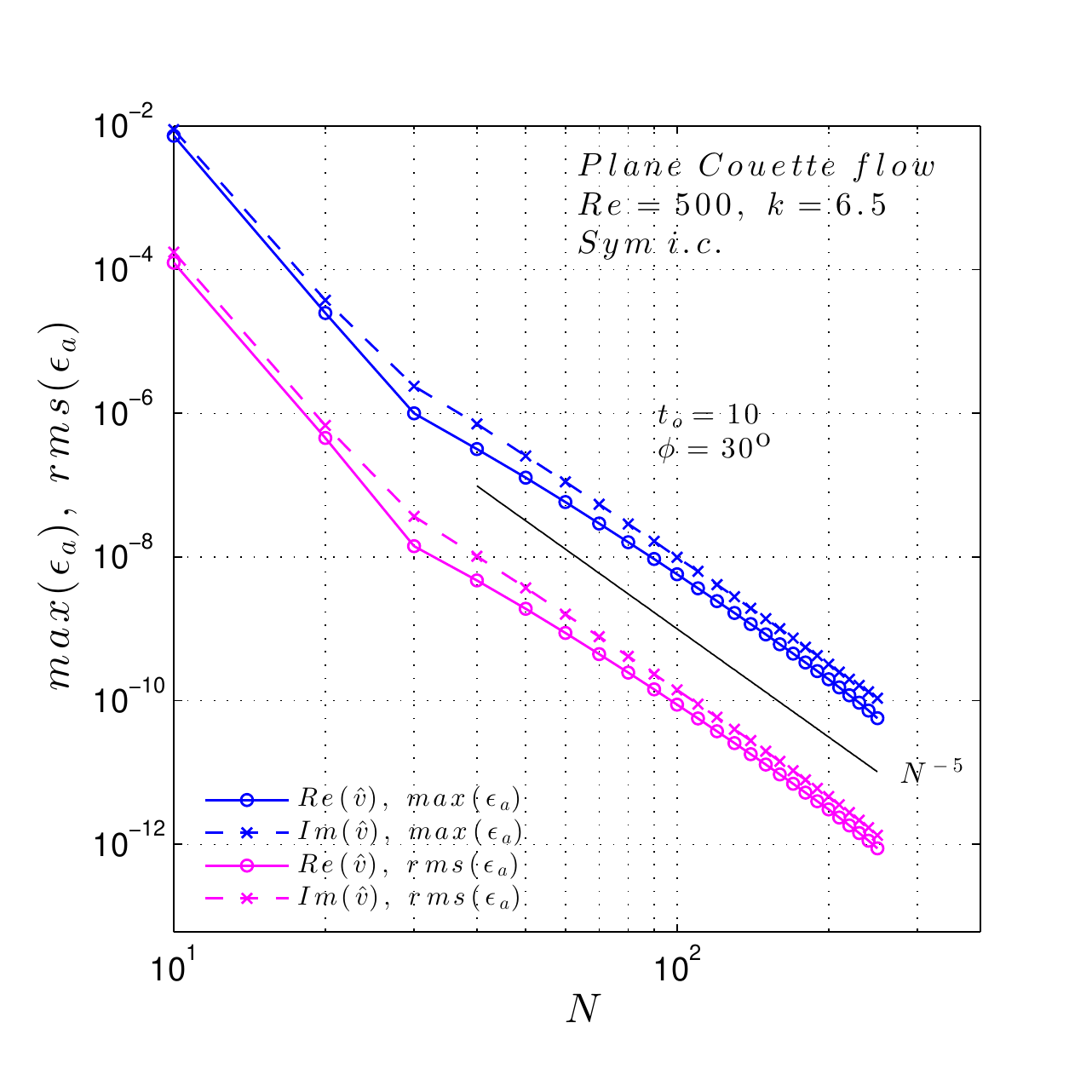}
	\subcaption{$t_0=10$, $Re=500$, $k=6.5$, $\phi=30^\circ$}
	 \end{subfigure}
        \begin{subfigure}{0.6\textwidth}
        \centering 
	\includegraphics[width=9.5cm]{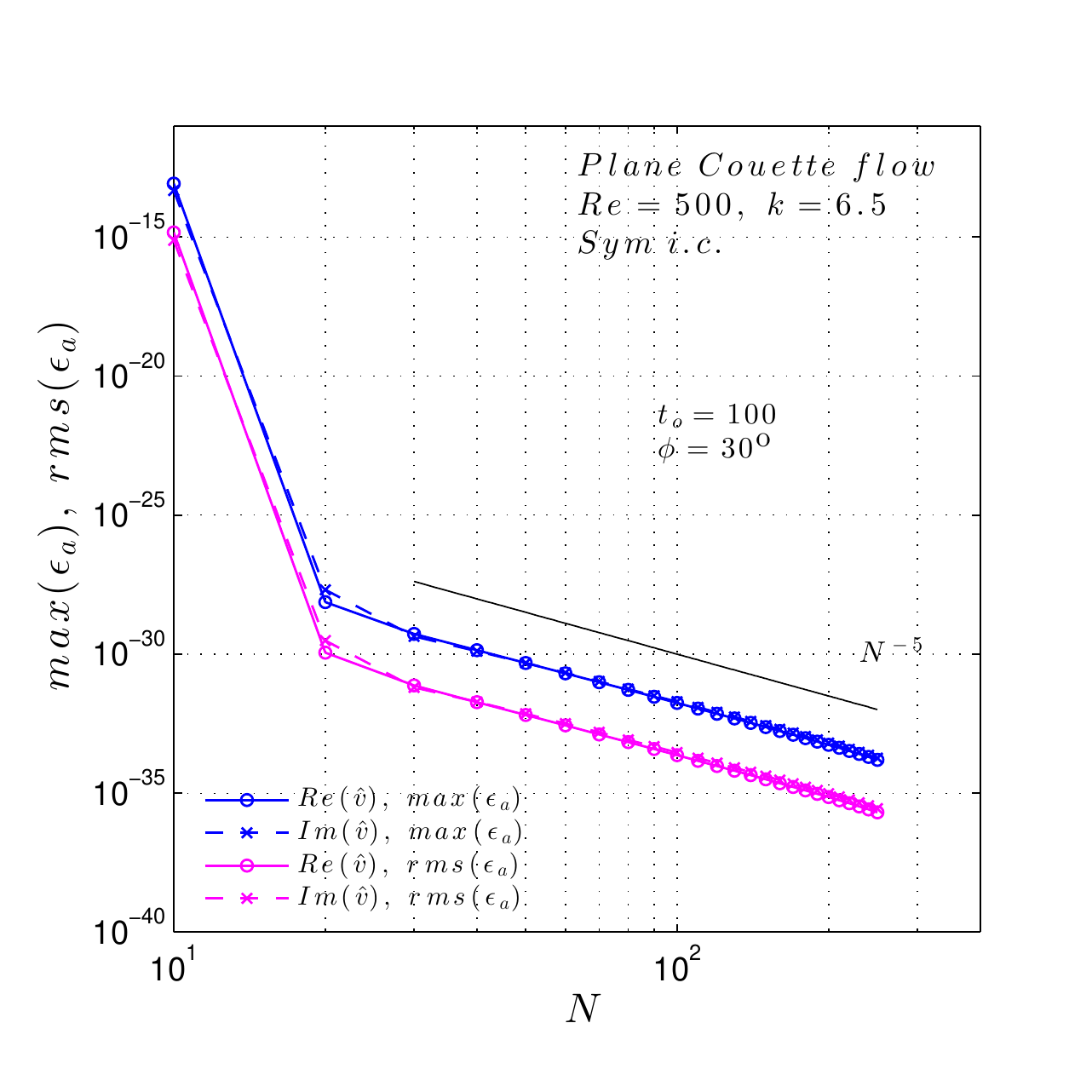}
	\subcaption{$t_0=100$, $Re=500$, $k=6.5$, $\phi=30^\circ$}
	 \end{subfigure}
	\caption{Maximum and rms of the absolute residual of $\hv$ for Plane
Couette flow and symmetrical (even) initial
condition. \textit{Continuous line}: real part;  \textit{dashed line}: imaginary part.
Comparison with $N^{-5}$.}
\label{fig:convergence_v_Co}
\end{figure}

\begin{figure}
        \centering
         \advance\leftskip-2.5cm
         \advance\rightskip-2cm
        \begin{subfigure}{0.6\textwidth}
        \centering 
	\includegraphics[width=9.5cm]{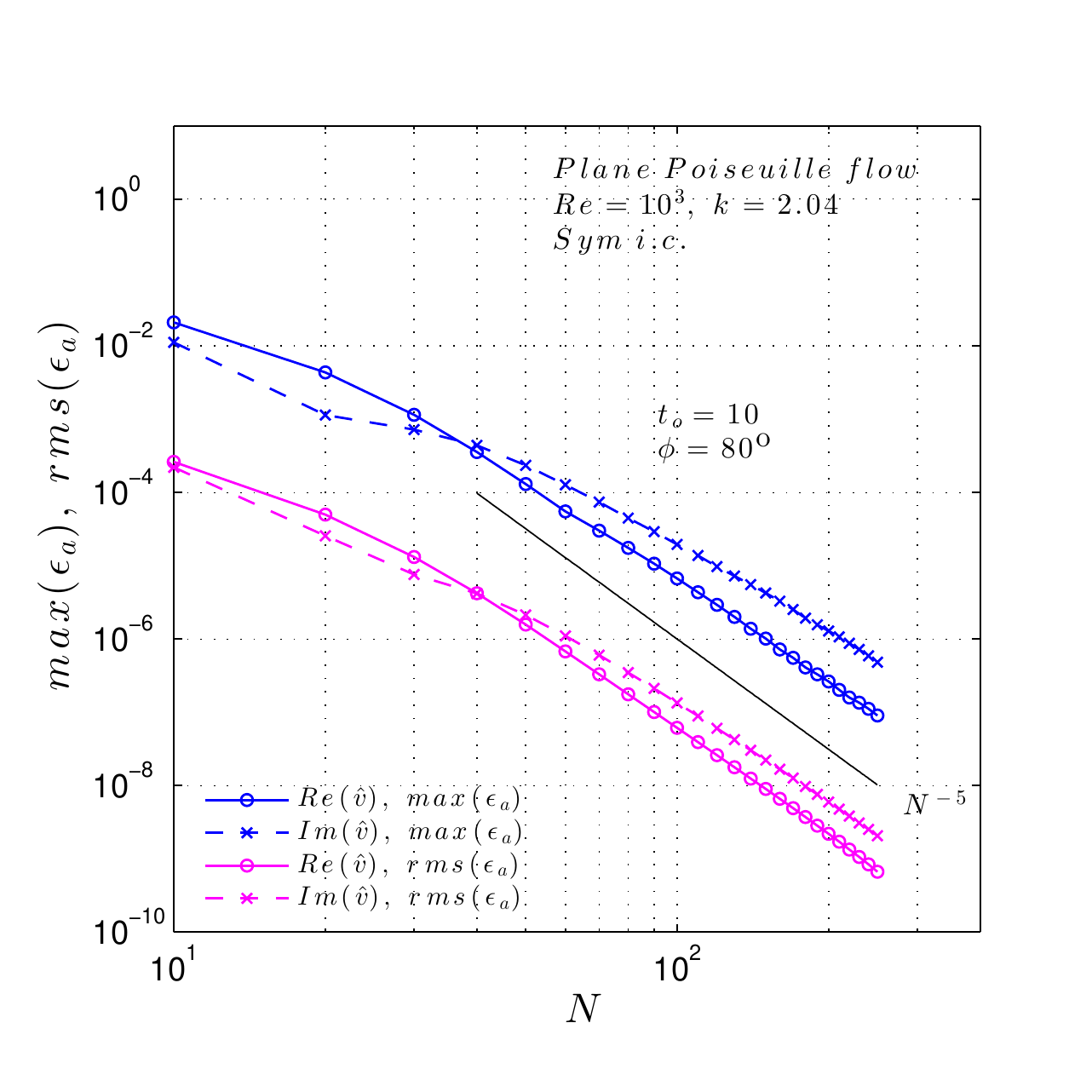}
	\subcaption{$t_0=10$, $Re=1000$, $k=2.04$, $\phi=80^\circ$}
	 \end{subfigure}
        \begin{subfigure}{0.6\textwidth}
        \centering 
	\includegraphics[width=9.5cm]{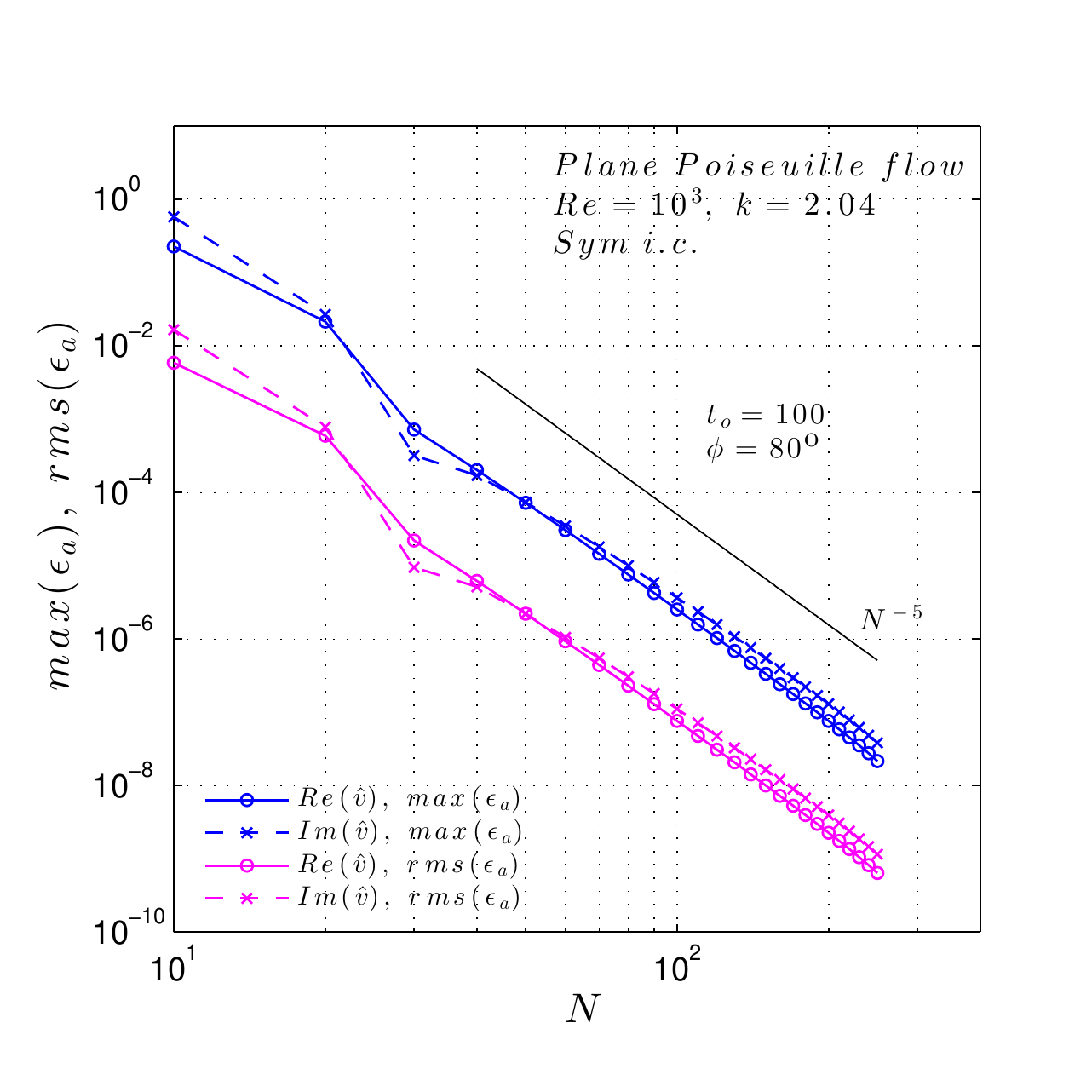}
	\subcaption{$t_0=100$, $Re=1000$, $k=2.04$, $\phi=80^\circ$}
	 \end{subfigure}
	\caption{Maximum and rms of the absolute residual of $\hv$ for Plane
Poiseuille flow and symmetrical (even) initial
condition. \textit{Continuous line}: real part; \textit{dashed line}: imaginary part.
Comparison with $N^{-5}$.}
\label{fig:convergence_v_Po}
\end{figure}

The error as function of the number of eigenfunctions is represented in a
bilogarithmic plane, so that the slope represent directly the order of the
method. It can be noticed from
\figref{fig:convergence_v_Co} and
\figref{fig:convergence_v_Po} that the error behaves almost like $1/N^5$ as 
$N\to \infty$ for different choices of the parameters and in both norms.
Moreover, differences between the imaginary and the real part are little.  
\FloatBarrier
\subsubsection{Convergence of $\he$}
\FloatBarrier
\begin{figure}
\vspace{-1cm}
        \centering
         \advance\leftskip-2.5cm
         \advance\rightskip-2cm
        \begin{subfigure}{0.6\textwidth}
        \centering 
	\includegraphics[width=9.5cm]{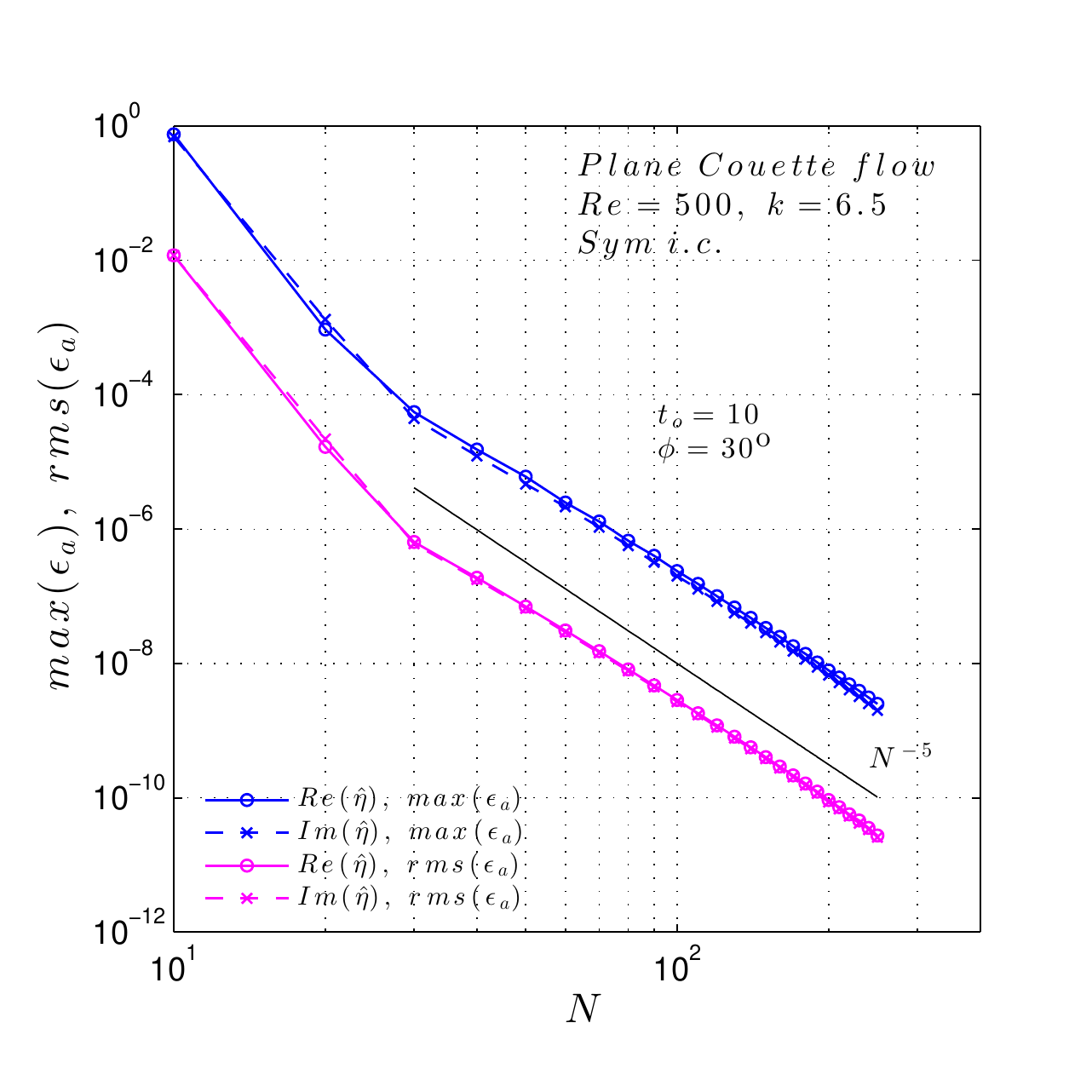}
	\subcaption{$t_0=10$, $Re=500$, $k=6.5$, $\phi=30^\circ$}
	 \end{subfigure}
        \begin{subfigure}{0.6\textwidth}
        \centering 
	\includegraphics[width=9.5cm]{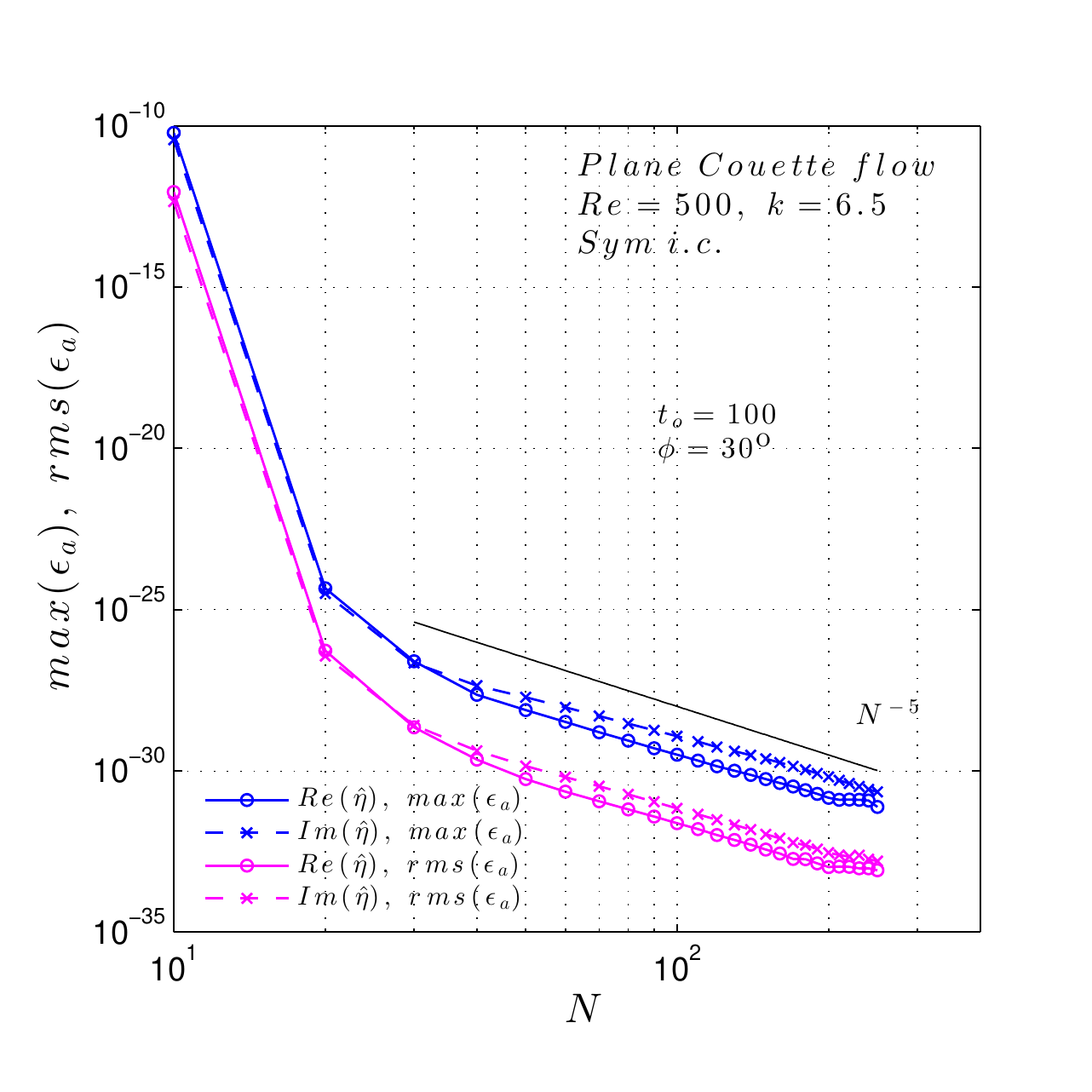}
	\subcaption{$t_0=100$, $Re=500$, $k=6.5$, $\phi=30^\circ$}
	 \end{subfigure}
	\caption{Maximum and rms of the absolute residual of $\he$ for Plane
Couette flow and symmetrical (even) initial
condition. \textit{Continuous line}: real part; \textit{dashed line}: imaginary part.
Comparison with $N^{-5}$.}

\label{fig:convergence_eta_Co}
\end{figure}
\vspace{-1cm}
\begin{figure}
        \centering
         \advance\leftskip-2.5cm
         \advance\rightskip-2cm
        \begin{subfigure}{0.6\textwidth}
        \centering 
	\includegraphics[width=9.5cm]{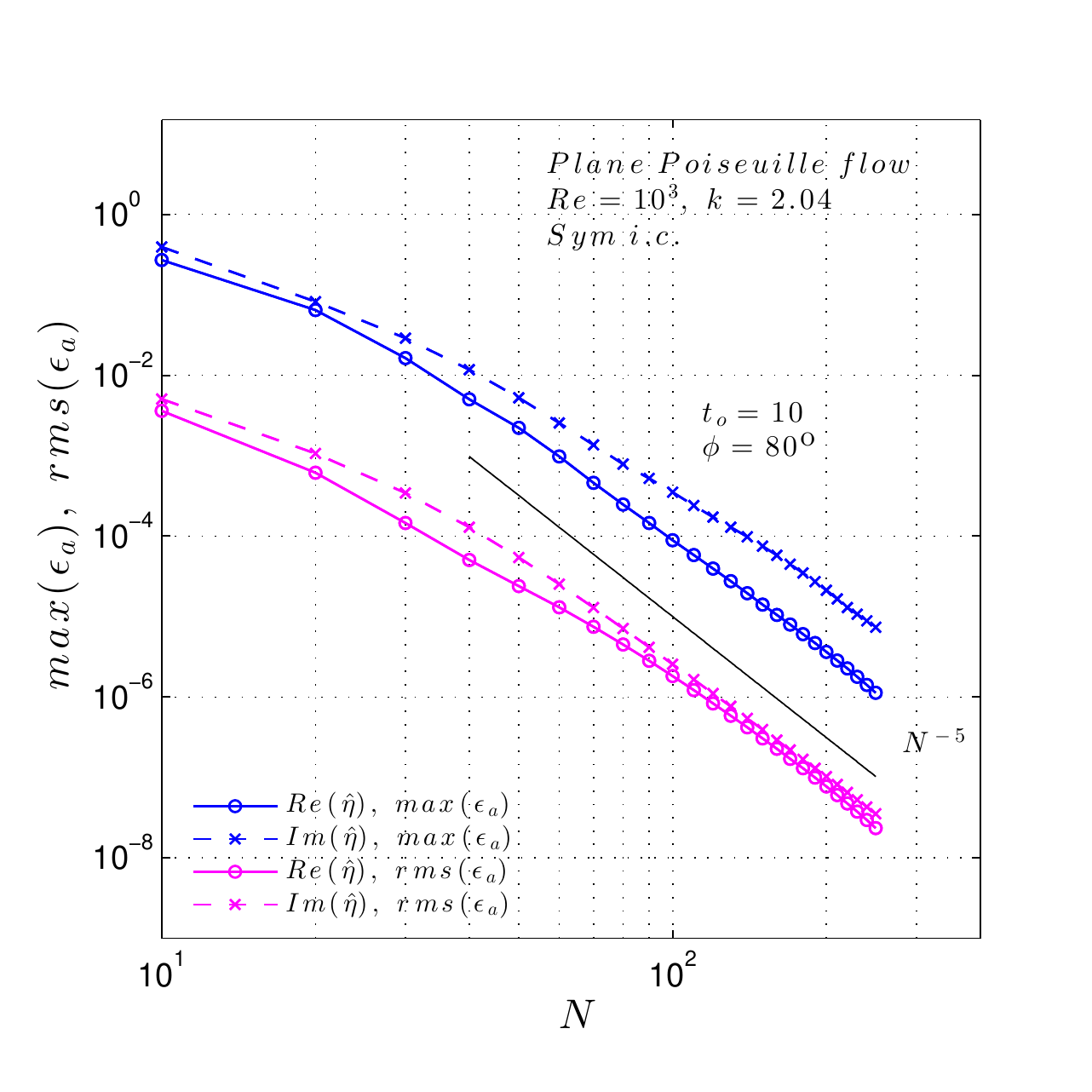}

	\subcaption{$t_0=10$, $Re=1000$, $k=2.04$, $\phi=80^\circ$}
	 \end{subfigure}
        \begin{subfigure}{0.6\textwidth}
        \centering 
\includegraphics[width=9.5cm]{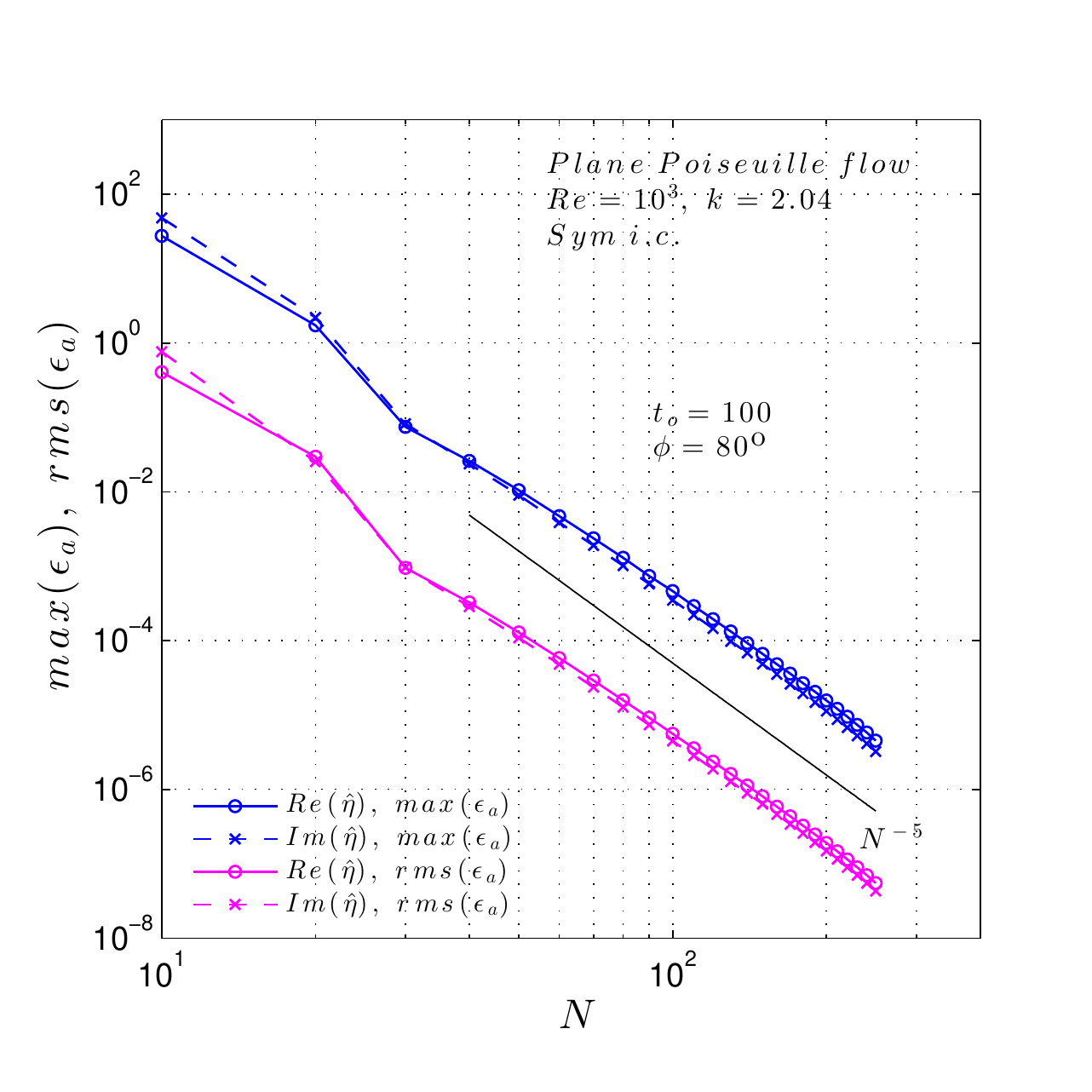}
	\subcaption{$t_0=100$, $Re=1000$, $k=2.04$, $\phi=80^\circ$}
	 \end{subfigure}
	\caption{Maximum and rms of the absolute residual of $\he$ for Plane
Poiseuille flow and symmetrical (even) initial
condition. \textit{Continuous line}: real part; \textit{dashed line}: imaginary part.
Comparison with $N^{-5}$.}
\label{fig:convergence_eta_Po}
\end{figure}
\FloatBarrier
\subsection{Termwise differentiation and convergence of derivatives}
\begin{figure}[h!]
        \centering
         \advance\leftskip-2.5cm
         \advance\rightskip-2cm
        \begin{subfigure}{0.6\textwidth}
        \centering 
	\includegraphics[width=9.0cm]{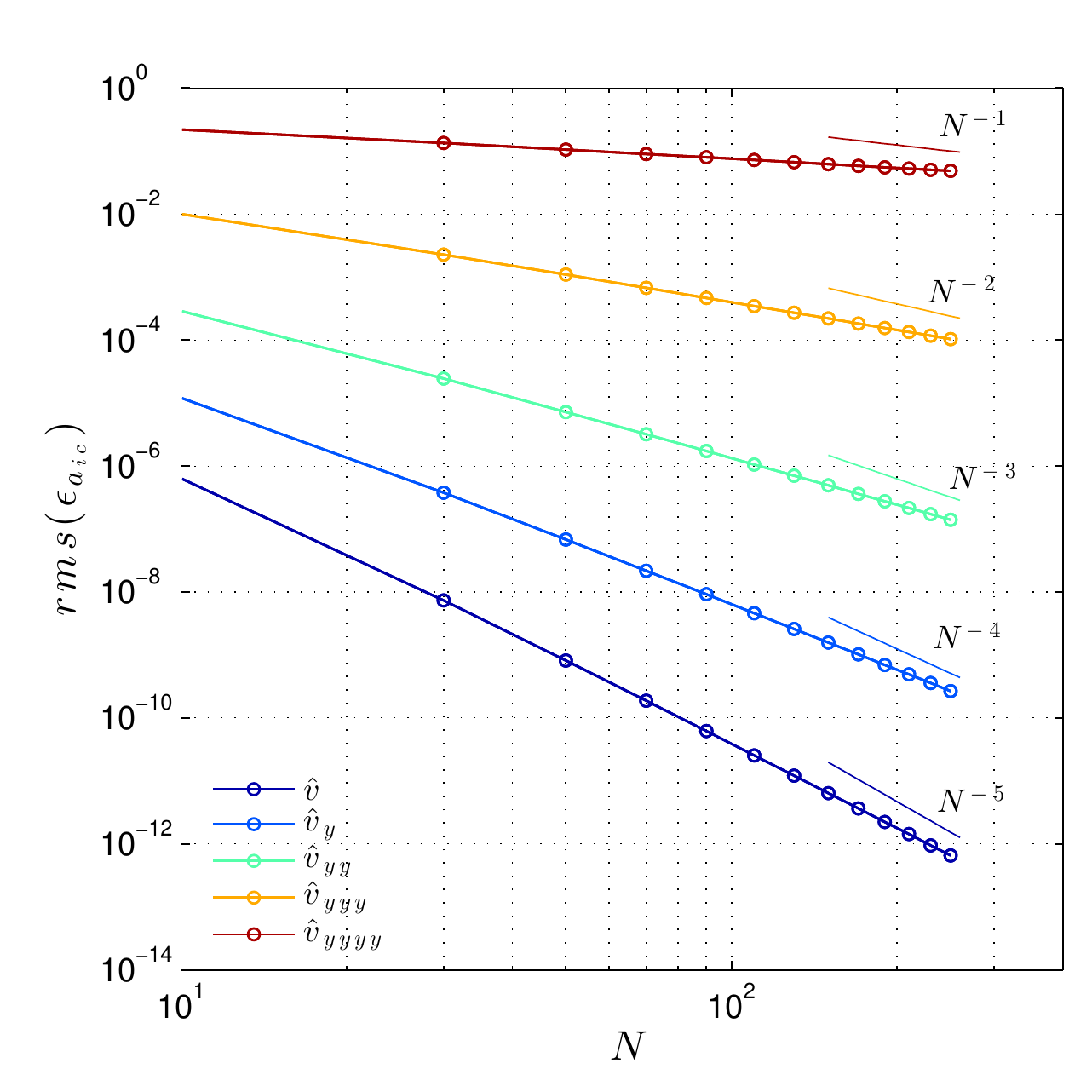}
	\vspace{0.5pt}
	\subcaption{Error $rms$}
	\label{v_der_rms}
	 \end{subfigure}
        \begin{subfigure}{0.6\textwidth}
        \centering 
\includegraphics[width=9.0cm]{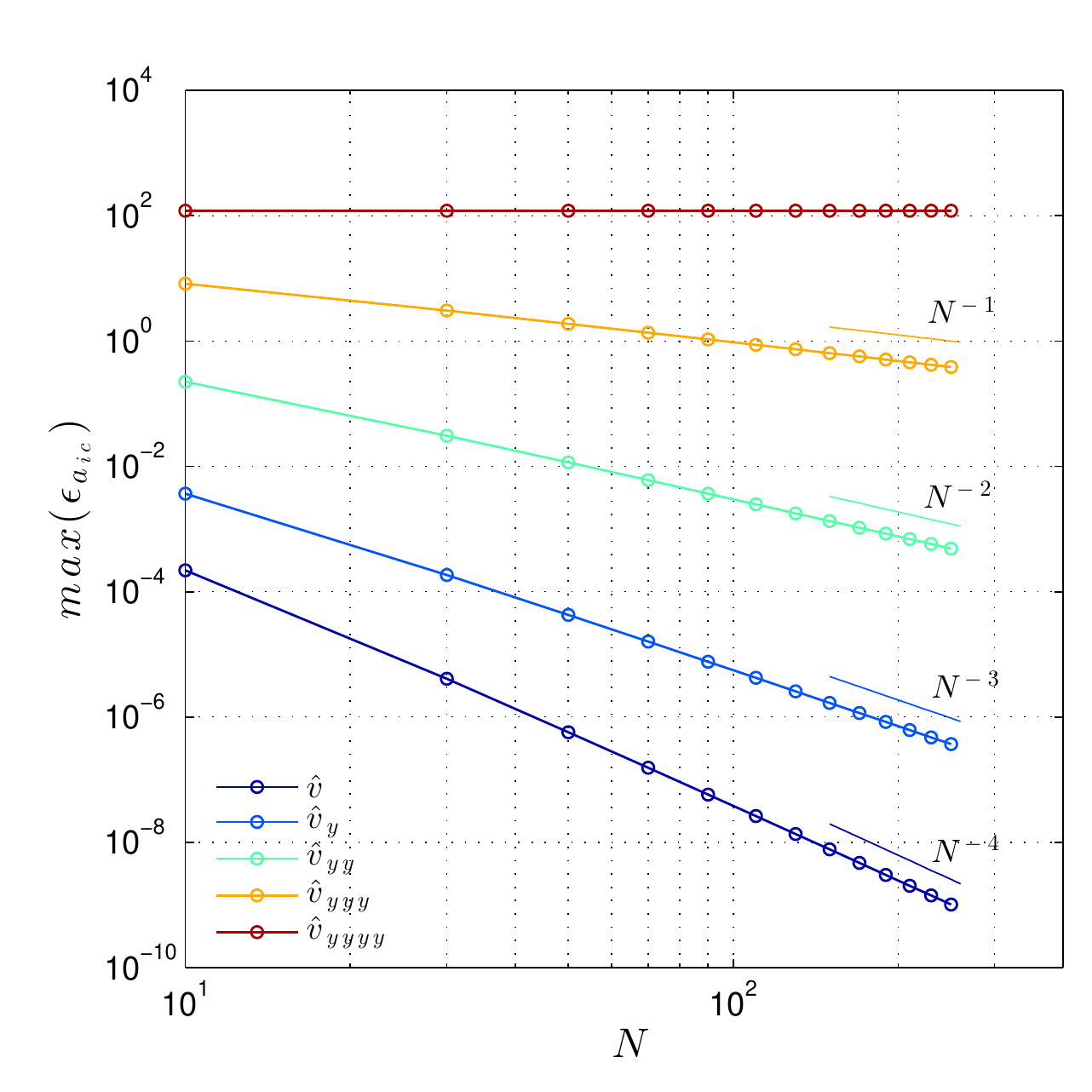}
	\vspace{0.5pt}
	\subcaption{Maximum error}
	\label{v_der_max}
	 \end{subfigure}
	\caption{Convergence of $\hv$ series and its derivatives to the velocity
profile defined by $ic=(1-y^2)^2$.}
\label{fig:convergence_deriv_v}
\end{figure}

\begin{figure}[h!]
        \centering
         \advance\leftskip-2.5cm
         \advance\rightskip-2cm
        \begin{subfigure}{0.6\textwidth}
        \centering 
	\includegraphics[width=9.0cm]{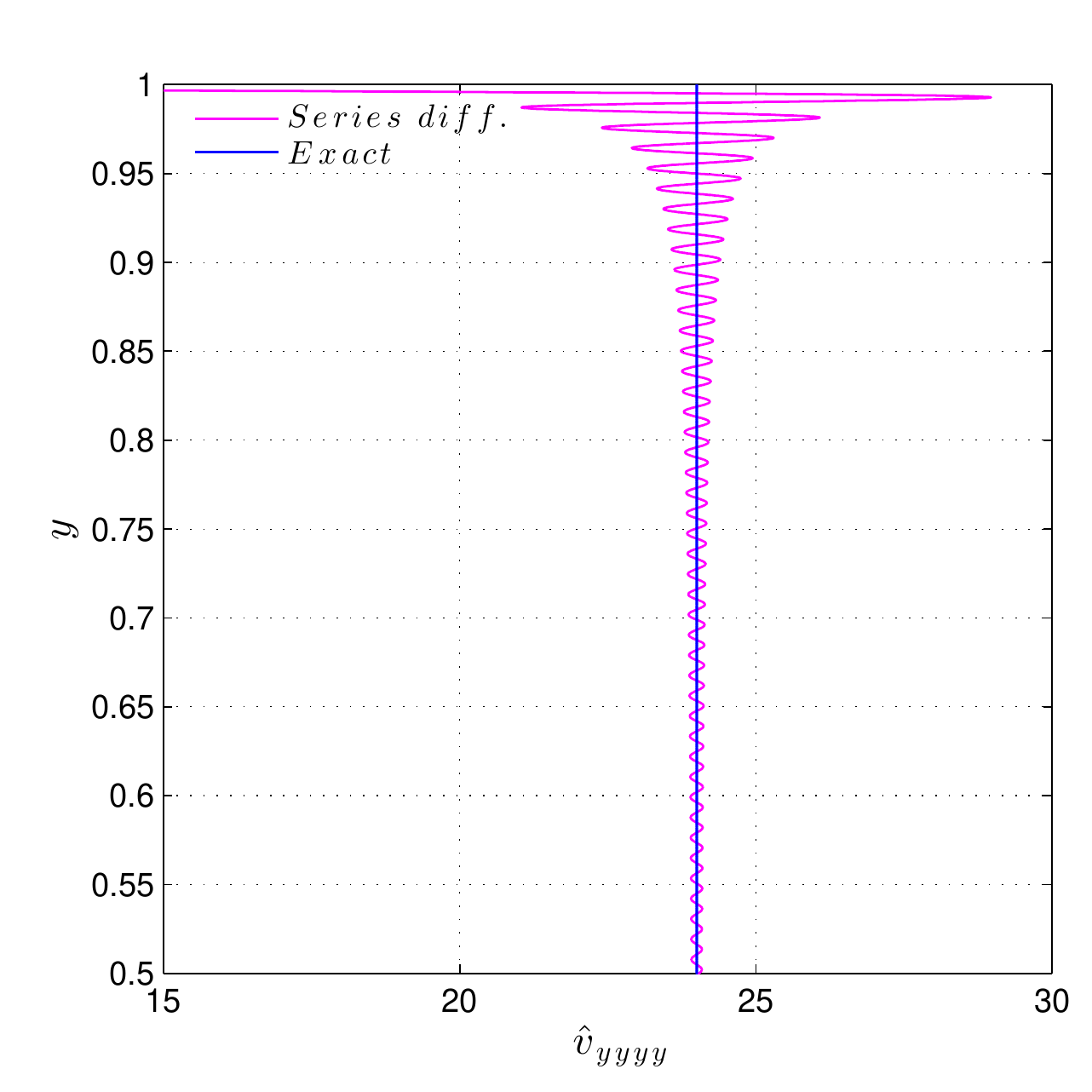}
	\vspace{0.5pt}
	\subcaption{}
	\label{fig:nonunif_v}
	 \end{subfigure}
        \begin{subfigure}{0.6\textwidth}
        \centering 
\includegraphics[width=9.0cm]{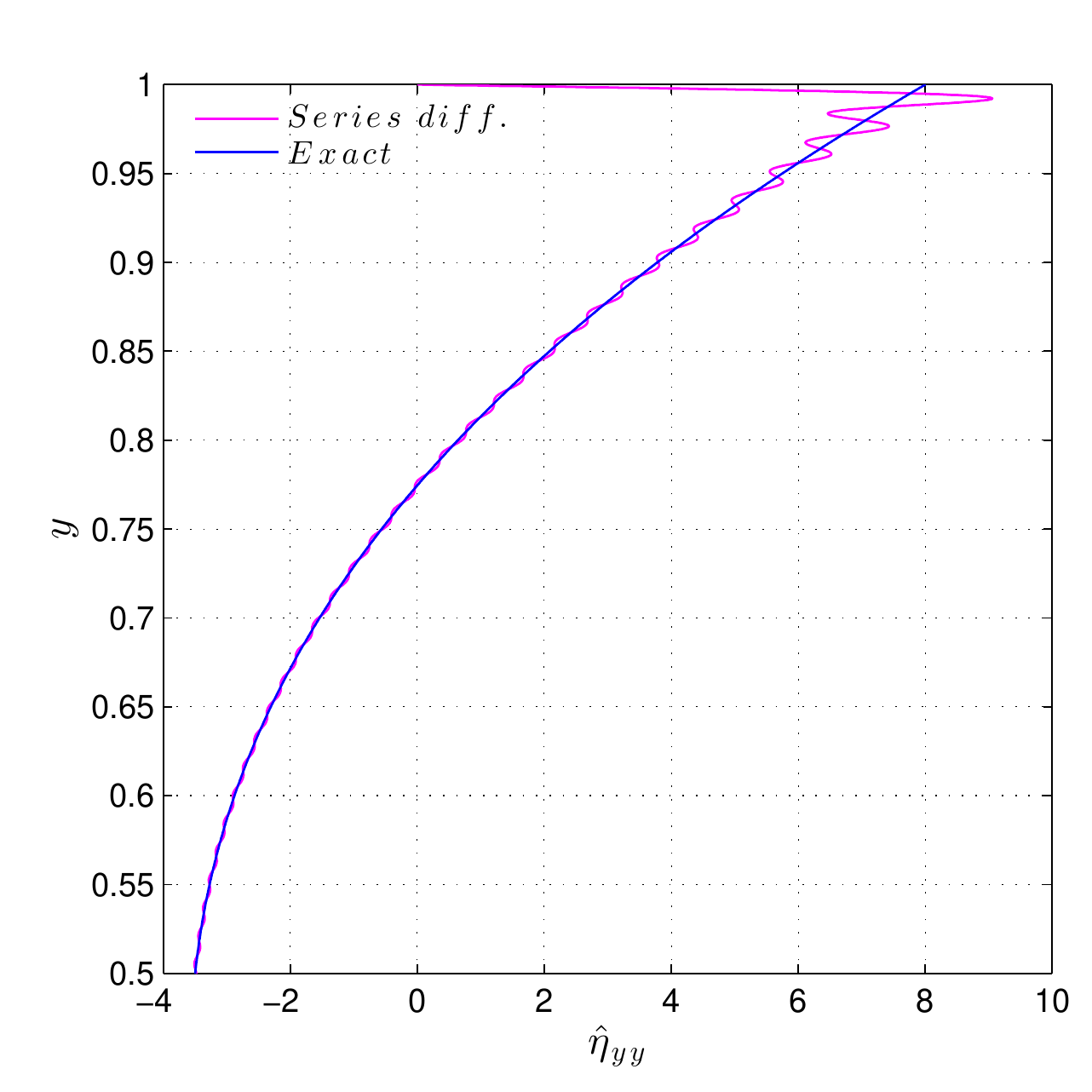}
	\vspace{0.5pt}
	\subcaption{}
	\label{fig:nonunif_eta}
	 \end{subfigure}
	\caption{Non-uniform convergence of the four-times
termwise differentiated $\hv$ series and of the two-times termwise
differentiated $\he$ series to the exact derivatives of $ic=(1-y^2)^2$. The
phenomenon is analogous to the Gibbs one, and subsists at the boundaries $y=\pm
1$. }
\label{fig:nonuniform}
\end{figure}

\FloatBarrier
The analytical expressions of the derivatives of the basis functions are known.
Thereby, it is easy to investigate the convergence of the series obtained by
termwise differentiation. In fact, it is known \citep[see][]{PDE_book} that
termwise differentiation is not always possible and if the derivatives of the
solution are needed, finite differences techniques may be required. The
convergence has been estimated applying the transform to a known function, i.e.
the symmetrical initial condition for the normal-velocity. The same profile is
used to estimate the convergence for the normal-vorticity, even if it won't be
used as initial condition in the further analysis; thus, in the following
figures, the absolute error is indicated as $\epsilon_{a_{ic}}$. \\
About the convergence of $\hv$, it can be noticed (see
\figref{fig:convergence_deriv_v}) that the series can be differentiated termwise
three times. The derivative of fourth order converges very slowly in
$L_2$-norm but
the maximum error remains almost constant, which means that the convergence is
non-uniform (see \figref{fig:nonunif_v}). The same results have been found by
\citet{Orszag1971}. The
derivatives of lower order converge in both norm to the exact expression, even
if the rate decreases with increasing order of derivation. The rate of
convergence of $\hv$ is now slightly less than $N^{5}$.\\

\begin{figure}[h!]
        \centering
         \advance\leftskip-2.5cm
         \advance\rightskip-2cm
        \begin{subfigure}{0.6\textwidth}
        \centering 
	\includegraphics[width=9.0cm]{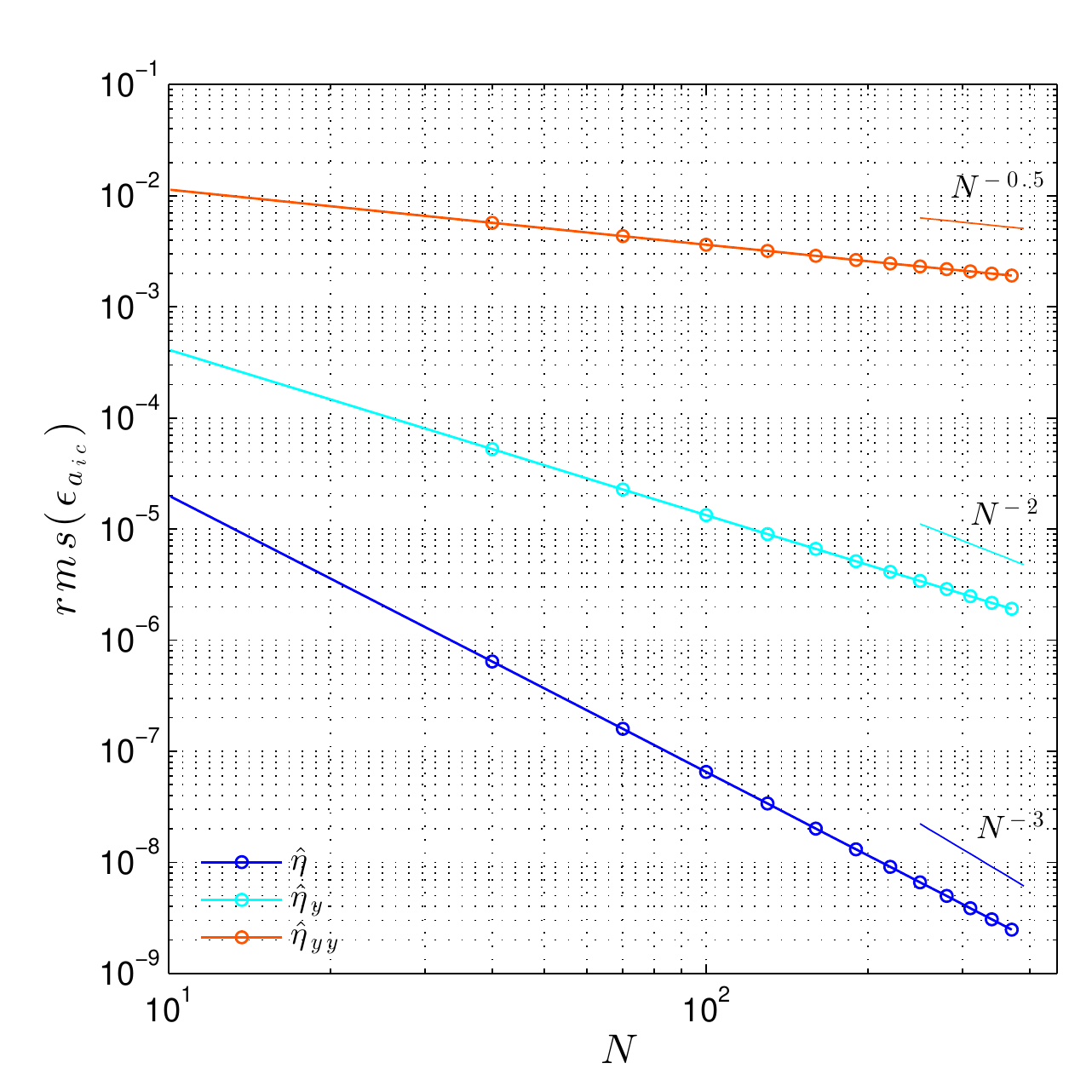}
	\vspace{0.5pt}
	\subcaption{Error $rms$}
	\label{eta_der_rms}
	 \end{subfigure}
        \begin{subfigure}{0.6\textwidth}
        \centering 
\includegraphics[width=9.0cm]{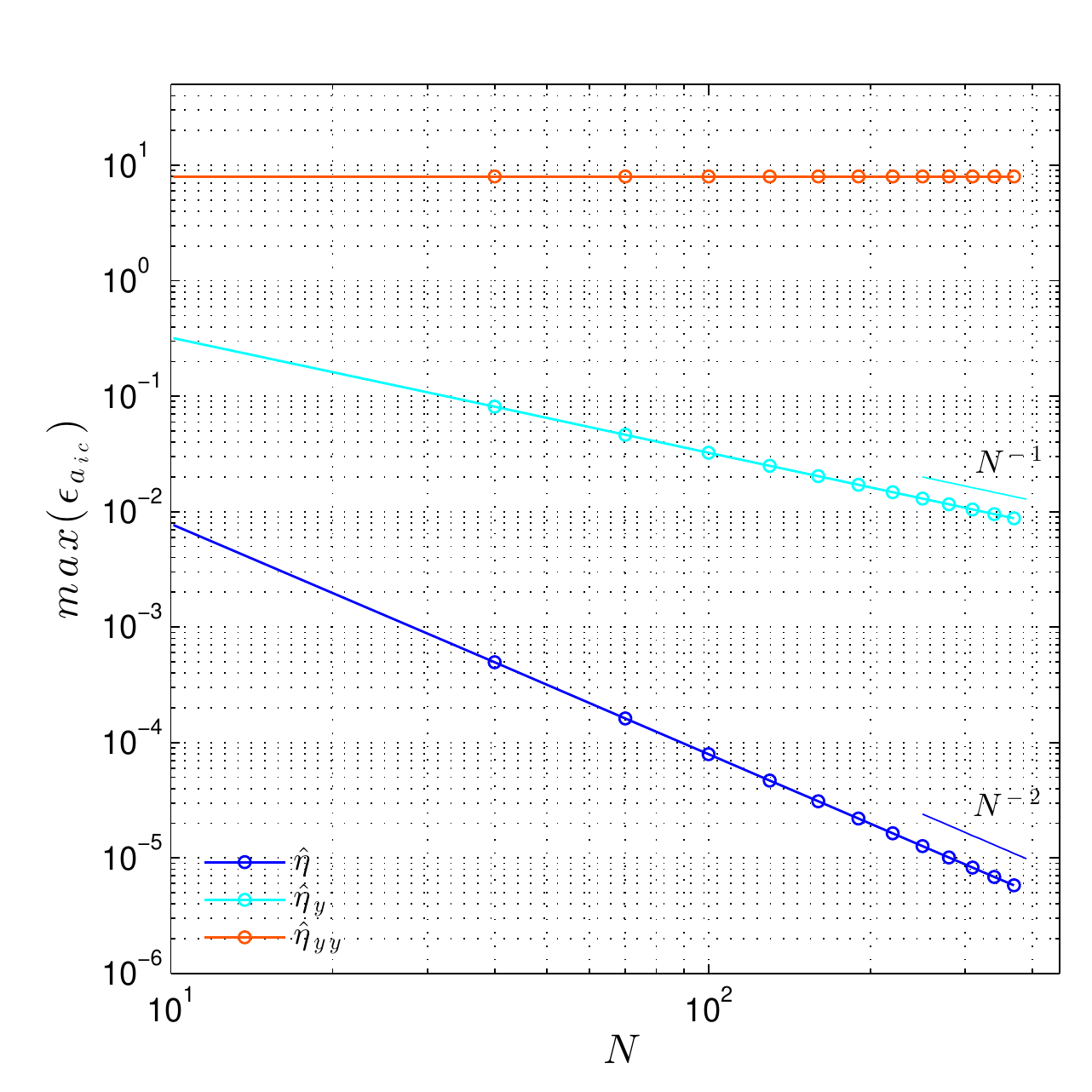}
	\vspace{0.5pt}
	\subcaption{Maximum error}
	\label{eta_der_max}
	 \end{subfigure}
	\caption{Convergence of $\he$ series and its derivatives to a profile
defined by $ic=(1-y^2)^2$.}
\label{fig:convergence_deriv_eta}
\end{figure}

About $\he$ series, it can be noticed from \figref{fig:convergence_deriv_eta}
that the termwise differentiation can be applied to obtain correct results up
to the first derivative. The second derivative converges slowly in norm two,
but non-uniformly (see \figref{fig:nonunif_eta}). In this case the error of
$\he$ seems to behave almost like
$N^{-3}$, so differently from what can be seen in
\figref{fig:convergence_eta_Co} and \figref{fig:convergence_eta_Po}, where the
$\he$ solution is forced by the normal-velocity $\hv$. 
\vspace{-1cm}

\chapter{Wave transient analysis: numerical results}\label{chap:wave_transient}
\section{Introduction}
In the present section, particular attention is given to the
 temporal evolution of the wave frequency and phase velocity. 
Recent studies \citep[see][]{Scarsoglio2012,Scarsoglio2009} have been pointing
out as through 
wave frequency investigations precious information can be obtained about the
different phases that characterize the spatio-temporal evolution of a
perturbation. \\ The importance of a better understanding of the transient live
of traveling waves relies, among other reasons, in the relation with rapid
transition to fluid turbulence. In fact, it is believed that the
exceptionally large algebraic growth which can occur in the disturbance
evolution before the asymptotic exponential mode is set, could promote a
phenomenon known as \textit{bypass transition} \citep[see e.g.][]{Henningson1994}.
Actually, it consists of a disturbance growth and breakdown on a timescale much
shorter than those typical for Tollmien-Schlichting (TS) waves. The term is used
to emphasize that these scenarios bypass the growth of two-dimensional waves and
their subsequent secondary instability \citep[see][]{Henningson1993}. It has
been shown that the transient algebraic growth can be significant even for
subcritical values of the Reynolds number, so that finite amplitudes can
rapidly be achieved, and nonlinear effects can enter into play. We will
discuss in the next chapter the consequent formation of turbulent spots. \par
The frequency temporal evolution has been poorly investigated, probably because
sheared incompressible flows are viewed as non-dispersive media. Nevertheless,
the frequency  transient behaviour revealed unexpected
phenomena, non predictable \textit{a priori}, that being related to the wave
phase velocity could have a remarkable influence on the main phase speed of a
group of waves, in particular on the early stage of a natural spot formation,
when the non-linear effects can be neglected, as shown by \citet{Cohen1991} in
the case of boundary-layer flow. Moreover, as pointed out by
\citet{Kachanov1994}, neither intermittence nor turbulent spots are observed in
K- or N-regimes of transition when the initial instability wave is strictly
periodic in time. The ``natural'' intermittence phenomena and the spot
formation are usually detected when the perturbations background is more
complicated and the instability wave has both amplitude and phase modulation in
time.\par
In this contest, the study of the frequency temporal evolution gains more and
more importance. From the latest works previously cited emerges that the
complexity of the frequency transient is mainly associated to jumps which
appear quite far along the temporal history. The normalized time at which the
jumps occur have been considered as the threshold between the first two
phases of a wave life, respectively the \textit{Early transient} and the
\textit{Intermediate transient}. The intermediate term lasts until the
asymptotic exponential energy growth/decay is reached (long term), and
appears to be the most probable state in a wave life, since on one hand its
temporal extension is at least one order of magnitude bigger than the early
term's
one and, on the other hand, at the end of this intermediate period the
disturbance will die or blowup.\par
In the following sections, an analysis of the phase speed time evolution
will be provided; The three phases of a wave life will emerge by taking into
account the frequency time history of both the normal-velocity and the
normal-vorticity; particularly, the existence of the intermediate period will be shown. Moreover, a relationship
between frequency jumps and the achievement of a self-similar asymptotic state
of the flow velocity and vorticity profiles will be shown.

\section{Wave frequency and phase velocity}
\subsection{Analysis of the $\hv$ component of flow
velocity}\label{sec:v_frequency}
 In the present section we take advantage of the solution method proposed
in \secref{sec:gal_v} to obtain the time evolution of the frequency
and phase speed of the wall-normal component of velocity $\hv$, varying the
three parameters that characterize the problem:
the Reynolds number, the obliquity angle and the polar wavenumber. The high
non-stationarity of the phenomenon will emerge, typically a jump
is observed at a certain time, which is considered to be the threshold between
the early period and the intermediate one. After this jump the frequency of
$\tilde v$ is characterized by a modulation about a constant mean value for Plane Couette flow, for
sufficiently high values of the polar wavenumber $k$. For all the
simulations performed, $N=250$ eigenfunctions are used for the solution
expansion.\par 
\subsubsection{Numerical computation}
The frequency of the perturbation is defined as the temporal derivative
of the unwrapped phase $\theta(y,t;\alpha,\beta)$, at a specific spatial point
along the $y$ coordinate. The wrapped phase
\begin{equation}
\theta_w(y,t;\alpha,\beta)=arg(\hv(y,t;\alpha,\beta))
\end{equation}
is a discontinuous function of $t$ in $[-\pi,+\pi]$, while the unwrapped phase
$\theta$ is continuous and it is obtained  by adding multiples of $\pm 2\pi$ when
absolute jumps  greater than or equal to $\pi$ radians occur. The wave
frequency is defined as
\begin{gather}
\omega(y,t;\alpha,\beta)=\bigg|\de{\theta(y,t;\alpha,\beta)}{t}\bigg|
\end{gather}
The phase velocity vector is given then by the dispersion relation
\begin{equation}
 \bm{c}=\frac{\omega}{k}\BU{\hat{k}}
\end{equation}
where $\BU{\hat{k}}=(cos(\phi),sin(\phi))$ is the unitary vector defining the
 polar wavenumber direction. The frequency of each signal can be numerically
computed at a fixed observation point $y=y_0$. In order to ensure a
high accuracy in the results, a fourth order centered finite-differences scheme
has been used to calculate the first temporal derivative.
The following scheme applies for the inner points of the defined time vector and
corresponding phase values $(t_i,\theta_i)$:
\begin{equation}
\omega(t_i;y_0,\alpha,\beta)=\omega_i=\dtot{\theta_i}{t}=\frac{\theta_{i-2}
-8\theta_{i-1}+8\theta_{i+1}-\theta_{i+2}}{12\Delta t}\ \ \ \ i=3,4,\hdots,N_t-2
\end{equation}
where $N_t$ is the total number of elements of the time and phase vector
and $\Delta t$ the time spacing. It is worth to underline that the proposed
method allows the user to define arbitrary time and space grids, differently
from Runge-Kutta routines, where the time step is free to
change accordingly to the stiffness of the problem and the needed accuracy. This
is actually an advantage, because the accuracy of the numerical estimated
derivatives is affected by the non-uniformity of the grid. Setting a uniform
time spacing we ensure that the finite-differences scheme is actually of the
fourth order \citep[see][]{CFD_book}. 
Since the scheme stencil is made of five points, for the first and the last two
points of the vector respectively a forward and backward fourth order finite-differences scheme is needed. Indeed, the
accuracy of the method for these
points could be lower. The following schemes are applied:
\begin{gather}
\omega_1=\dtot{\theta_1}{t}=\frac{-25\theta_{1}
+48\theta_{2}-36\theta_{3}+16\theta_{4}-3\theta_{5}}{12\Delta t}\\[10pt]
\omega_2=\dtot{\theta_2}{t}=\frac{-3\theta_{1}
-10\theta_{2}+18\theta_{3}-6\theta_{4}+\theta_{5}}{12\Delta t}
\end{gather}
\begin{gather}
\omega_{N_t-1}=\dtot{\theta_{N_t-1}}{t}=\frac{3\theta_{N_t}
+10\theta_{N_t-1}-18\theta_{N_t-2}+6\theta_{N_t-3}-\theta_{N_t-4}}{12\Delta
t}\\[10pt]
\omega_{N_t}=\dtot{\theta_{N_t}}{t}=\frac{+25\theta_{N_t}
-48\theta_{N_t-1}+36\theta_{N_t-2}-16\theta_{N_t-3}+3\theta_{N_t-4}}{12\Delta t}
\end{gather}
\vspace{1cm}
\FloatBarrier
\subsubsection{Phase velocity temporal evolution}
\begin{figure}[h!]
        \centering
         \advance\leftskip-2.5cm
         \advance\rightskip-2cm
        \begin{subfigure}{0.6\textwidth}
        \centering 
	\includegraphics[width=9.0cm]{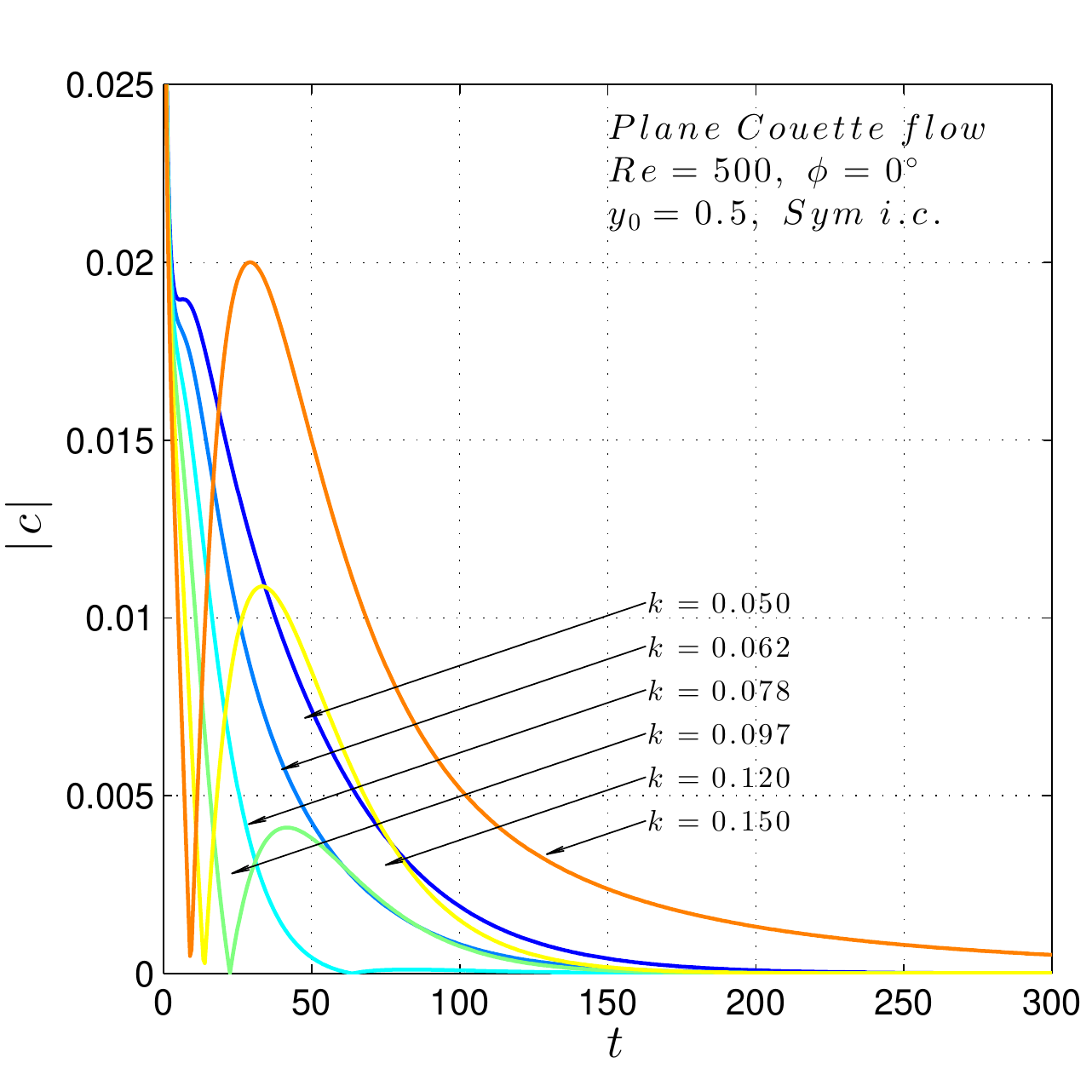}
	\vspace{0.5pt}
	\subcaption{}
	\label{fig:c_CO_Re500_phi0_variok_1}
	 \end{subfigure}
        \begin{subfigure}{0.6\textwidth}
        \centering 
\includegraphics[width=9.0cm]{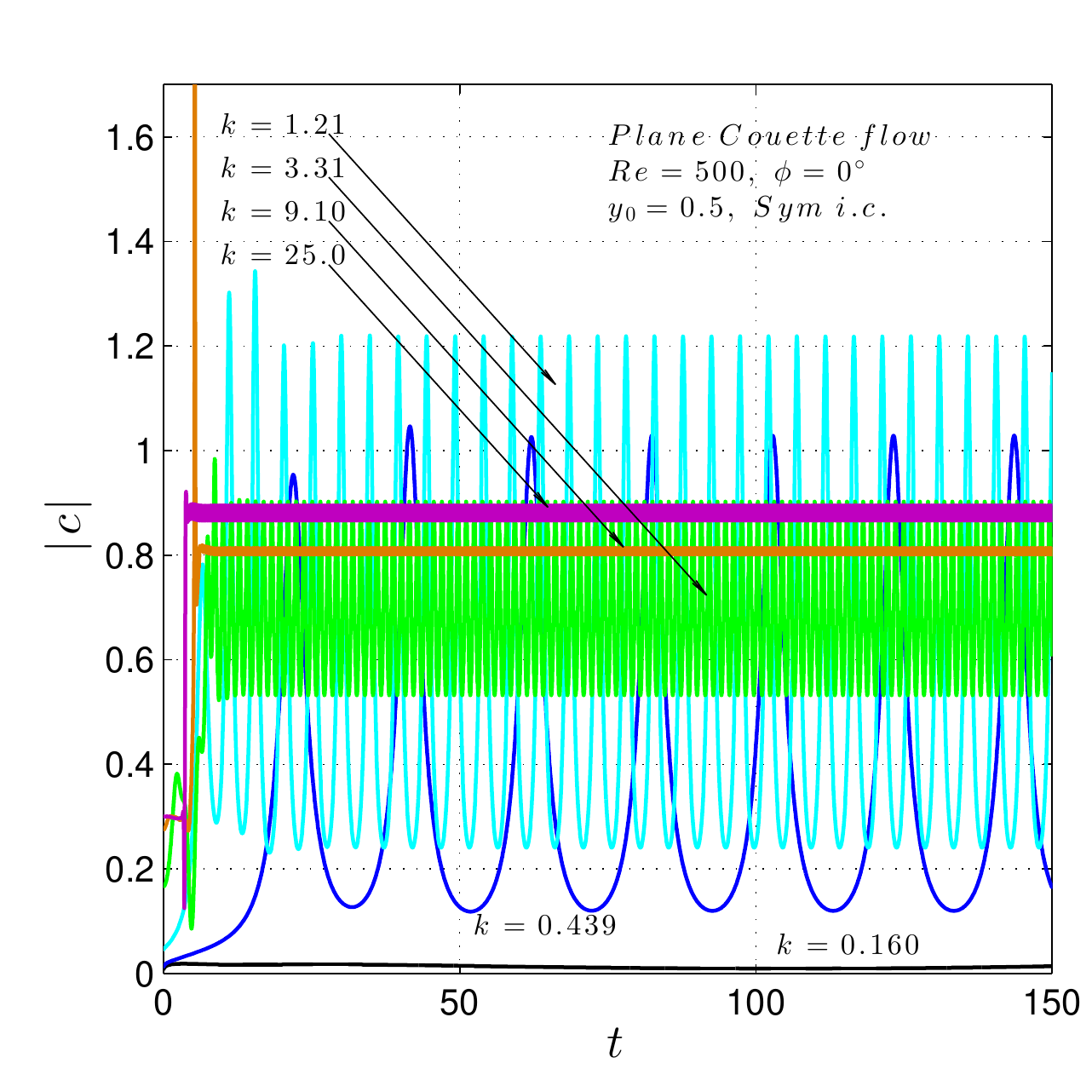}
	\vspace{0.5pt}
	\subcaption{ }
	\label{fig:c_CO_Re500_phi0_variok_2}
	 \end{subfigure}
	\caption{Temporal evolution of the absolute value of the phase
velocity, calculated from $\hv$, for PCf at $Re=500$, $\phi=0^{\circ}$ and
\textit{sym.} initial condition, defined by $\hv_0=(1-y^2)^2$. The polar
wavenumber covers the range $k \in[0.05, 25]$, uniformly distributed in the
logarithmic space. The fixed observation point is $y_0=0.5$.}
\label{fig:c_CO_Re500_phi0_variok}
\end{figure}

\begin{figure}[h!]
        \centering
         \advance\leftskip-2.5cm
         \advance\rightskip-2cm
        \begin{subfigure}{0.6\textwidth}
        \centering 
	\includegraphics[width=9.0cm]{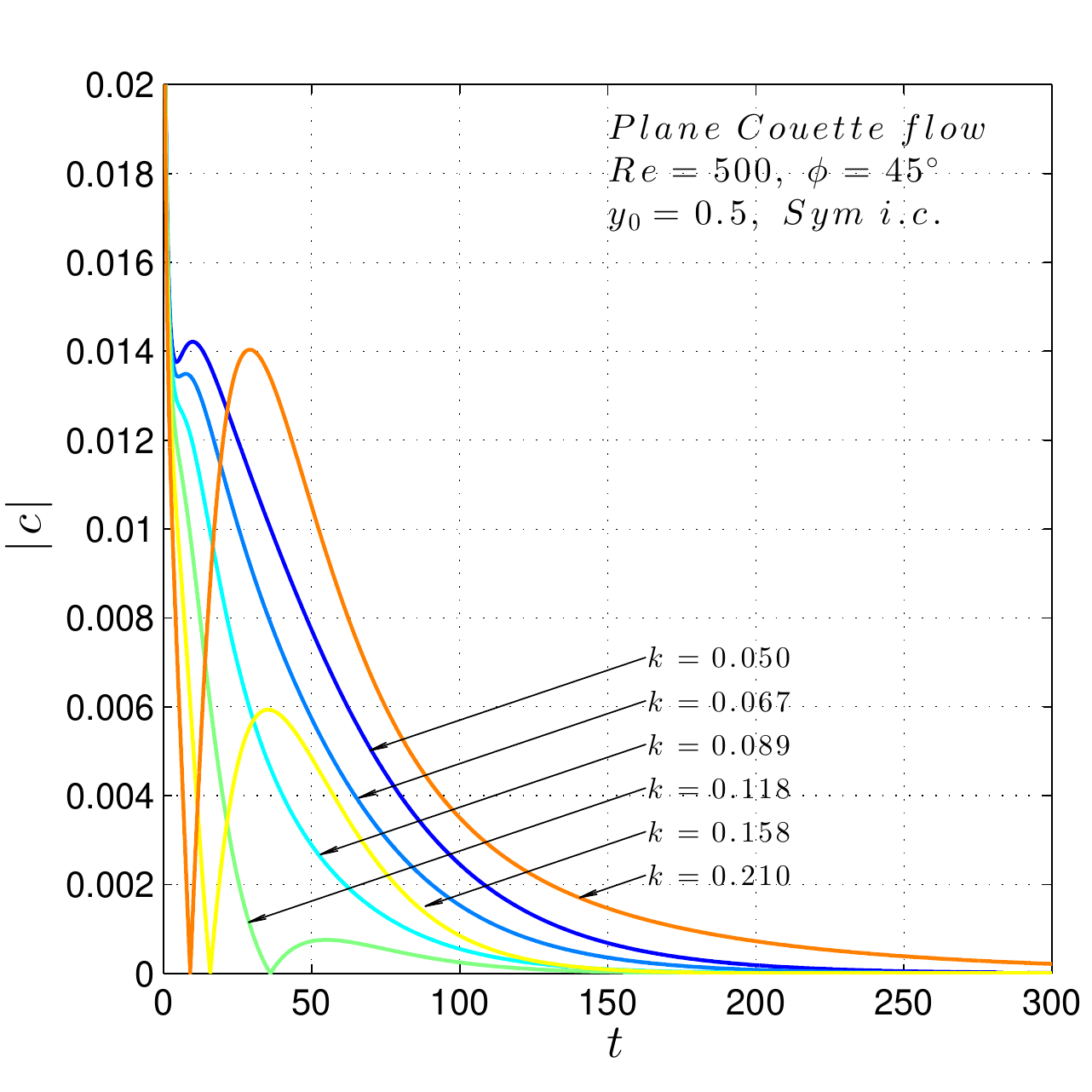}
	\vspace{0.5pt}
	\subcaption{}
	\label{fig:c_CO_Re500_phi45_variok_1}
	 \end{subfigure}
        \begin{subfigure}{0.6\textwidth}
        \centering 
\includegraphics[width=9.0cm]{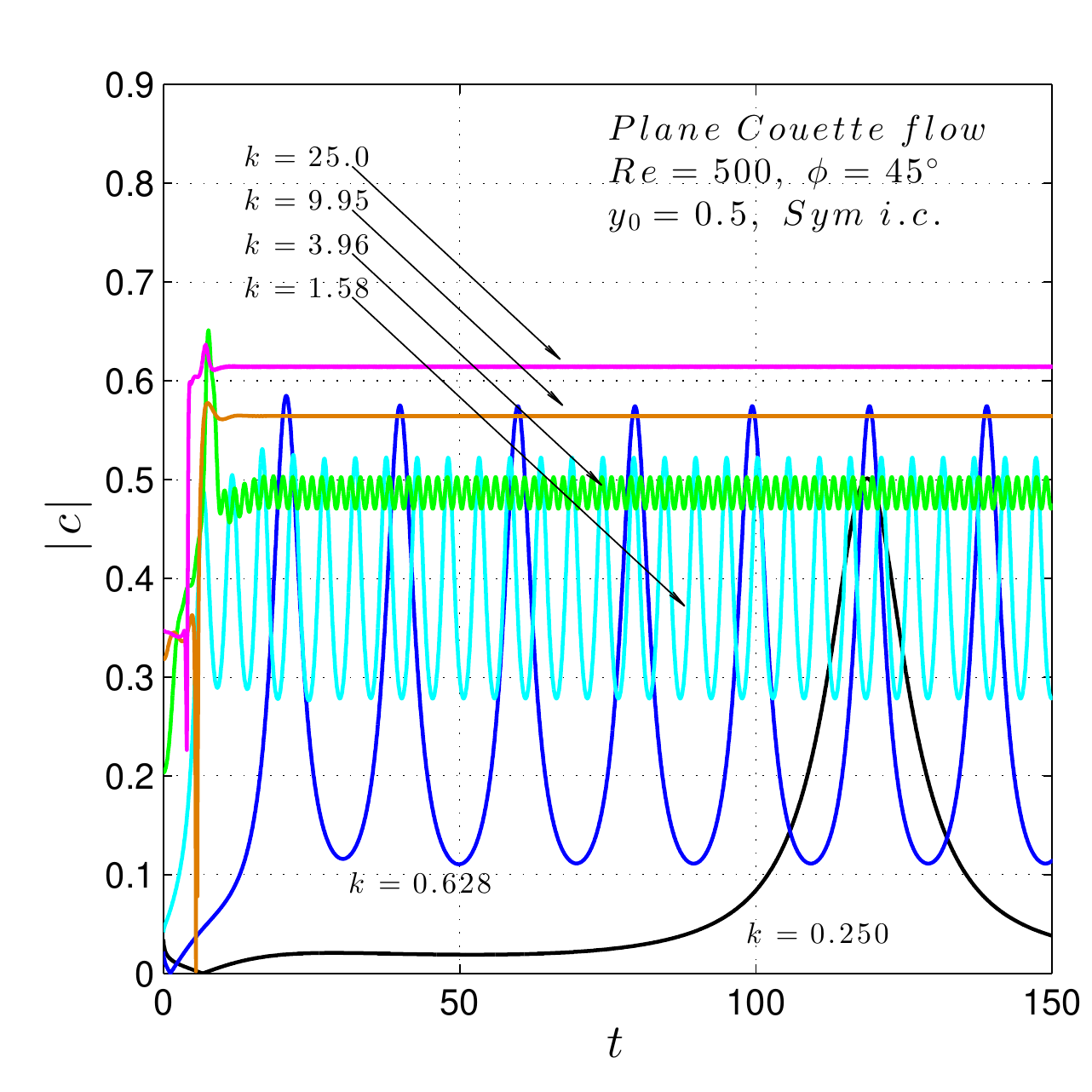}
	\vspace{0.5pt}
	\subcaption{}
	\label{fig:c_CO_Re500_phi45_variok_2}
	 \end{subfigure}
	\caption{Temporal evolution of the absolute value of the phase
velocity, calculated from $\hv$, for PCf at $Re=500$, $\phi=45^{\circ}$ and
\textit{sym.} initial condition. The
polar wavenumber covers the range $k \in[0.05, 25]$, uniformly
distributed in the logarithmic space. The fixed observation point is
$y_0=0.5$.}
\label{fig:c_CO_Re500_phi45_variok}
\end{figure}

\begin{figure}[h!]
        \centering
         \advance\leftskip-2.5cm
         \advance\rightskip-2cm
        \begin{subfigure}{0.6\textwidth}
        \centering 
	\includegraphics[width=9.0cm]{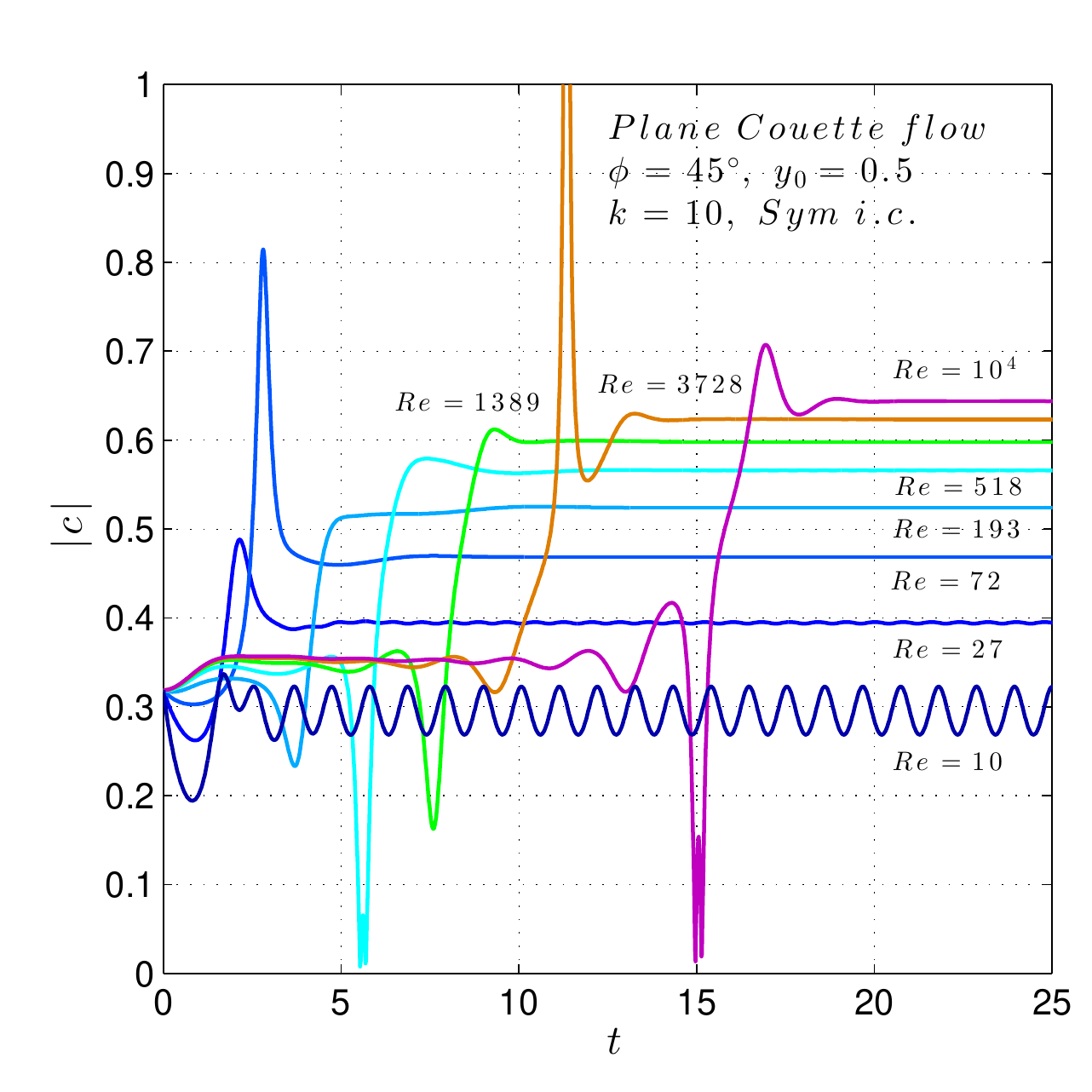}
	\vspace{0.5pt}
	\subcaption{}
	\label{fig:c_CO_phi45_varioRe_1}
	 \end{subfigure}
        \begin{subfigure}{0.6\textwidth}
        \centering 
\includegraphics[width=9.0cm]{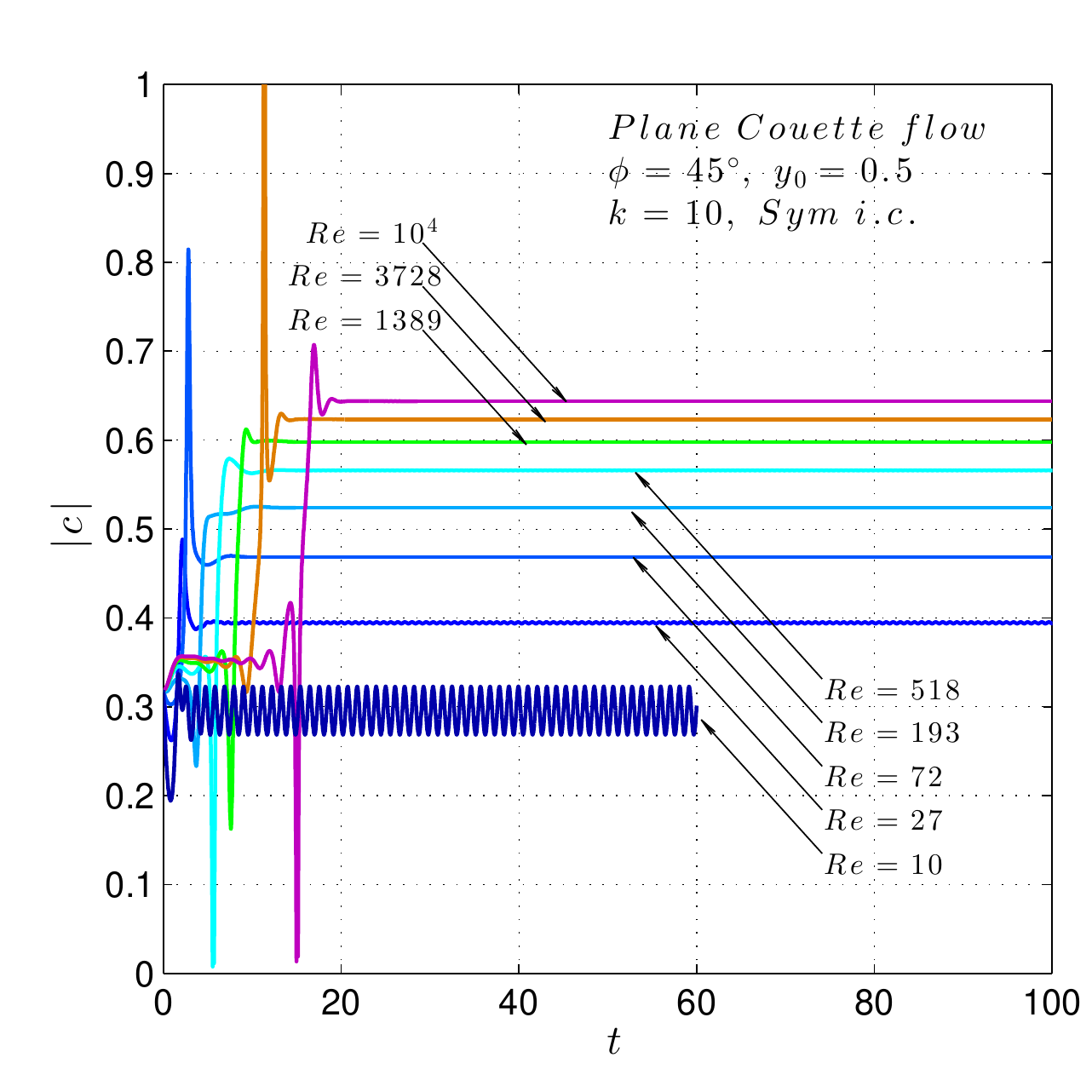}
	\vspace{0.5pt}
	\subcaption{ }
	\label{fig:c_CO_Re500_phi45_varioRe_2}
	 \end{subfigure}
	\caption{Temporal evolution of the absolute value of the phase
velocity, calculated from $\hv$, for PCf for $k=10$, $\phi=45^{\circ}$ and
\textit{sym.} initial condition. The Reynolds number covers the range $Re
\in[10, 10000]$, uniformly distributed in the logarithmic space. The fixed
observation point is
$y_0=0.5$. (a) Detail of the \textit{Early transient} and frequency jumps; (b)
\textit{Intermediate transient}.}
\label{fig:c_CO_Re500_phi45_varioRe}
\end{figure}

\begin{figure}[htb]
        \centering
        \vspace{-0.4cm}
         \advance\leftskip-2.5cm
         \advance\rightskip-2cm
        \begin{subfigure}{0.6\textwidth}
        \centering 
	\includegraphics[width=9.0cm]{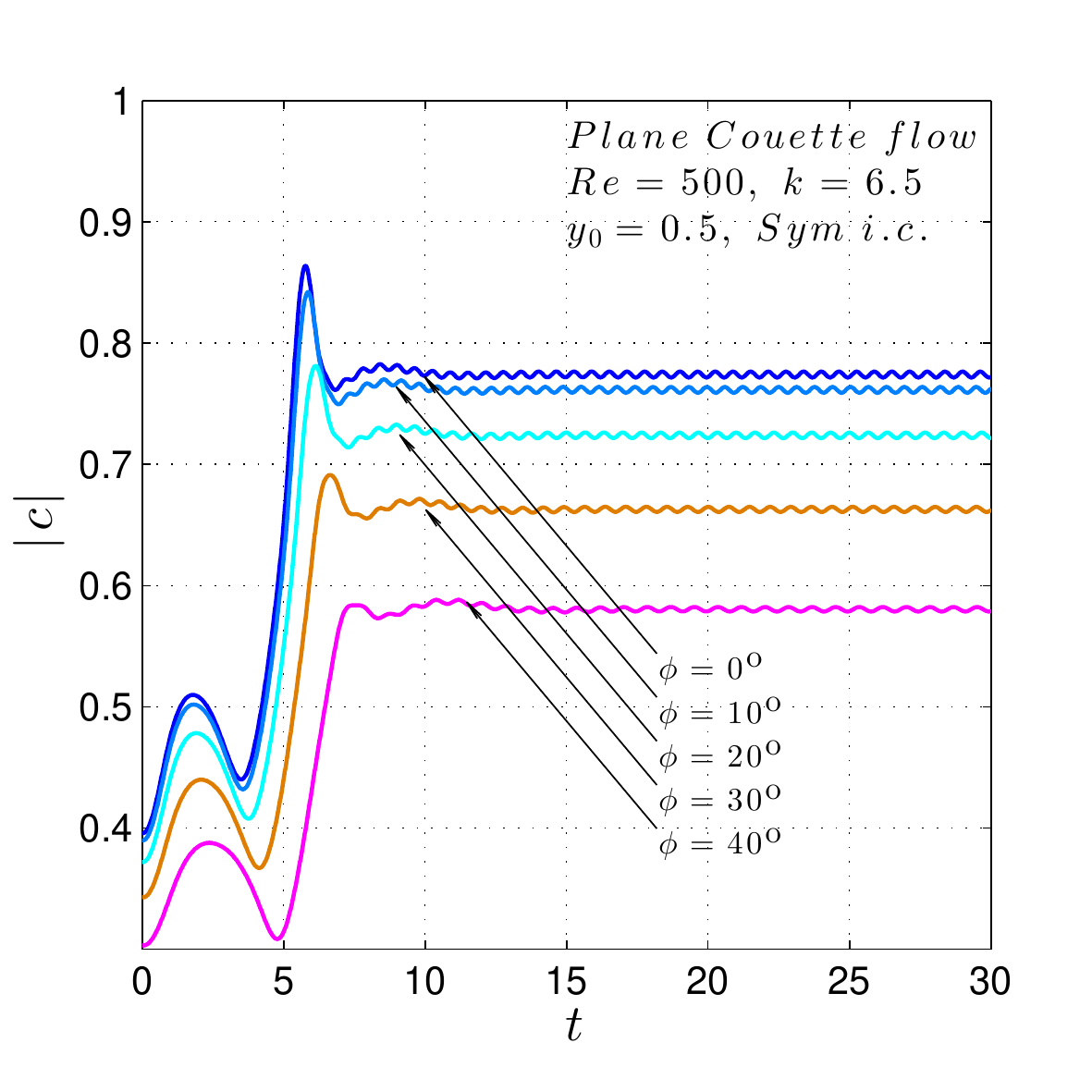}
	\vspace{0.5pt}
	\subcaption{}
	\label{fig:c_CO_Re500_k6p5_variophi_1}
	 \end{subfigure}
        \begin{subfigure}{0.6\textwidth}
        \centering 
\includegraphics[width=9.0cm]{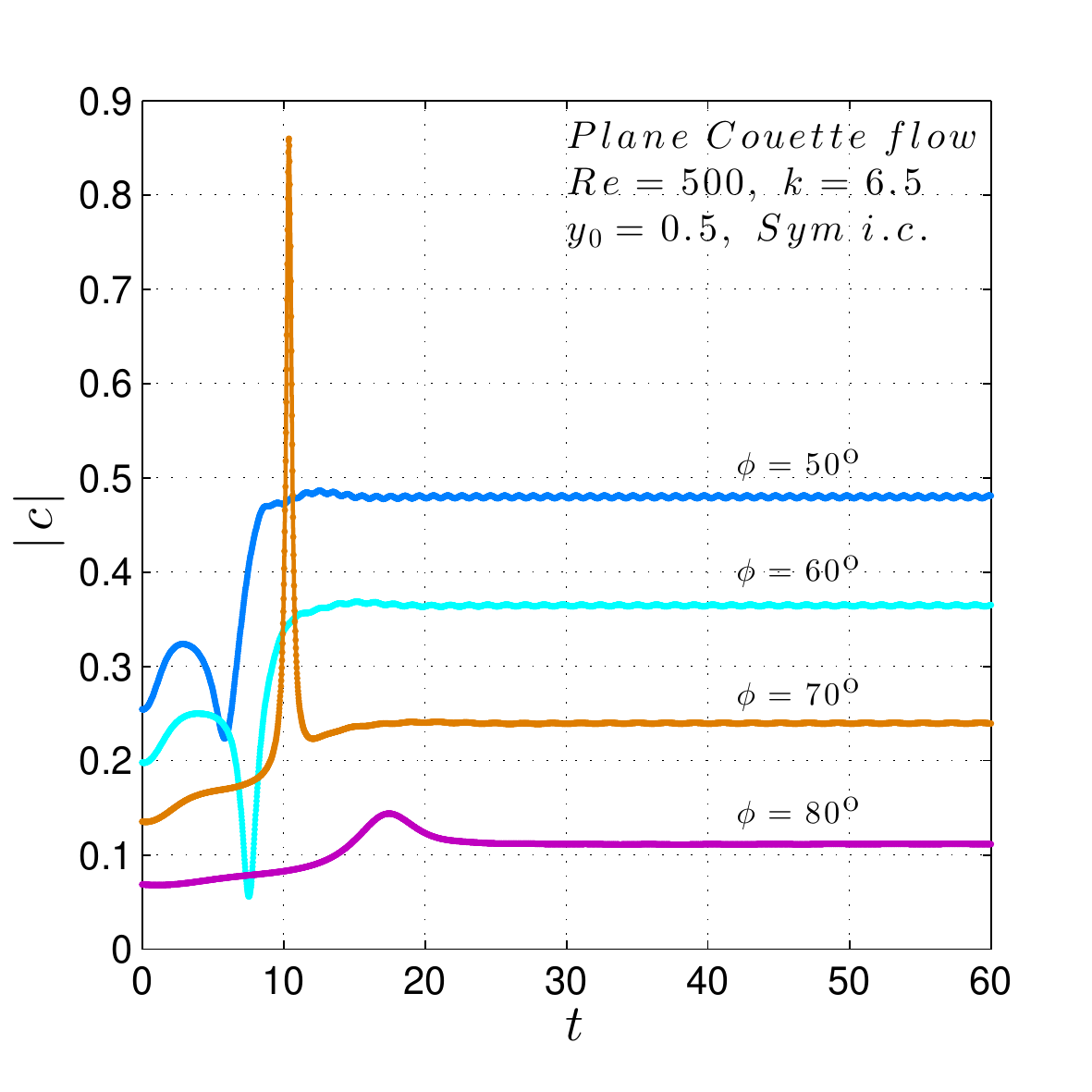}
	\vspace{0.5pt}
	\subcaption{ }
	\label{fig:c_CO_Re500_k6p5_variophi_2}
	 \end{subfigure}
	\caption{Temporal evolution of the absolute value of the phase
velocity, calculated from $\hv$, for PCf for $k=6.5$, $Re=500$ and \textit{sym.}
initial condition. The obliquity perturbation angle covers the range $\phi
\in[0^{\circ}, 90^{\circ}]$, uniformly distributed. The fixed observation point
is $y_0=0.5$. The case $\phi=90^{\circ}$ is not represented since the
orthogonal wave is stationary for all $k$, $Re$ and initial condition.}
\label{fig:c_CO_Re500_k6p5_variophi}
\end{figure}

\FloatBarrier
In the following, the temporal history of the absolute value of the phase
velocity of the $\tilde v$ component, $|c(t)|$, for the Plane Couette flow is presented, for different
combinations of the parameters. As one can be notice from figures
\ref{fig:c_CO_Re500_phi0_variok}, \ref{fig:c_CO_Re500_phi45_variok},
\ref{fig:c_CO_Re500_phi45_varioRe}, \ref{fig:c_CO_Re500_k6p5_variophi}, the
problem, though linear, offers a complex highly non-stationary scenario, which
is hardly possible to estimate \textit{a priori}. Two different periods for the
phase velocity (and so the frequency) temporal evolution can be observed, the
\textit{Early term} and the \textit{Intermediate term}. The last one ends when
the perturbation energy growth factor reaches the exponential asymptotic trend, and will be investigated in detail in
the following section.
The transition between the early and the intermediate transient appears to happen in a
narrow time window, and  it is often characterized by an abrupt jump to a higher
mean value which is maintained throughout the rest of the perturbation's life,
indicated as $\bar{c}$ or $\bar{\omega}$ for the phase velocity and the wave
frequency respectively.\par
Another important observation is that the asymptotic value generally is not a
constant one for PCf, but a modulation characterized by a specific period $T_c$ is
present. For this reason it is suitable to refer to frequency asymptotic mean
values. A motivation of this fact will be provided
in the next paragraph. As it clearly appears from
\figref{fig:c_CO_Re500_phi0_variok_2} and
\figref{fig:c_CO_Re500_phi45_variok_2}, the oscillations amplitude
 can be significant with respect to the mean value, specially for low
wavenumbers. With increasing
$k$, the frequency temporal evolution appears to shift from
peaks-characterized  to sinusoidal.\par
Moreover, it has been verified that exists
a certain threshold in the wavenumber, denoted with
$k_j$,
below which neither jump occurs nor phase velocity modulation is
observed, but the wave frequency experiences a monotonic decay to the zero
value, after a transient evolution (see
\figref{fig:c_CO_Re500_phi0_variok_1},
\ref{fig:c_CO_Re500_phi45_variok_1} and \tabref{tab:k_nojump}).\\
In \tabref{tab:Tj_Re500_sym} and \tabref{tab:Tj_Re5000_sym}  values of
nondimensional time at which the jump occurs are reported. Since the jump is
spread within a certain time window, a strict definition of $T_j$ is not
provided, so the normalized time corresponding to the frequency
peak that typically occurs after the jump is considered, in the present work,
as an index of the end of the early transient. Even if the temporal evolution
of the wave frequency varies with respect to the simulation parameters and the
``jump''  itself shows different shapes, some general trends can be observed.
In fact, $T_j$ seems to increase with increasing Reynolds number (\figref{fig:c_CO_phi45_varioRe_1}) and obliquity angle
(\figref{fig:c_CO_Re500_k6p5_variophi_1}), while it decreases
with increasing polar wavenumber (\figref{fig:c_CO_Re500_phi45_varioRe_2}).

\vspace{-1cm}


\FloatBarrier

\begin{table}[h!]
\centering
  \begin{tabular}{cccccc}
 \hline
  \rule[-0.3cm]{0mm}{0.8cm}
  \boldmath ${k  \backslash \phi}$  & \boldmath ${0^{\circ}}$  & \boldmath
${20^{\circ}}$ & \boldmath ${40^{\circ}}$ & \boldmath ${60^{\circ}}$ & \boldmath
${80^{\circ}}$\\ 
  \hline \rule[0 cm]{0mm}{0.5cm}   
  \boldmath $0.80$  & 10.1 & 10.9 & 13.8 & 23.5 & 25.0  \\
  \boldmath $1.37$  & 13.2 & 14.2 & 18.1 & 31.2 & 96.1  \\ 
  \boldmath $2.34$  & 9.00 & 9.65 & 12.2 & 15.3 & 38.3  \\
  \boldmath $4.00$  & 5.65 & 6.02 & 7.10 & 10.5 & 26.2  \\
  \boldmath $6.84$  & 5.50 & 5.84 & 7.15 & 14.3 & 16.8  \\
  \boldmath $11.7$  & 4.47 & 4.72 & 5.68 & 6.75 & 32.5  \\
  \boldmath $20.0$  & 3.88 & 4.14 & 4.28 & 7.51 & 22.5  \\
  \boldmath $34.2$  & 3.05 & 3.10 & 3.90 & 4.60 & 15.5  \\ 
  \boldmath $58.5$  & 2.15 & 2.27 & 2.52 & 3.40 & 11.5  \\ 
  \boldmath $100$  & 1.41 & 1.51 & 1.74 & 2.32 & 8.20  \\ \hline
 \end{tabular}
\caption{Frequency jump nondimensional time $T_j$ for various combination of the
simulation parameters, for $Re=500$ and \textit{sym.} initial condition. 
$T_j$ is considered as the time at which the frequency maximum value, typically
located just after the jump, occurs.}
\label{tab:Tj_Re500_sym}
\end{table}

\begin{table}[h!]
\centering
  \begin{tabular}{cccccc}
 \hline
  \rule[-0.3cm]{0mm}{0.8cm}
  \boldmath ${k  \backslash \phi}$  & \boldmath ${0^{\circ}}$  & \boldmath
${20^{\circ}}$ & \boldmath ${40^{\circ}}$ & \boldmath ${60^{\circ}}$ & \boldmath
${80^{\circ}}$\\ 
  \hline \rule[0 cm]{0mm}{0.5cm}   
  \boldmath $0.80$  & 23.8 & 25.5 & 31.8 & 50.7 & 132  \\
  \boldmath $1.37$  & 13.6 & 14.5 & 18.0 & 28.2 & 90.0  \\ 
  \boldmath $2.34$  & 8.16 & 8.72 & 10.7 & 16.5 & 49.4  \\
  \boldmath $4.00$  & 12.5 & 13.4 & 16.5 & 18.3 & 56.0  \\
  \boldmath $6.84$  & 11.6 & 12.3 & 15.3 & 19.2 & 44.0  \\
  \boldmath $11.7$  & 10.2 & 11.0 & 11.9 & 16.0 & 32.4  \\
  \boldmath $20.0$  & 7.99 & 8.48 & 9.56 & 13.2 & 26.3  \\
  \boldmath $34.2$  & 6.47 & 6.90 & 7.61 & 10.5 & 20.4  \\ 
  \boldmath $58.5$  & 4.52 & 4.62 & 5.39 & 7.51 & 14.5  \\ 
  \boldmath $100$  & 3.13 & 3.33 & 3.75 & 5.18 & 10.1 \\ \hline
 \end{tabular}
\caption{Frequency jump nondimensional time $T_j$ for various combination of the
simulation parameters, for $Re=5000$ and \textit{sym.} initial condition.}
\label{tab:Tj_Re5000_sym}
\end{table}
\FloatBarrier
\subsubsection{Intermediate Term and Long Term behaviour}
\begin{figure}[h!]
        \centering
         \advance\leftskip-2.5cm
         \advance\rightskip-2cm
        \begin{subfigure}{0.6\textwidth}
        \centering 
	\includegraphics[width=9.0cm]{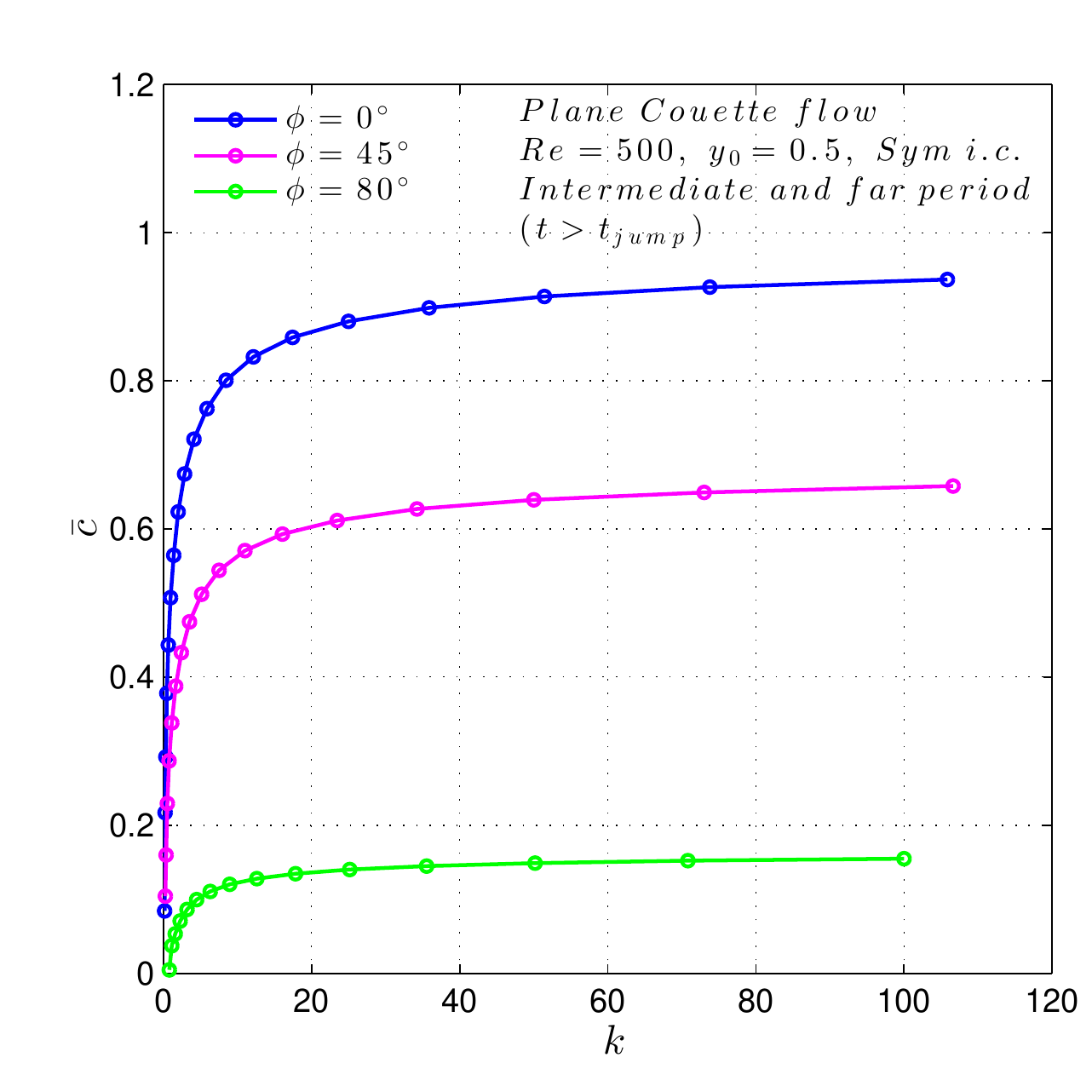}
	\vspace{0.5pt}
	\subcaption{}
	\label{fig:c_mean_variophi}
	 \end{subfigure}
        \begin{subfigure}{0.6\textwidth}
        \centering 
\includegraphics[width=9.0cm]{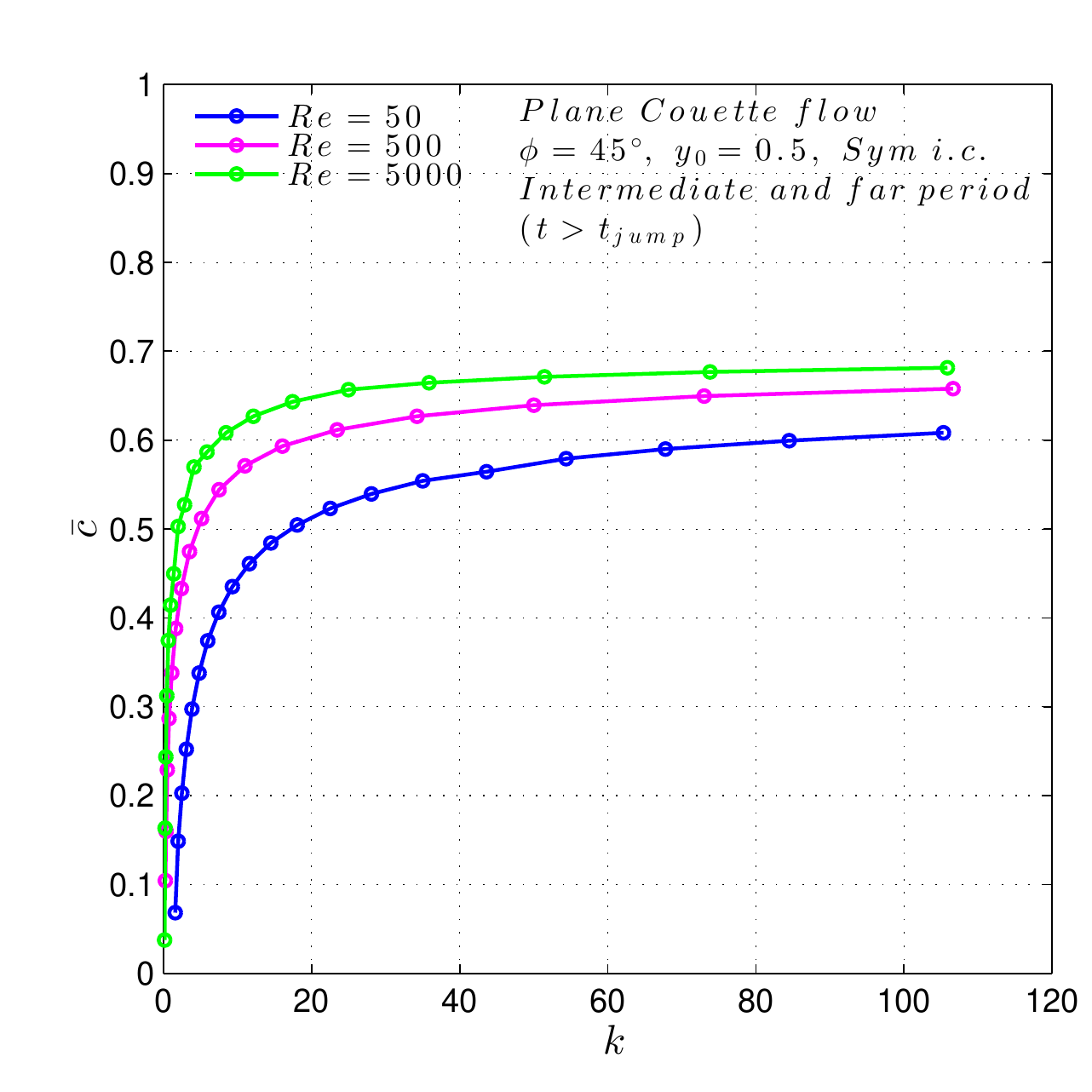}
	\vspace{0.5pt}
	\subcaption{ }
	\label{fig:c_mean_varioRe}
	 \end{subfigure}
	\caption{(a) Trends of the absolute mean value of phase
velocity, calculated from $\hv$, for PCf for $Re=500$, $\phi=\{0^{\circ},
45^{\circ}, 80^{\circ}\}$ and
\textit{sym.} initial condition. The polar wavenumbers are uniformly distributed
in the logarithmic space. The fixed
observation point is
$y_0=0.5$. (b)  Absolute mean values of $\bar{c}$, calculated for
PCf for  $\phi=45^{\circ}$, $Re=\{50, 500, 5000\}$ and
\textit{sym.} initial condition. These asymptotic results are independent on
both the initial condition and $y_0$.}
\label{fig:c_asymptotic}
\end{figure}

\begin{figure}[h!]
        \centering
         \advance\leftskip-2.5cm
         \advance\rightskip-2cm
        \begin{subfigure}{0.6\textwidth}
        \centering 
	\includegraphics[width=9.0cm]{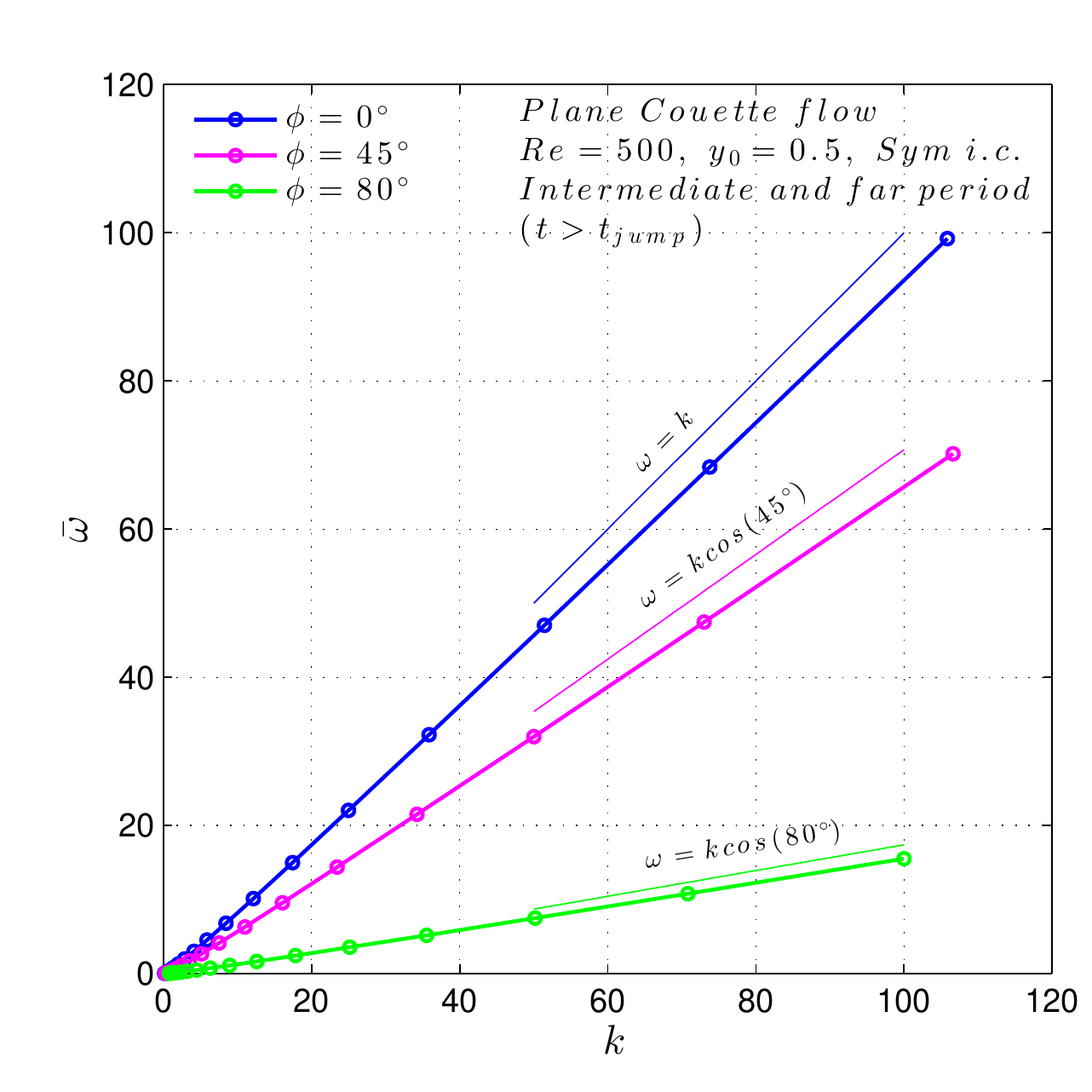}
	\vspace{0.5pt}
	\subcaption{}
	\label{fig:O_mean_variophi}
	 \end{subfigure}
        \begin{subfigure}{0.6\textwidth}
        \centering 
\includegraphics[width=9.0cm]{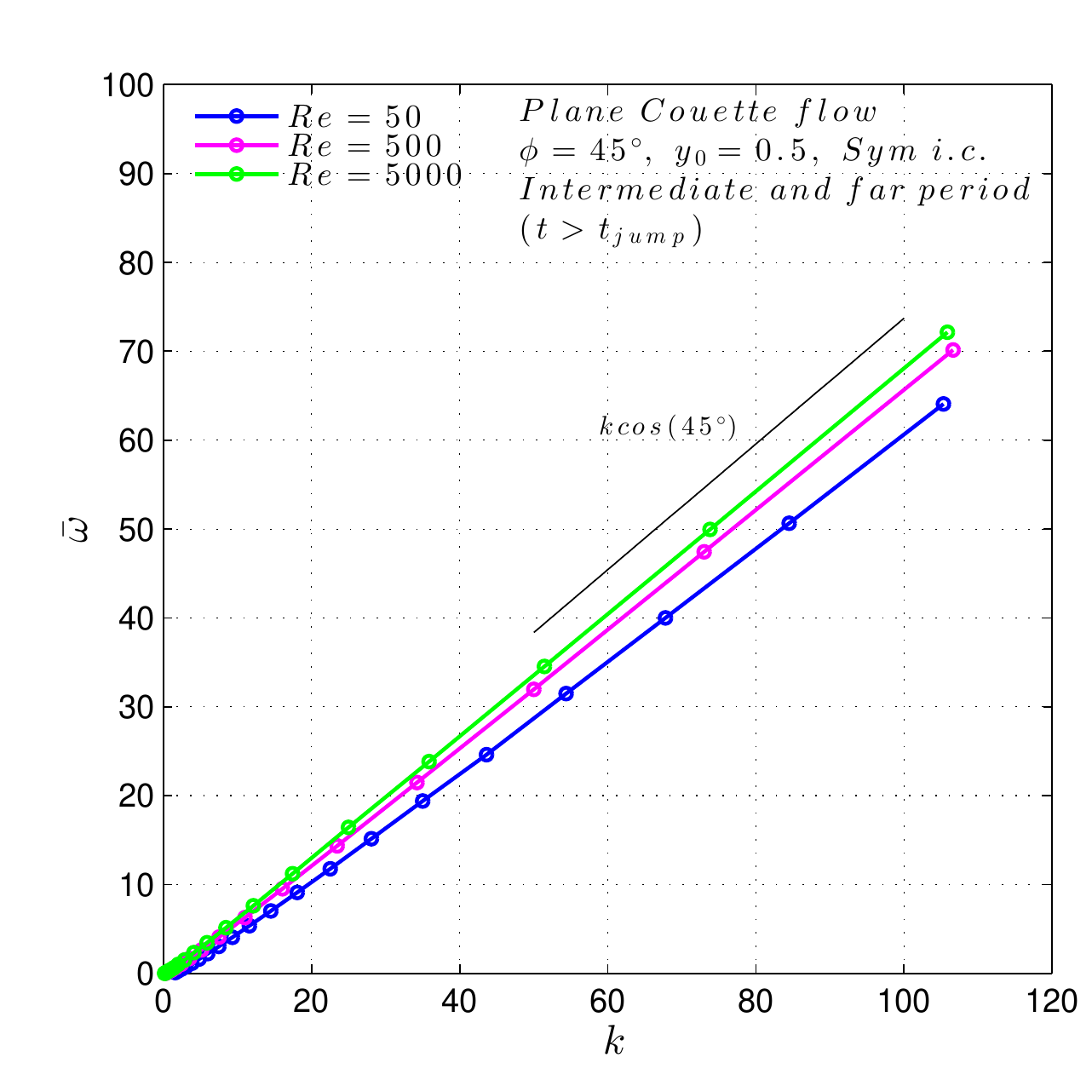}
	\vspace{0.5pt}
	\subcaption{ }
	\label{fig:O_mean_varioRe}
	 \end{subfigure}
	\caption{(a) Trends of the absolute mean value of the frequency,
calculated from $\hv$, for PCf for $Re=500$, $\phi=\{0^{\circ},
45^{\circ}, 80^{\circ}\}$ and
\textit{sym.} initial condition. The polar wavenumbers are uniformly distributed
in the logarithmic space. The fixed
observation point is
$y_0=0.5$. (b)  Absolute mean values of $\bar{\omega}$, calculated for
PCf for  $\phi=45^{\circ}$, $Re=\{50, 500, 5000\}$ and
\textit{sym.} initial condition. These asymptotic results are independent on
both the initial condition and $y_0$.}
\label{fig:O_asymptotic}
\end{figure}

\FloatBarrier
An interesting aspect of the nonmodal analysis, hardly ever considered in the
past, is the possibility to investigate on how the asymptotic state is
reached. In the present and in the following sections some new results will be
presented.
The trends of the mean values of the asymptotic phase
velocity for various combinations of the simulation parameters are reported in
\figref{fig:c_mean_variophi} and \figref{fig:c_mean_varioRe}. 
Accordingly to known results from the cited literature, the phase velocity
asymptotic mean value increases with increasing $k$, $Re$ and decreasing
$\phi$. The asymptotic value doesn't depend neither on the initial condition nor
on the observation point $y_0$, but only on the spectrum of $\hv$ for that
particular parameters combination. In fact, the frequency corresponds to the
real part of the least damped eigenvalue of the Orr-Sommerfeld operator, while
the damping is given by the imaginary part. Even if the present method is
developed to study the temporal evolution of perturbations from an initial
value problem, the spectra af $\hv$ can
easily be obtained by computing the eigenvalues of the matrix $\BU{A}$ (see
\secref{sec:gal_eta}).\\
About the mean values of the asymptotic frequency $\bar{\omega}$, trends are
shown in \figref{fig:O_mean_variophi} and \figref{fig:O_mean_varioRe}. It can
be noticed that for high values of $k$, the general tendency can be
approximated by the relation $\bar{\omega}=kcos(\phi)$. This trend was observed
for Plane Poiseuile flow and for Wake flow, as well, by \citet{Scarsoglio2012}.

\newpage

\begin{figure}[htb]
        \centering
	\includegraphics[width=9.0cm]{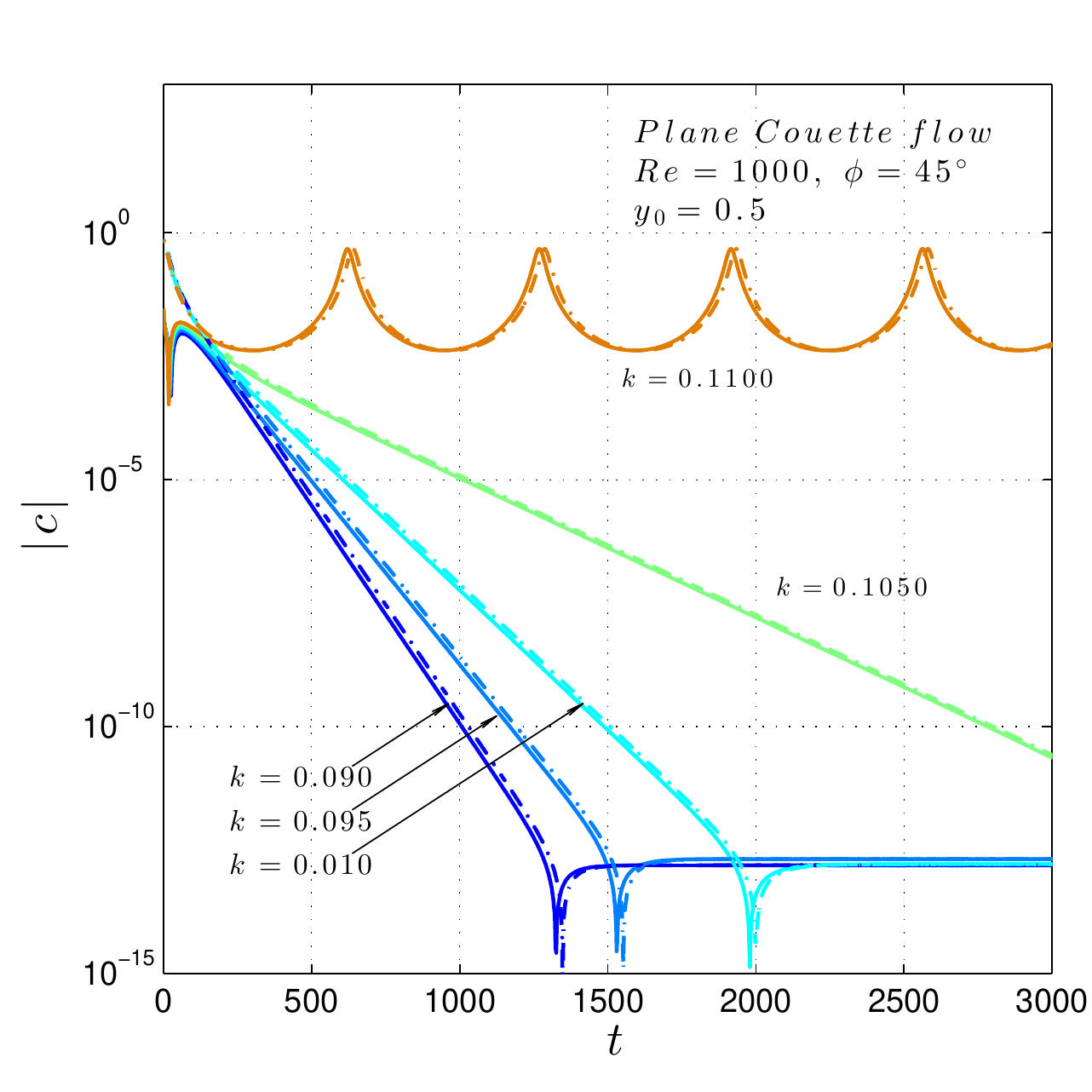}	
\caption{Temporal
evolution of the absolute value of the phase
velocity, calculated from $\hv$, for PCf for $\phi=45$, $Re=1000$. An abrupt
variation in the temporal evolution trend can be observed once $k=0.105$ is exceeded.
Dot-dashed line: \textit{asym} initial condition; continuous line: \textit{sym}
i.c.}
\label{fig:time_evol_semilog}
\end{figure}

\begin{figure}[h!]
        \centering
         \advance\leftskip-2.5cm
         \advance\rightskip-2cm
        \begin{subfigure}{0.6\textwidth}
        \centering 
	\includegraphics[width=9.0cm]{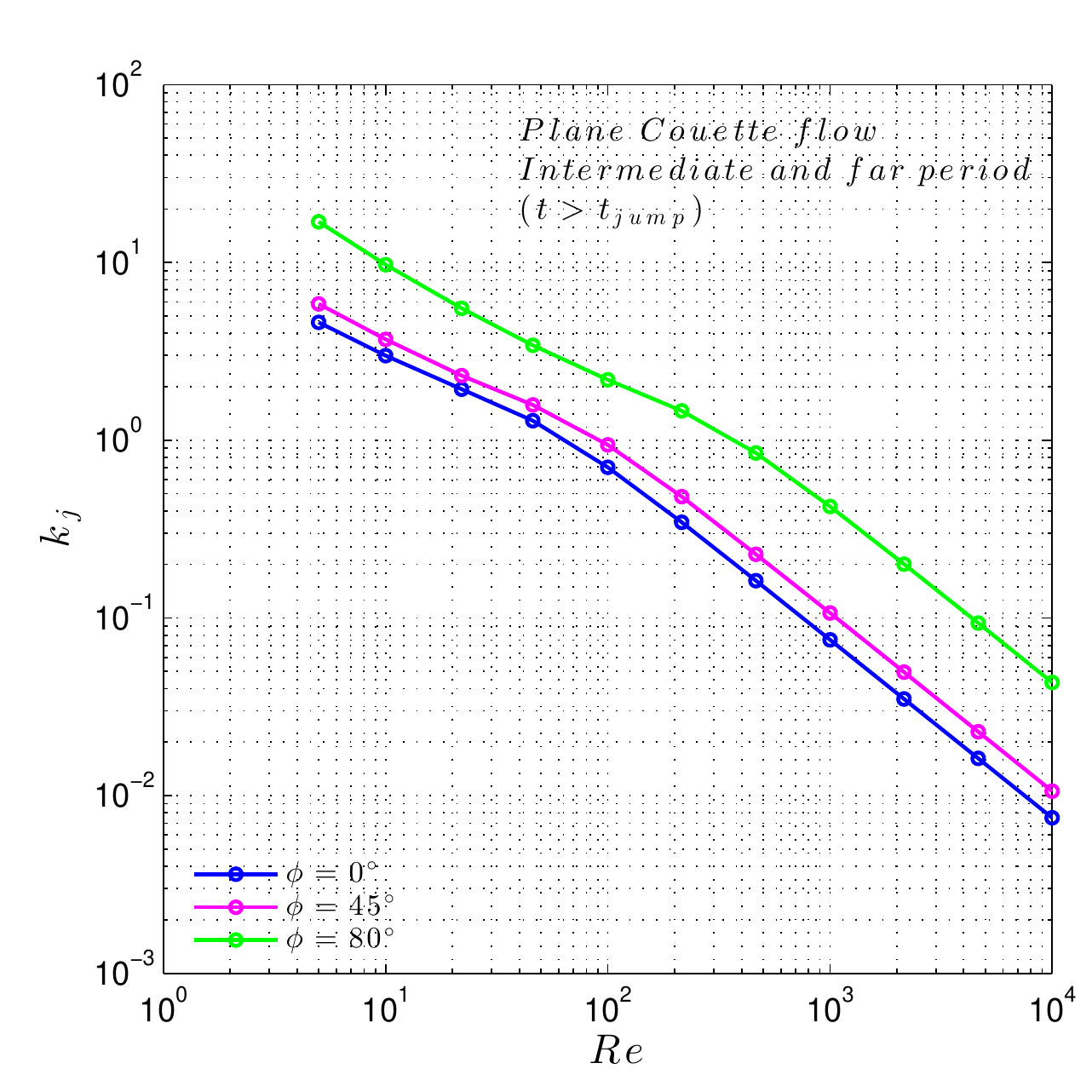}
	\vspace{0.5pt}
	\subcaption{}
	\label{fig:KJ_Re}
	 \end{subfigure}
        \begin{subfigure}{0.6\textwidth}
        \centering 
\includegraphics[width=9.0cm]{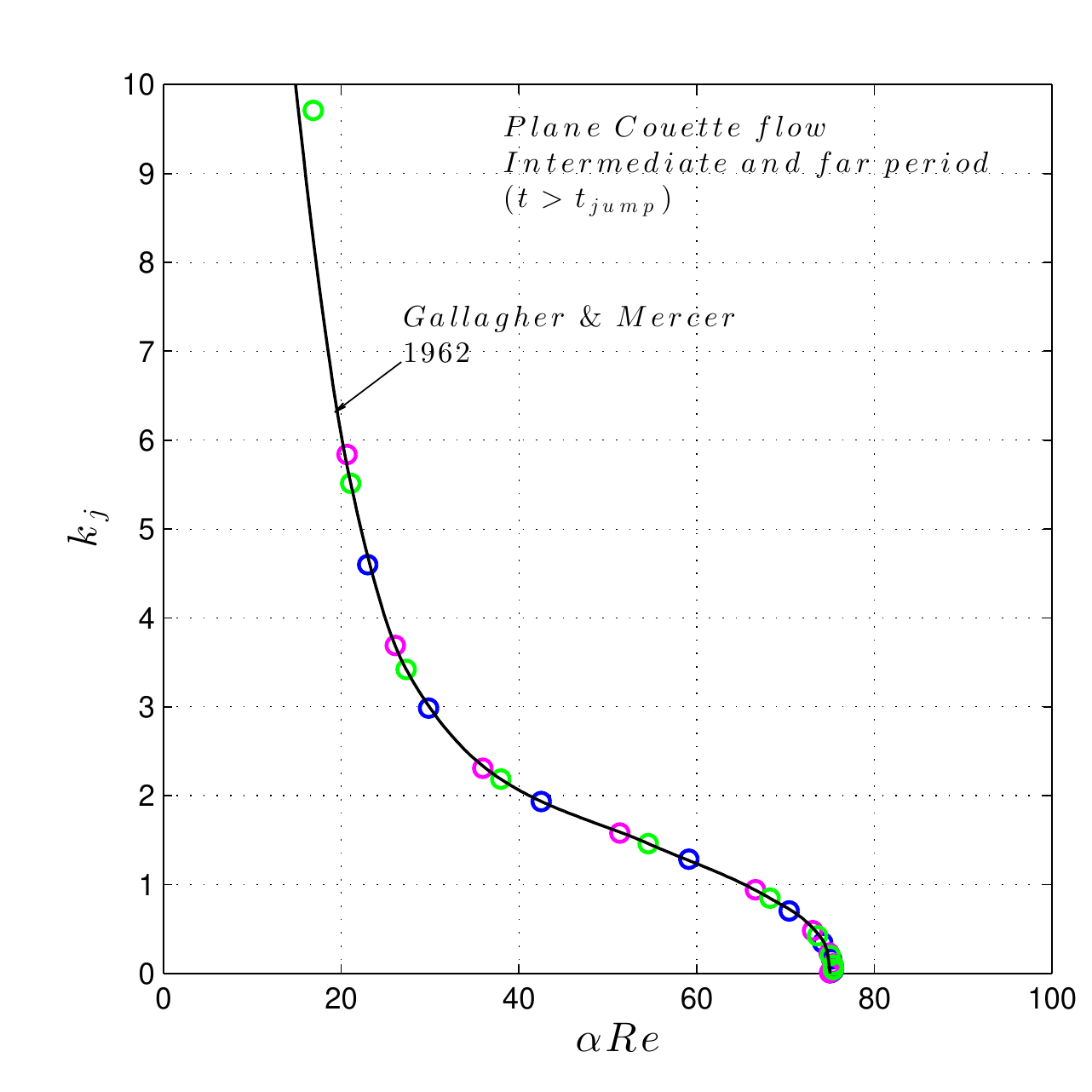}
	\vspace{0.5pt}
	\subcaption{ }
	\label{fig:KJ_alfaRe}
	 \end{subfigure}
	\caption{Trends of the threshold wavenumber $k_j$ for PCf, as a
function of $Re$ for $\phi=\{0^{\circ},45^{\circ},80^{\circ}\}$ (a).
Comparison with  results found in literature (Gallagher \& Mercer, 1962) for
the bidimensional case ($\phi=0$). It is worth to notice that here the same
trend is found for the three-dimensional cases.}
\label{fig:KJ}
\end{figure}

As seen in the previous section, there exists a threshold for $k$ below which
the temporal evolution is characterized by a frequency decay to zero, after an
early transient. To better observe these conditions, in
\figref{fig:time_evol_semilog} the phase velocity
evolution has been traced on a semilogarithmic plane. Also for these cases
there is no influence of the initial condition on the far
periods. Moreover it can be noticed that for $k<k_j$ there is a phase of
exponential decay leading to a stationary state.\\
The phenomenon has been observed by \citet{Gallagher1962} through a modal
analysis (2D). They discovered that below a certain value of $\alpha Re$
all the eigenvalues were real. Increasing $Re$, a threshold level is reached,
where the least damped eigenvalues are real and coincident; they split into a
complex conjugate pair for larger values of the Reynolds number. The authors
pointed out the abruptness of the transition, as well. 
Here various simulations have been performed in order to verify the precision
of the method and of the numerical code for the asymptotic solution computation.
The results have been compared to those of the cited authors, and an excellent
agreement has been found. In addition, a generalization for the
three-dimensional case is shown (see \figref{fig:KJ_alfaRe}).\\
Another comparison have been done with the results obtained by
\citet{Orszag1971}, who developed a method for the modal analysis
based on a Chebyshev polynomials expansion. For Plane Poiseuille flow at
$Re=10000,\ \alpha=1$ and $\phi=0^{\circ}$ he found the following value
for the unstable eigenvalue: $c=0.23752649+0.00373967i$. With N=250 we
obtain the value of $c=0.23752629+0.00373964i$ (with $\Delta
y=10^{-3}$, $N=250$). The same accuracy is assured for the the other
eigenvalues. 

\begin{table}[h!]
\centering
  \begin{tabular}{cccc}
 \hline
  \rule[-0.3cm]{0mm}{0.8cm}
  \boldmath ${Re  \backslash \phi}$  & \boldmath ${0^{\circ}}$  & \boldmath
${45^{\circ}}$ & \boldmath ${80^{\circ}}$\\ 
  \hline \rule[0 cm]{0mm}{0.5cm}   
  \boldmath $10$     & 2.9840 & 3.6900 & 9.7110  \\
  \boldmath $22$     & 1.9330 & 2.3100 & 5.5150   \\ 
  \boldmath $46$     & 1.2850 & 1.5790 & 3.4200  \\
  \boldmath $100$    & 0.7040 & 0.9420 & 2.1865   \\
  \boldmath $215$    & 0.3450 & 0.4805 & 1.4610 \\
  \boldmath $464$    & 0.1620 & 0.2283 & 0.8469  \\
  \boldmath $1000$   & 0.0754 & 0.1065 & 0.4239  \\
  \boldmath $2154$   & 0.0350 & 0.0495 & 0.2006  \\ 
  \boldmath $4642$   & 0.0162 & 0.0229 & 0.0935   \\ 
  \boldmath $10000$  & 0.0075 & 0.0106 & 0.0434  \\ \hline
 \end{tabular}
\caption{Values of threshold wavenumber $k_j$ below which the asymptotic
frequency tends to zero, i.e the wave tends to a stationary (damped) state at
high
times. Here $k_j$ is reported for three different values of obliquity angle and
for ten values of Reynolds number, spanning four decades, uniformly distributed
in the logarithmic space. }
\label{tab:k_nojump}
\end{table}

\begin{figure}[h!]
        \centering
         \advance\leftskip-2.5cm
         \advance\rightskip-2cm
        \begin{subfigure}{0.6\textwidth}
        \centering 
	\includegraphics[width=9.0cm]{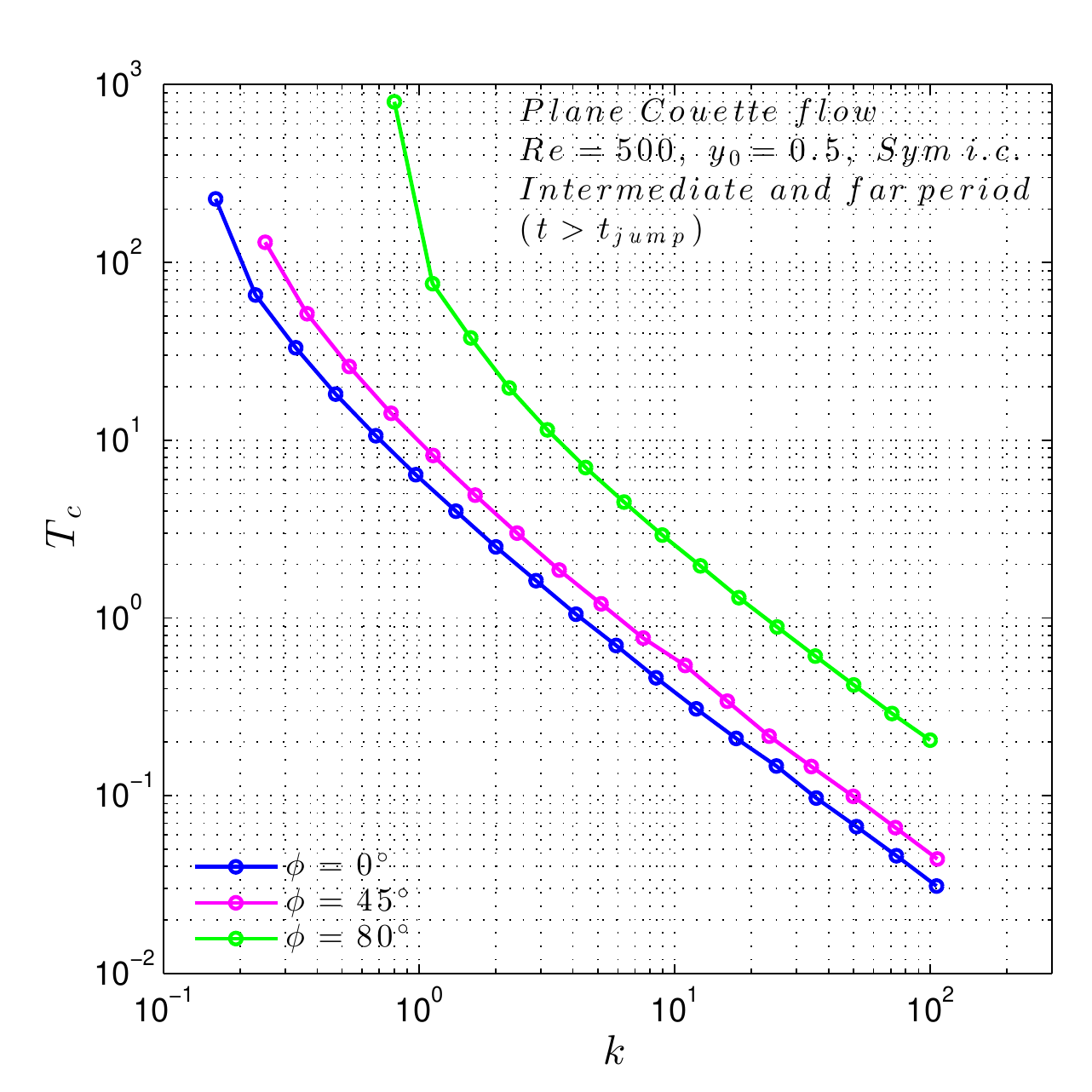}
	\vspace{0.5pt}
	\subcaption{}
	\label{fig:c_freq_variophi}
	 \end{subfigure}
        \begin{subfigure}{0.6\textwidth}
        \centering 
\includegraphics[width=9.0cm]{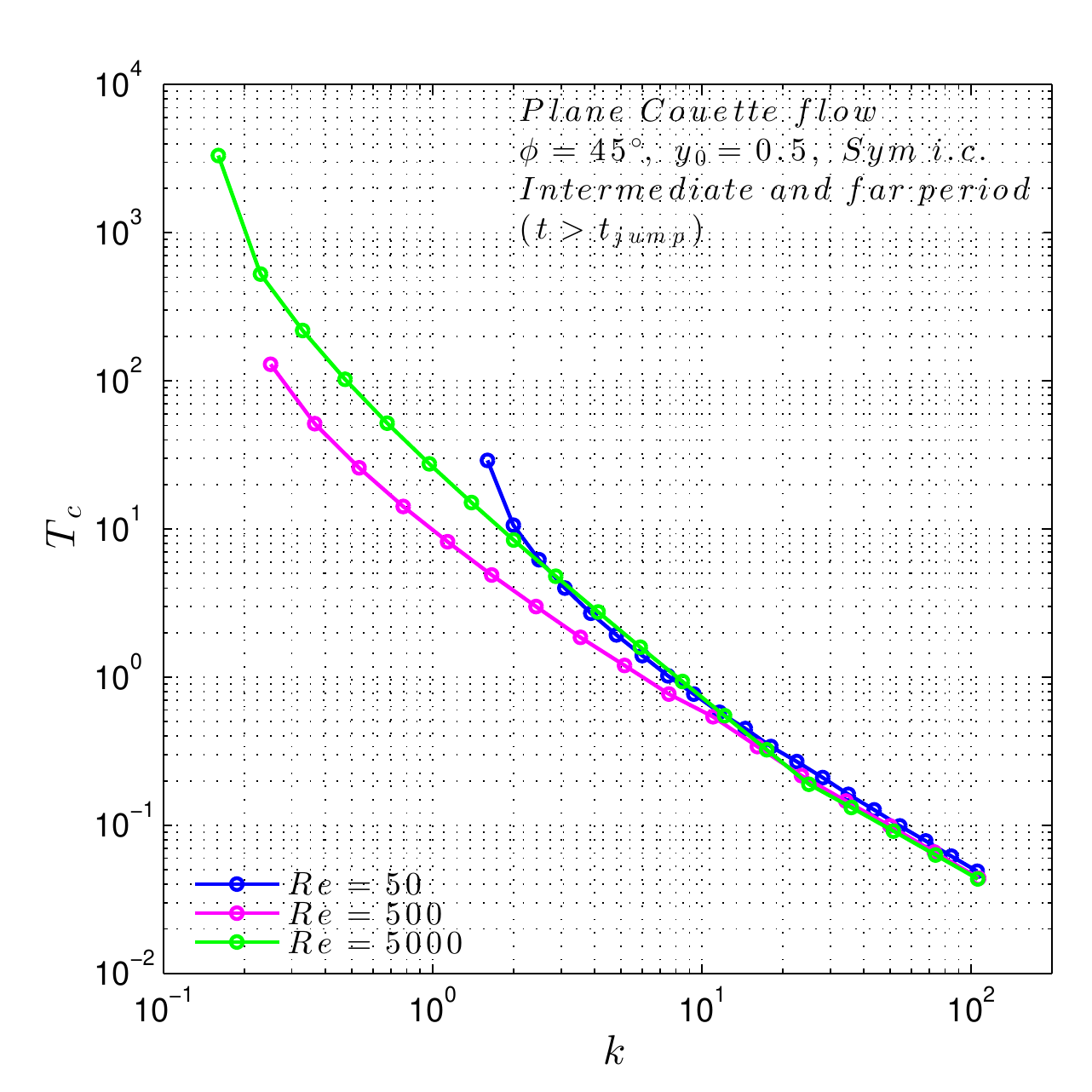}
	\vspace{0.5pt}
	\subcaption{ }
	\label{fig:c_freq_varioRe}
	 \end{subfigure}
	\caption{Trends of the phase velocity asymptotic period $T_c$ for PCf,
as a function of $k$ for $\phi=\{0^{\circ},45^{\circ},80^{\circ}\}$ and $Re=500$
(a). $Re=\{50,500,5000\}$ and $\phi=45^{\circ}$ (b). The observation point $y_0$
and the initial condition does not influence the period of the modulation.} 
\label{fig:c_freq}
\end{figure}
\FloatBarrier
We find interesting to investigate the trend and the reasons of the
phase velocity asymptotic modulation, since no detailed literature is found
about this topic. The modulation is characterized by a period
$T_c$ that decreases with increasing polar wavenumber, according to an
exponential law, for sufficiently high values of $k$
(\figref{fig:c_freq_variophi}). $T_c$ increases with increasing obliquity angle
as well, while the influence of $Re$ is weak at high $k$. \\
Even if further investigations are needed to verify these results, we
underline the total agreement with the results obtained by direct numerical
integration of \eqref{Orr-Somm} and \eqref{Squire} with the methods of lines and Runge-Kutta ODE solver.
In the following we try to provide a  motivation about their
non-contradictory nature with the modal theory. 
The spectrum of channel flows (for sufficiently high values of $k$) is generally
composed by three branches,whose label $A,P,S$ were given by \citet{Mack1976}.
In the case of Plane Poiseuille flow all the three branches are present and
correspond respectively to wall modes ($c_r\to 0$), center modes ($c_r\to 1$)
and highly damped modes ($c_r\to 2/3$). In the case of Plane Couette flow, the
spectrum does not contain a $P$ branch but it has two $A$ branches, composed by
complex conjugate eigenvalues.\par
\begin{figure}[h!]
        \centering
         \advance\leftskip-2.5cm
         \advance\rightskip-2cm
        \begin{subfigure}{0.6\textwidth}
        \centering
\includegraphics[width=9.0cm]{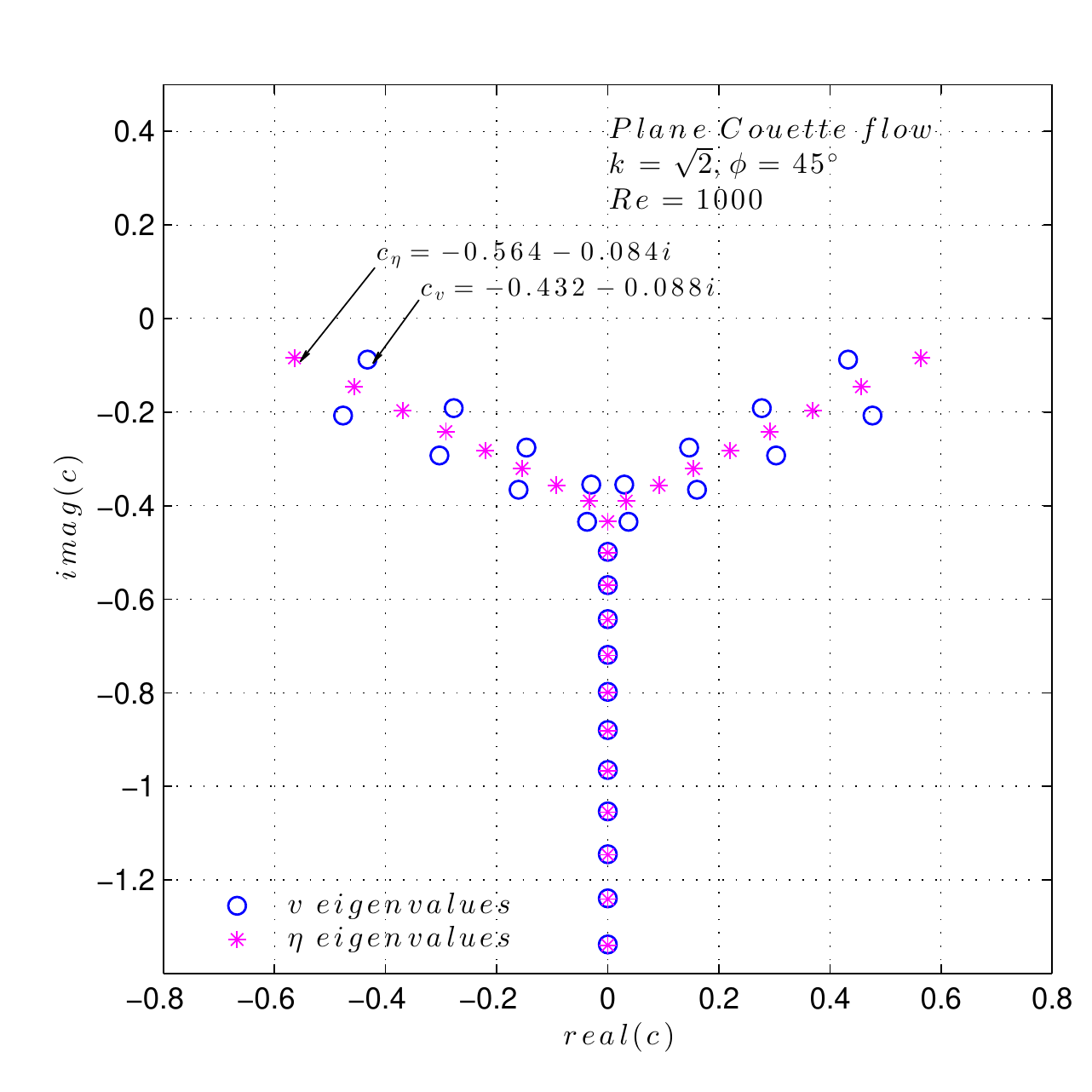}
	\vspace{0.5pt}
	\subcaption{}
	\label{fig:Spectrum_PCf_Re1000_k_1p41_phi_45}
	 \end{subfigure}
        \begin{subfigure}{0.6\textwidth}
        \centering 
\includegraphics[width=9.0cm]{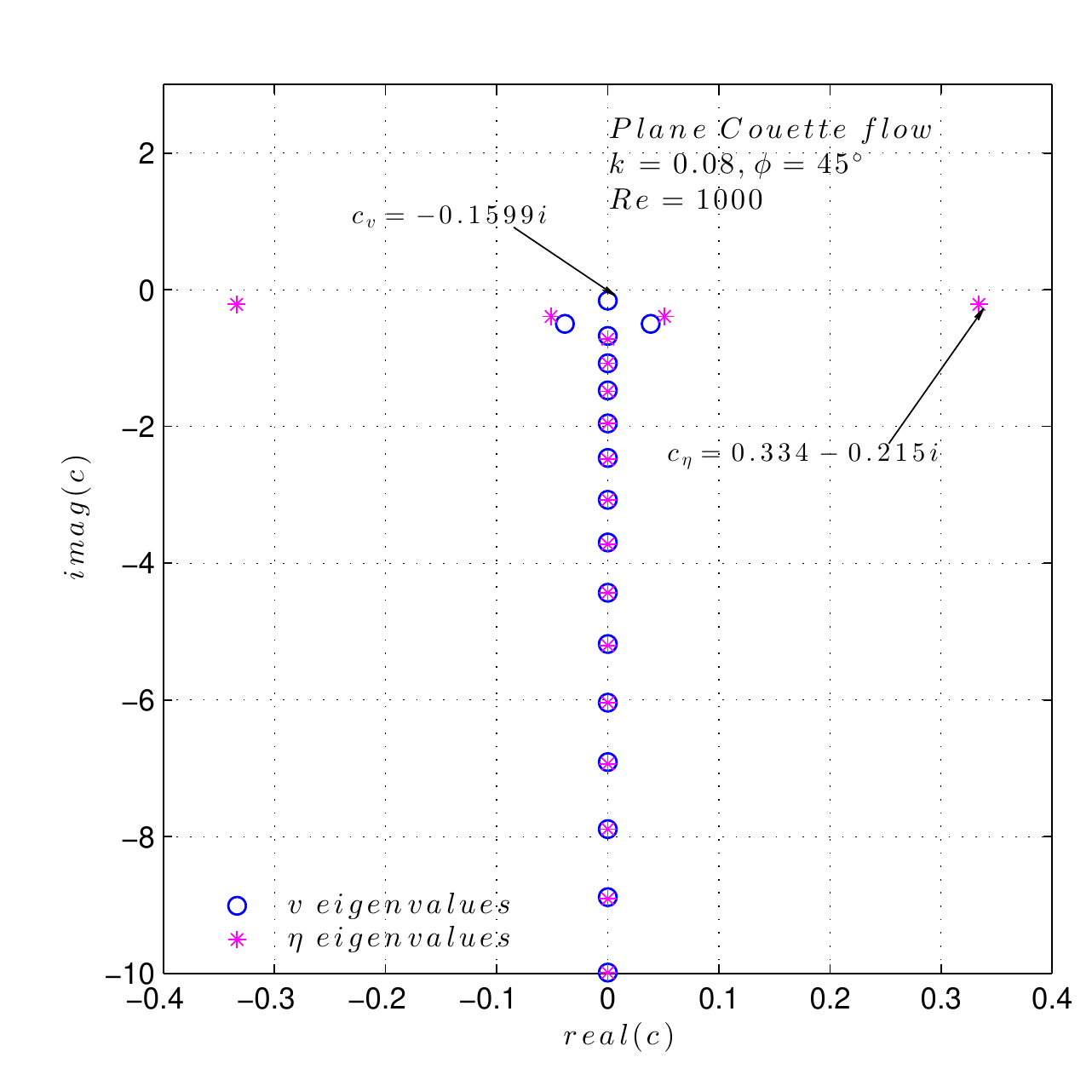}
	\vspace{0.5pt}
	\subcaption{ }
	\label{fig:Spectrum_PCf_Re1000_k_0p08_phi_45}
	 \end{subfigure}
	\caption{(a) Spectrum of both the Orr-Sommerfeld and the Squire
operators for PCf at $Re=1000$, $\phi=45^{\circ}$ and $k=\sqrt{2}$. The least
damped eigenvalue, respectively for $\hv$ and for $\he$, is explicitly
reported. (b) Spectra for PCf at $Re=1000$, $\phi=45^{\circ}$ and $k=0.08$. Here
the polar wavenumber is below the critical value, in fact we observe that the
least damped eigenvalue is real. This means that $\hv$ tends to a stationary
damped state as $t\to\infty$.}
\label{fig:Spectra_1}
\end{figure}
\begin{figure}[t!]
        \centering
         \advance\leftskip-2.5cm
         \advance\rightskip-2cm
        \begin{subfigure}{0.6\textwidth}
        \centering
\includegraphics[width=9.0cm]{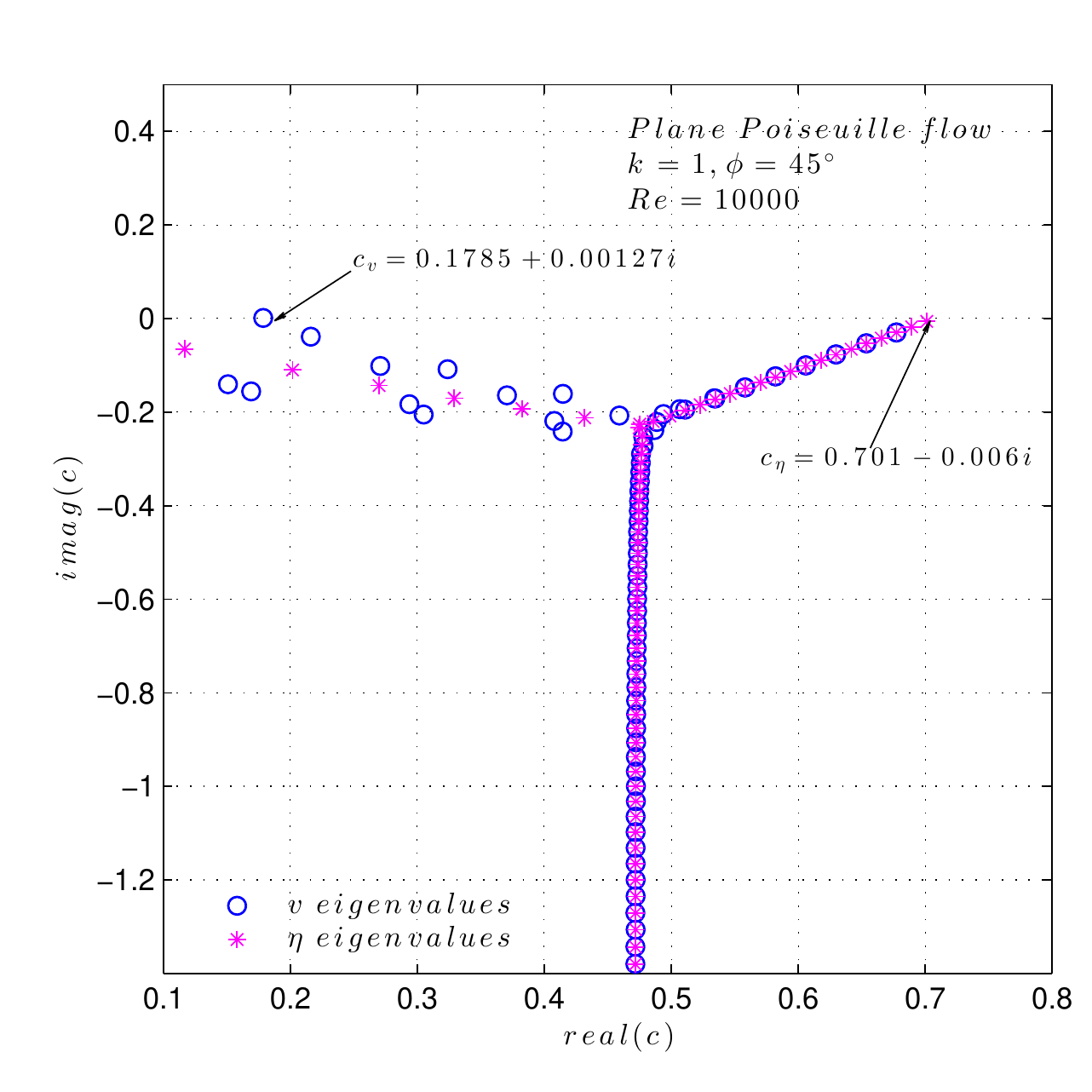}
	\vspace{0.5pt}
	\subcaption{}
	\label{fig:Spectrum_PPf_Re1000_k_1_phi_45}
	 \end{subfigure}
        \begin{subfigure}{0.6\textwidth}
        \centering 
\includegraphics[width=9.0cm]{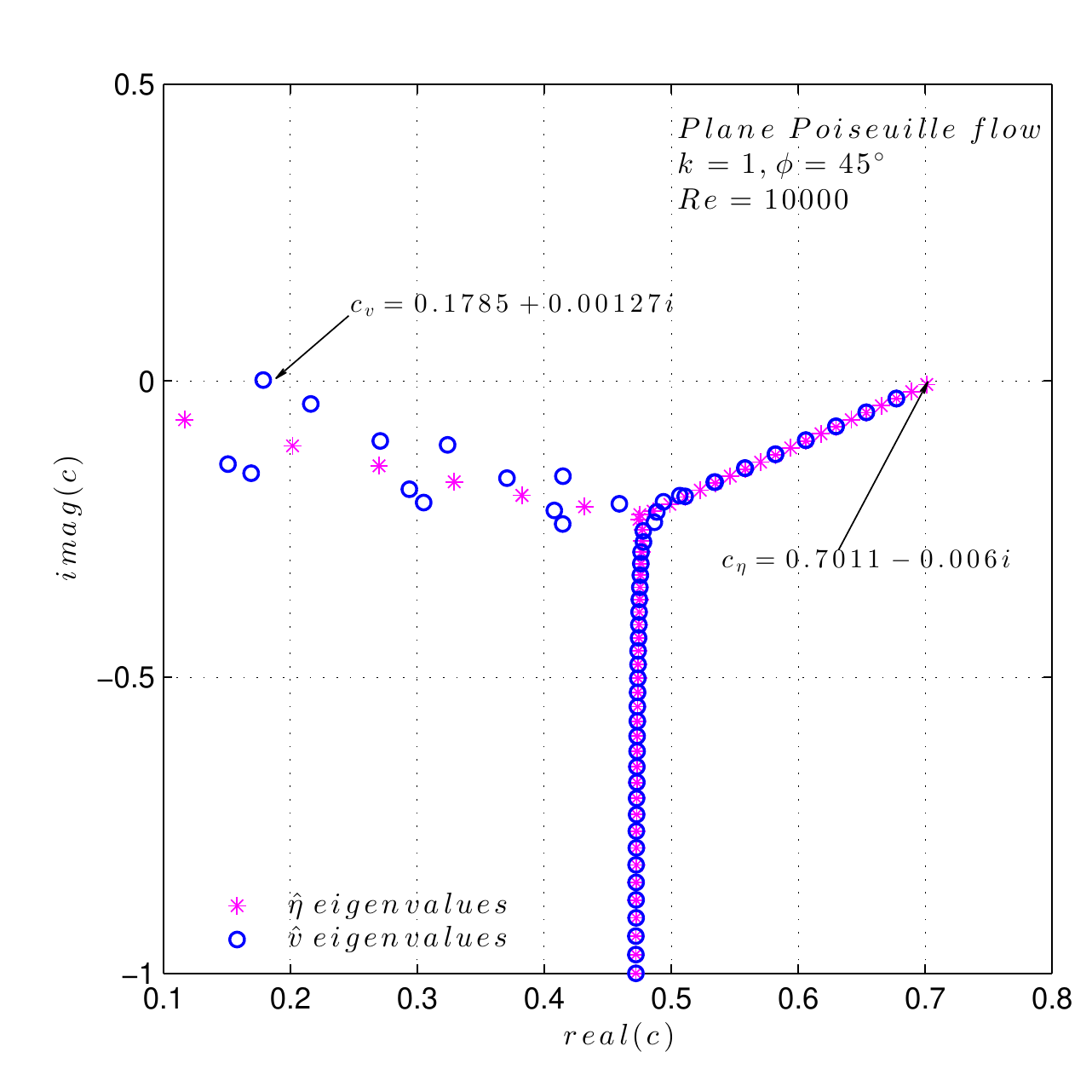}
	\vspace{0.5pt}
	\subcaption{ }
	\label{fig:Spectrum_PCf_Re1000_k_10_phi_45}
	 \end{subfigure}
	\caption{(a) Spectrum of both the Orr-Sommerfeld and the Squire
operators for PPf at $Re=1000$, $\phi=45^{\circ}$ and $k=1$. This is an
instable configuration: an eigenvalue with positive imaginary part
exists. (b) Spectra for PCf at $Re=1000$, $\phi=45^{\circ}$ and $k=10$. With
increasing $k$, or $Re$, the eigenvalues exact computation becomes difficult.
The issue  concerns the sensitivity of the spectrum to small perturbations
(e.g. the computer finite precision). This is a property of the
linear operator rather than a property of the numerical scheme, and the
junction point of the three branches exhibit the largest sensitivity
\citep[see][]{Schmid_book}. However, the validity of the solution is not
compromised, since the problem doesn't affect the first eigenvalues.}
\label{fig:Spectra_2}
\end{figure}

The solution of the initial value problem generally contains the
contribute of all the frequency components, as can be clearly seen from the
solution of the ODE-reduced velocity equation \eqref{ODE_solution2}. Considering
the spectrum of PCf, we observe that there are two least damped, complex conjugate, eigenvalues (with the same damping
rate and opposite real frequency). It
has been observed, as well, that the coefficients $\BU{h_0}$ are complex conjugate;
the same does not apply, however, to the final solution where the coefficients
are mixed up by the matrix $\BU{L}$ and eventually multiplied by the
corresponding shape function $X_n$. In this case the phase of the final solution
results oscillating and so the frequency. The mean value of the
asymptotic frequency corresponds exactly to the real part of the least damped
eigenvalues pair.\\
To support this motivation,  simulations have been performed for PPf, for
parameters combinations where the least damped eigenvalue is unique. As
expected, for these cases no phase velocity oscillations are observed. In the
following, some spectra examples for both Plane Couette flow and Plane
Poiseuille flow are provided. We remind that in the classical modal analysis
the solution, e.g. the normal velocity, is expressed as 

\begin{equation}
 \hv(y,t)=\hat{\hv}(y)e^{-i\omega t}
\end{equation}

where $\omega=\omega_r+i\omega_i$ is the complex  eigenvalue, so the real
part $\omega_r$ represent the wave frequency, while the imaginary component
$\omega_i$ is damping rate.\newpage
 According to the convention adopted in the present work (see the
exponential terms in \eqref{ODE_solution2}), the damping rates of the single
modes are given by the real part of the eigenvalues, $\mu_r=\omega_i$ and the
frequencies correspond to the imaginary part with changed sign $\mu_i=-\omega_r$. In order not to
be confusing, we will express the spectra  with the classical convention found in
literature, in
terms of phase velocity. Only the least damped eigenvalues are represented
in \figref{fig:Spectra_1} and \figref{fig:Spectra_2}.

\FloatBarrier

As one can notice, the spectra of the Orr-Somerfeld ($\hv$) and Squire ($\he$)
operators are usually different, and for Plane Couette flow in many cases the
least damped eigenvalue belong to the set of the Squire operator (see for
example \figref{fig:Spectrum_PCf_Re1000_k_1p41_phi_45}). this fact is found to
have an influence on the dynamic of the system which has not been taken in
account yet, as shown in the next section.

\FloatBarrier
\ 
\newpage
\subsection{Behaviour of the vorticity component $\he$ and global
considerations}
In order to understand the behavior of the complete solution, the normal
vorticity must be considered or, alternatively, the other components of
perturbation velocity $\hu$ and $\hw$. In the present section the phase
velocity of the vorticity signal is investigated and some new results are presented. The same fourth order
finite-difference scheme introduced in \secref{sec:v_frequency} is used for the
computation of the phase first derivative.

\subsubsection{Temporal evolution of the $\he$ phase velocity}
As seen in the previous section, the non-modal analysis allows to observe
the complete life of a perturbation, from the early transient to the asymptotic
state predicted by the modal analysis. However, in the past the same
frequency for both the normal velocity and the normal vorticity (or,
similarly, for the three components of velocity) was usually considered by the
authors dedicated to the modal analysis. As pointed out by
\citet{Schmid_book}, only the particular solution of the modal Squire equation
has the same frequency of $\tilde v$. In the following analysis, the role of the
homogeneous part $\he_h$ in the frequency temporal evolution of $\tilde \eta$ is shown.
Moreover in the following section we will focus on the evolution of the velocity
and vorticity profiles along the $y$ coordinate, and their correlation with the
frequency time history.

\begin{figure}[htb]
        \centering
        \vspace{-0.5pt}
         \advance\leftskip-2.5cm
         \advance\rightskip-2cm
        \begin{subfigure}{0.6\textwidth}
        \centering
\includegraphics[width=9.0cm]{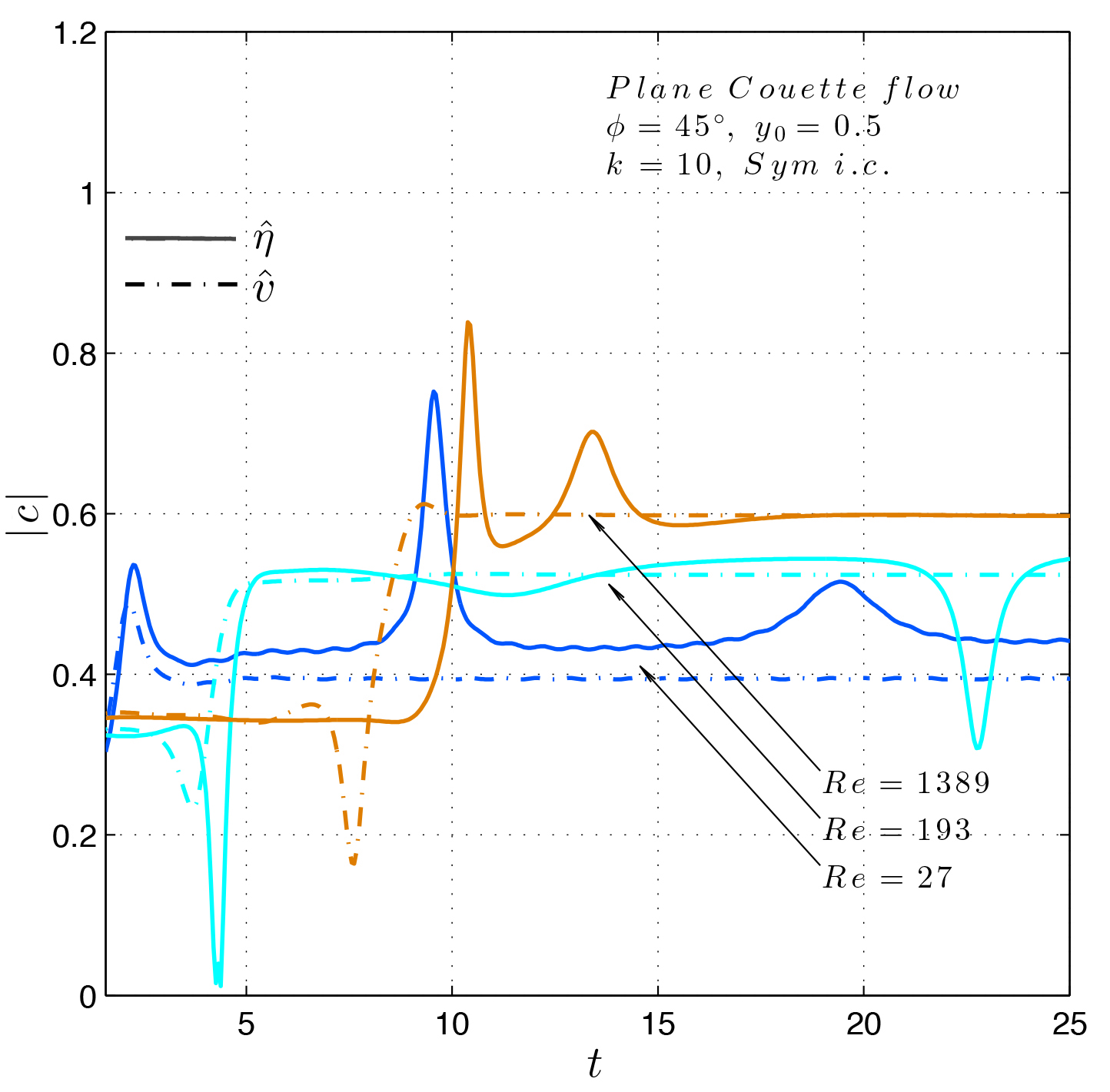}
	\vspace{0.5pt}
	\subcaption{}
	\label{fig:c_om_varioRe_1}
	 \end{subfigure}
        \begin{subfigure}{0.6\textwidth}
        \centering 
\includegraphics[width=9.0cm]{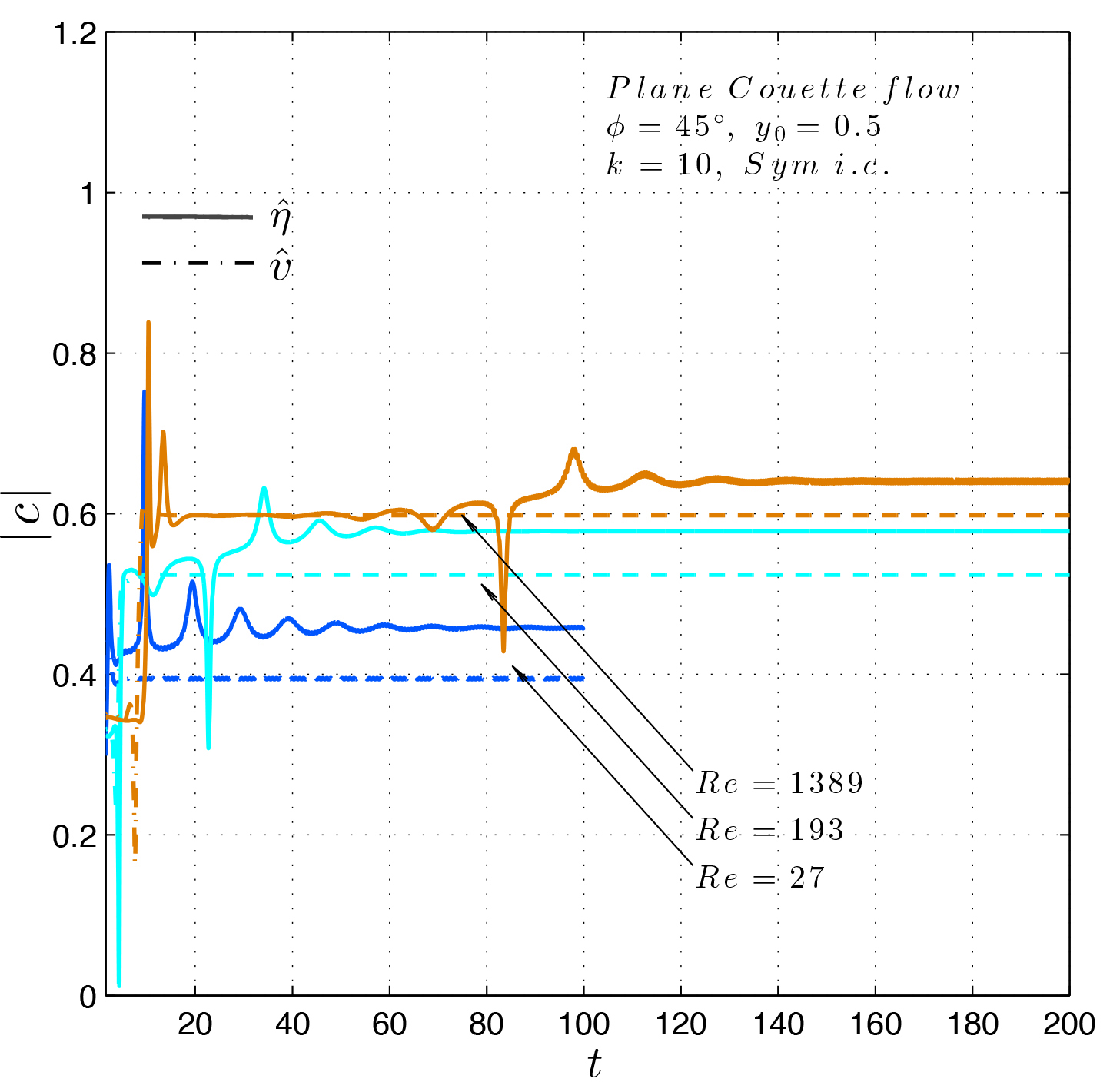}
	\vspace{0.5pt}
	\subcaption{ }
	\label{fig:c_om_varioRe_2}
	 \end{subfigure}
	\caption{Temporal evolution of the absolute value of the phase
velocity, calculated from $\he$ (continuous line), for PCf for $k=10$,
$\phi=45^{\circ}$, $Re=\{27,193,1389\}$ and
\textit{sym.} initial condition.The fixed observation point is
$y_0=0.5$. (a) Detail of the early transient and first frequency jump;
(b) Intermediate and far transient: the second jump can be observed. The phase
velocity of $\tilde v$ is shown with dot-dashed lines.}
\label{fig:c_om_varioRe}
\end{figure}

As shown in \figref{fig:c_om_varioRe_1}, for sufficiently high values of $Re$,
the phase velocity of $\tilde \eta$ ($c_\eta$, in the following) has approximately
the same evolution of the phase velocity of $\tilde v$ analized in the previous section, $c_v$ in the following, even if
it is clear that they are not coincident.
To be more precise, a certain lag in the first jump time $T_j$ is observed.\\
Figure \ref{fig:c_om_varioRe_2} reveals an interesting aspect: the
frequency of $\tilde \eta$ experiences a second jump, after which it reaches the
asymptote predicted by the modal theory. In the time window between the these
two jumps the mean value of $c_\eta$ is about the one of $c_v$, and this phase
of the wave life can also last several time units, depending on the
parameters. In fact, the time at which the second jump, $T_{j2}$, occurs
increases with increasing $Re$ and with decreasing $k$, while about the
influence of the obliquity angle, we observe that $T_{j2}$ increases with
increasing $\phi$ for high $k$, but the opposite trend occurs at lower
wavenumbers (see \tabref{tab:Tj2}). 

\begin{table}[h!]
\vspace{1cm}
\centering
  \begin{tabular}{cccccc}
 \hline
  \rule[-0.3cm]{0mm}{0.8cm}
  \boldmath ${k  \backslash \phi}$  & \boldmath ${20^{\circ}}$ & \boldmath
${40^{\circ}}$ & \boldmath ${60^{\circ}}$ & \boldmath
${80^{\circ}}$\\ 
  \hline \rule[0 cm]{0mm}{0.5cm}   
  \boldmath $0.80$  & 1030 & 860 & 710 & 475  \\
  \boldmath $1.37$  & 420 & 352 & 334 & 332  \\ 
  \boldmath $2.34$  & 176 & 164 & 170 & 194  \\
  \boldmath $4.00$  & 96.2 & 97.0 & 112 & 162  \\
  \boldmath $6.84$  & 45.5 & 53.5 & 58.0& 94.0  \\
  \boldmath $11.7$  & 38.8 & 46.0 & 50.8 & 85.4  \\
  \boldmath $20.0$  & 34.8 & 41.5 & 45.5 & 79.0  \\
  \boldmath $34.2$  & 23.9 & 37.6 & 42.2 & 75.3  \\ \hline
 \end{tabular}
\caption{Frequency jump nondimensional time $T_{j2}$ for various combination
of the
simulation parameters, for $Re=500$ and \textit{sym.} initial condition. Since
the transition to the asymptotic value of $c_\eta$ can be more or less smooth,  
$T_{j2}$ is considered as the time at which the frequency peak, typically
located just after the jump, occurs.}
\label{tab:Tj2}
\end{table}
\FloatBarrier
\ \newpage
\subsubsection{Intermediate Term and Long Term behaviour}
\begin{figure}[htbf]
        \centering
         \advance\leftskip-2.5cm
         \advance\rightskip-2cm
        \begin{subfigure}{0.6\textwidth}
        \centering
\includegraphics[width=9.6cm]{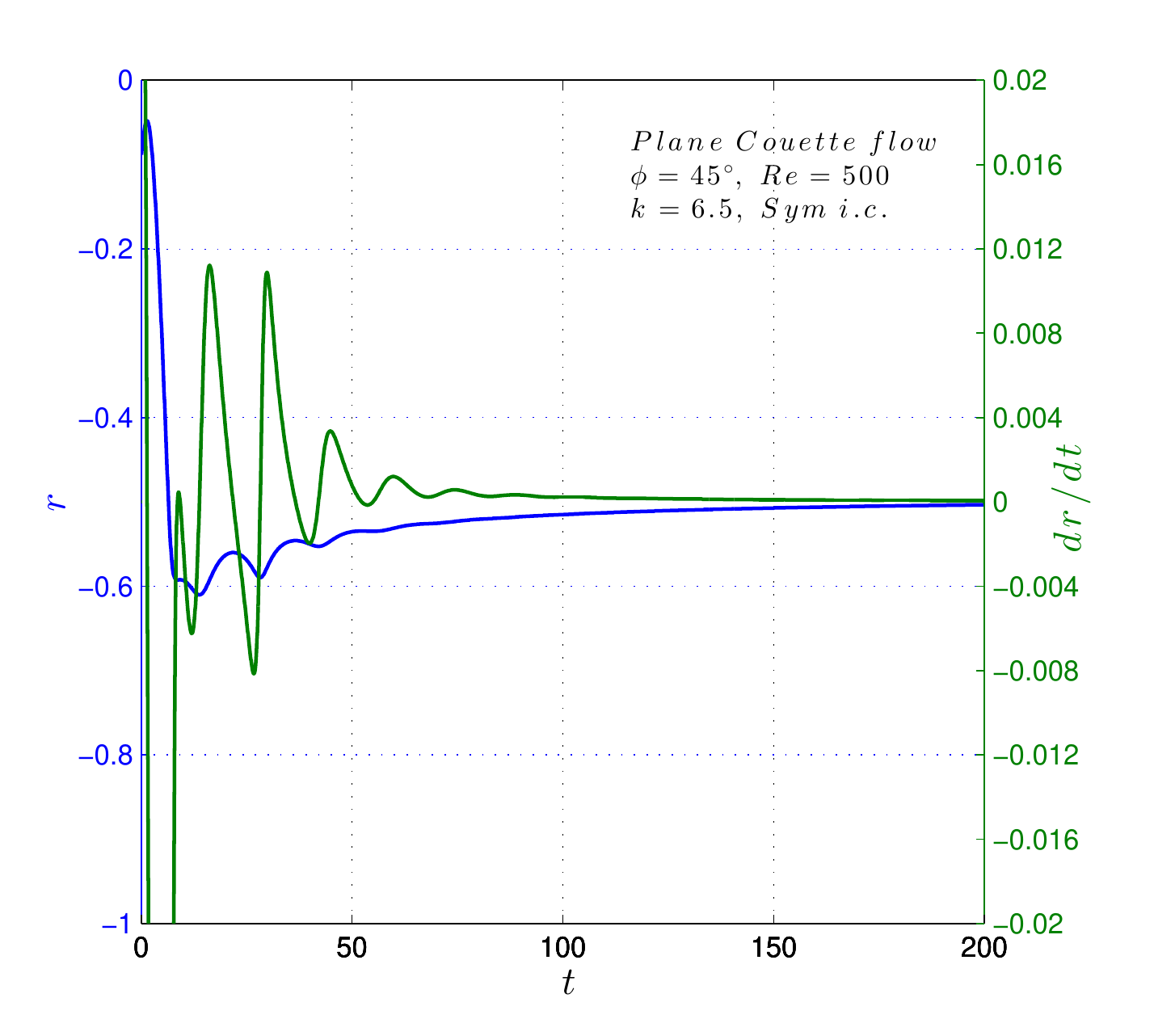}
	\vspace{0.5pt}
	\subcaption{}
	\label{fig:r_dr_CO_500_k6p5}
	 \end{subfigure}
        \begin{subfigure}{0.6\textwidth}
        \centering 
\includegraphics[width=9.6cm]{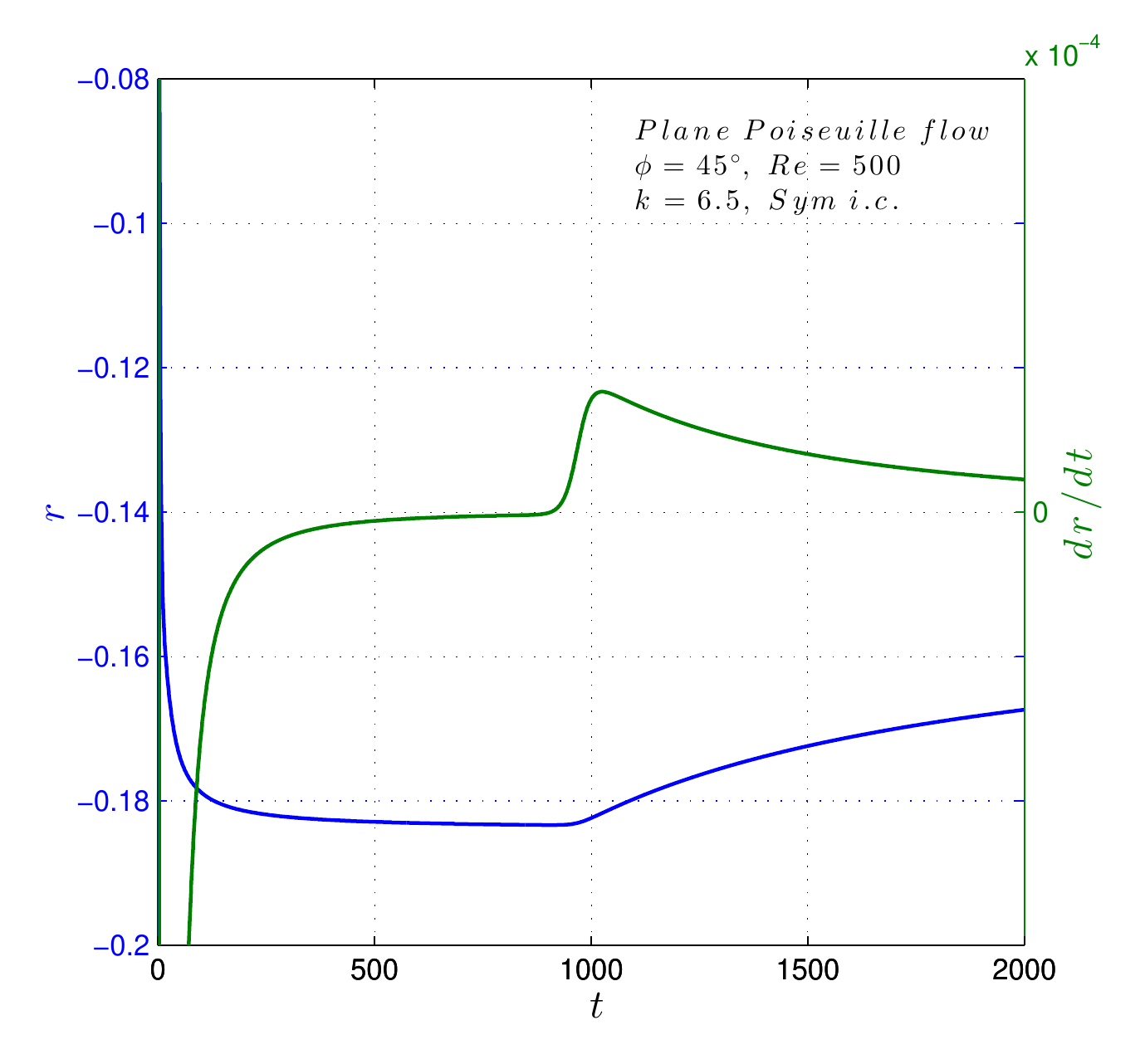}
	\vspace{0.5pt}
	\subcaption{ }
	\label{fig:r_dr_PO_500_k6p5}
	 \end{subfigure}	
	\caption{Temporal evolution of the kinetic energy growth rate (\textit{blue line}
) and its derivative (\textit{green line}) for PCf with
$k=6.5$, $\phi=45^{\circ}$, $Re=500$ and \textit{sym.} initial condition. It is
evident a correlation with the frequency jumps shown in the following plots. }
\label{fig:r_dr}
\end{figure}

In order to understand the physical reasons why $c_\eta$ experiences two jumps during
its temporal evolution, we take advantage of the mathematical formulation
introduced in \secref{sec:v_solution} and \secref{sec:eta_solution}. The Squire
equation \eqref{Squire} is forced by the solution $\hv$ of the Orr-Sommerfeld
PDE \eqref{Orr-Somm}. The general solution can be expressed as
$\he=\he_h+\he_p$ as shown in \secref{sec:gal_eta}. The particular solution
contains the same eigenvalues spectrum of the forcing $\hv$, while the spectrum
of the Squire operator (the homogeneous part) $\he_h$ is different. Thinking
about
a generic forced linear system, it is clear that the  asymptotic solution has the
same frequency of the forcing term if its amplitude is constant. If the forcing
term itself is damped, the asymptotic frequency depends on the damping of both
the forcing term and the homogeneous solution. If the damping rate of the
forcing term is higher than the one of the homogeneous operator, the
frequency for $t\to\infty$ will be the ``natural pulsation'' of the system.\\
Here the system is far more complicated but the same phenomenon is observed;
for several configurations of the parameters, looking at the spectra (e.g.
\figref{fig:Spectra_1}, \figref{fig:Spectra_2}) one can
notice that the least damped eigenvalue belongs to the Squire set. In these
cases, the second jump of $c_\eta$ occurs. $T_{j2}$ depends on the initial
coefficients of the series $\BU{h_0}$ (i.e. on the initial condition) and on the
ratio of the real part of the eigenvalues $\mu$, to $\mu^*$, and can be qualitatively considered as the end of the
intermediate term and the beginning of the asymptote, as can be seen from the trends of the kinetic energy growth rate
in \figref{fig:r_dr}.

\begin{figure}[h!]
        \centering
         \advance\leftskip-2.5cm
         \advance\rightskip-2cm
        \begin{subfigure}{0.6\textwidth}
        \centering
\includegraphics[width=9.0cm]{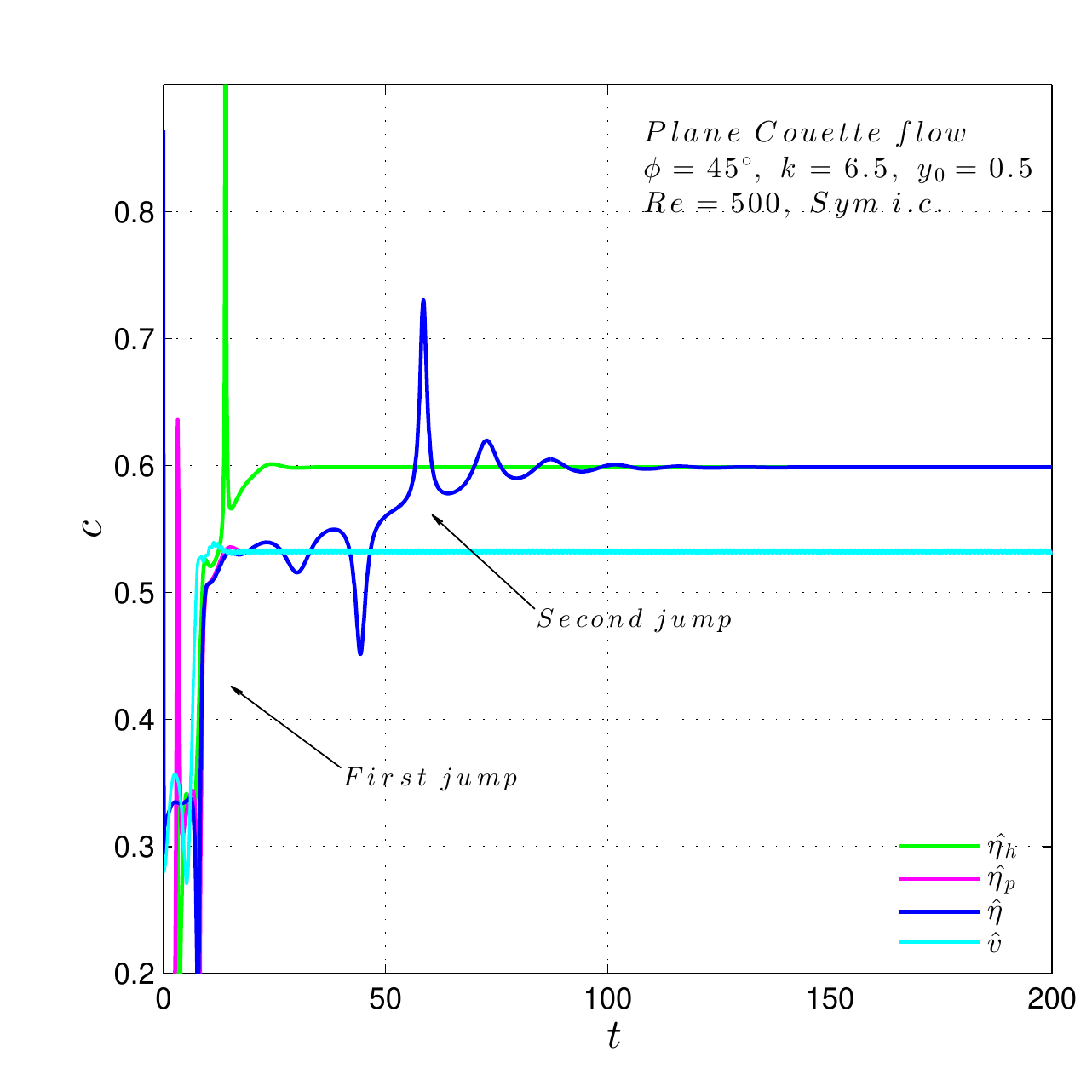}
	\vspace{-1cm}
	\label{fig:c_tutte_CO_Re500_k6p5_phi45}
	 \end{subfigure}
        \begin{subfigure}{0.6\textwidth}
        \centering 
\includegraphics[width=9.0cm]{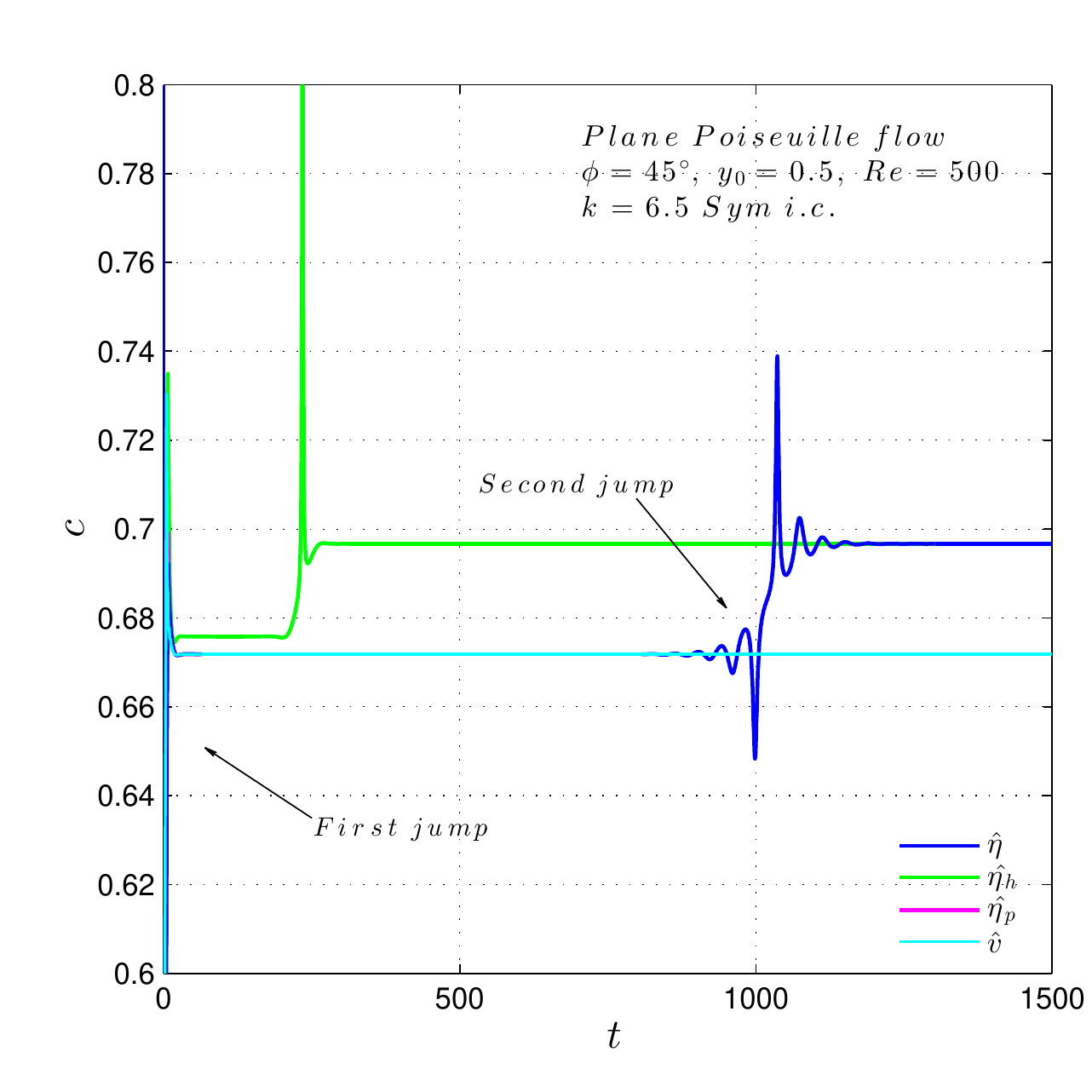}
	\vspace{-1cm}
	\label{fig:c_tutte_PO_Re500_k6p5_phi45}
	 \end{subfigure}
        \begin{subfigure}{0.6\textwidth}
        \centering
\includegraphics[width=9.0cm]{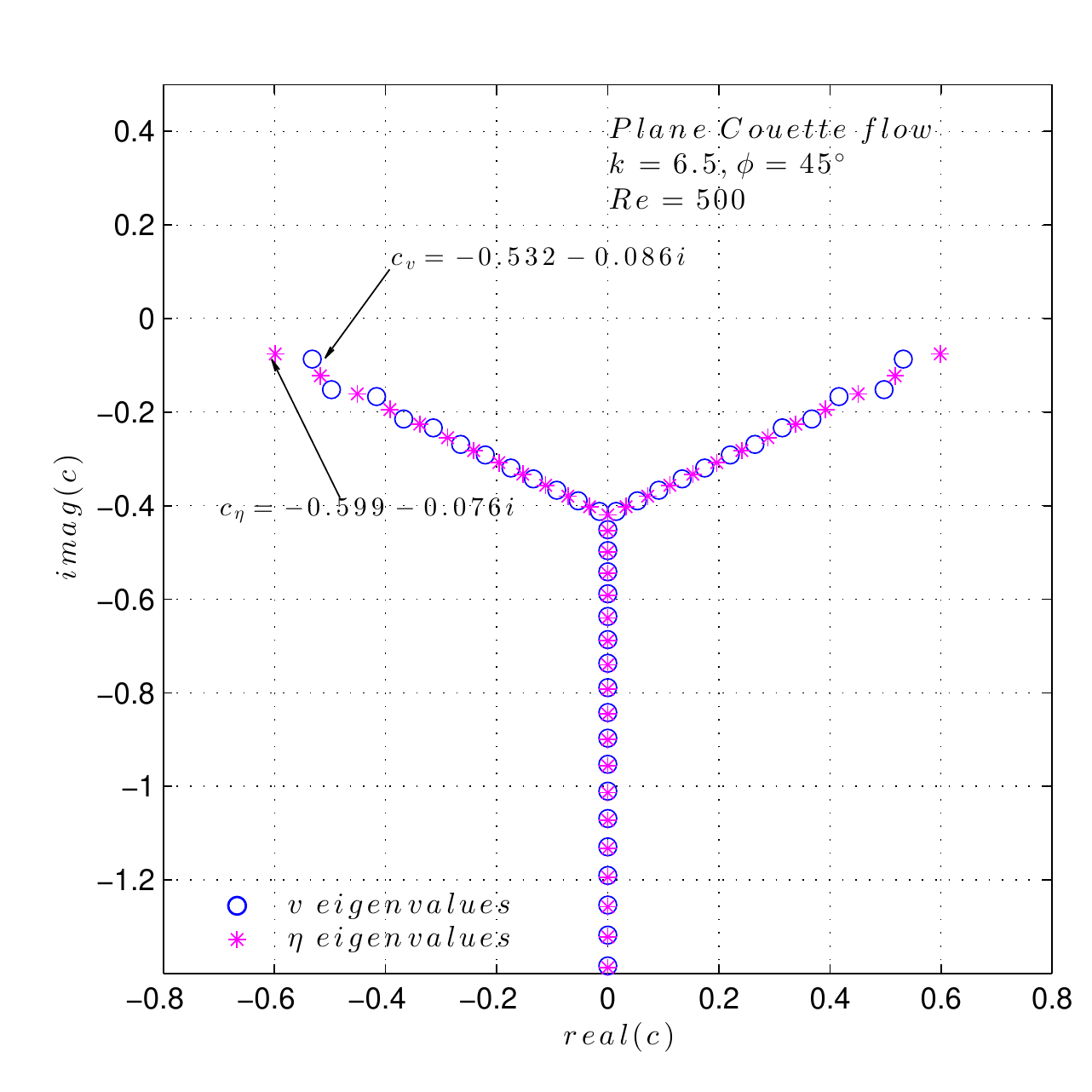}
	\vspace{0.5pt}
	\subcaption{}
	\label{fig:Spectrum_PCf_Re500_k_6p5_phi_45}
	 \end{subfigure}
        \begin{subfigure}{0.6\textwidth}
        \centering 
\includegraphics[width=9.0cm]{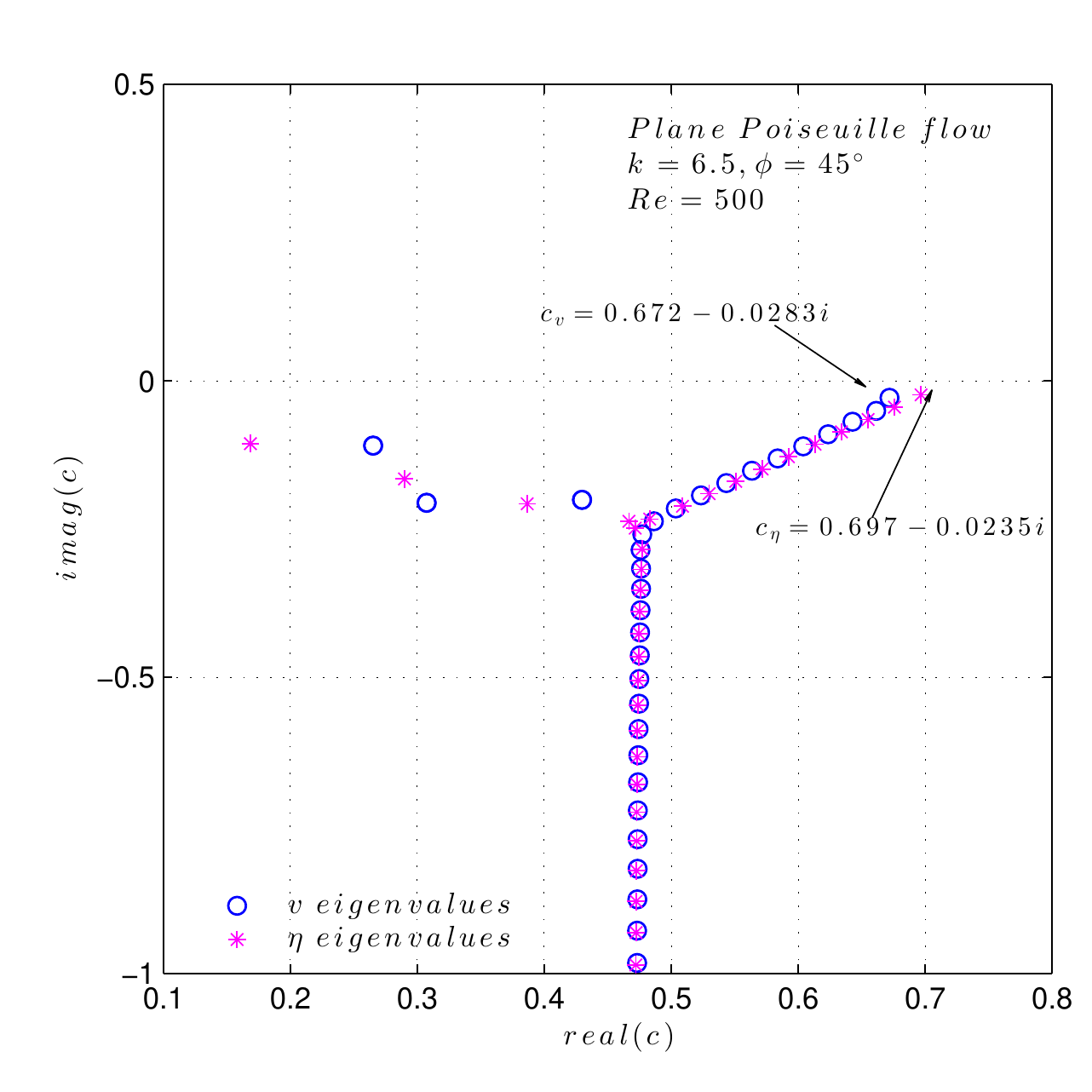}
	\vspace{0.5pt}
	\subcaption{ }
	\label{fig:Spectrum_PPf_Re500_k_6p5_phi_45}
	 \end{subfigure}	
	\caption{(a) Temporal evolution of the absolute value of the phase
velocity of $\tilde \eta$, $\tilde{\eta}_h$, $\tilde{\eta}_p$ and comparison with $c_v$ for PCf with
$k=6.5$, $\phi=45^{\circ}$, $Re=500$ and \textit{sym.} initial condition
(upper plot). In the lower plot the spectrum for the same configuration is shown.
(b) Phase velocity for Plane Poiseuille flow, same parameters configuration. In
both cases the least damped eigenvalue belongs to the Squire set.}
\label{fig:c_tutte}
\end{figure}
\begin{figure}[h!]
        \centering
         \advance\leftskip-2.5cm
         \advance\rightskip-2cm
        \begin{subfigure}{0.6\textwidth}
        \centering
\includegraphics[width=9.0cm]{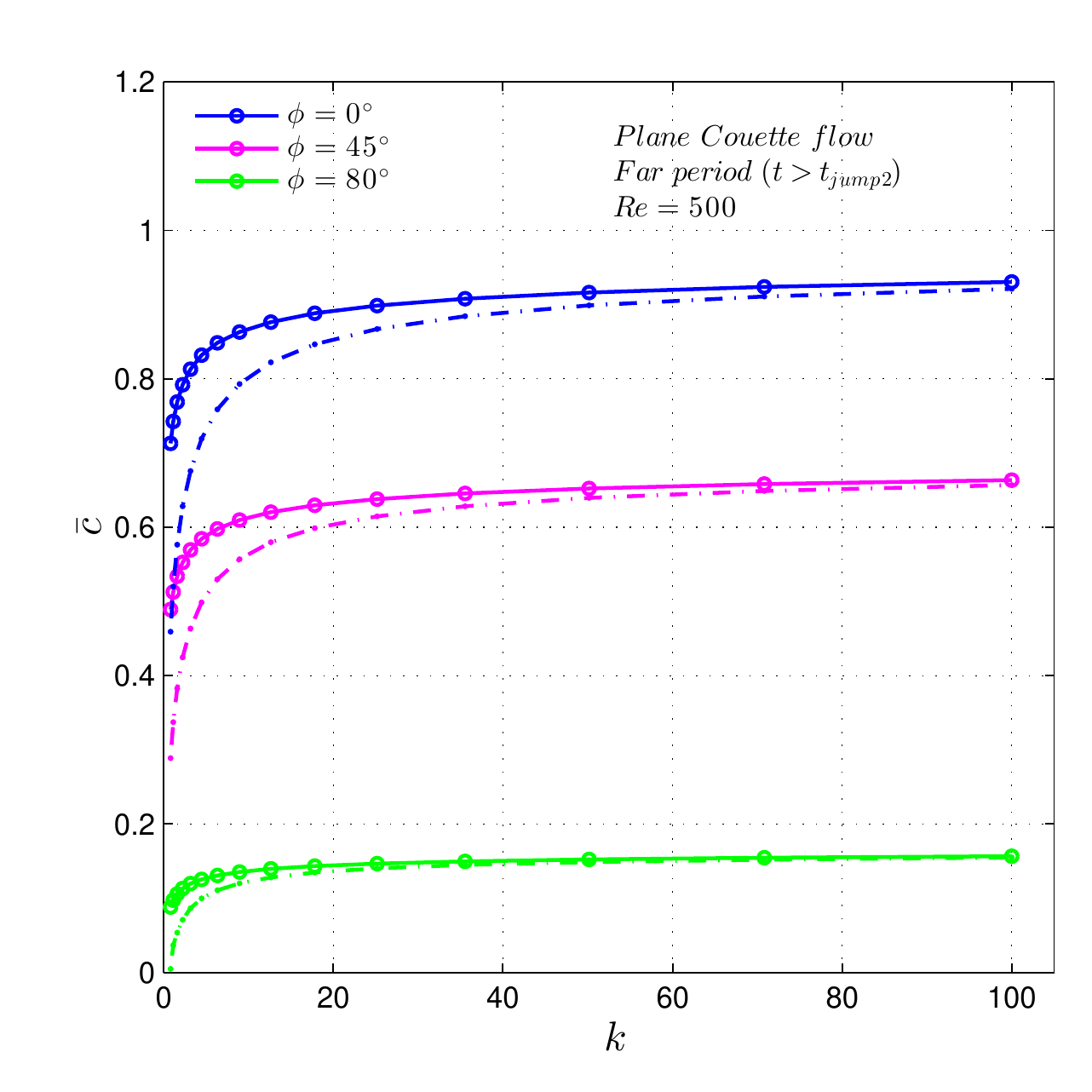}
	\vspace{0.5pt}
	\subcaption{}
	\label{fig:c_mean_variophi_eta}
	 \end{subfigure}
        \begin{subfigure}{0.6\textwidth}
        \centering 
\includegraphics[width=9.0cm]{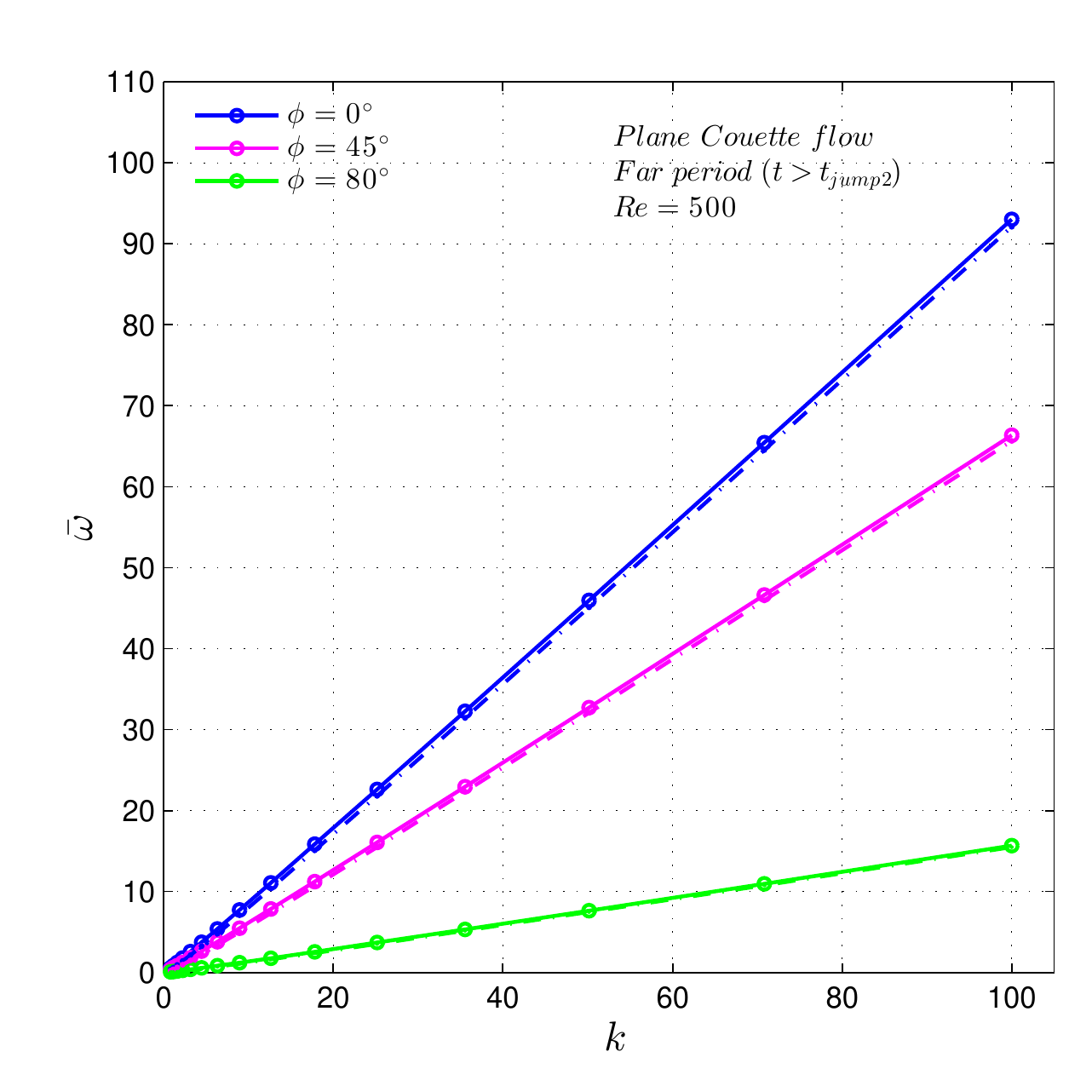}
	\vspace{0.5pt}
	\subcaption{ }
	\label{fig:O_mean_variophi_eta}
	 \end{subfigure}	
	\caption{(a) Asymptotic absolute values of $c_\eta$,
for PCf with $Re=500$, $\phi=\{10^{\circ},
45^{\circ}, 80^{\circ}\}$. The polar wavenumbers are uniformly distributed
in the logarithmic space. (b) Asymptotic absolute values of ${\omega_\eta}$.
Comparison with the trends of $\bar{c}_v,\ \bar{\omega}_v$, plotted with  dot-dashed line.
}
\label{fig:mean_variophi_eta}
\end{figure}

\FloatBarrier
\ \newpage
For Plane Couette flow, the same modulation observed in the phase velocity of
$\tilde v$ is found in the $\tilde \eta$ component, even if the characteristic amplitude and
the period are generally different. The same motivation discussed in the
previous
section applies, the eigenvalues of $\he$ for this type of flow are complex
conjugate, indeed. The frequency of this modulation appears to be generally
higher than the one of $c_v$ , supporting the
fact that this modulation is related to the imaginary part of the least damped
$\mu_i$ (for $c_v$) or $\mu^*_i$ (for $c_\eta$), as can be inferred by looking
at the spectra, since usually $|\Re(c_v)|<|\Re(c_\eta)|$ for the least damped.
The trend of the asymptotic frequency is reported in
\figref{fig:c_mean_variophi_eta} and \figref{fig:O_mean_variophi_eta}, where
one can see that as $k\to\infty$ the difference between $c_v$ and $c_\eta$
tends to vanish. Anyway the general trend of the two frequencies, varying the
parameters, is approximately the same.

\FloatBarrier
\ \newpage

\section{Velocity and vorticity profiles, similarity
considerations and solutions in the physical space}
\subsection{Profiles of $\hv$, $\he$ and their similarity properties}\label{sec:trasf_profiles}
\begin{figure}[h!]
        \centering
        \advance\leftskip-1.2cm
        \begin{subfigure}{1\textwidth}
        \centering
\includegraphics[width=16.0cm]{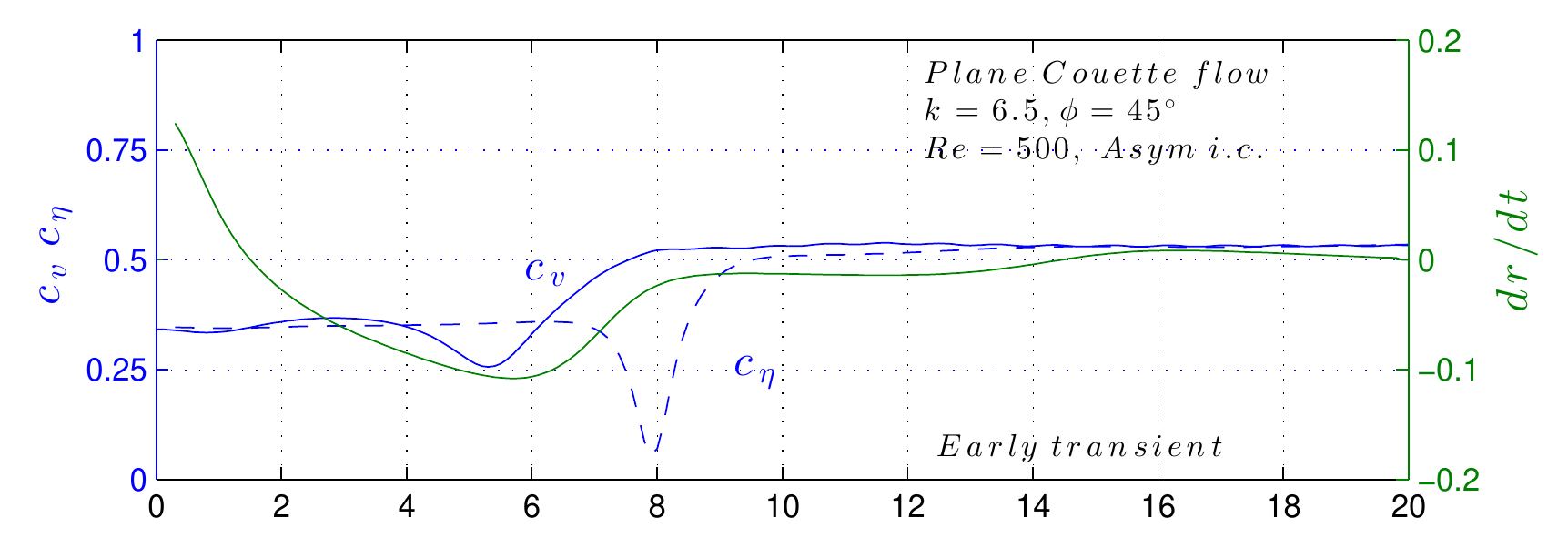}
	\vspace{-0.5cm}
	\label{fig:prof_CO_Re500_k6p5_early_1}
	 \end{subfigure}
        \begin{subfigure}{1\textwidth}
        \centering 
\includegraphics[width=16cm]{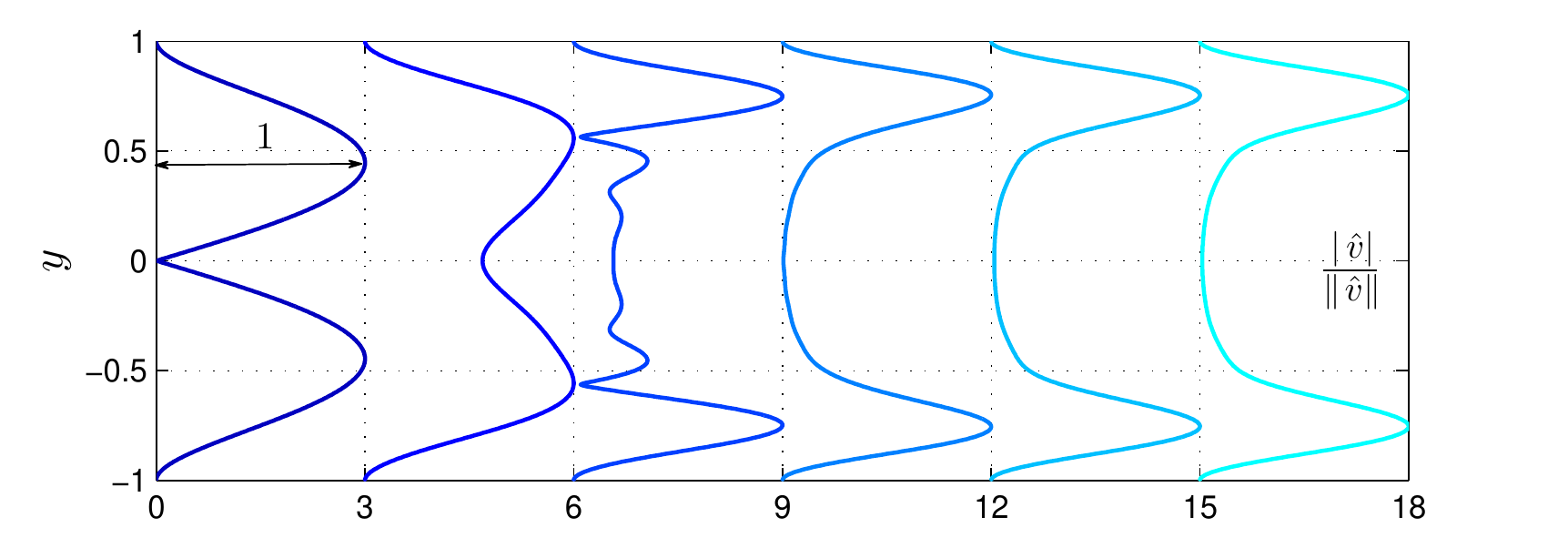}
	\vspace{-0.5cm}
	\label{fig:prof_CO_Re500_k6p5_early_2}
	 \end{subfigure}
	\begin{subfigure}{1\textwidth}
        \centering
\includegraphics[width=16.0cm]{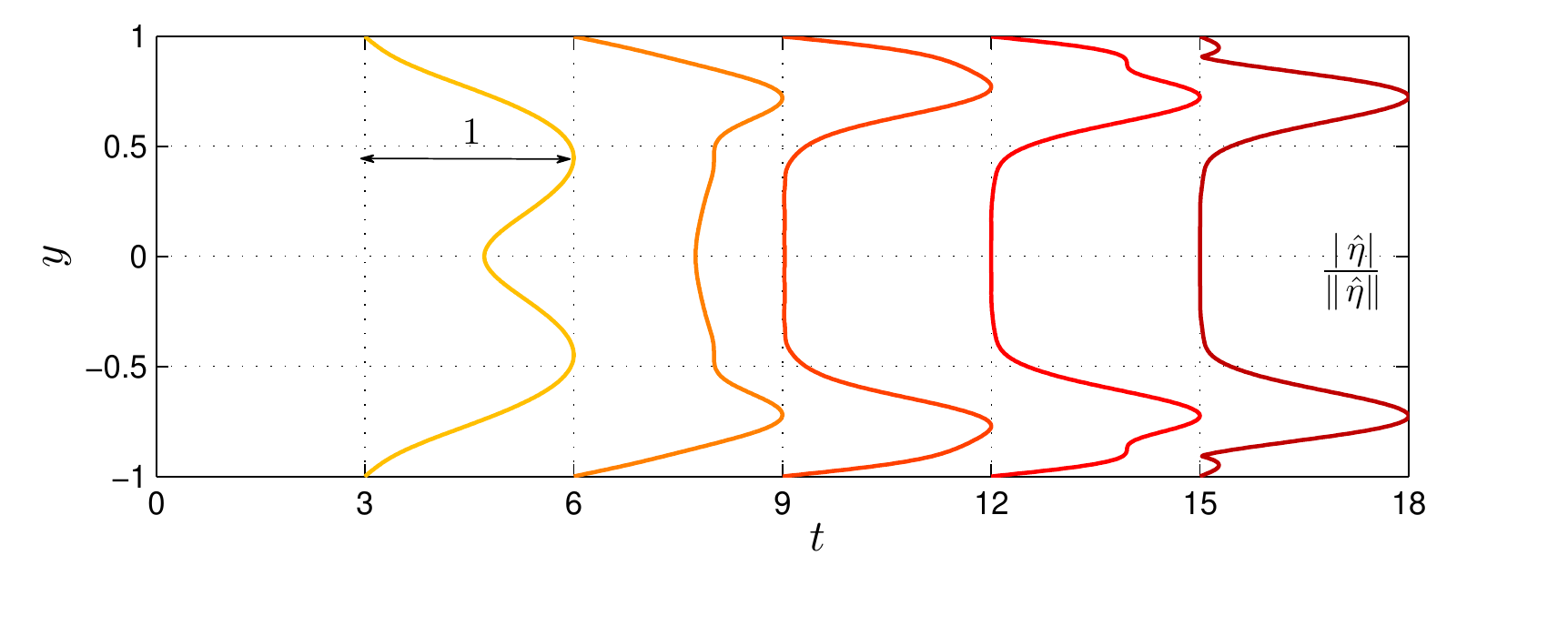}
	\vspace{-1.0cm}
	\label{fig:prof_CO_Re500_k6p5_early_3}
	 \end{subfigure}
	\caption{Plane Couette flow early transient for $Re=500$,
$\phi=45^{\circ}$, $k=6.5$ and \textit{asym.} initial condition.
Top: phase velocity temporal evolution for the $\hv$ and $\he$ disturbance
(respectively, blue continuous line and blue dashed line) and first derivative
of the kinetic energy growth rate (green line). Middle: profiles of the modulus
of $\hv$, normalized with respect to the maximum ($L_\infty$-norm). Bottom:
profiles of
$|\he|/\| \he \|_\infty$.}
\label{fig:prof_CO_Re500_k6p5_early}
\end{figure}

\begin{figure}[h!]
        \centering
        \advance\leftskip-1.2cm
        \begin{subfigure}{1\textwidth}
        \centering
\includegraphics[width=16.0cm]{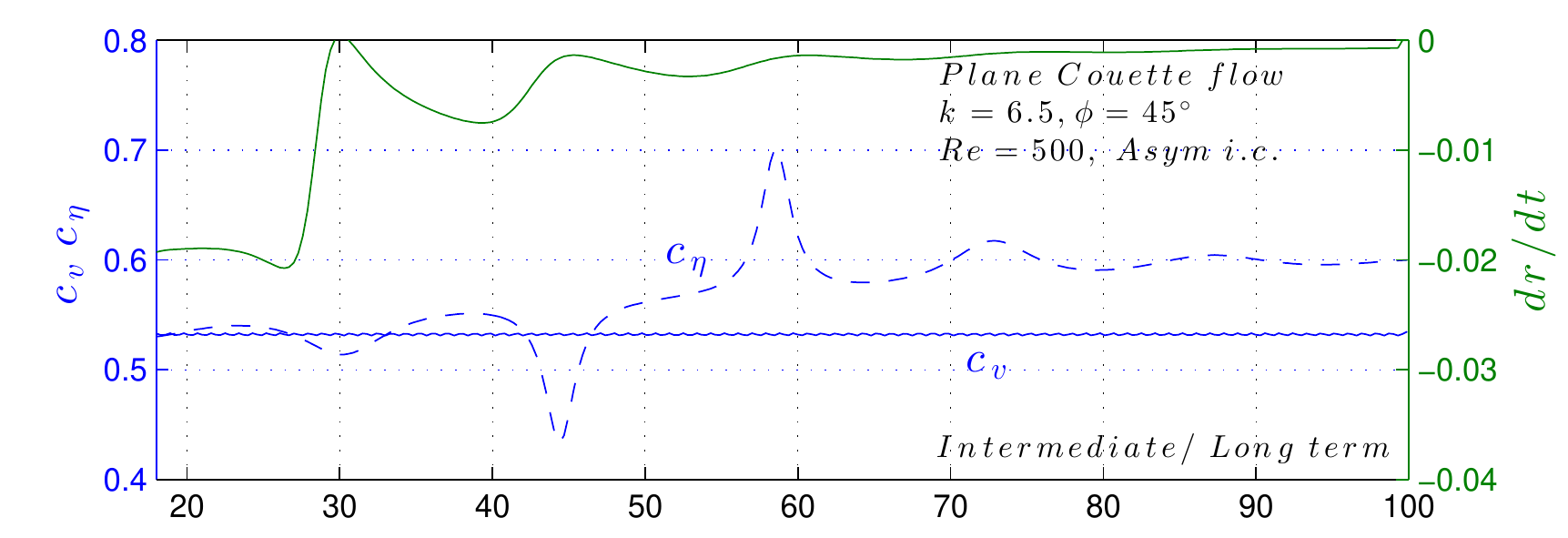}
	\vspace{-0.5cm}
	\label{fig:prof_CO_Re500_k6p5_far_1}
	 \end{subfigure}
        \begin{subfigure}{1\textwidth}
        \centering 
\includegraphics[width=16cm]{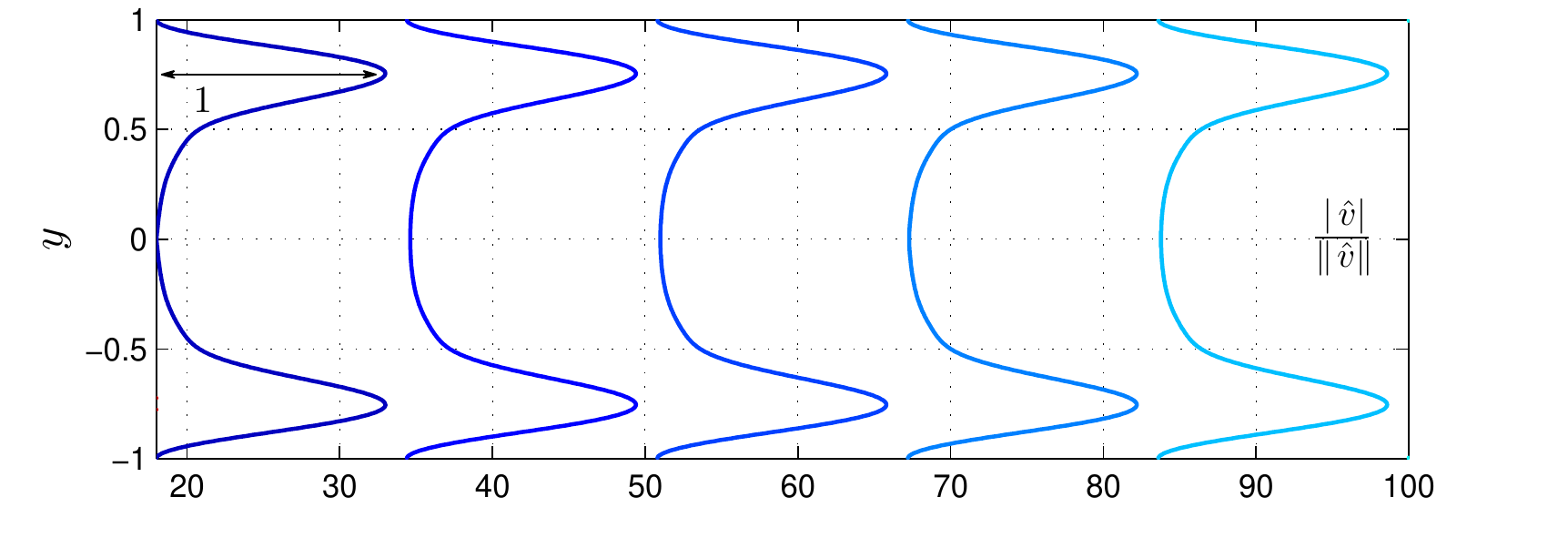}
	\vspace{-0.5cm}
	\label{fig:prof_CO_Re500_k6p5_far_2}
	 \end{subfigure}
	\begin{subfigure}{1\textwidth}
        \centering
\includegraphics[width=16.0cm]{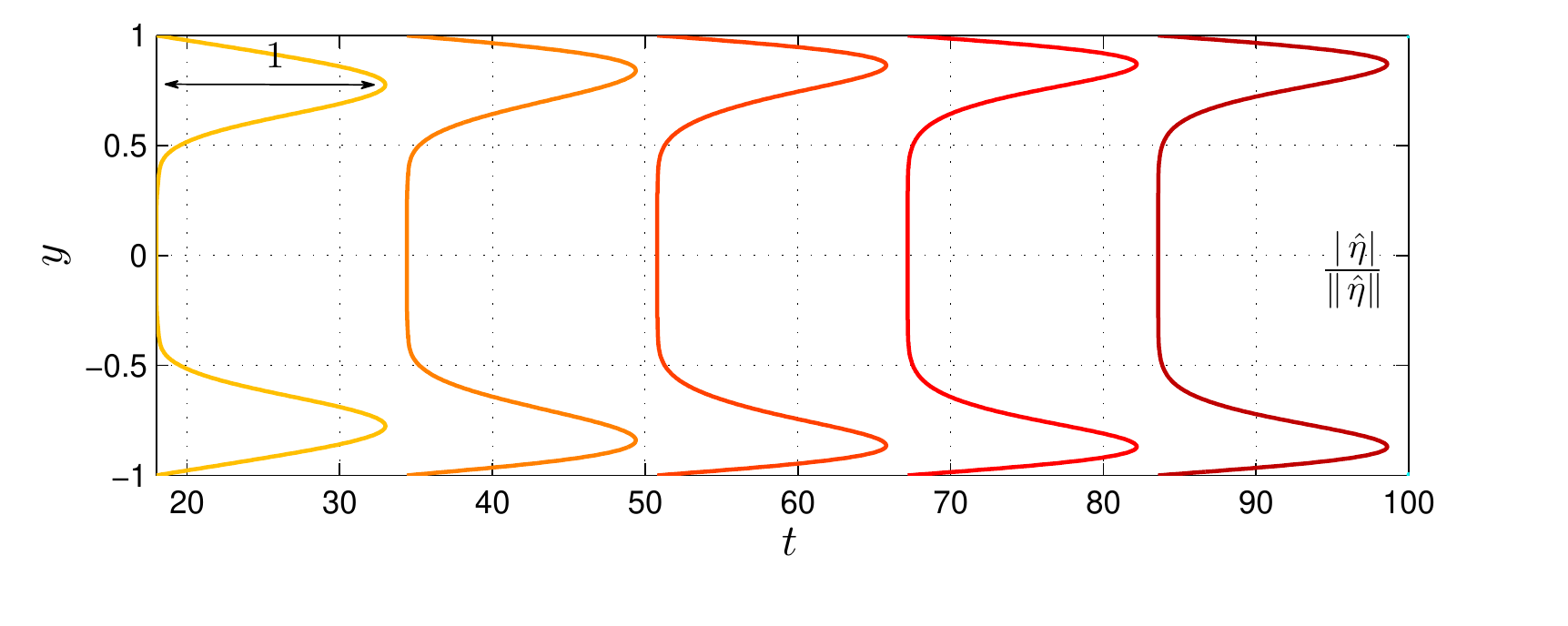}
	\vspace{-1.0cm}
	\label{fig:prof_CO_Re500_k6p5_far_3}
	 \end{subfigure}
	\caption{Plane Couette flow intermediate and far term for $Re=500$,
$\phi=45^{\circ}$, $k=6.5$ and \textit{asym.} initial condition.
Top: phase velocity temporal evolution for the $\hv$ and $\he$ disturbance
(respectively, blue continuous line and blue dashed line) and first derivative
of the kinetic energy growth rate (green line). Middle: profiles of the modulus
of $\hv$, normalized with respect to the maximum ($L_\infty$-norm). Bottom:
profiles of
$|\he|/\| \he \|_\infty$. Note that $|\hv|$ is fully self-similar after the
first
jump, while $|\he|$ achieves gradually the similarity after the second jump.  }
\label{fig:prof_CO_Re500_k6p5_far}
\end{figure}

\begin{figure}[h!]
        \centering
        \advance\leftskip-1.2cm
        \begin{subfigure}{1\textwidth}
        \centering
\includegraphics[width=16.0cm]{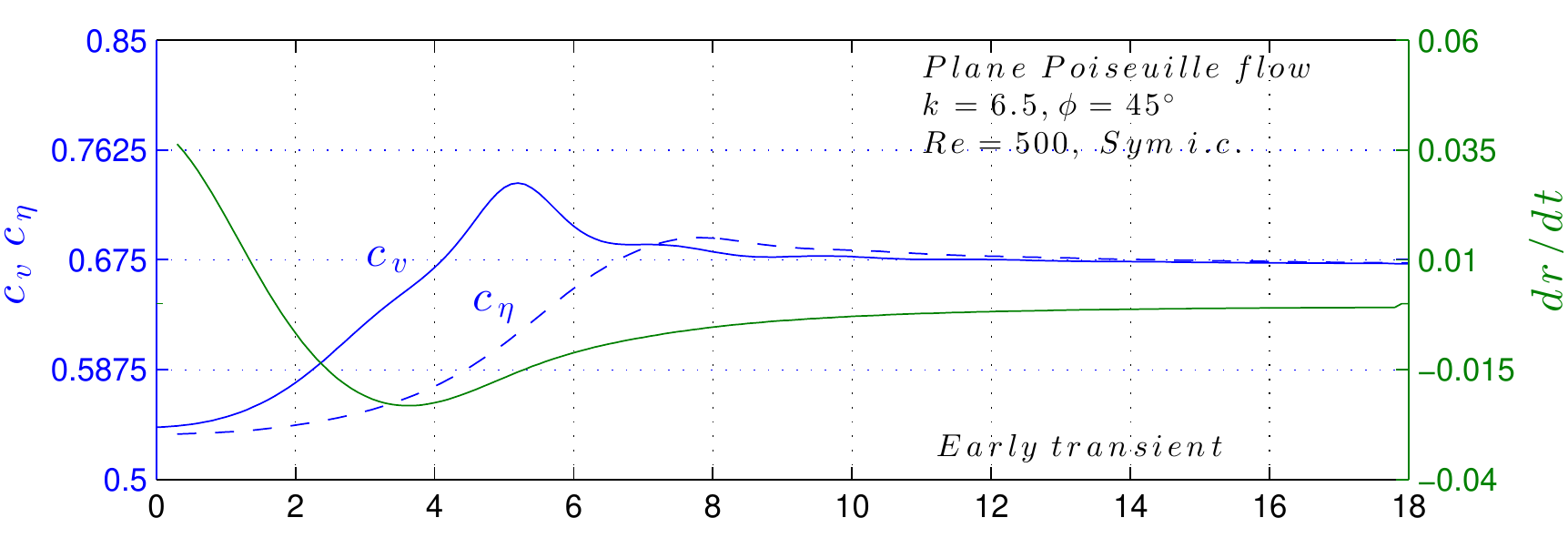}
	\vspace{-0.5cm}
	\label{fig:prof_PO_Re500_k6p5_early_1_sym}
	 \end{subfigure}
        \begin{subfigure}{1\textwidth}
        \centering 
\includegraphics[width=16cm]{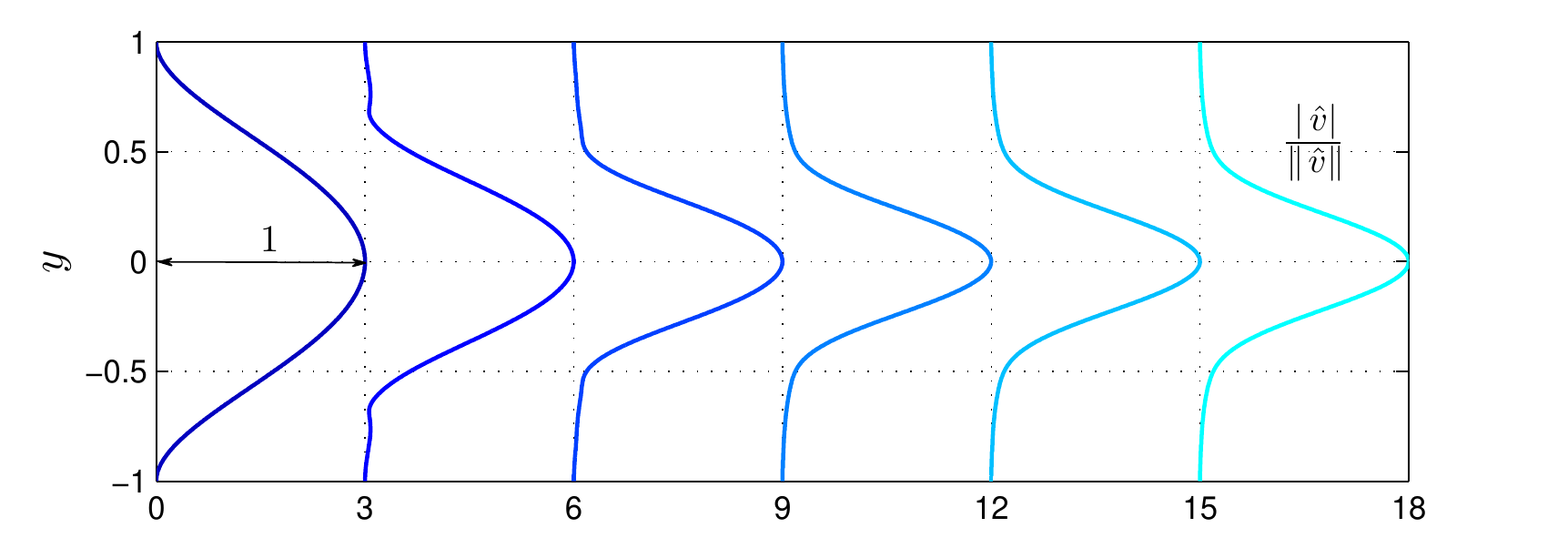}
	\vspace{-0.5cm}
	\label{fig:prof_PO_Re500_k6p5_early_2_sym}
	 \end{subfigure}
	\begin{subfigure}{1\textwidth}
        \centering
\includegraphics[width=16.0cm]{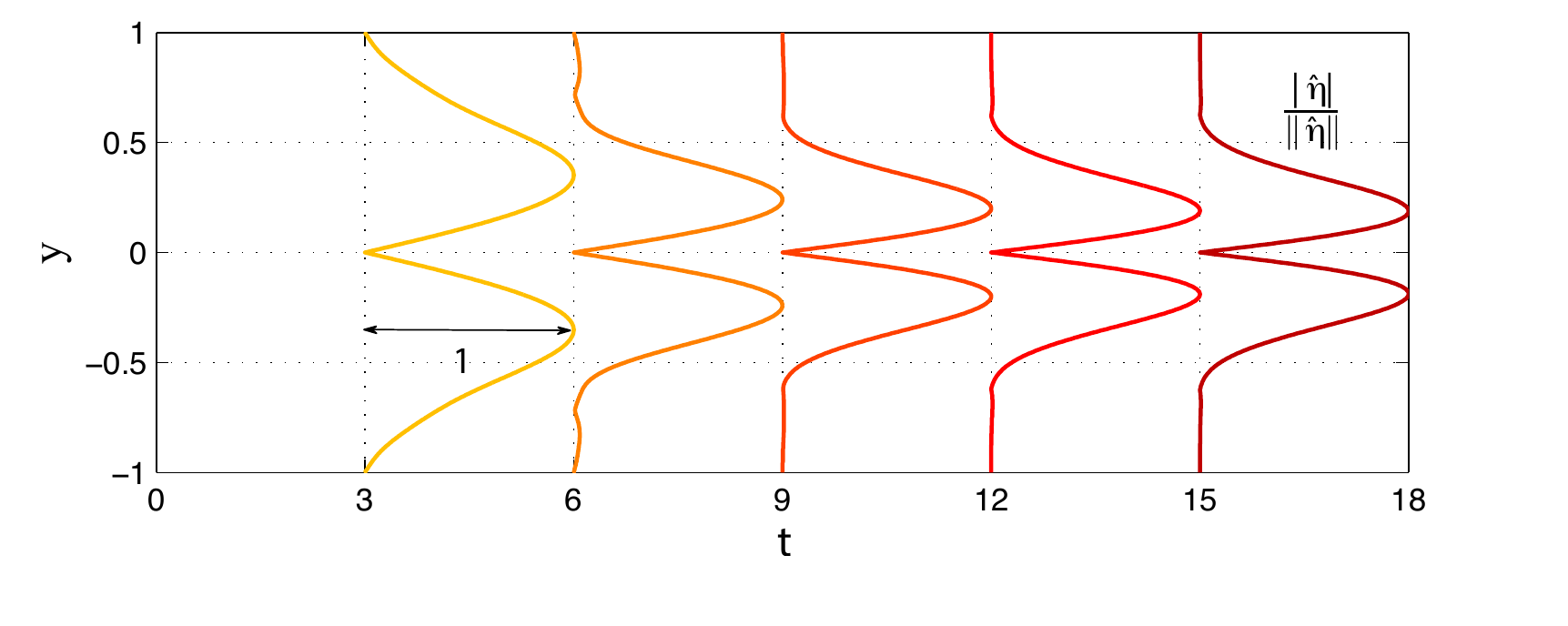}
	\vspace{-1.0cm}
	\label{fig:prof_PO_Re500_k6p5_early_3_sym}
	 \end{subfigure}
	\caption{Plane Poiseuille flow early transient for
$Re=500$, $\phi=45^{\circ}$, $k=6.5$ and \textit{sym.} initial condition.
Top: phase velocity temporal evolution for the $\hv$ and $\he$ disturbance
(respectively, blue continuous line and blue dashed line) and first derivative
of the kinetic energy growth rate (green line). Middle: profiles of the modulus
of $\hv$, normalized with respect to the maximum ($L_\infty$-norm). Bottom:
profiles of
$|\he|/\| \he \|_\infty$. }
\label{fig:prof_PO_Re500_k6p5_early_sym}
\end{figure}

\begin{figure}[h!]
        \centering
        \advance\leftskip-1.2cm
        \begin{subfigure}{1\textwidth}
        \centering
\includegraphics[width=16.0cm]{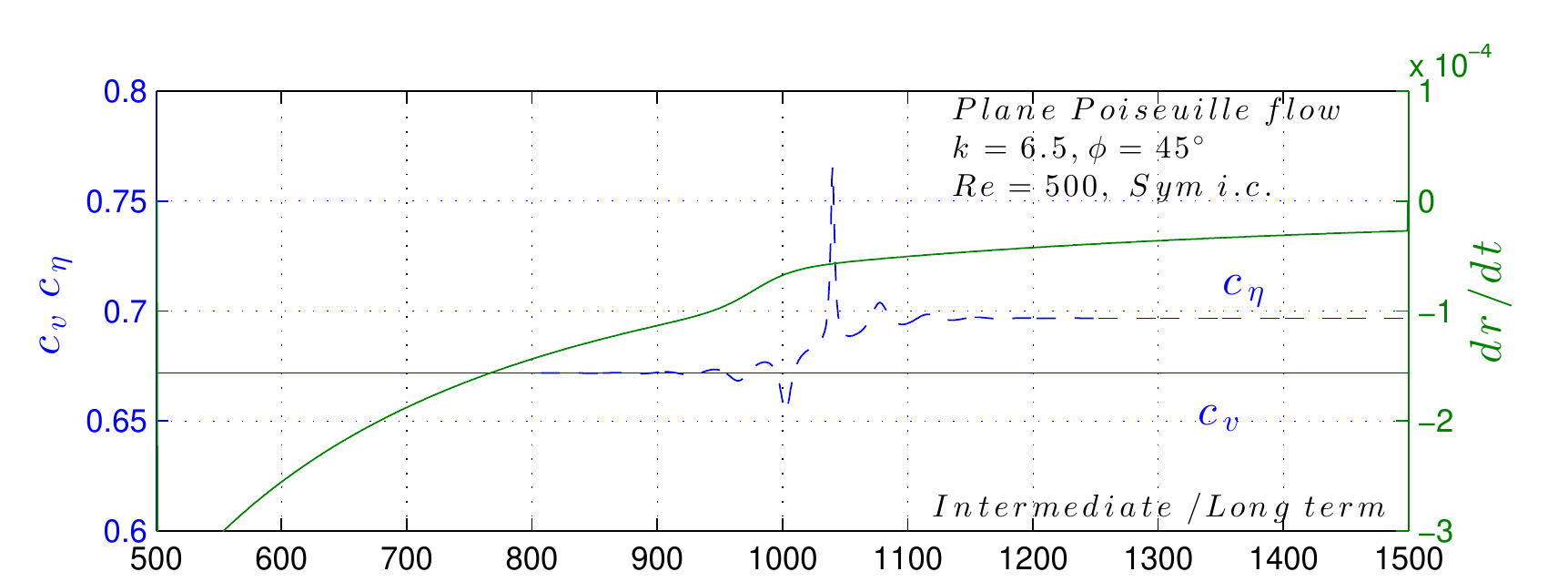}
	\vspace{-0.5cm}
	\label{fig:prof_PO_Re500_k6p5_far_1_sym}
	 \end{subfigure}
        \begin{subfigure}{1\textwidth}
        \centering 
\includegraphics[width=16cm]{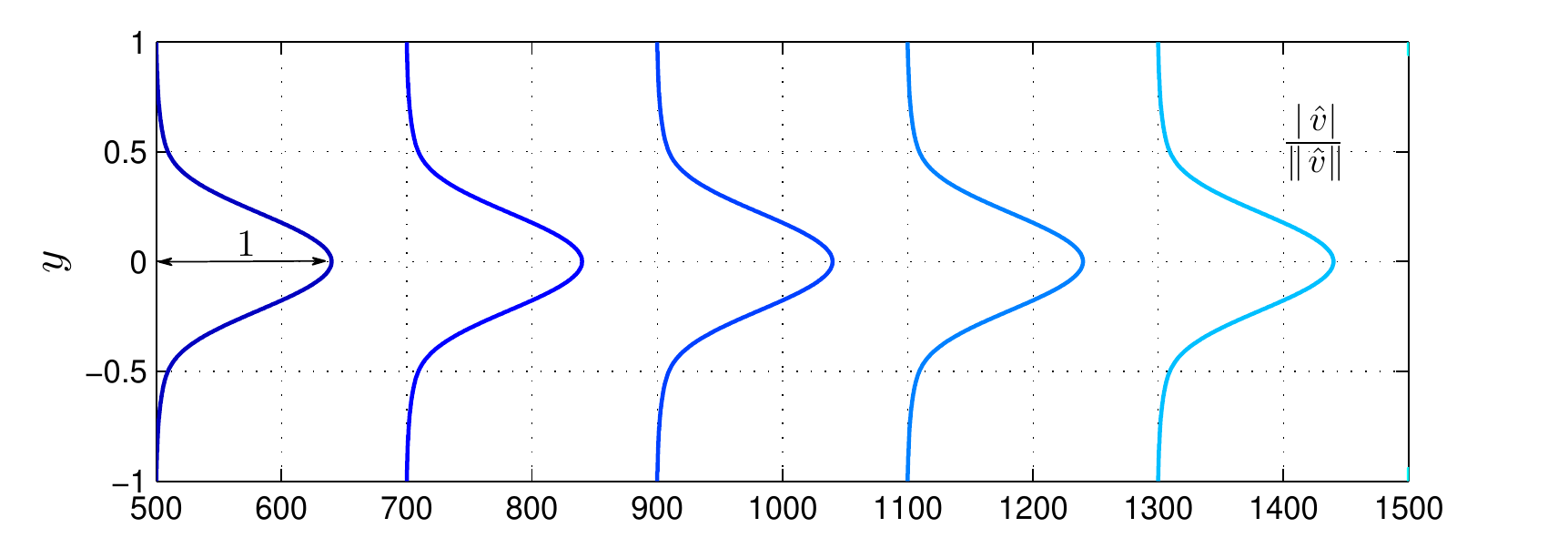}
	\vspace{-0.5cm}
	\label{fig:prof_PO_Re500_k6p5_far_2_sym}
	 \end{subfigure}
	\begin{subfigure}{1\textwidth}
        \centering
\includegraphics[width=16.0cm]{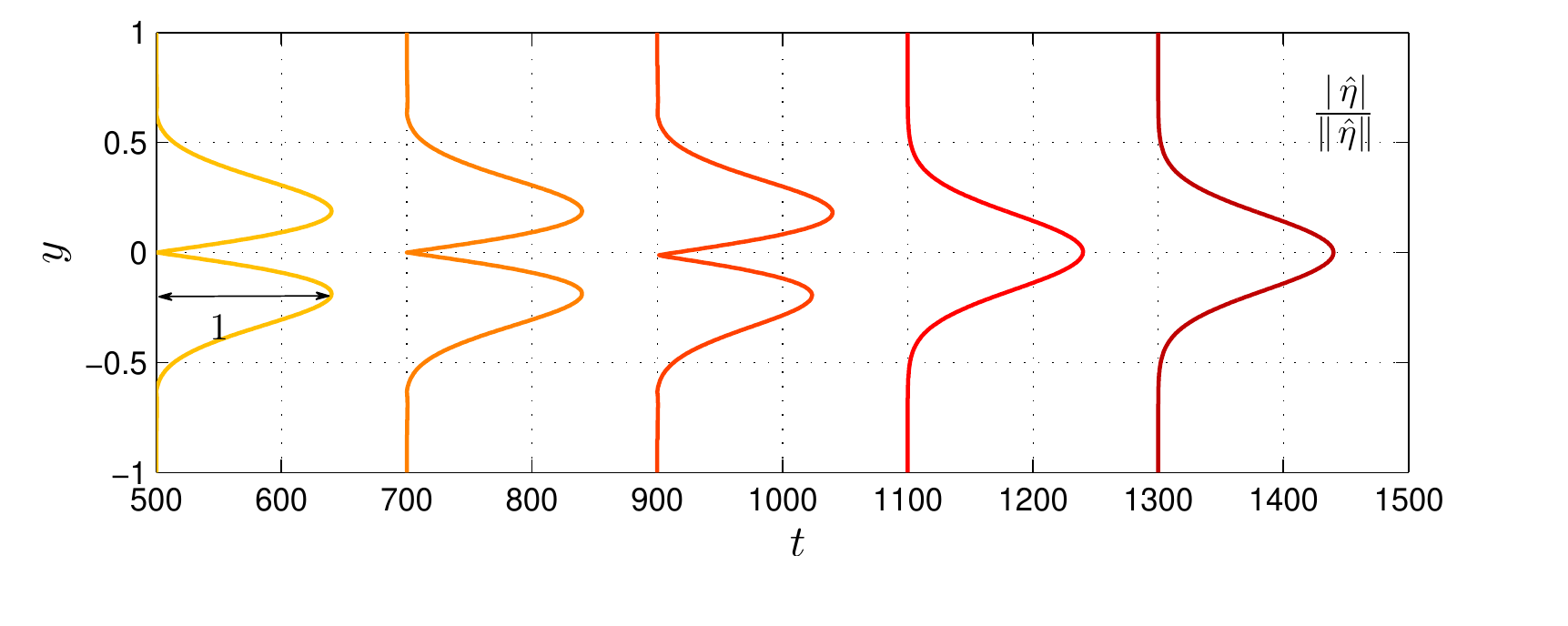}
	\vspace{-1.0cm}
	\label{fig:prof_PO_Re500_k6p5_far_3_sym}
	 \end{subfigure}
	\caption{Plane Poiseuille flow intermediate and far term for $Re=500$,
$\phi=45^{\circ}$, $k=6.5$ and \textit{sym.} initial condition.
Top: phase velocity temporal evolution for the $\hv$ and $\he$ disturbance
(respectively, blue continuous line and blue dashed line) and first derivative
of the kinetic energy growth rate (green line). Middle: profiles of the modulus
of $\hv$, normalized with respect to the maximum ($L_\infty$-norm). Bottom:
profiles of
$|\he|/\| \he \|_\infty$. It is interesting to notice the abrupt transition
to the
final state of the vorticity profile, which occurs in correspondence to 
the second frequency jump. Both the states before and after $T_{j2}$ seem to
have similarity properties.}
\label{fig:prof_PO_Re500_k6p5_far_sym}
\end{figure}

\begin{figure}[h!]
        \centering
        \advance\leftskip-1.2cm
        \begin{subfigure}{1\textwidth}
        \centering
\includegraphics[width=16.0cm]{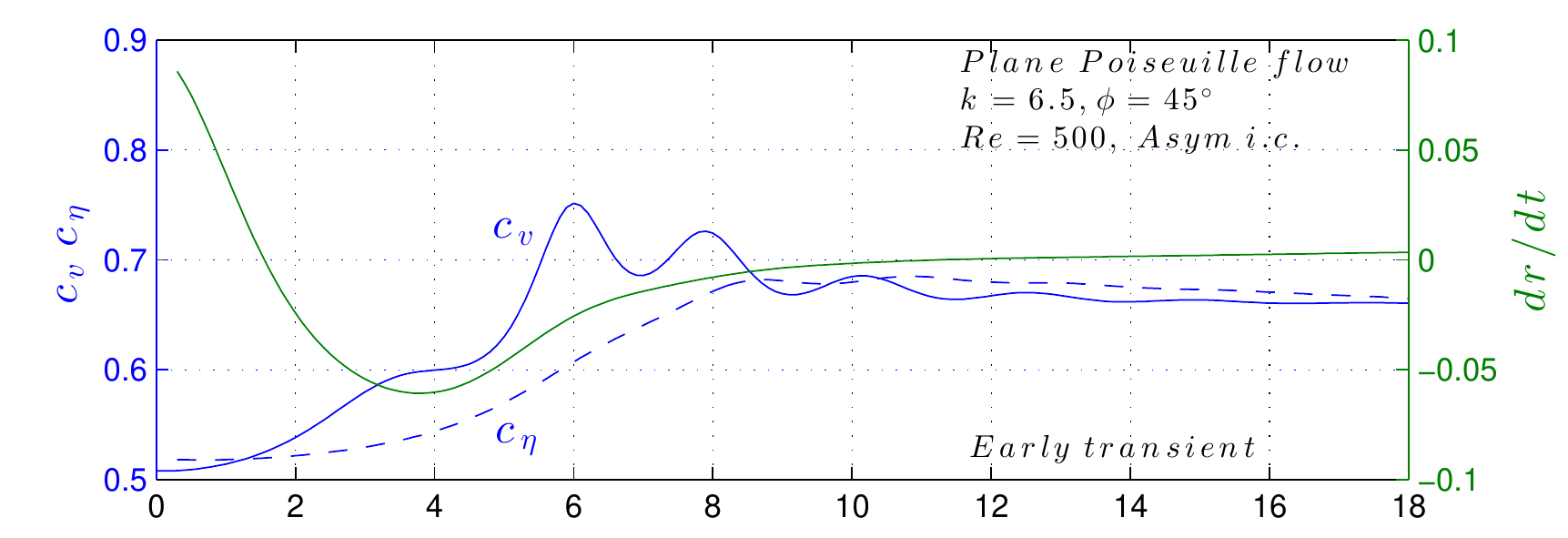}
	\vspace{-0.5cm}
	\label{fig:prof_PO_Re500_k6p5_early_1}
	 \end{subfigure}
        \begin{subfigure}{1\textwidth}
        \centering 
\includegraphics[width=16cm]{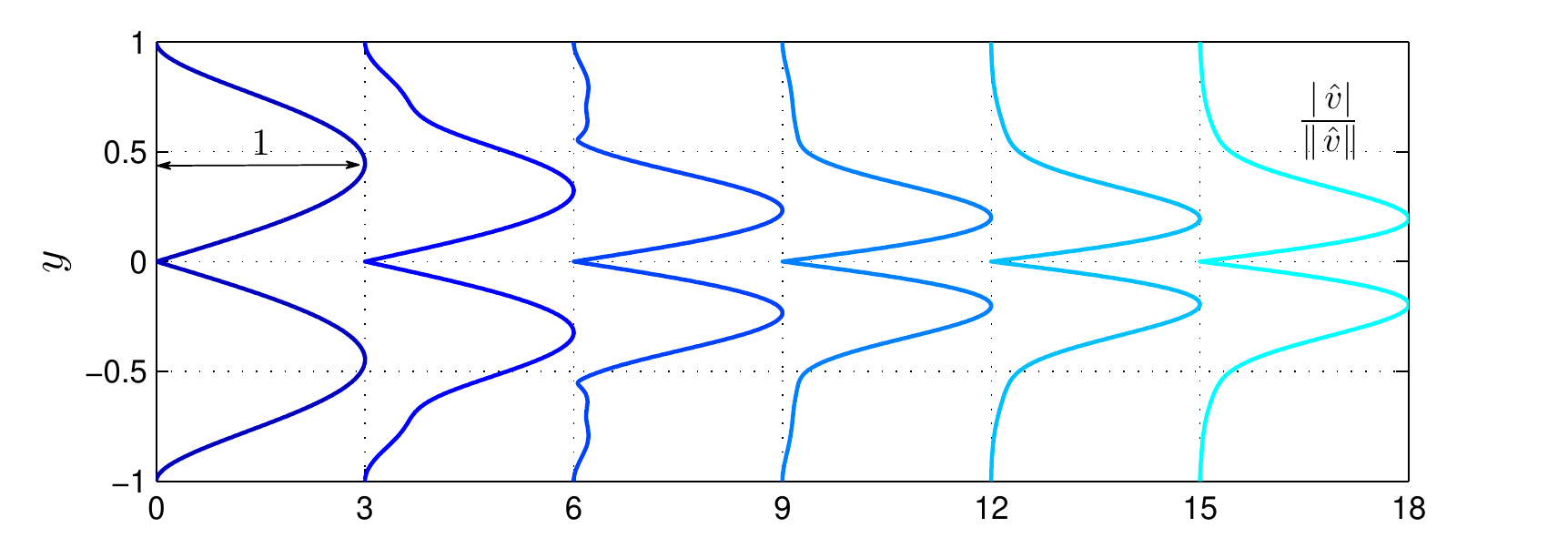}
	\vspace{-0.5cm}
	\label{fig:prof_PO_Re500_k6p5_early_2}
	 \end{subfigure}
	\begin{subfigure}{1\textwidth}
        \centering
\includegraphics[width=16.0cm]{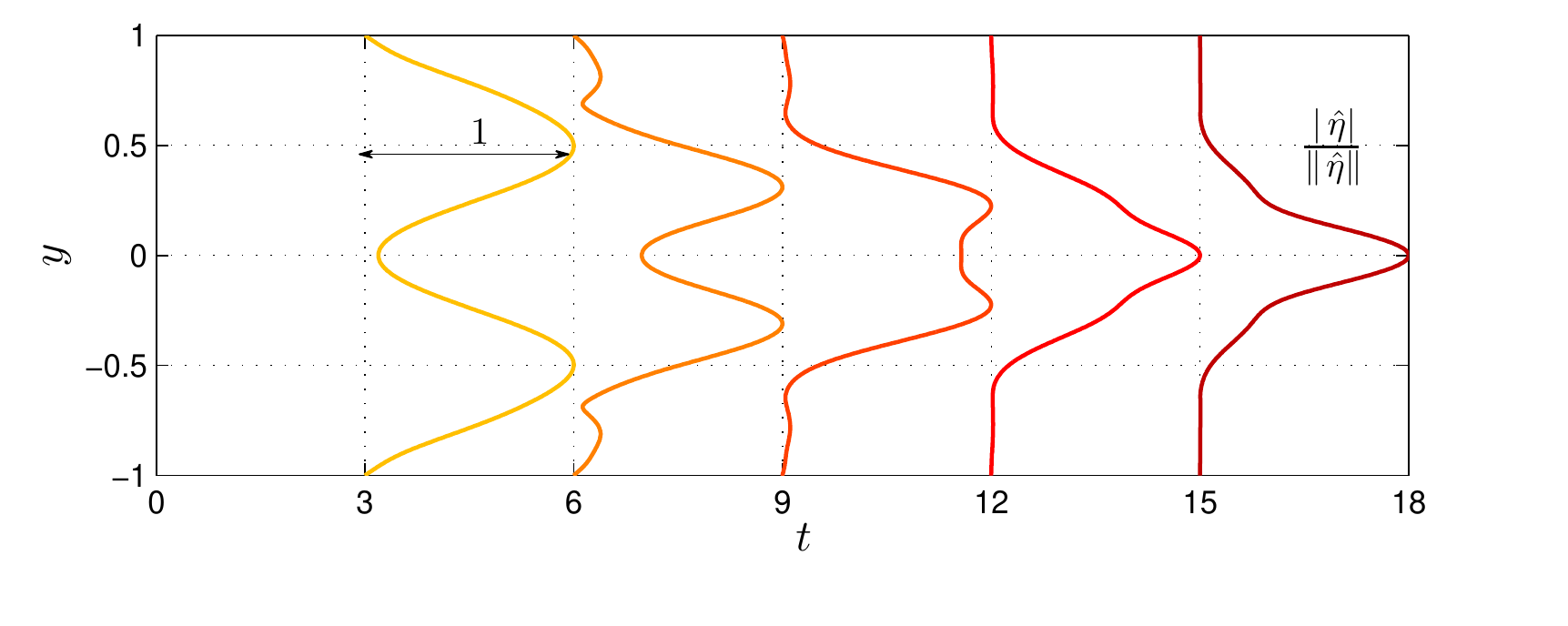}
	\vspace{-1.0cm}
	\label{fig:prof_PO_Re500_k6p5_early_3}
	 \end{subfigure}6
	\caption{Plane Poiseuille flow early transient, for
$Re=500$, $\phi=45^{\circ}$, $k=6.5$ and \textit{asym.} initial condition.
Top: phase velocity temporal evolution for the $\hv$ and $\he$ disturbance
(respectively, blue continuous line and blue dashed line) and first derivative
of the kinetic energy growth rate (green line). Middle: profiles of the modulus
of $\hv$, normalized with respect to the maximum ($L_\infty$-norm). Bottom:
profiles of
$|\he|/\| \he \|_\infty$. }
\label{fig:prof_PO_Re500_k6p5_early}
\end{figure}

\begin{figure}[h!]
        \centering
        \advance\leftskip-1.2cm
        \begin{subfigure}{1\textwidth}
        \centering
\includegraphics[width=16.0cm]{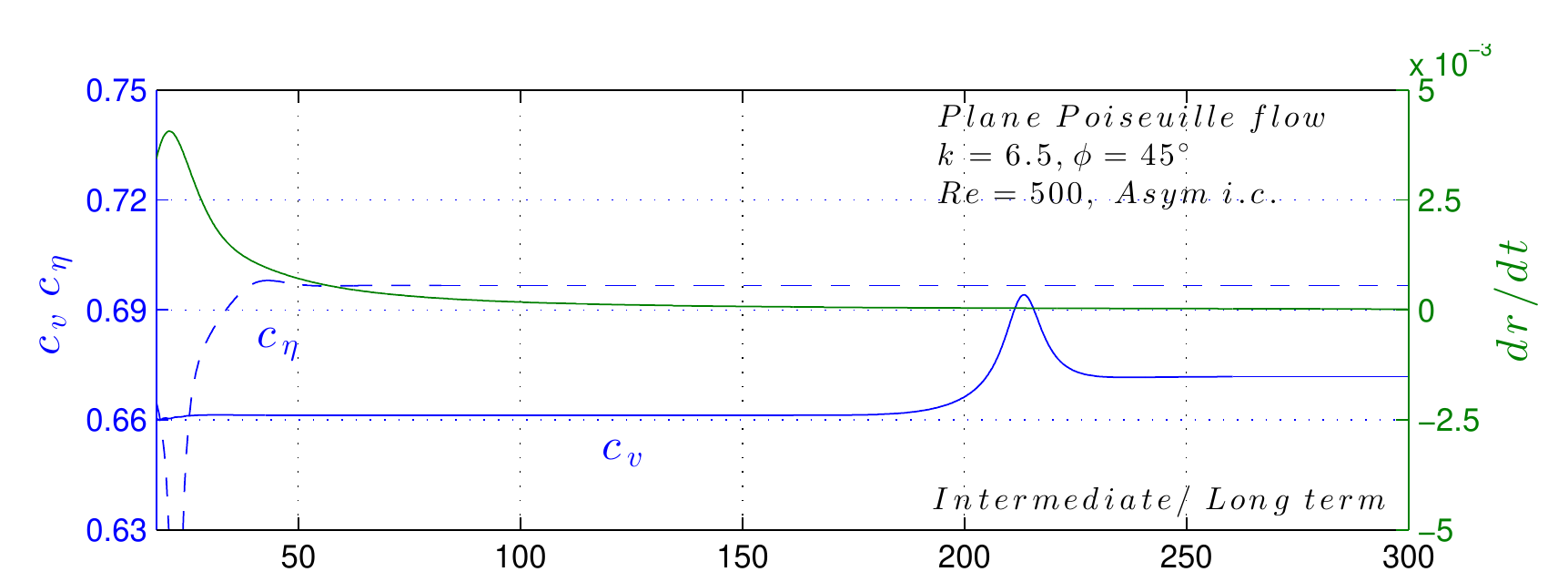}
	\vspace{-0.5cm}
	\label{fig:prof_PO_Re500_k6p5_far_1}
	 \end{subfigure}
        \begin{subfigure}{1\textwidth}
        \centering 
\includegraphics[width=16cm]{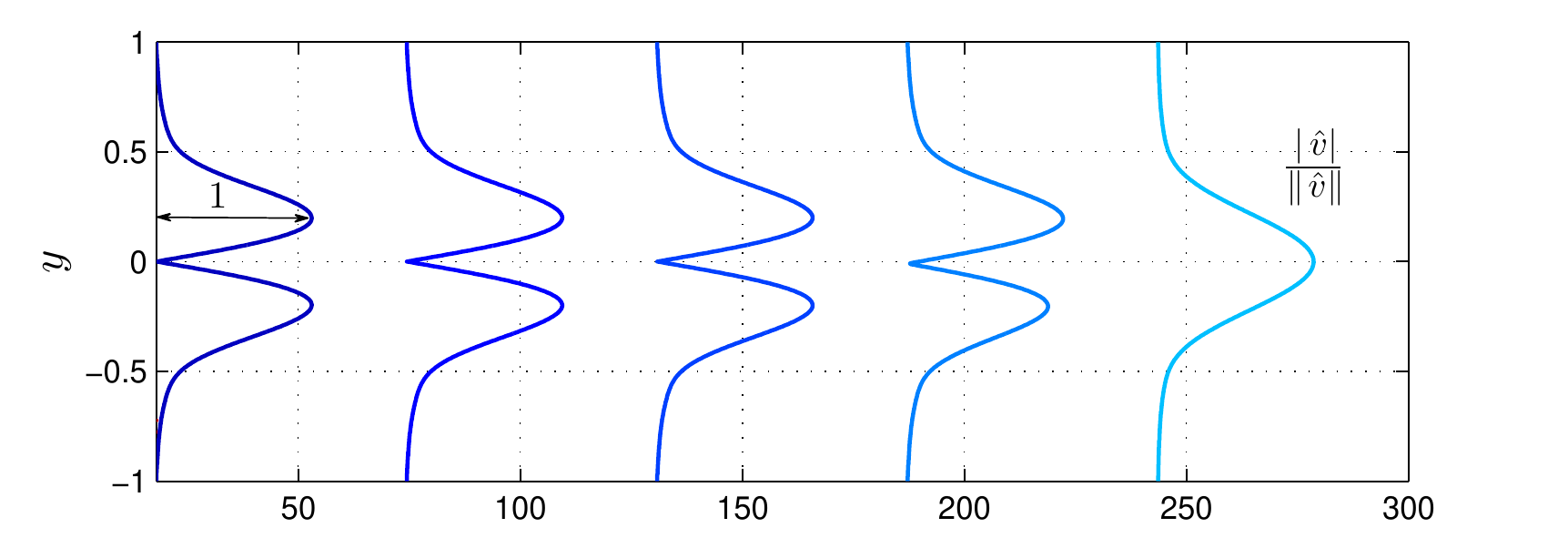}
	\vspace{-0.5cm}
	\label{fig:prof_PO_Re500_k6p5_far_2}
	 \end{subfigure}
	\begin{subfigure}{1\textwidth}
        \centering
\includegraphics[width=16.0cm]{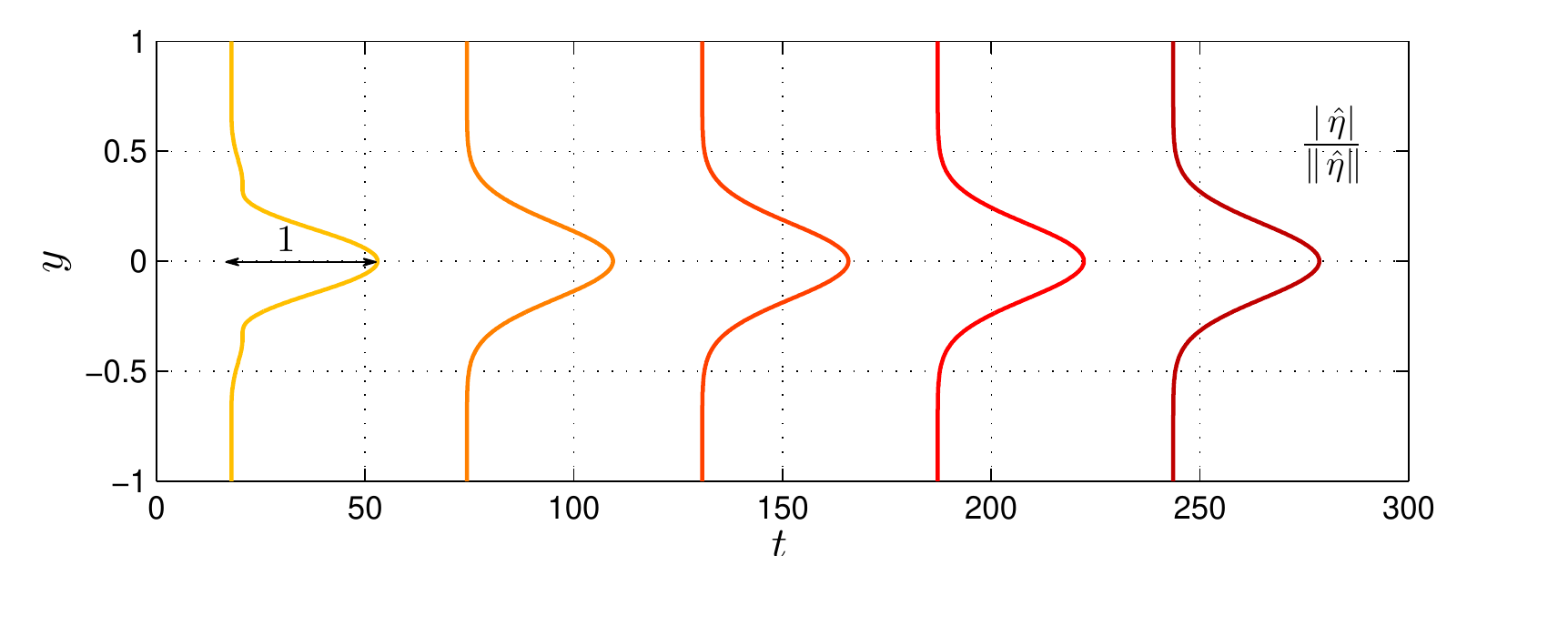}
	\vspace{-1.0cm}
	\label{fig:prof_PO_Re500_k6p5_far_3}
	 \end{subfigure}
	\caption{Plane Poiseuille flow intermediate and far term for $Re=500$,
$\phi=45^{\circ}$, $k=6.5$ and \textit{asym.} initial condition.
Top: phase velocity temporal evolution for the $\hv$ and $\he$ disturbance
(respectively, blue continuous line and blue dashed line) and first derivative
of the kinetic energy growth rate (green line). Middle: profiles of the modulus
of $\hv$, normalized with respect to the maximum ($L_\infty$-norm). Bottom:
profiles of
$|\he|/\| \he \|_\infty$. The difference from the case of
$\figref{fig:prof_PO_Re500_k6p5_far_sym}$ should be noticed: indeed, here
$|\hv|$ stabilizes to the final self-similar state after the vorticity
component. This state is announced by a the second transition of $c_v$ rather
than the ``usual'' one of $c_\eta$. This is likely due to the combination of the
symmetry properties of both the initial condition (here antisymmetrical) and the
asymptotic states.}
\label{fig:prof_PO_Re500_k6p5_far}
\end{figure}

\begin{figure}[h!]
        \centering
        \advance\leftskip-1.2cm
        \begin{subfigure}{1\textwidth}
        \centering
\includegraphics[width=16.0cm]{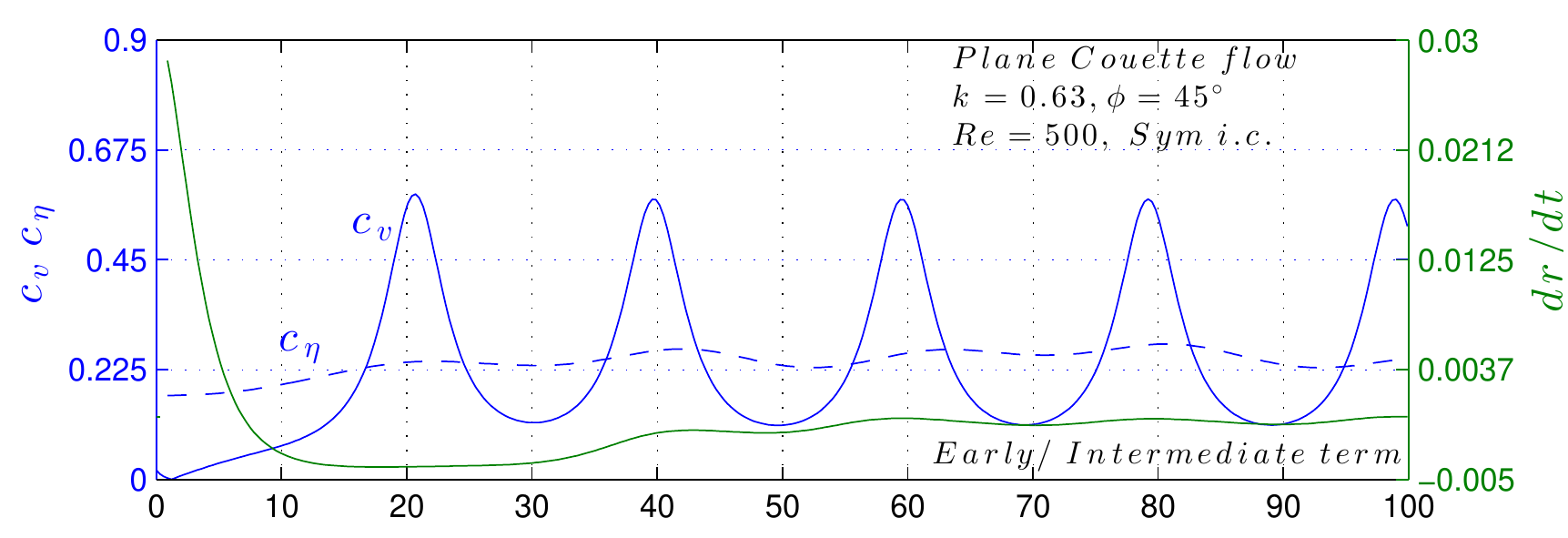}
	\vspace{-0.5cm}
	\label{fig:prof_CO_Re500_k0p63_far_1}
	 \end{subfigure}
        \begin{subfigure}{1\textwidth}
        \centering 
\includegraphics[width=16cm]{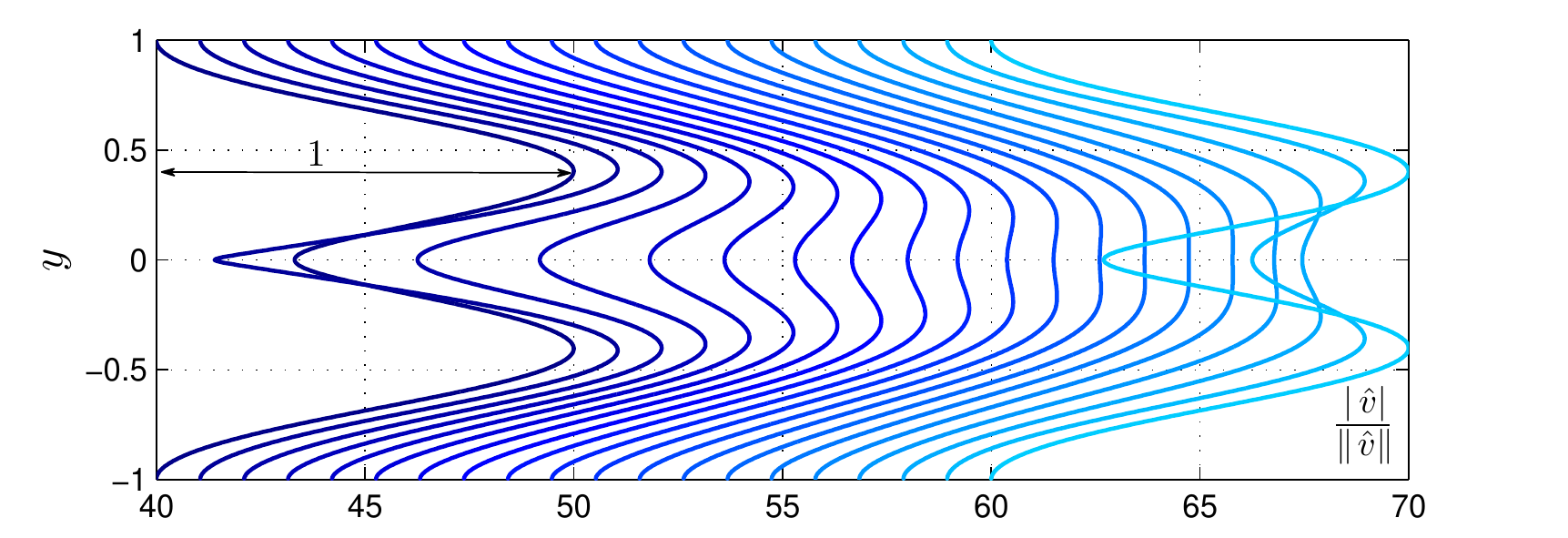}
	\vspace{-0.5cm}
	\label{fig:prof_CO_Re500_k0p63_far_2}
	 \end{subfigure}
	\begin{subfigure}{1\textwidth}
        \centering
\includegraphics[width=16.0cm]{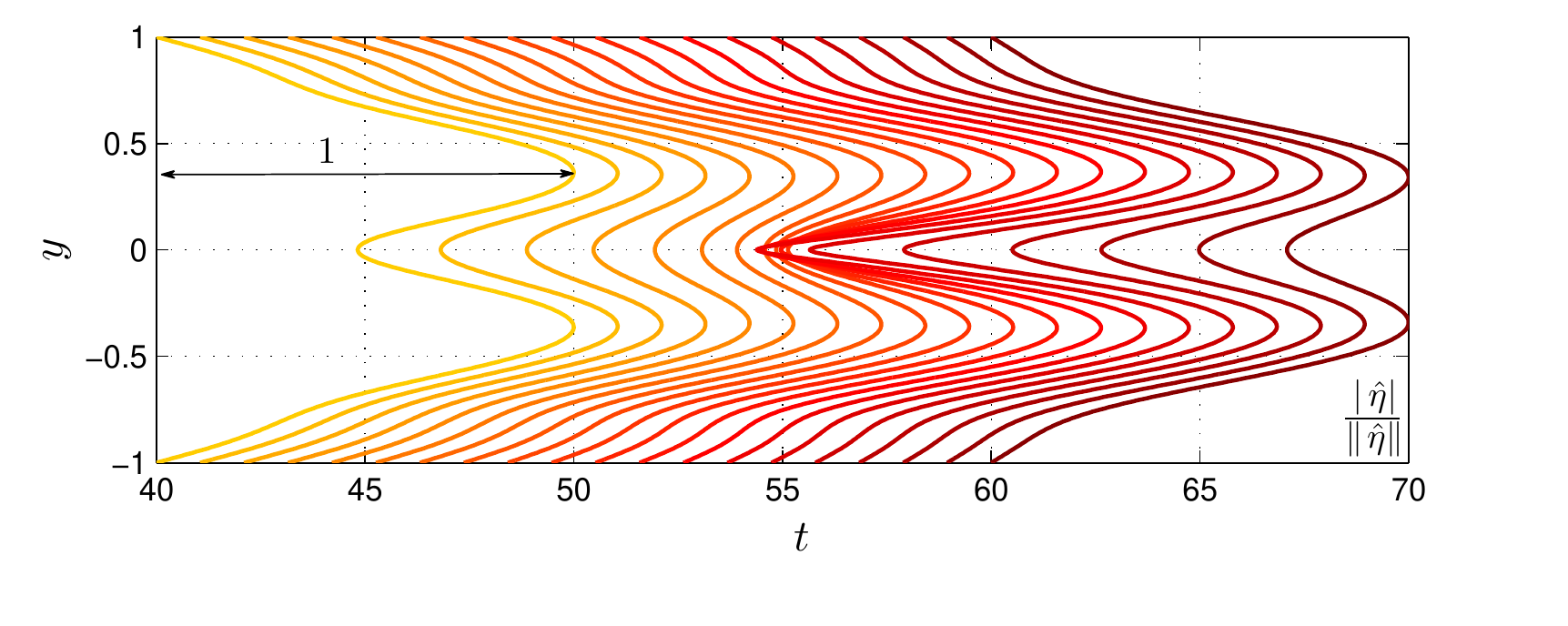}
	\vspace{-1.0cm}
	\label{fig:prof_CO_Re500_k0p63_far_3}
	 \end{subfigure}
	\caption{Plane Couette flow intermediate and far term for $Re=500$,
$\phi=45^{\circ}$, $k=0.63$ and \textit{sym.} initial condition.
Top: phase velocity temporal evolution for the $\hv$ and $\he$ disturbance
(respectively, blue continuous line and blue dashed line) and first derivative
of the kinetic energy growth rate (green line). Middle: profiles of the modulus
of $\hv$, normalized with respect to the maximum ($L_\infty$-norm). Bottom:
profiles of
$|\he|/\| \he \|_\infty$. This is an interesting case characterized by large
frequency oscillations; the modulation of phase velocity is related to a
periodic variation of the normalized profiles of both $|\hv|$ and $|\he|$. }
\label{fig:prof_CO_Re500_k0p63_far}
\end{figure}

In this section, the temporal evolution of the normal vorticity and velocity
profiles along the $y$ coordinate is investigated. We observe that the
frequency jumps previously introduced are strictly related to the spatial
distribution of the solutions $\hv$ and $\he$, i.e. to the distribution of the
complete flow field, in the wavenumber space. In figures
\ref{fig:prof_CO_Re500_k6p5_early}-\ref{fig:prof_CO_Re500_k0p63_far} the
velocity and vorticity profiles are reported, together with the phase velocity
time history and the evolution of the first derivative of the kinetic energy
growth rate. Actually, the quantities analyzed in the following are the modules
of the complex-valued solutions $\hv$ and $\he$
\begin{equation}
 |\hv|=\sqrt{\Re^2{\hv}+\Im^2{\hv}}\hspace{1.5 cm} 
|\he|=\sqrt{\Re^2{\he}+\Im^2{\he}}
\end{equation}
The module of the general quantity  in the wavenumber space can be related to
the solution in the physical space. In fact, taking advantage of linearity, the
inverse transform for a single wave reads \citep{Criminale_book}
\begin{gather}
\label{eq:invtrans_1}
\tilde{v}(x,y,z,t)=\frac{1}{2}\left[\hv(y,t)e^{i\alpha x+i\beta
z}+\hv^*(y,t)e^{-i\alpha x-i\beta z}\right] \\
\label{eq:invtrans_2}
\tilde{\eta}(x,y,z,t)=\frac{1}{2}\left[\he(y,t)e^{i\alpha x+i\beta
z}+\he^*(y,t)e^{-i\alpha x-i\beta z}\right]
\end{gather}
where the * sign represents the complex conjugate. Hence, the sum of the first
complex quantity at right hand side and its conjugate represents the real
disturbance quantity in the physical space; the same applies for $\tilde u$ and $\tilde w$,
derived from \eqref{eq:sol_u} and \eqref{eq:sol_w}. Since the complex conjugate
values can be easily obtained once $\hv$ and $\he$ are computed, this is a
convenient way to express the solution.\\
The explicit relation between the real and imaginary part of the solutions and
the quantities in the physical space is derived from the above expressions
\begin{gather}
\tilde{v}(x,y,z,t)=\Re \hv\ cos(\alpha x+\beta z)-\Im\hv\ sin(\alpha x+\beta z)
 \\
\tilde{\eta}(x,y,z,t)=\Re \he\ cos(\alpha x+\beta z)-\Im\he\ sin(\alpha x+\beta
z) 
\end{gather}
The profile along the coordinate $y$ of the module $|\hv|(y,t)$ or $|\he|(y,t)$
indicates the envelope of the maxima of $\tilde v$, or $\tilde \eta$, at a fixed
point $(x,z)$ 
\begin{gather}
 |\hv|(y,t_0)=\max_{x,z}\{\tilde{v}(x,y,z,t_0)\}\\
 |\he|(y,t_0)=\max_{x,z}\{\tilde{\eta}(x,y,z,t_0)\}
\end{gather}
The following figures show how the temporal evolution of the disturbance phase
velocity is closely related to the spatial distribution. To be more precise,
the solution in terms of modules seems to achieve a self-similarity in time,
when the frequency becomes constant. In fact, in these conditions the profiles
coincide if normalized with their $L_\infty$-norm (the maximum along $y$) or,
similarly,
with the $L_2$-norm. This means that the space-dependent and time-dependent
parts
of the solution are separable.  
 \begin{equation}
  \frac{|\hv|(y,t)}{(\max_y\hv)(t)}=f(y)\hspace{1.5cm}Self-similarity 
 \end{equation}
Usually the component of normal velocity is found to achieve this condition
after $T_j$, the time at which the first frequency jump occurs, as shown in
\figref{fig:prof_CO_Re500_k6p5_early}-\ref{fig:prof_CO_Re500_k6p5_far} for
Plane Couette flow,
and \figref{fig:prof_PO_Re500_k6p5_early_sym}-\ref{fig:prof_PO_Re500_k6p5_far}
for 
Plane Poiseuille flow. The vorticity profile continues to evolve until the
second phase velocity transition occurs, for $t=T_{j2}$. For the cited cases, we
observe that for PCf the $|\he|$ spatial distribution varies quite smoothly
(\figref{fig:prof_CO_Re500_k6p5_far}) while for PPf an abrupt variation in the
parity of the profile occurs (the double hump of the modules correspond to
odd profiles in the physical plane), as shown in
\figref{fig:prof_PO_Re500_k6p5_far_sym}.\par
An interesting case is shown
in \figref{fig:prof_PO_Re500_k6p5_early}-\ref{fig:prof_PO_Re500_k6p5_far}, where
the sudden profile change, and the associated second frequency jump are
experienced by the velocity component rather than the vorticity one. This is
probably due to the influence of the antisymmetrical initial condition on the early
and intermediate wave transient. This influence may be related to the parity of
the asymptotic state, which is independent on the initial condition.
It is also interesting to notice that the intermediate phase, starting after the
first jump, is usually very close to similarity conditions; in this term, the
phase velocities of the two signals are nearly coincident. Moreover, we
underline that the intermediate transient is, in addition to the early period,
the most relevant term in a perturbation's life. Indeed, in the introduced
cases $T_{j2}$  occurs when the wave kinetic energy is extremely small
(the last jump represents the beginning of the asymptotic conditions).\par
A connection between the periodic frequency modulation observed for Plane
Couette flow in \secref{sec:v_frequency} and the spatial distribution is pointed
out in \figref{fig:prof_CO_Re500_k0p63_far}; here it should be noticed that a
periodic continuous variation in the (normalized) profiles of $|\hv|$ and
$|\he|$ occurs. Actually, the periodic change happens in the channel
central region, while the near-wall region remains  unchanged and self-similar. The corresponding case in the physical
space is reported in \figref{fig:anti_CO_Re500_k0p63_phi45_sym_v}.

\FloatBarrier
\subsection{Maxima of kinetic energy for Plane Couette flow}
\begin{figure}[h!]
        \centering
         \begin{subfigure}{1\textwidth}
        \centering
\includegraphics[width=11.0cm]{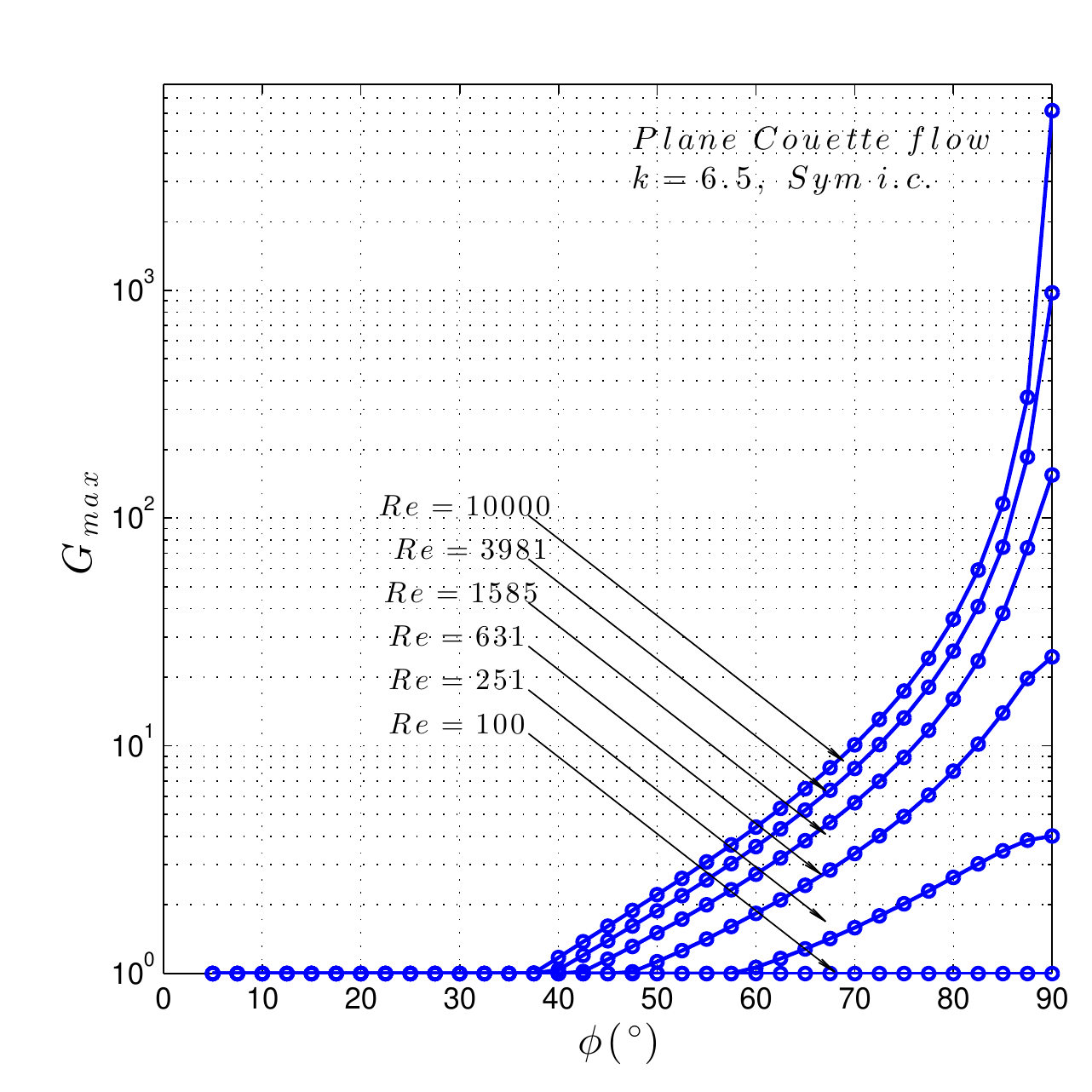}
	\vspace{-0.6cm}
	 \end{subfigure}
	  \begin{subfigure}{1\textwidth}
        \centering
\includegraphics[width=11.0cm]{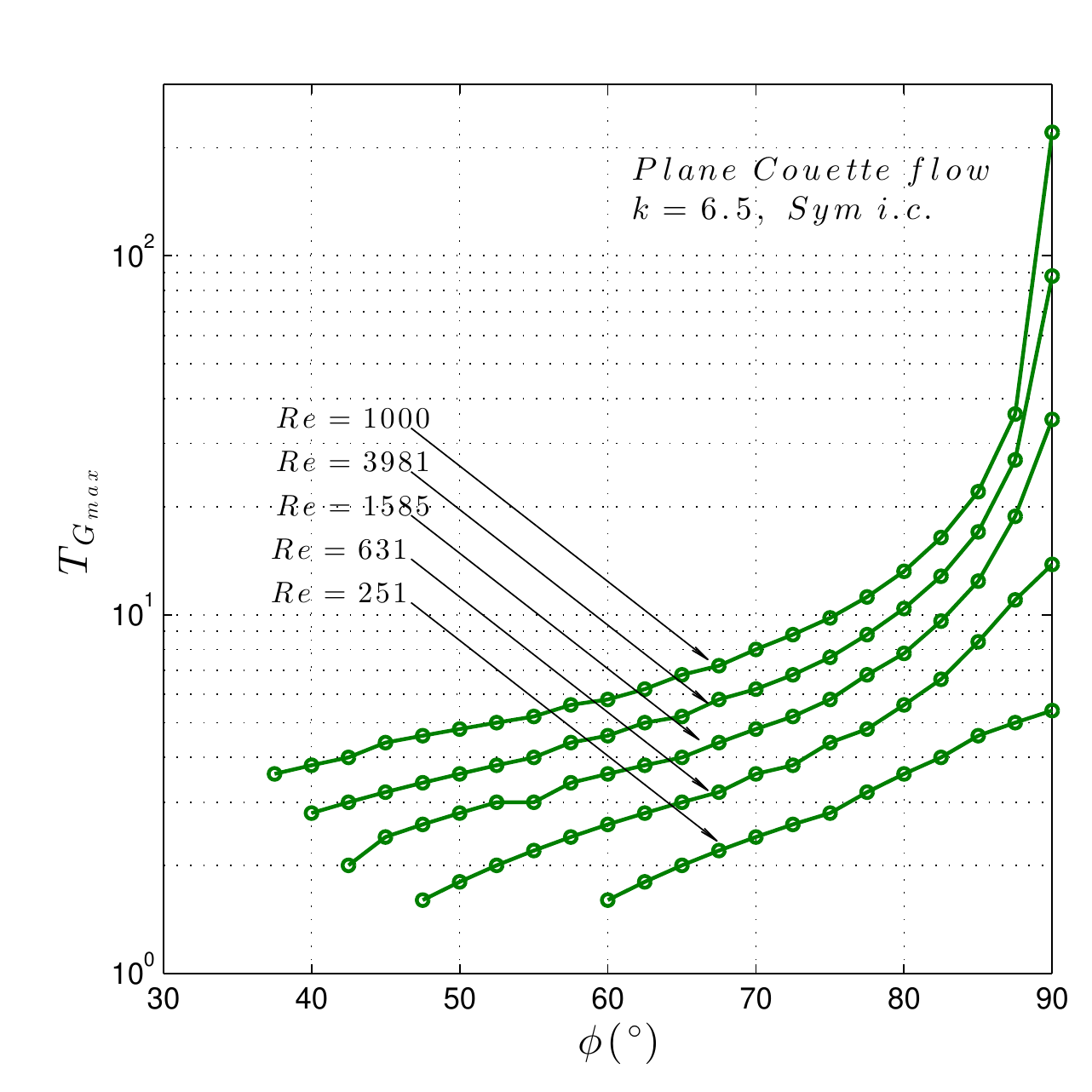}
	 \end{subfigure}
	 \caption{Upper plot: maxima of  the kinetic energy density $G$ as a function of the obliquity angle $\phi$,
parametrized with $Re$, for Plane Couette flow with $k=6.5$ and antisymmetrical initial condition. Lower plot:
nondimensional times corresponding the the maxima of $G$. The values of Reynolds number are uniformly distributed in the
logarithmic space.}
\label{fig:Gmax_CO_k6p5_sym_varioRe}
\end{figure}

\begin{figure}[h!]
        \centering
         \begin{subfigure}{1\textwidth}
        \centering
\includegraphics[width=11.0cm]{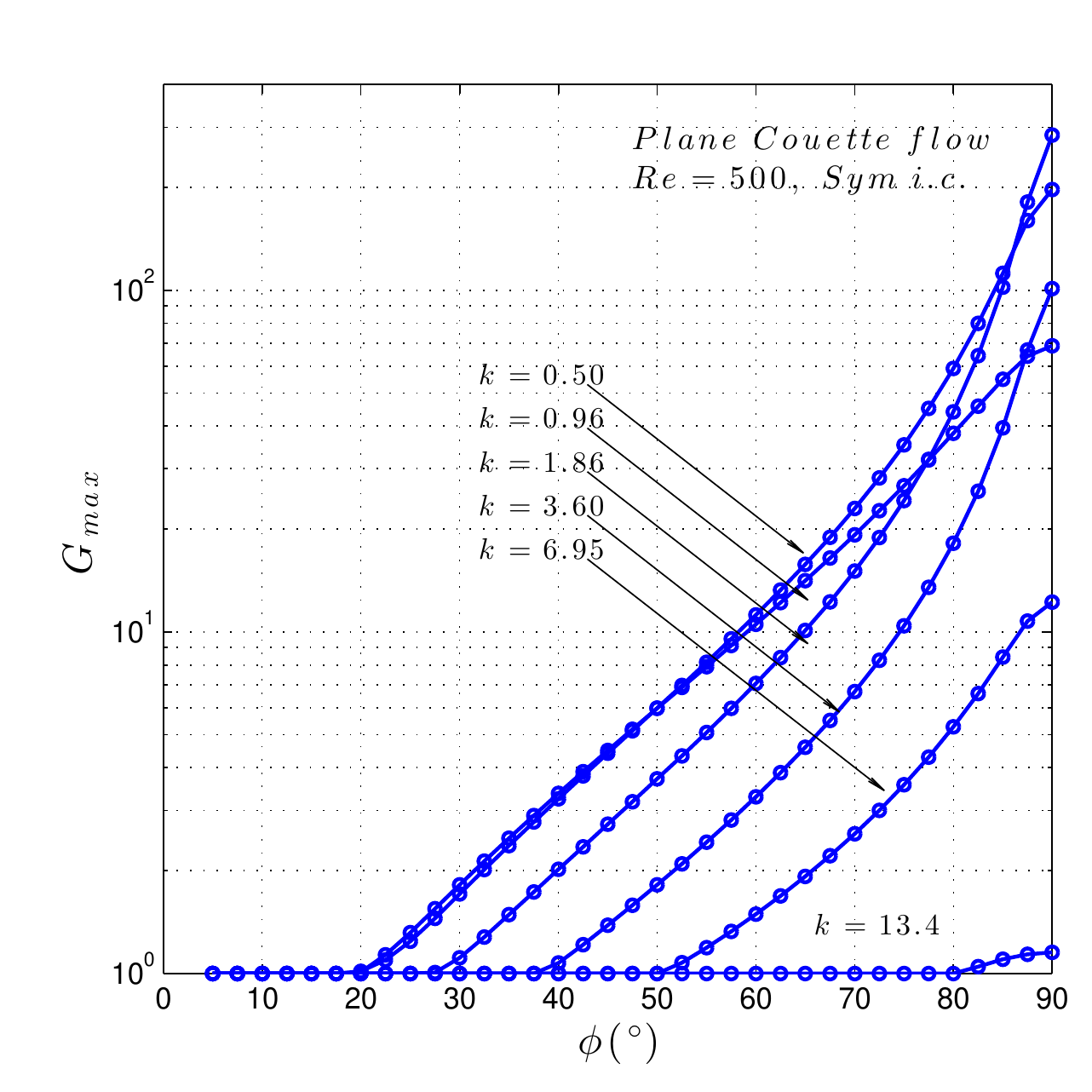}
	\vspace{-0.6cm}
	 \end{subfigure}
	  \begin{subfigure}{1\textwidth}
        \centering
\includegraphics[width=11.0cm]{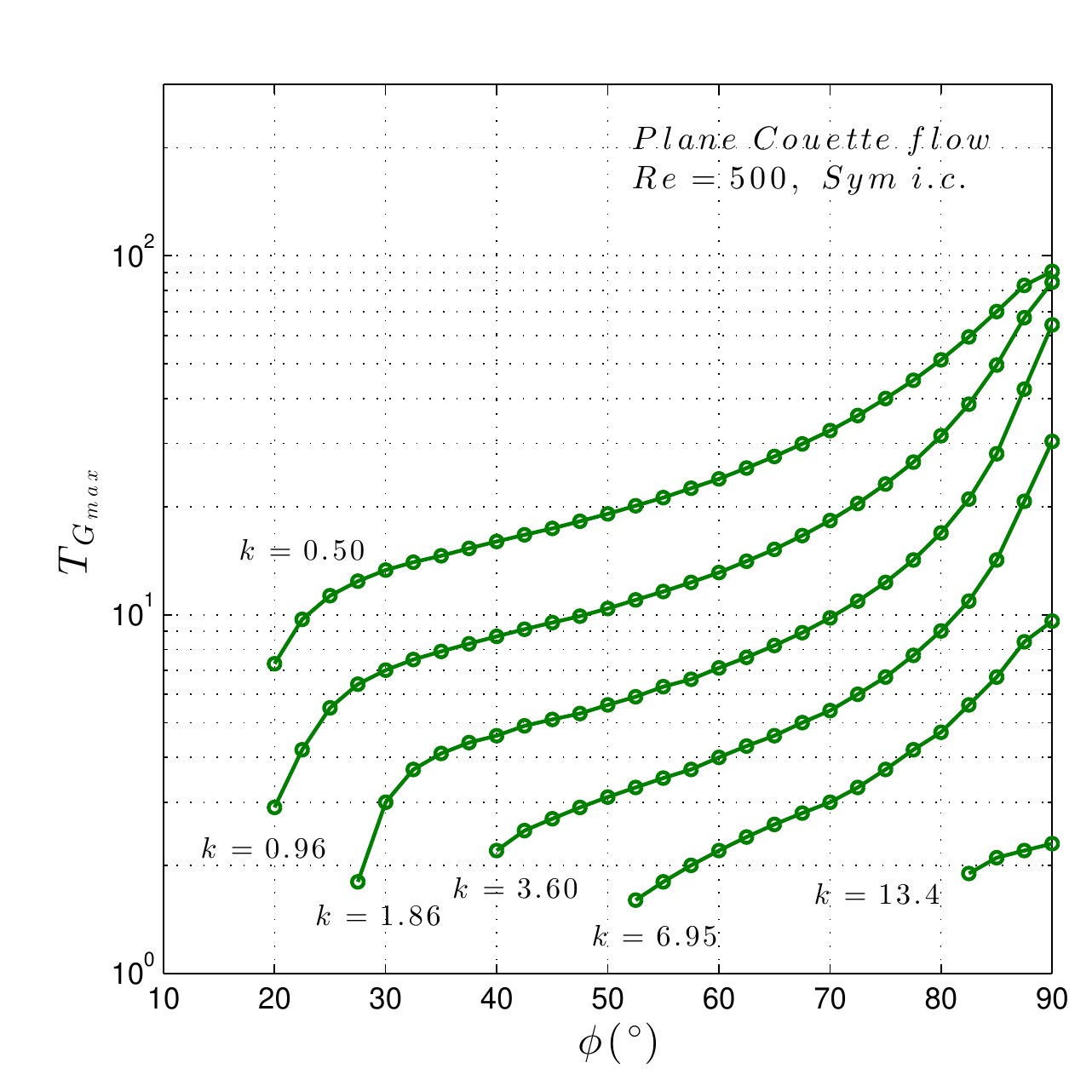}
	 \end{subfigure}
	 \caption{Upper plot: maxima of  the kinetic energy density $G$ as a function of the obliquity angle $\phi$,
parmetrized with $k$, for Plane Couette flow with $Re=500$ and symmetrical initial condition. Lower plot:
nondimensional times corresponding the the maxima of $G$. The values of polar wavenumber are uniformly distributed in
the
logarithmic space.}
\label{fig:Gmax_CO_k6p5_sym_variok}
\end{figure}

\begin{figure}[h!]
        \centering
         \begin{subfigure}{1\textwidth}
        \centering
\includegraphics[width=11.0cm]{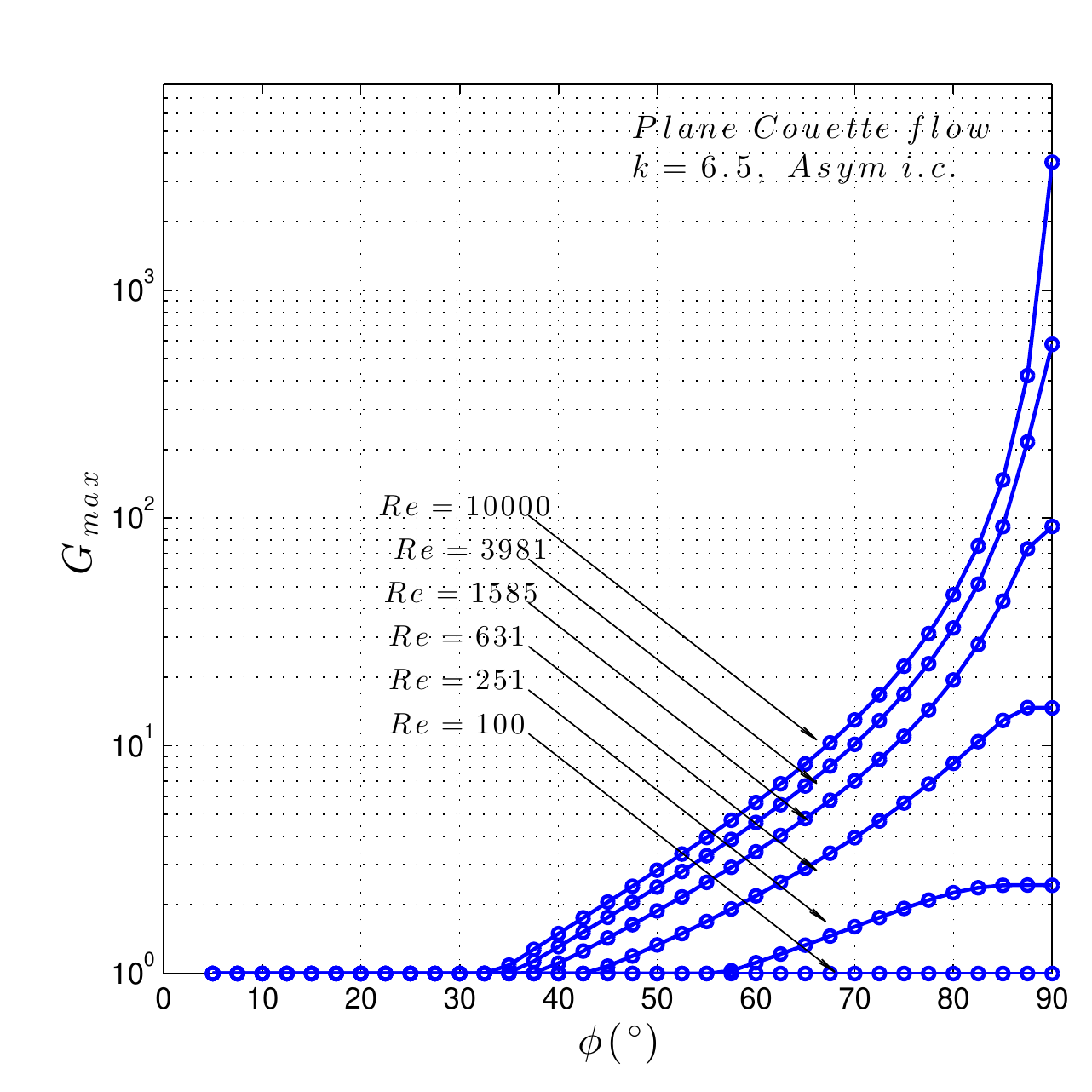}
	\vspace{-0.6cm}
	 \end{subfigure}
	  \begin{subfigure}{1\textwidth}
        \centering
\includegraphics[width=11.0cm]{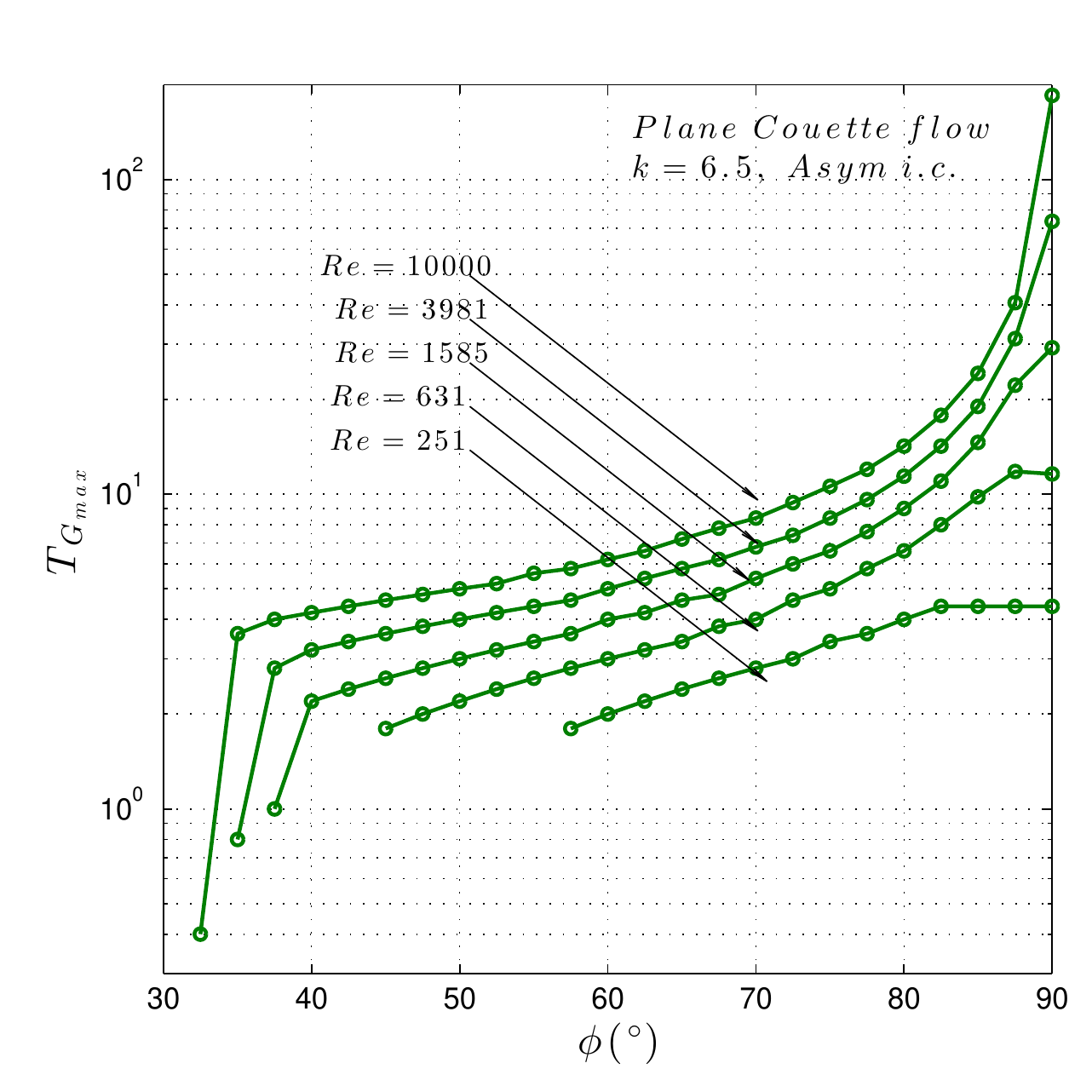}
	 \end{subfigure}
	 \caption{Upper plot: maxima of  the kinetic energy density $G$ as a function of the obliquity angle $\phi$,
parametrized with $Re$, for Plane Couette flow with $k=6.5$ and antisymmetrical initial condition. Lower plot:
nondimensional times corresponding the the maxima of $G$. The values of Reynolds number are uniformly distributed in the
logarithmic space.}
\label{fig:Gmax_CO_k6p5_asym_varioRe}
\end{figure}

\begin{figure}[h!]
        \centering
         \begin{subfigure}{1\textwidth}
        \centering
\includegraphics[width=11.0cm]{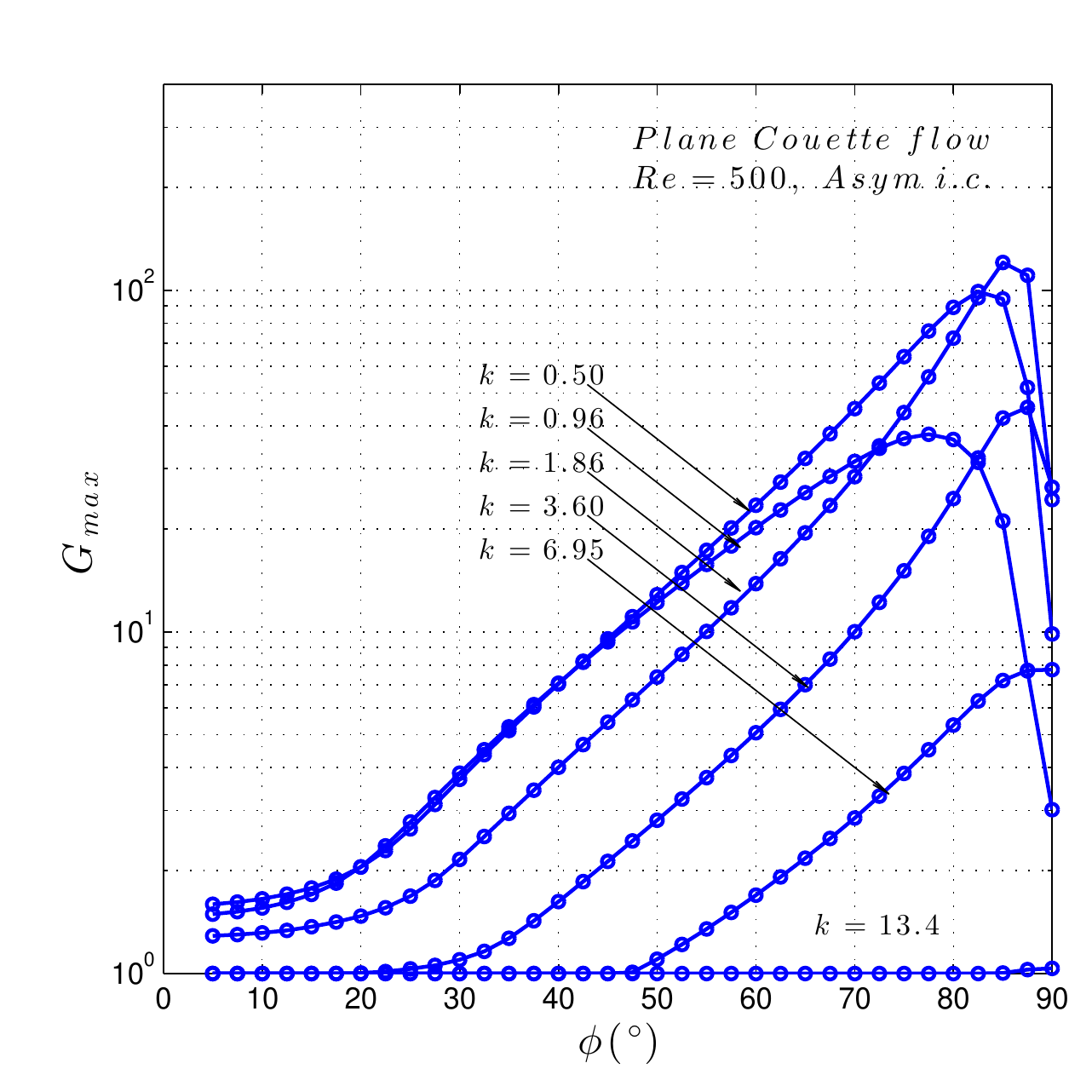}
	\vspace{-0.6cm}
	 \end{subfigure}
	  \begin{subfigure}{1\textwidth}
        \centering
\includegraphics[width=11.0cm]{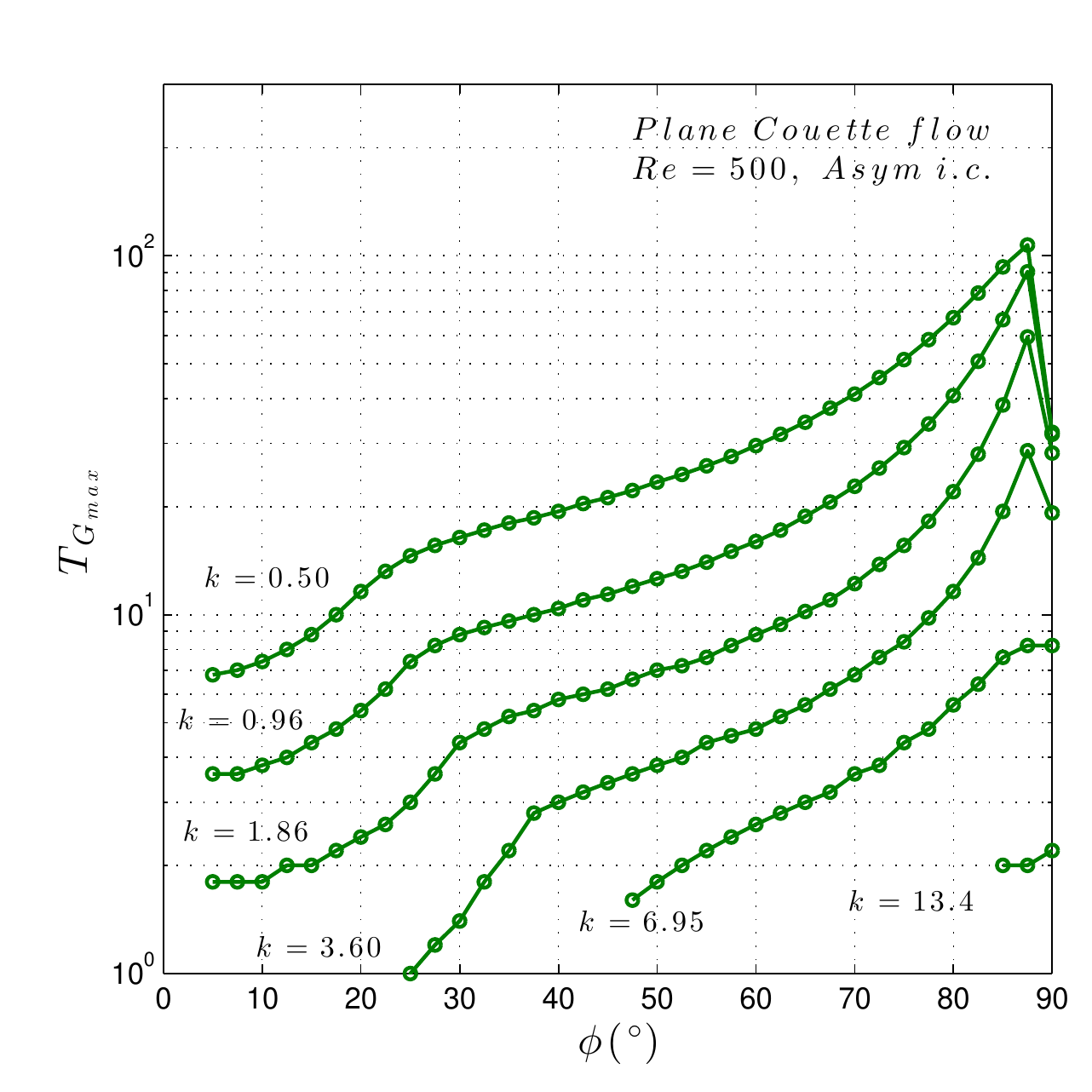}
	 \end{subfigure}
	 \caption{Upper plot: maxima of  the kinetic energy density $G$ as a function of the obliquity angle $\phi$,
parmetrized with $k$, for Plane Couette flow with $Re=500$ and antisymmetrical initial condition. Lower plot:
nondimensional times corresponding the the maxima of $G$. The values of polar wavenumber are uniformly distributed in
the
logarithmic space.}
\label{fig:Gmax_CO_k6p5_asym_variok}
\end{figure}

Even if the focus of the chapter is on the wave frequency and
the similarity properties of the velocity profiles, it is thought to be appropriate to include 
this little paragraph about the maxima gained by the kinetic energy during the perturbation's life. The maps of figures
\ref{fig:Gmax_CO_k6p5_sym_varioRe}-\ref{fig:Gmax_CO_k6p5_asym_variok}, together with the evolution of the
real normalized velocity and vorticity fields introduced in the next paragraph, contribute to gain understanding of the
complete scenario.\par
It is known that in the early and intermediate terms even large transient growths can be experienced by the components
of flow velocity, vorticity, and by the kinetic energy. The normalized kinetic energy density $G$ defined in
\secref{sec:G} can effectively measure the transient growth for a perturbation with prescribed initial condition.
Following the definition by \citet{Criminale_book}, an asymptotically stable configuration is called
\textit{algebraically unstable} if $G>0$ for some $t>0$; \textit{algebraically stable} if $G<0$ for all time;
\textit{algebraically neutral} if $G=0$ for all time. The reasons for the algebraic growth are mainly three. First, the
non-orthogonality of the eigenfunctions, as shown by \citet{Schmid_book}. Secondly, a possible resonance between the
Orr-Sommerfeld and the Squire damped exponential modes can occur, as shown by \citet{Benney1981}. However, the
resonance does not occur for the boundary layer. The last reason deals with the presence of a continuous spectrum (so,
it only applies to unbounded flows), see the work by \citet{Criminale1990}.\par
In the following, the maxima of $G$ are traced as a function of the obliquity angle. The nondimensional time at which
the maxima occurs is reported as well. Curves for six values of Reynolds number (\figref{fig:Gmax_CO_k6p5_sym_varioRe}
and \figref{fig:Gmax_CO_k6p5_asym_varioRe}) and polar wavenumber (\figref{fig:Gmax_CO_k6p5_sym_variok}
and \figref{fig:Gmax_CO_k6p5_asym_variok}) are shown, for both the symmetrical (\figref{fig:Gmax_CO_k6p5_sym_varioRe}
and \figref{fig:Gmax_CO_k6p5_sym_variok}) and the antisymmetrical initial condition
(\figref{fig:Gmax_CO_k6p5_asym_varioRe}
and \figref{fig:Gmax_CO_k6p5_asym_variok}). It is interesting to notice that, for fixed $Re$ and $k$ , it is not
generally true that the maximum occurs for $\phi=90^{\circ}$. This is evident from
\figref{fig:Gmax_CO_k6p5_asym_variok}.

\FloatBarrier

\ \newpage
\FloatBarrier
\subsection{Wave solutions in the physical space}
\begin{figure}[h!]
        \centering
        \advance\leftskip-1.2cm
        \begin{subfigure}{1\textwidth}
        \centering
\includegraphics[width=16.0cm]{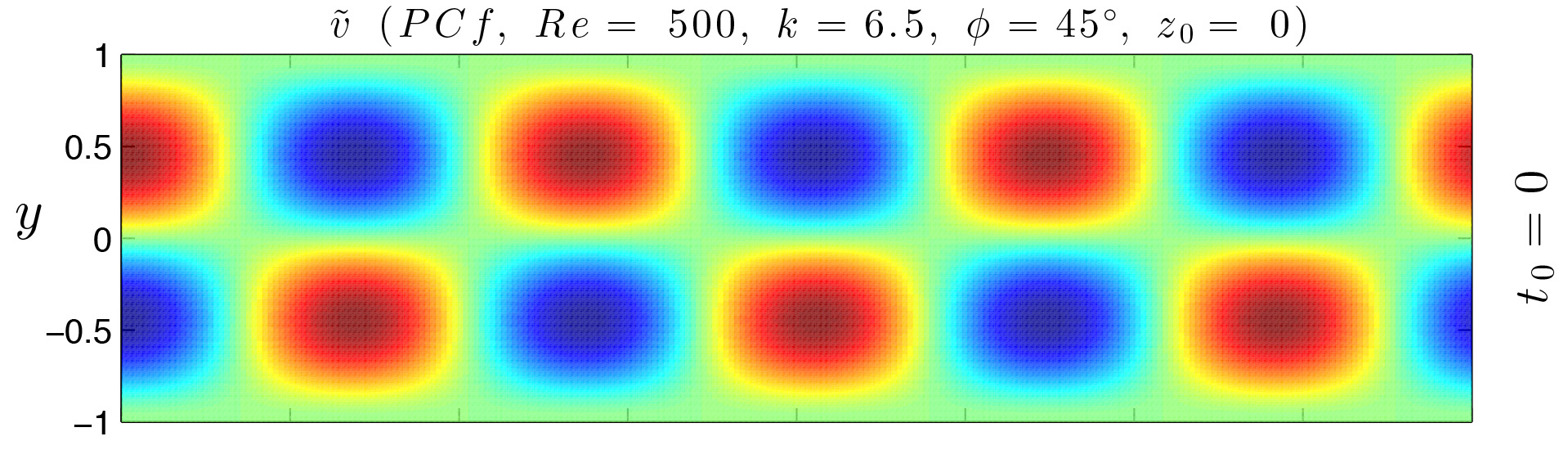}
	\vspace{-0.9cm}
	 \end{subfigure}
        \begin{subfigure}{1\textwidth}
        \centering 
\includegraphics[width=16cm]{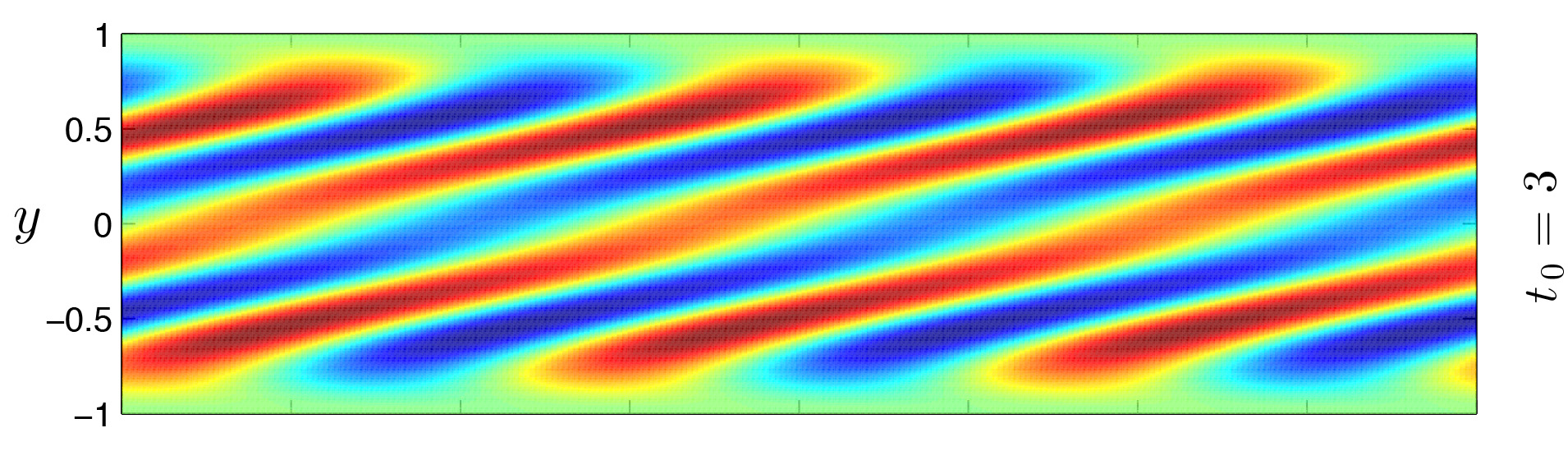}
	\vspace{-0.9cm}
	 \end{subfigure}
	\begin{subfigure}{1\textwidth}
        \centering
\includegraphics[width=16.0cm]{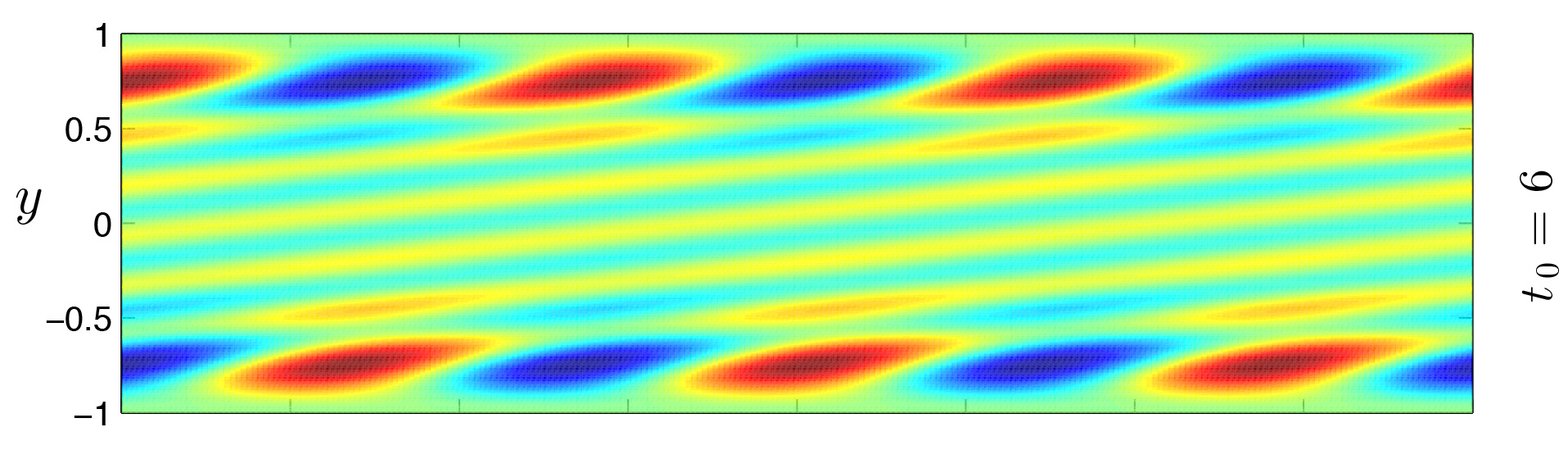}
	\vspace{-0.6cm}
	 \end{subfigure}
	 \begin{subfigure}{1\textwidth}
        \centering
\includegraphics[width=16.0cm]{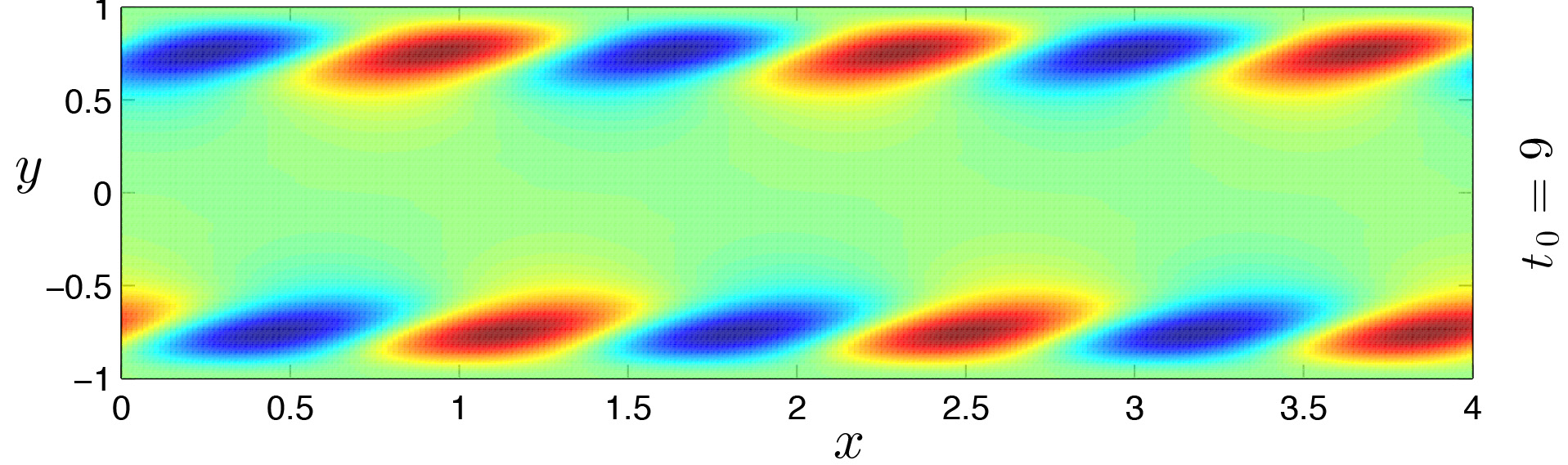}
	\vspace{-0.6cm}
	 \end{subfigure}
	\caption{Visualization of the wall-normal perturbation velocity $\tilde
v(x,y,z_0,t_0)$, $xy$ plane, for Plane Couette flow with the same parameters of
\figref{fig:prof_CO_Re500_k6p5_early}. It should be noticed that the
$x$-component of the phase velocity takes the same sign of the base flow. Remind that for the adopted
conventions the base flow is oriented as the longitudinal axis $x$; with reference to the figure, the upper wall moves
to the right, the lower to the left. \textit{Red}: maximum (positive); \textit{blue}: minimum (negative).}
\label{fig:anti_CO_Re500_k6p5_phi45_asy}
\end{figure}

\begin{figure}[h!]
        \centering
        \advance\leftskip-1.2cm
        \begin{subfigure}{1\textwidth}
        \centering
\includegraphics[width=16.0cm]{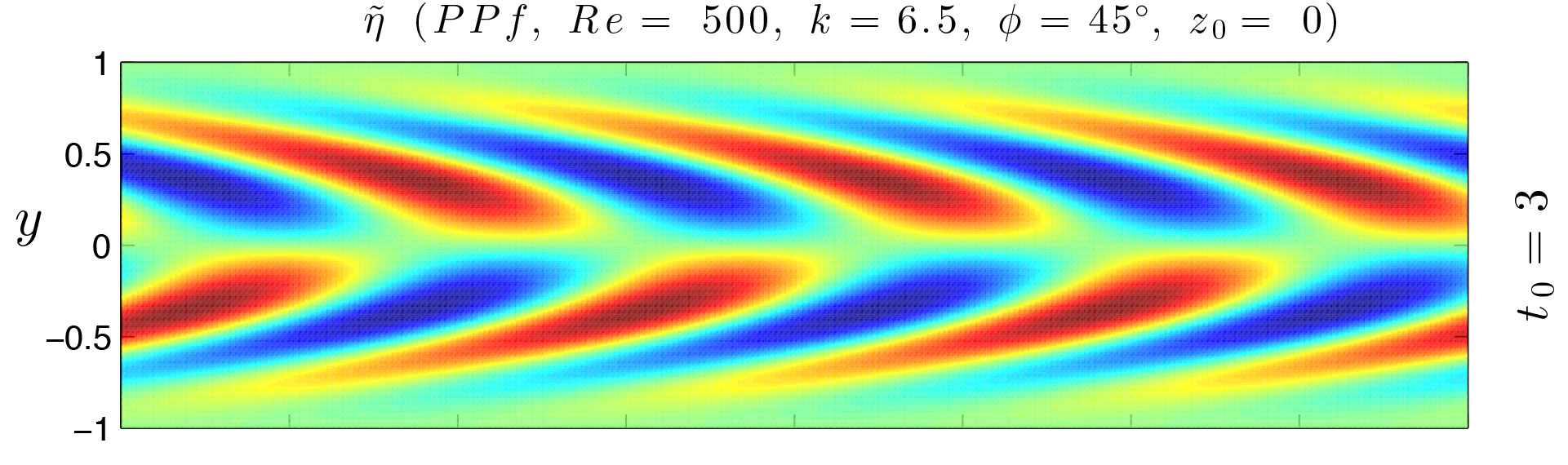}
	\vspace{-0.9cm}
	 \end{subfigure}
        \begin{subfigure}{1\textwidth}
        \centering 
\includegraphics[width=16cm]{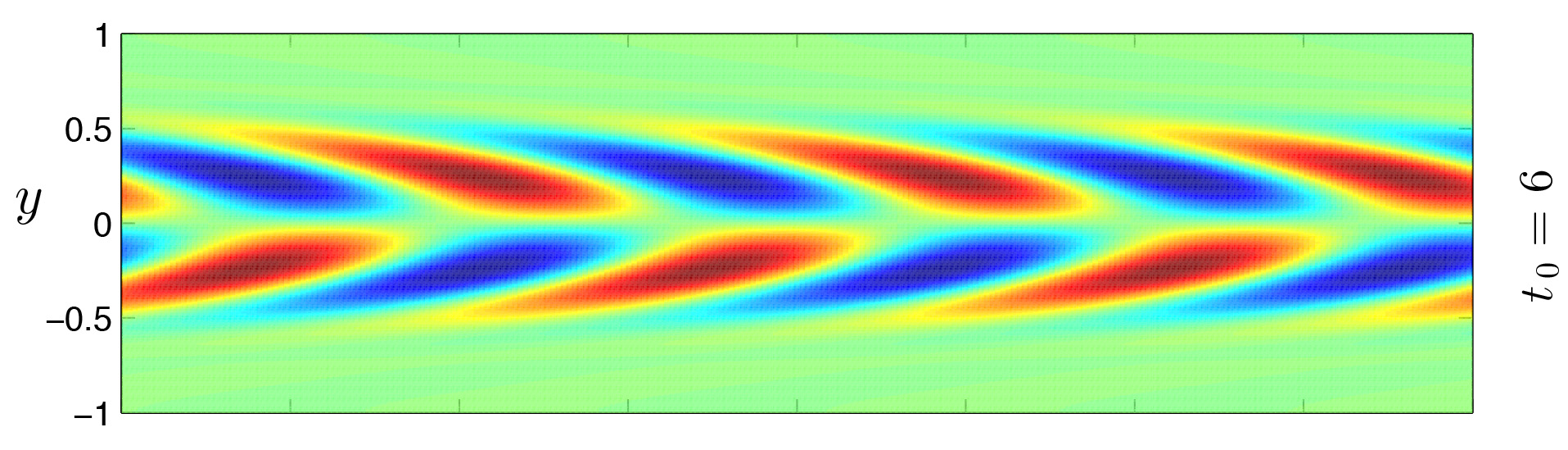}
	\vspace{-0.9cm}
	 \end{subfigure}
	\begin{subfigure}{1\textwidth}
        \centering
\includegraphics[width=16.0cm]{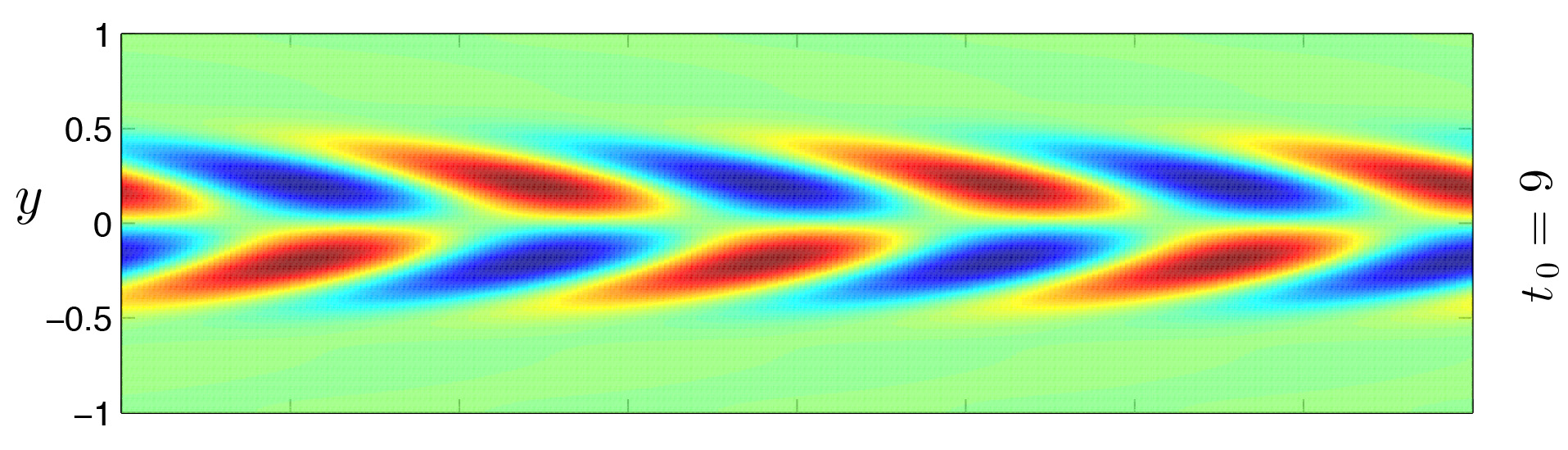}
	\vspace{-0.6cm}
	 \end{subfigure}
	 \begin{subfigure}{1\textwidth}
        \centering
\includegraphics[width=16.0cm]{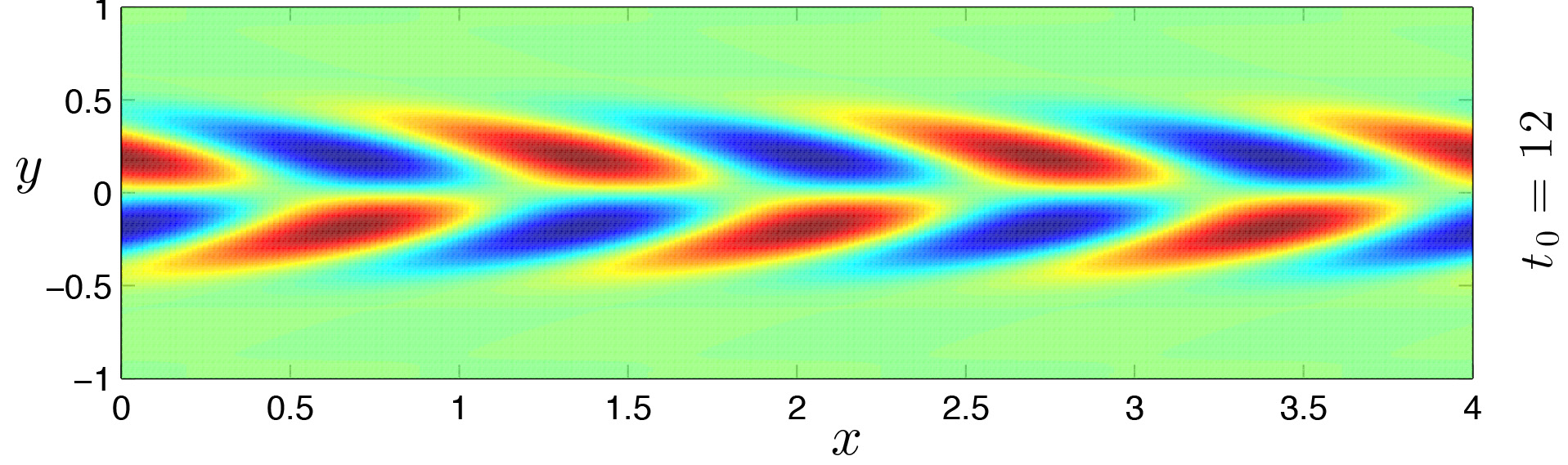}
	\vspace{-0.6cm}
	 \end{subfigure}
	\caption{Visualization of the wall-normal perturbation vorticity $\tilde
\eta(x,y,z_0,t_0)$, $xy$ plane, for Plane Poiseuille flow. The same parameters
of \figref{fig:prof_PO_Re500_k6p5_early} are set, to allow a comparison. From these slices, the disturbance seems to
move in the streamwise direction, meaning that the $x$-component of phase velocity has the same sign of the base flow,
as observed in Couette flow. Remind that the disturbance direction is defined by the polar wavenumber vector.
\textit{Red}:
maximum (positive); \textit{blue}: minimum (negative). }
\label{fig:anti_PO_Re500_k6p5_phi45_asy_eta_early}
\end{figure}

\begin{figure}[h!]
        \centering
        \advance\leftskip-1.2cm
        \begin{subfigure}{1\textwidth}
        \centering
\includegraphics[width=16.0cm]{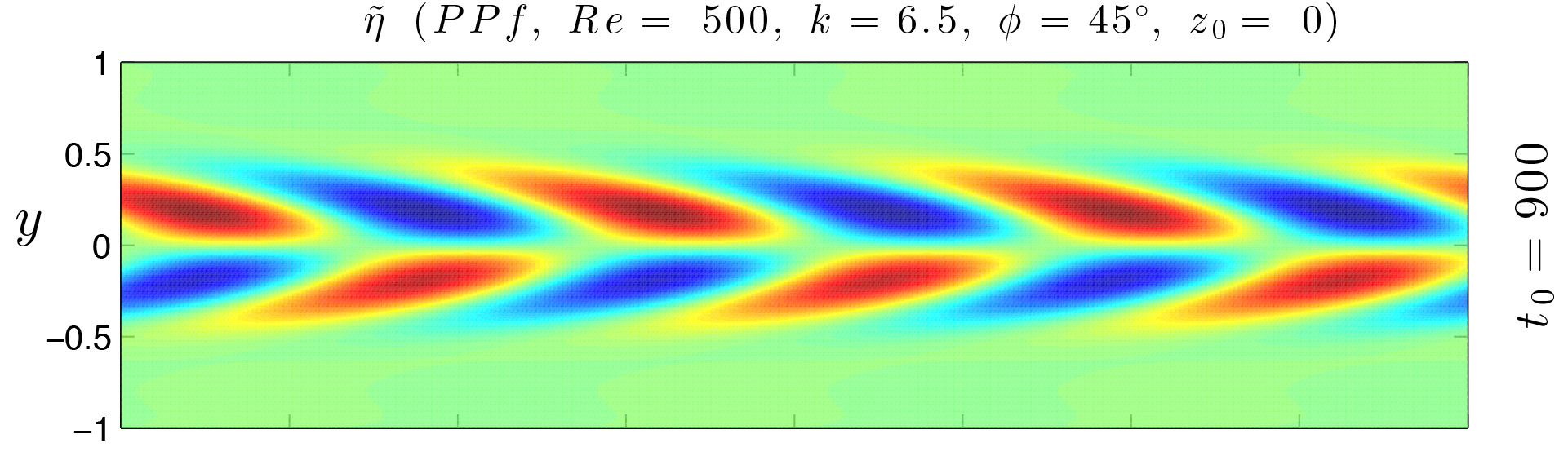}
	\vspace{-0.9cm}
	 \end{subfigure}
        \begin{subfigure}{1\textwidth}
        \centering 
\includegraphics[width=16cm]{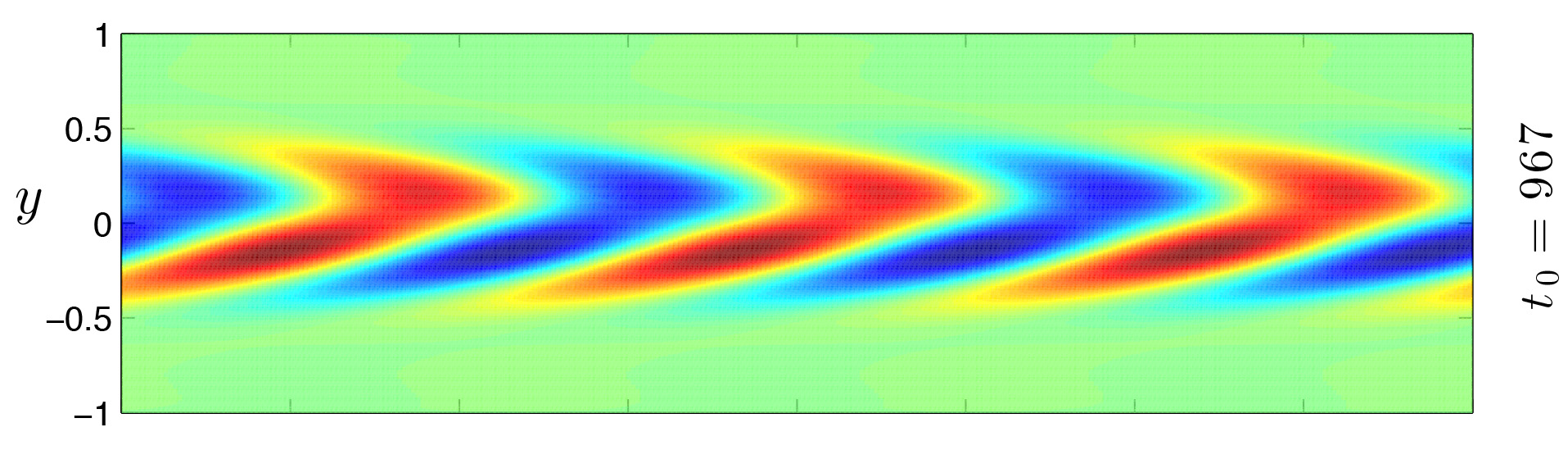}
	\vspace{-0.9cm}
	 \end{subfigure}
	\begin{subfigure}{1\textwidth}
        \centering
\includegraphics[width=16.0cm]{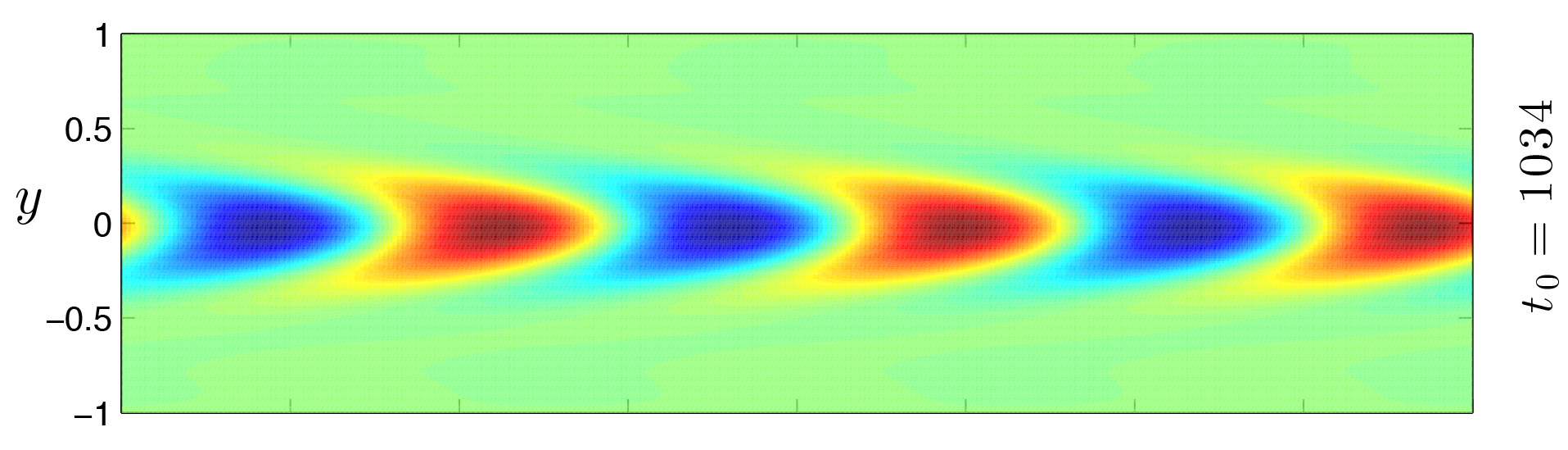}
	\vspace{-0.6cm}
	 \end{subfigure}
	 \begin{subfigure}{1\textwidth}
        \centering
\includegraphics[width=16.0cm]{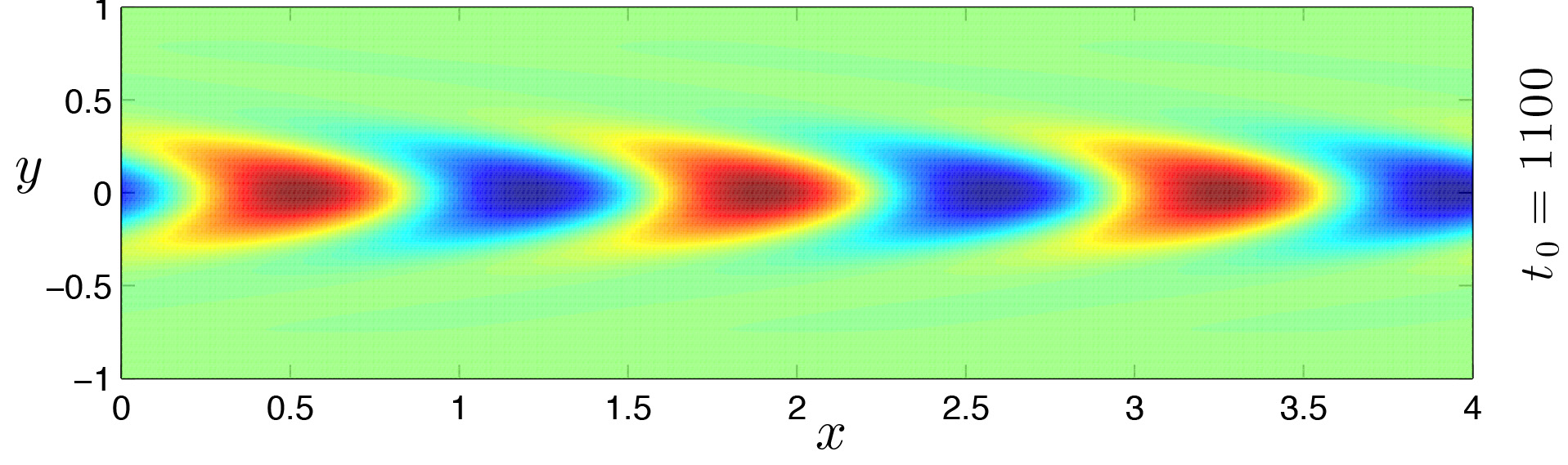}
	\vspace{-0.6cm}
	 \end{subfigure}
	\caption{Visualization of the wall-normal perturbation vorticity $\tilde
\eta(x,y,z_0,t_0)$, $xy$ plane, for Plane Poiseuille flow. This case correspond to \figref{fig:prof_PO_Re500_k6p5_far},
and the effects of the transition to the asymptotic conditions, through the second frequency jump of $\omega_\eta$, is
shown. \textit{Red}: maximum (positive); \textit{blue}: minimum (negative). }
\label{fig:anti_PO_Re500_k6p5_phi45_asy_eta_far}
\end{figure}

\begin{figure}[h!]
        \centering
        \advance\leftskip-1.2cm
        \begin{subfigure}{1\textwidth}
        \centering
\includegraphics[width=16.0cm]{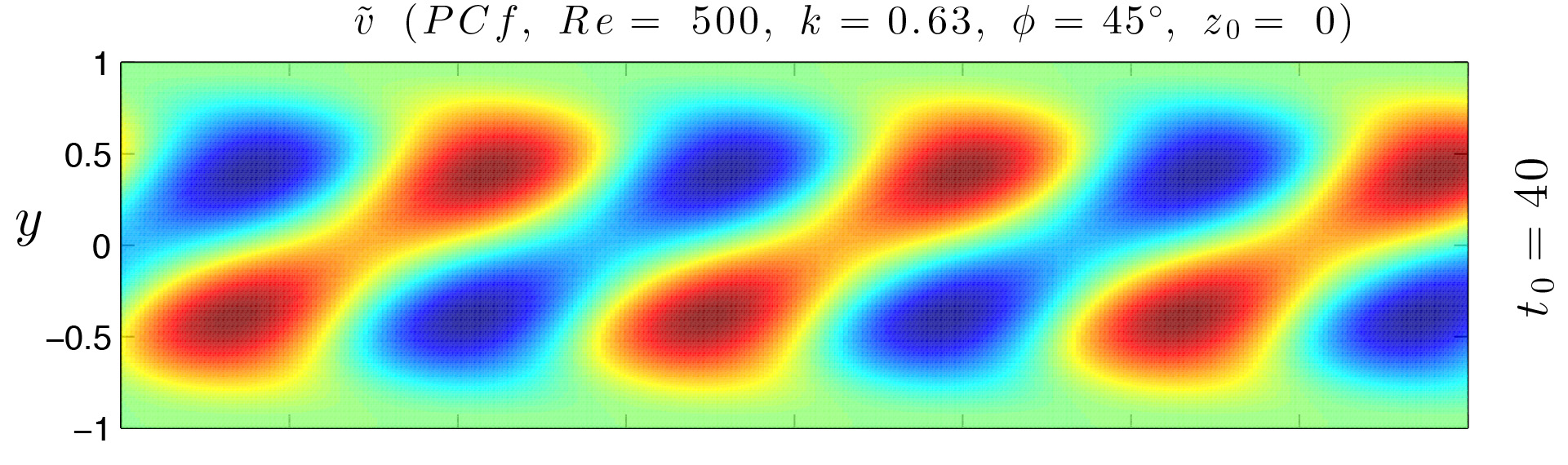}
	\vspace{-0.9cm}
	 \end{subfigure}
        \begin{subfigure}{1\textwidth}
        \centering 
\includegraphics[width=16cm]{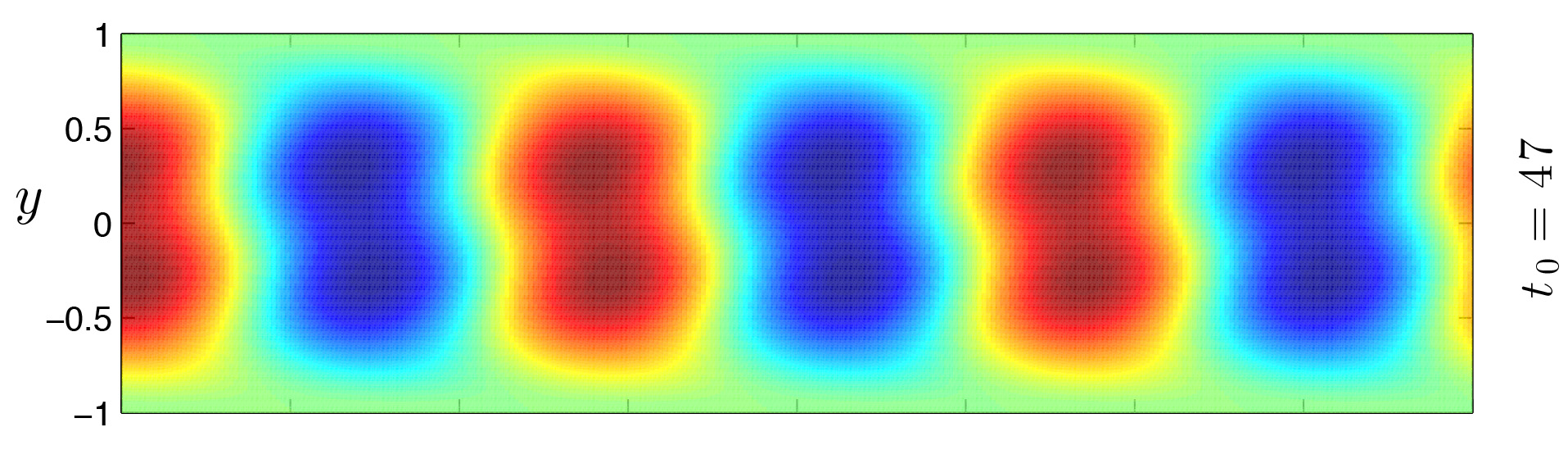}
	\vspace{-0.9cm}
	 \end{subfigure}
	\begin{subfigure}{1\textwidth}
        \centering
\includegraphics[width=16.0cm]{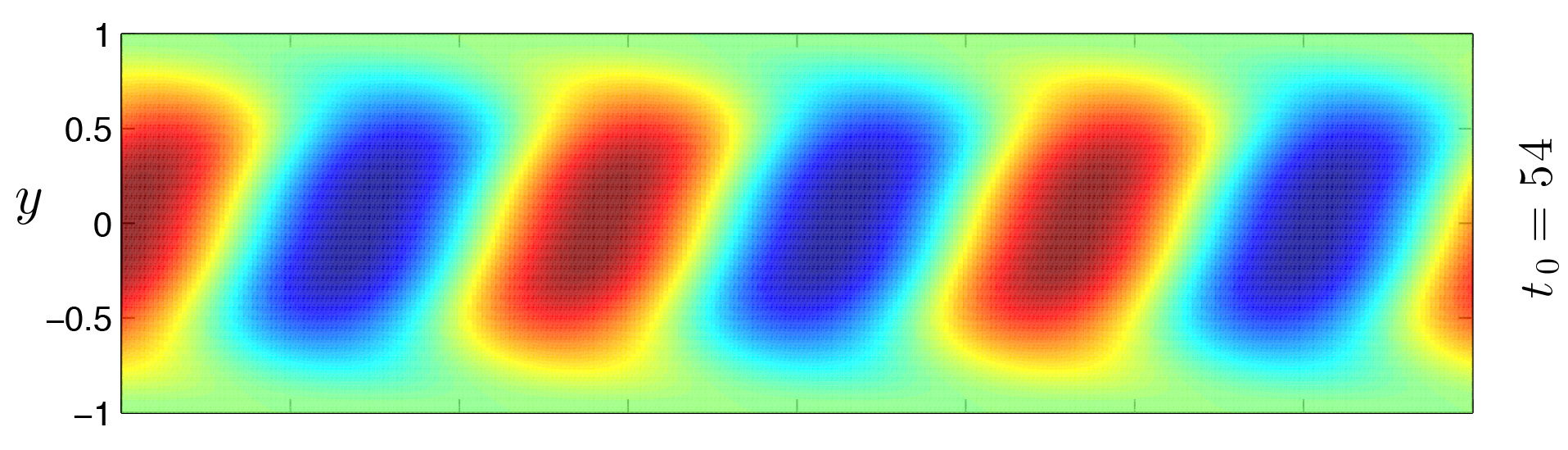}
	\vspace{-0.6cm}
	 \end{subfigure}
	 \begin{subfigure}{1\textwidth}
        \centering
\includegraphics[width=16.0cm]{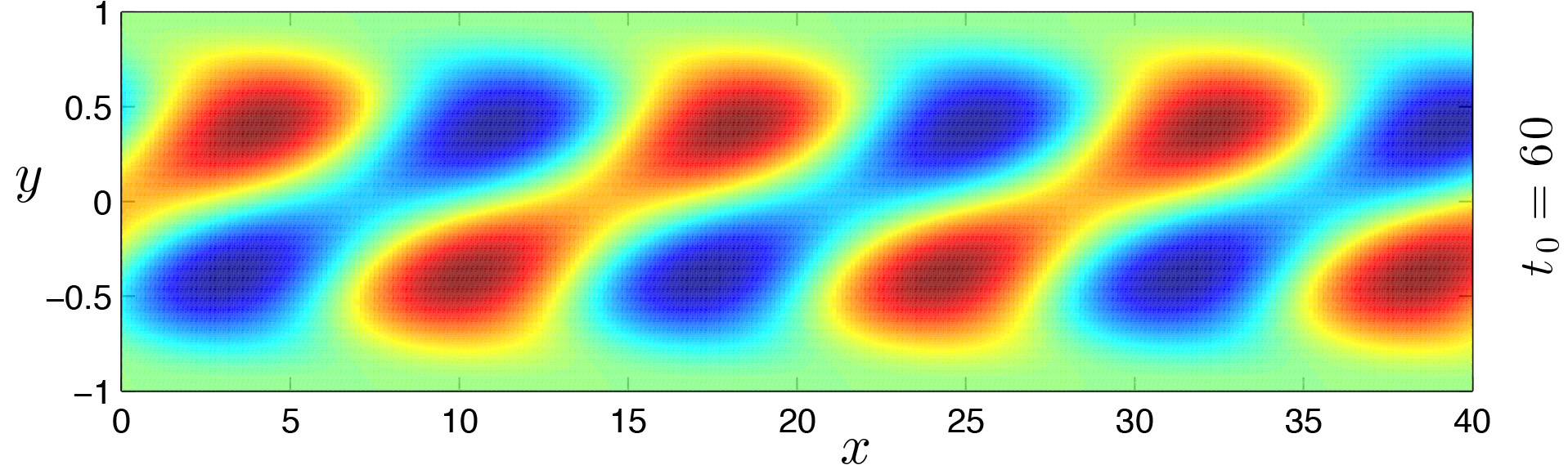}
	\vspace{-0.6cm}
	 \end{subfigure}
	\caption{Visualization of the wall-normal perturbation velocity $\tilde
v(x,y,z_0,t_0)$, $xy$ plane, for Plane Couette flow. This low-wavenumber condition was introduced and analyzed
in \figref{fig:prof_CO_Re500_k0p63_far}. From the physical space, the high non-stationarity of the whole scenario is
evident. The two opposite moving layers, where the largest disturbances reside, appears to experience a periodic 
attachment and consecutive detachment. The images sequence represents one period in the modulation of $c_v$, the
maximum of phase velocity occurs in the conditions (a) and (d). \textit{Red}: maximum (positive); \textit{blue}: minimum
(negative). }
\label{fig:anti_CO_Re500_k0p63_phi45_sym_v}
\end{figure}

In the following, some solutions among those introduced in 
\secref{sec:trasf_profiles} are inverse-transformed using  the relations \eqref{eq:invtrans_1} and
\eqref{eq:invtrans_2} to obtain the quantities in the physical space. A few
visualizations in the $xy$ plane are here reported (figures \ref{fig:anti_CO_Re500_k6p5_phi45_asy} to
\ref{fig:anti_CO_Re500_k0p63_phi45_sym_v}), for the same cases
introduced at the end of the previous section, in order to clarify the physical
meaning of the module of the complex quantities in the wavenumber space, and to observe the behavior of the
flow quantities in the real three-dimensional space. For all the following flow visualizations a variable color scale is
adopted to represent at all times the solution, that consequently has to be intended as normalized to its maximum value.

\chapter{Wave packets linear evolution}\label{chap:wave_packets}
\section{Introduction}
In the present chapter the evolution of linear wave packets is investigated. The aim of this study, as stated in the
Introduction, is to emphasize the role of the linear mechanisms in a scenario preceding the breakdown and the
transition to turbulence. To be more
precise, the focus will be on the Plane Couette flow, extensively studied in the previous chapters, and on Blasius
boundary layer flow (Bbl, in the following). The wave solutions for the latter are obtained with a Runge-Kutta code by
numerical integration of the Orr-Sommerfeld and Squire PDE equations by the method of lines
\citep[see][]{Ames_book}.\par
\subsubsection{Bypass transition and turbulent spots}
Although the Plane Couette flow is stable to infinitesimal perturbations for all values of the Reynolds number,
experimental evidences showed that for sufficiently high values of $Re$ the flow becomes turbulent. This process 
is observed in bounded flows and in Bbl as well, and it is known as \textit{Bypass transition}. The term is due to the
fact 
that this scenario bypasses the growth of two-dimensional waves and their secondary instability. Since this
laminar-turbulent transition is observed even for values of the Reynolds number lower than the critical one, obtained by
the modal stability theory, many shear flows fall in the class of \textit{subcritical} transitional flows. The general
scenario is the following. The transition does not occur simultaneously in the whole domain, but through nucleation
and growth of organized patches of turbulent flow, called \textit{turbulent spots}, that eventually fill the space. The
first observation was made by \citet{Emmons1951} in a water table flow. The PCf has been extensively studied in the
past, probably because its zero mean advection speed allows easier tracking of the spots. Experimental investigations
has shown the existence of a threshold Reynolds number $Re_c$ below which the spots keep a finite probability to
relaminarize; among these, we remind the works by \citet{Daviaud1992} ($Re_c=370\pm10$), \citet{Tillmark1992}
($Re_c=360\pm10$), \citet{Hegseth1996} ($Re_c=325$). The most common experimental apparatus consists of a
counter-translating belt driven by two rotating cylinders, the working fluid is water and a finite-amplitude disturbance
is triggered by fluid injection. Among the nonlinear direct numerical simulations, we report  more recent works by
\citet{Lagha1_2007}, \citet{Duguet2010} ($Re_c=324\pm1$), \citet{Duguet2011} ($Re_c=325$). Usually an germ-like initial
condition, (typically two counter-rotating vortices), is given in the physical plane. In these cases the authors showed
that the turbulent region exhibits elongated flow structures, called \textit{streaks}, and that for PCf the spot shape
is elliptical. When the transition is natural the \textit{streaky} structure remains, the spots nucleate randomly in
space and
their shape is more irregular, or oblique bands are found \citep[see][]{Manneville2011}. The typical distance between
two \textit{streaks} is found to be of the same order of magnitude of the channel half-height, $\lambda_s\approx
O(h)-O(3h)$. In
addition, other typical characteristics of the spot are its propagation speed and spreading rates which depends on the
base flow and the Reynolds number. However, for channel flows the spreading  of a turbulent spot is quite rapid, if
compared to the typical turbulent diffusion: this mechanism is known as ``growth by destabilization'' of the surrounding
laminar flow. Analyzing the structure of the Couette spot, \citet{Lundbladth1991} and \citet{Dauchot1995} classified it
as a case between the Poiseuille and the boundary layer spot.\par
About the latter, a boundary layer spot is characterized by a horseshoe structure. The complete process of transition
on a flat plate with zero pressure gradient has been subject to extensive studies since the beginning of the past
century with the work of \citet{Burgers1924} and successively by \citet{Tollmien1929} and \citet{Schlichting1933}.
Several features distinguish the Bbl transition from the one occurring in internal flows. From the leading edge of the
plate, as $x$ increases, the laminar flow is destabilized  until the transition zone is reached, where arrowhead
turbulent spots appear. A complete description of the structure and the evolution of spots can be found in the
experimental works by \citet{Cantwell1978} and  \citet{Gad-El-Hak1981}. The former also
provided  beautiful visualizations  of both the lateral side  of the spot and
its bottom side (the sublayer) taking advantage of the glass walls of the water channel. For a complete description of
all the boundary layer mechanisms of transition and the onset of turbulence see the review by \citet{Kachanov1994}.

\subsubsection{Wave packets and role of the linear stages in the transition process}
Comparing to the amount of studies about the non-linear stages of transition and the descriptions of the turbulent
spots, few investigations are found about the role of the linear evolution of small disturbances in the transitional
process. The results of the modal analysis have probably been overestimated, and only recently a renewed interest in the
transient evolution of linear three-dimensional disturbances arised. The importance of three-dimensionality and so the
spanwise variation of the velocity components was firstly pointed out by the boundary layer experiments by
\citet{Klebanoff1961}. The instability of this oblique wave develops in $\Lambda$-vortices (K-transition).
\citet{Zang1989} demonstrated that the growth of the oblique waves is correlated with the existence of a mode with
$\alpha=0$, i.e. an orthogonal mode. The fact that the presence of this mode is a prerequisite for the rising of
secondary instability was confirmed in earlier investigations. \citet{Schmid1992} looked at small amplitude wave
pairs, and \citet{Henningson1993} devoted to investigating a possible mechanism for bypass transition, pointing out
the role of the linear phase and arguing that the mechanism for energy transfer is primarily linear. In fact, the
disturbances with no streamwise dependence ($\alpha=0$) are usually those which experience the most rapid growth
(see
\figref{fig:Gmax_CO_k6p5_sym_varioRe}-\ref{fig:Gmax_CO_k6p5_asym_variok}). \citet{Henningson1994} argued the necessity
of linear growth mechanisms for subcritical growth of arbitrary amplitude perturbations. Indeed, almost all the Fourier
components are contained in a generic initial condition, and those corresponding to the spanwise wavenumber axis are
found to be rapidly excited due to linear mechanism. Moreover, even if the initial condition is poor in those
components, rapid growth still occurs when non-linear interactions transfer energy in that area of the wavenumber
space. The same happens for finite amplitude disturbances: \citet{Henningson1993} pointed out that the energy growth is
only caused by the linear mechanism, leading to the \textit{streaky} horizontal velocity pattern. For subcritical flows
this means that the transient growth effect must operate for transition to take place. \par
The  \textit{streaky} structure, typical of various flow configurations,  is also related to the evolution of optimal
(in
a linear
sense) disturbances, which can arise and bring to nonlinearity \citep[see e.g.][]{Brandt2003}. In fact, the wall-normal
shape of linearly optimal disturbances determined by \citet{Andersson1999} is surprisingly similar to the measured
$u_{rms}$ values. We should also cite the work by \citet{Cherubini2010}, who looked for optimal initial
conditions and also made a comparison between a linear and a nonlinear analysis of a spot evolution in Bbl. They shows
that the \textit{streaky} structure and the general shape of the spot are already determined by the linear analysis, due
to the
kinetic energy transient growth. Only in the following nonlinear phase, secondary instability of the \textit{streaks}
occurs and
the spot central region becomes turbulent.\newpage

\section{Linear spot in Plane Couette flow}
\begin{figure}[h!]
        \centering
        \hspace{-0.5cm}
         \advance\leftskip-2.5cm
         \advance\rightskip-2cm
        \begin{subfigure}{0.6\textwidth}
        \centering
\includegraphics[width=9.0cm]{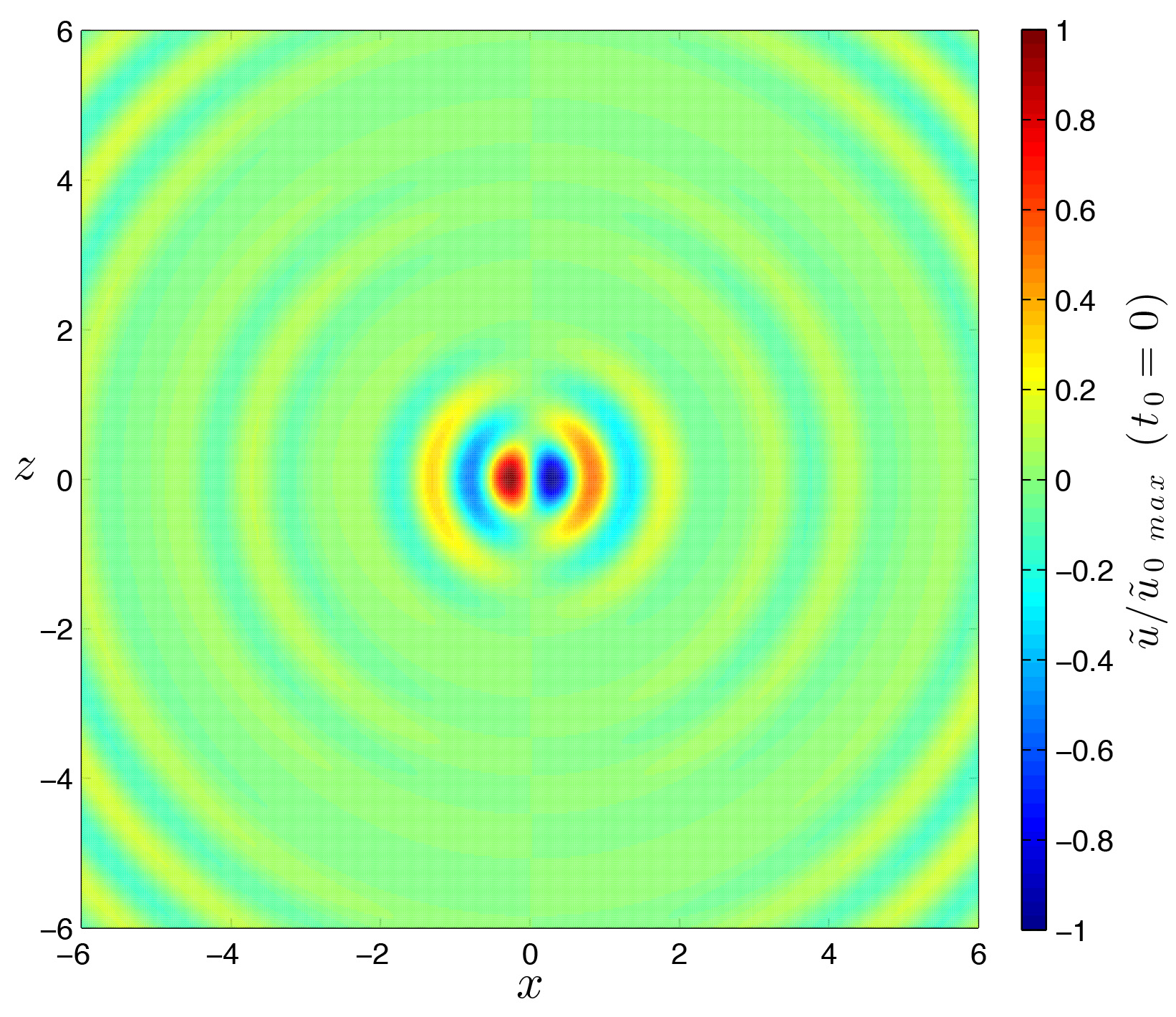}
	\vspace{0cm}
	\label{fig:CO_Re500_localized_U_1}
	 \end{subfigure}
        \begin{subfigure}{0.6\textwidth}
        \centering 
\includegraphics[width=9.0cm]{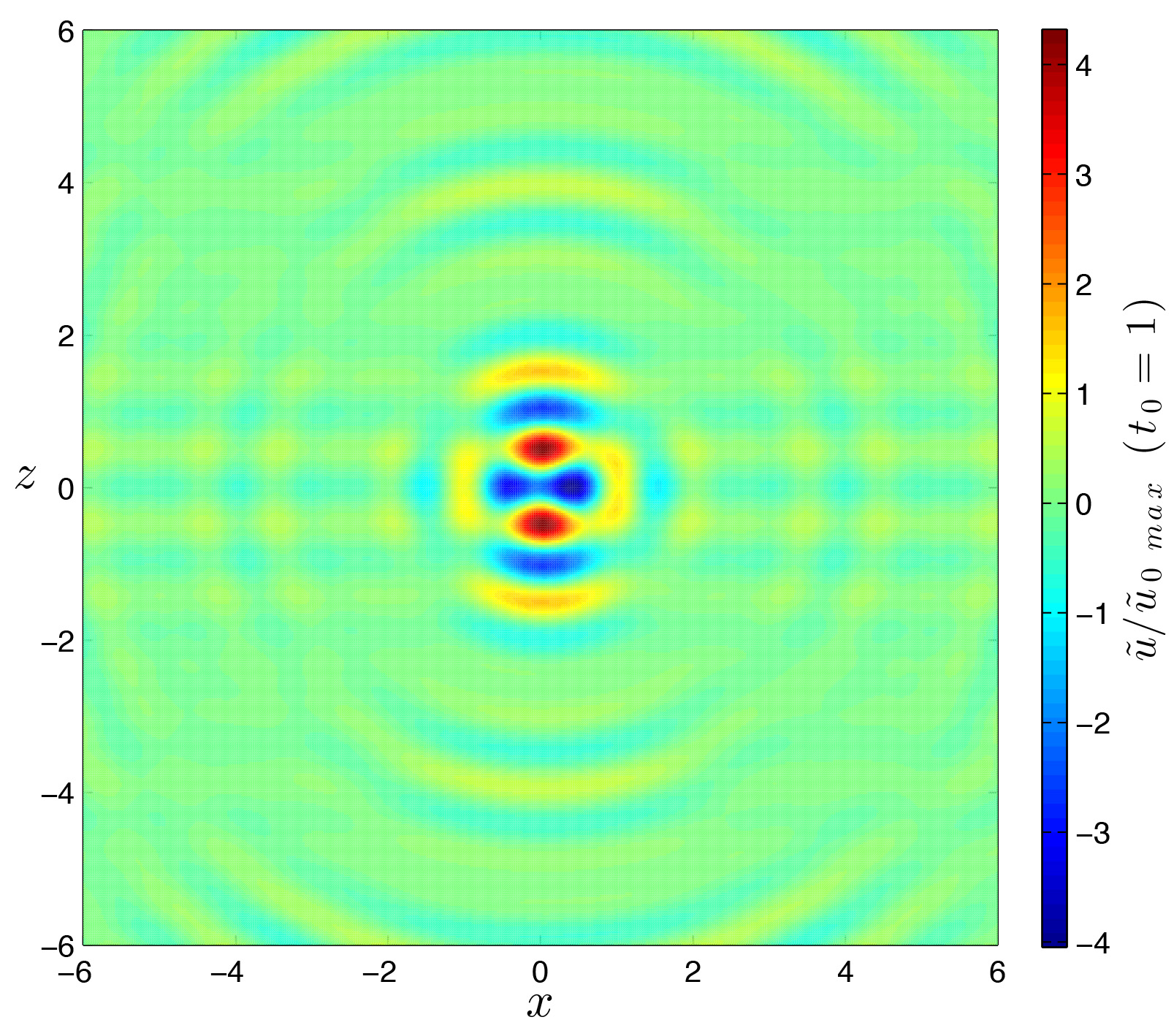}
	\vspace{0cm}
	\label{fig:CO_Re500_localized_U_2}
	 \end{subfigure}
        \begin{subfigure}{0.6\textwidth}
        \centering
\includegraphics[width=9.0cm]{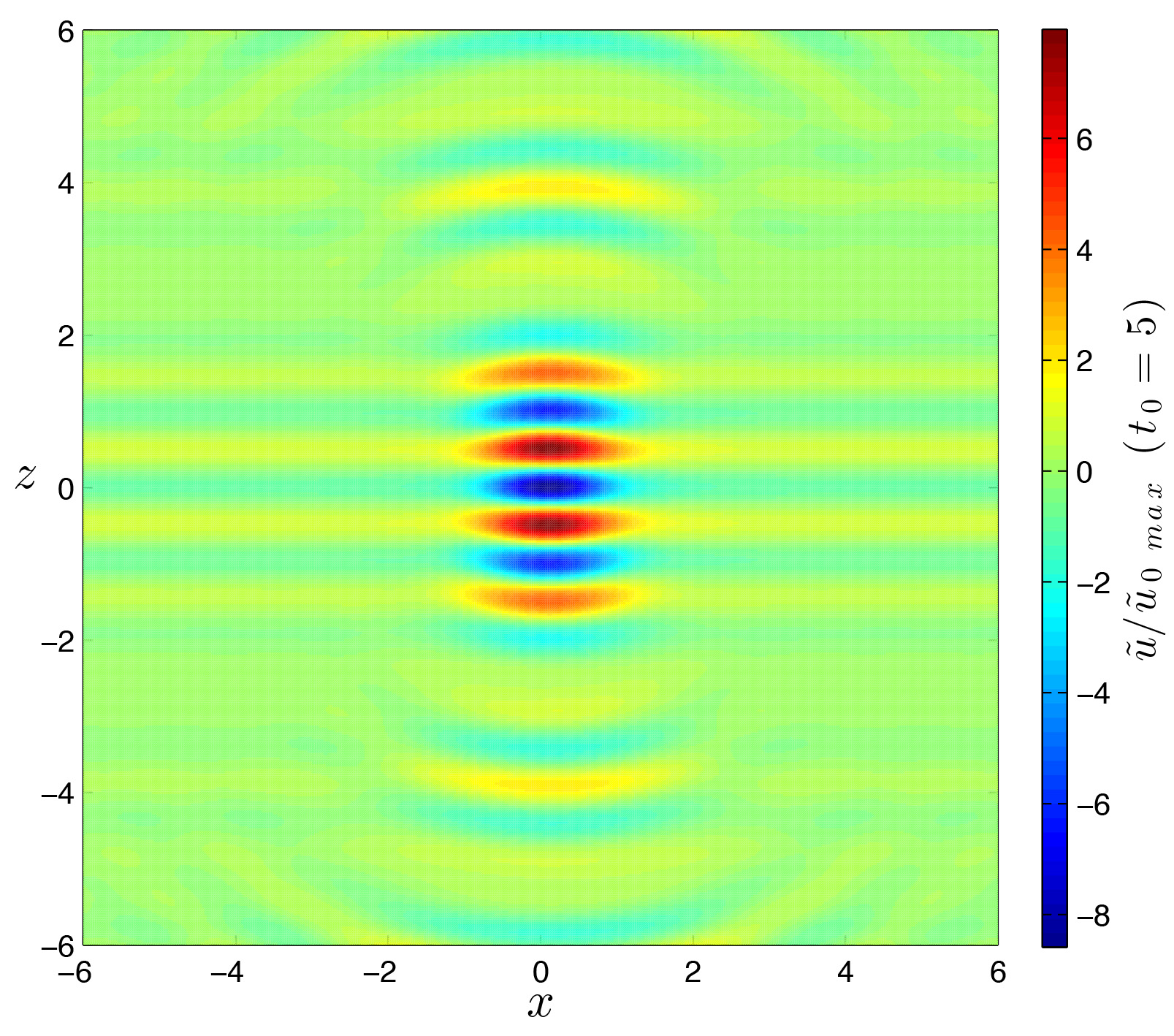}
	\vspace{0.5pt}
	\label{fig:CO_Re500_localized_U_3}
	 \end{subfigure}
        \begin{subfigure}{0.6\textwidth}
        \centering 
\includegraphics[width=9.0cm]{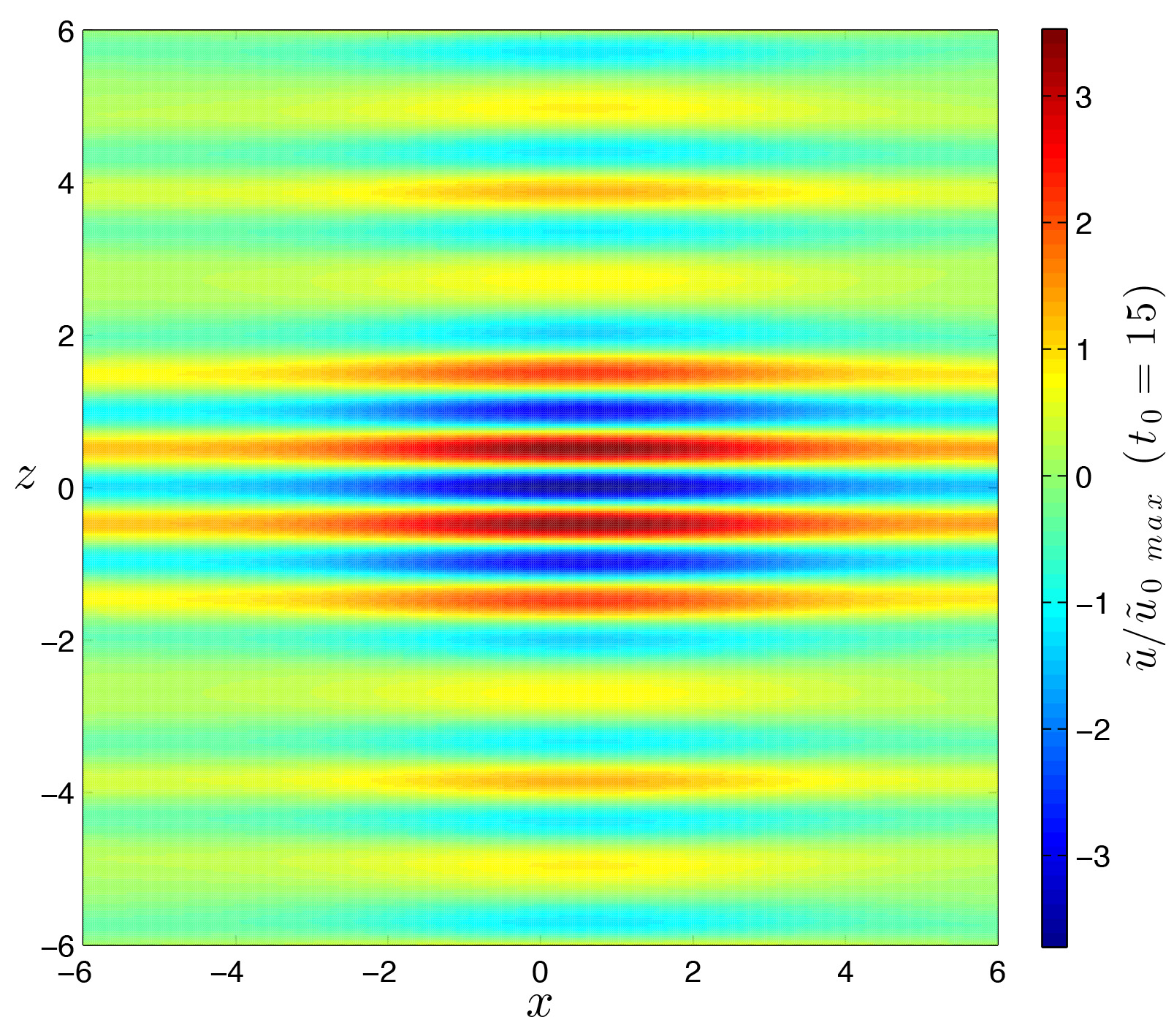}
	\vspace{0.5pt}
	\label{fig:CO_Re500_localized_U_4}
	 \end{subfigure}	
	\caption{Visualization of longitudinal velocity $\tilde u$ for PCf with $Re=500$. Views of $xz$ plane at the
channel symmetry plane at different times. The evolution of a localized perturbation is obtained by superposition of 220
waves with polar wavenumber $k=\{5.7,\ 6.5,\ 7.3\}$, obliquity angle spanning the  full circle, $\phi\in\{-90^\circ,\
+90^\circ\}$, with both \textit{sym} and \textit{asym} initial conditions. The values are normalized with respect to the
maximum  at time $t=0$.}
\label{fig:CO_Re500_localized_U} 
\end{figure}

\begin{figure}[h!]
        \centering
         \advance\leftskip-2.5cm
         \advance\rightskip-2cm
        \begin{subfigure}{0.6\textwidth}
        \centering
\includegraphics[width=9.0cm]{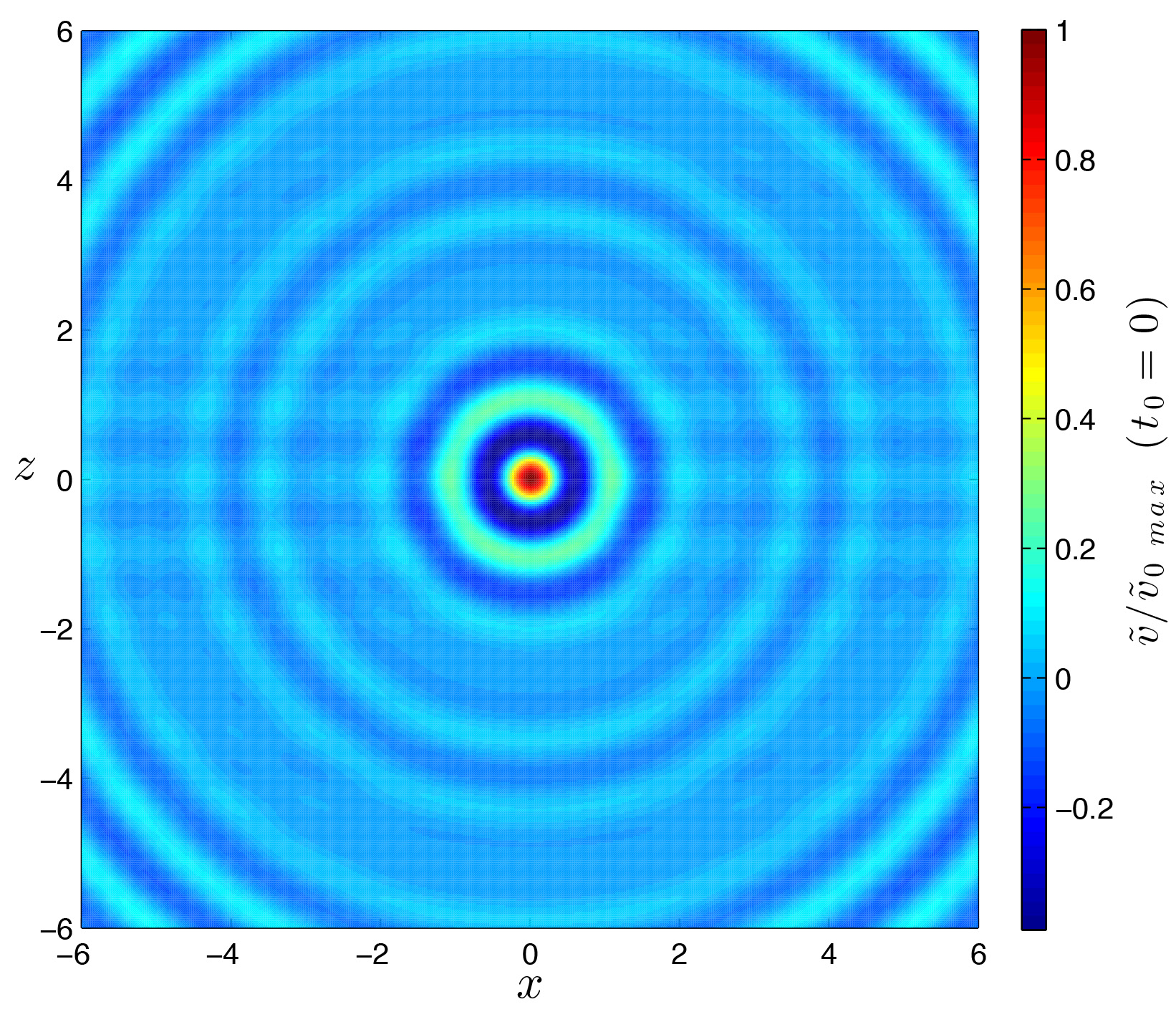}
	\vspace{0cm}
	 \end{subfigure}
        \begin{subfigure}{0.6\textwidth}
        \centering 
\includegraphics[width=9.0cm]{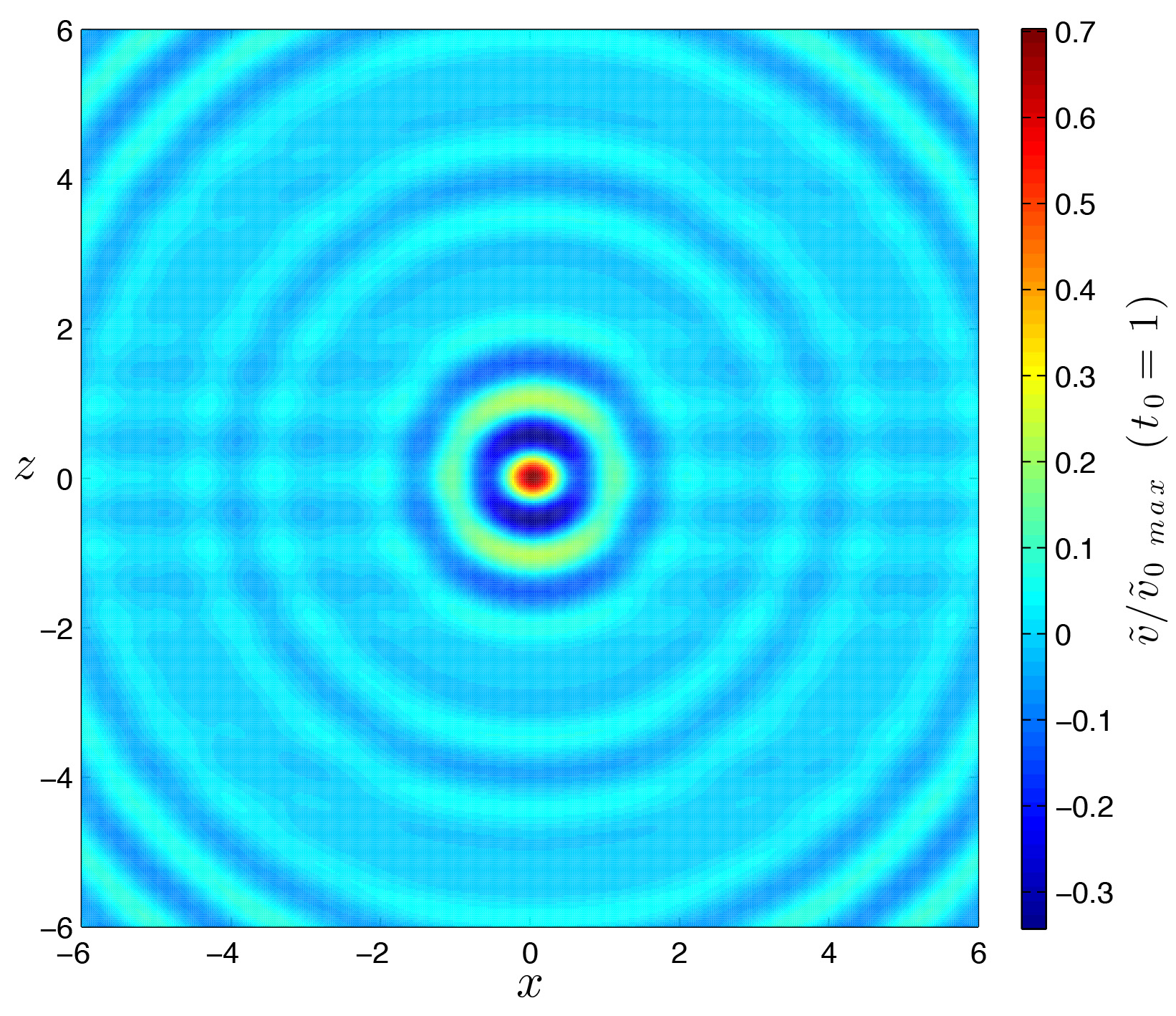}
	\vspace{0cm}
	 \end{subfigure}
        \begin{subfigure}{0.6\textwidth}
        \centering
\includegraphics[width=9.0cm]{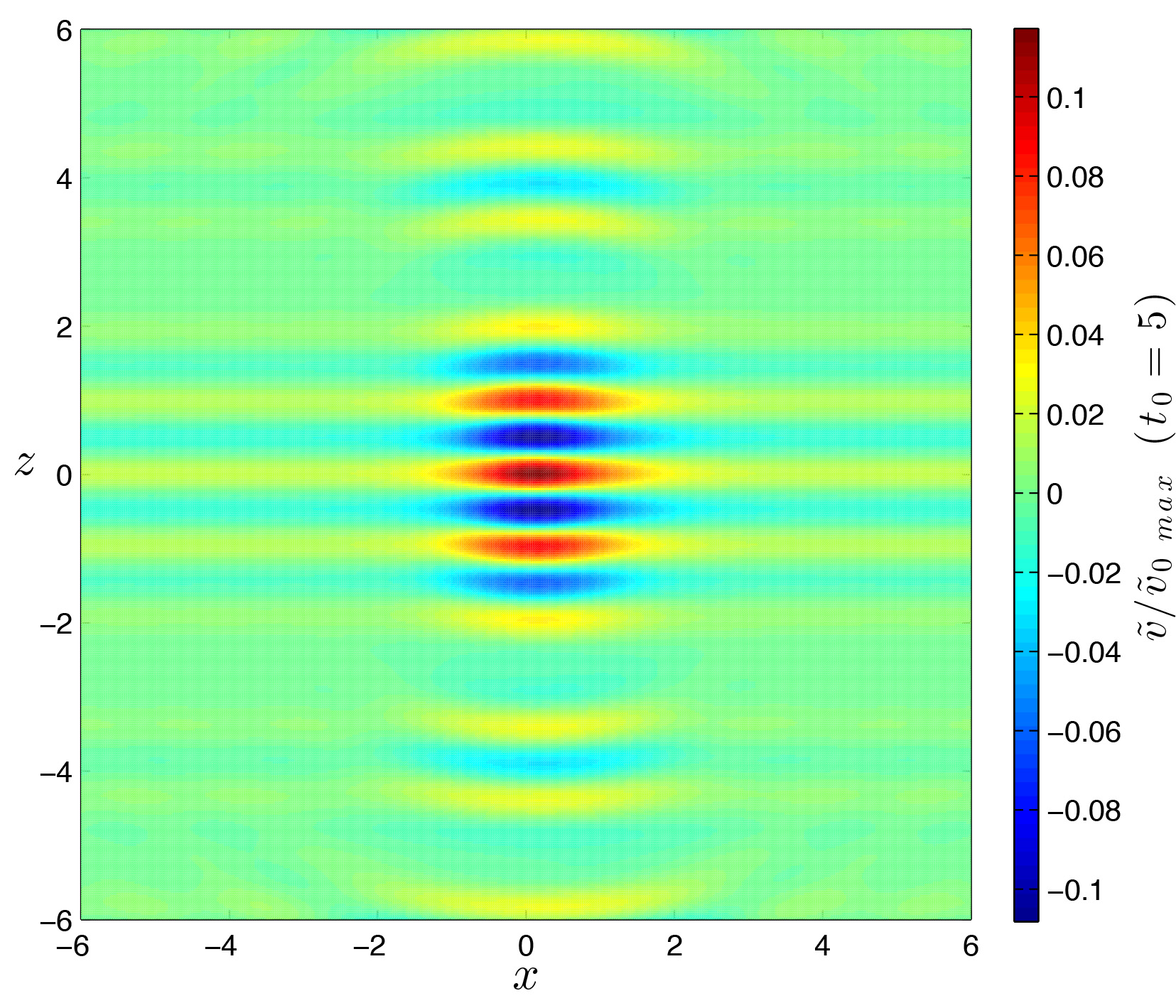}
	\vspace{0.5pt}
	 \end{subfigure}
        \begin{subfigure}{0.6\textwidth}
        \centering 
\includegraphics[width=9.0cm]{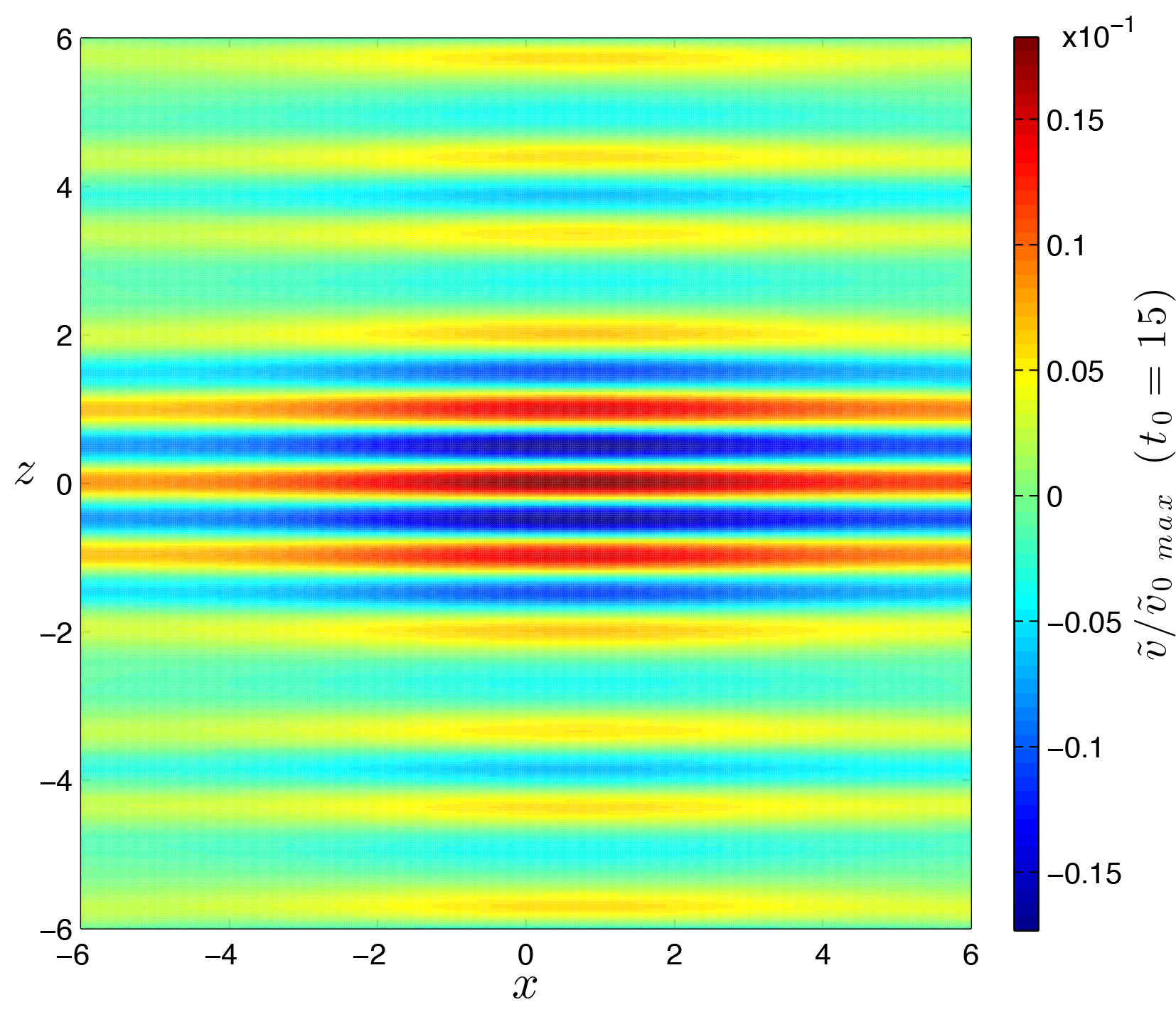}
	\vspace{0.5pt}
	 \end{subfigure}	
	\caption {Visualization of the wall-normal velocity $\tilde v$ for PCf with $Re=500$.  Views of $xz$ plane at
the channel symmetry plane at different times. The evolution of a localized perturbation is obtained by superposition of
220 waves with polar wavenumber $k=\{5.7,\ 6.5,\ 7.3\}$, obliquity angle spanning the  full circle,
$\phi\in\{-90^\circ,\ +90^\circ\}$,
with both \textit{sym} and \textit{asym} initial conditions. The values are normalized with respect to the maximum at
time $t=0$.}
\label{fig:CO_Re500_localized_V}
\end{figure}

\begin{figure}[h!]
        \centering
         \advance\leftskip-2.5cm
         \advance\rightskip-2cm
        \begin{subfigure}{0.6\textwidth}
        \centering
\includegraphics[width=9.0cm]{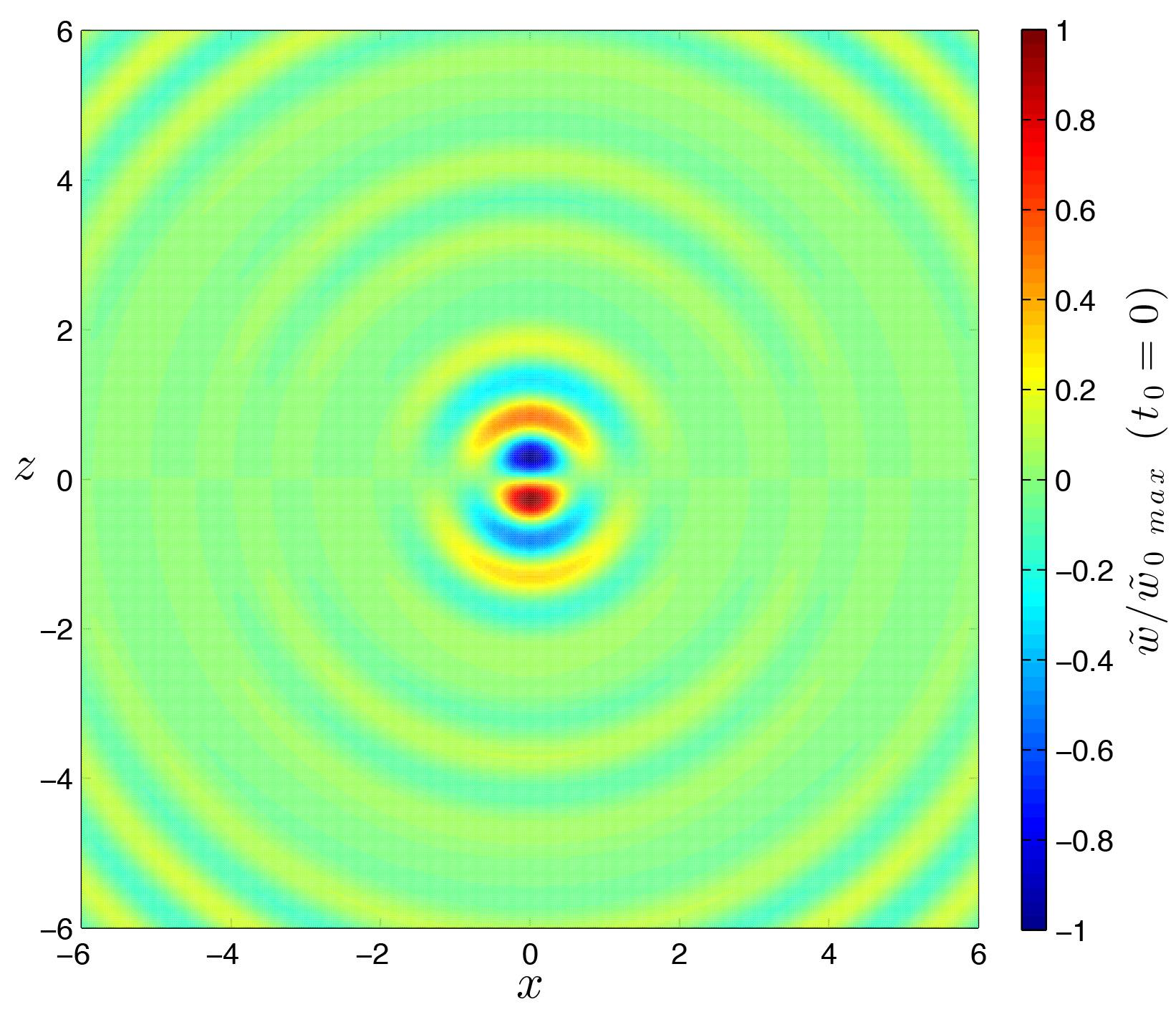}
	\vspace{0cm}
	 \end{subfigure}
        \begin{subfigure}{0.6\textwidth}
        \centering 
\includegraphics[width=9.0cm]{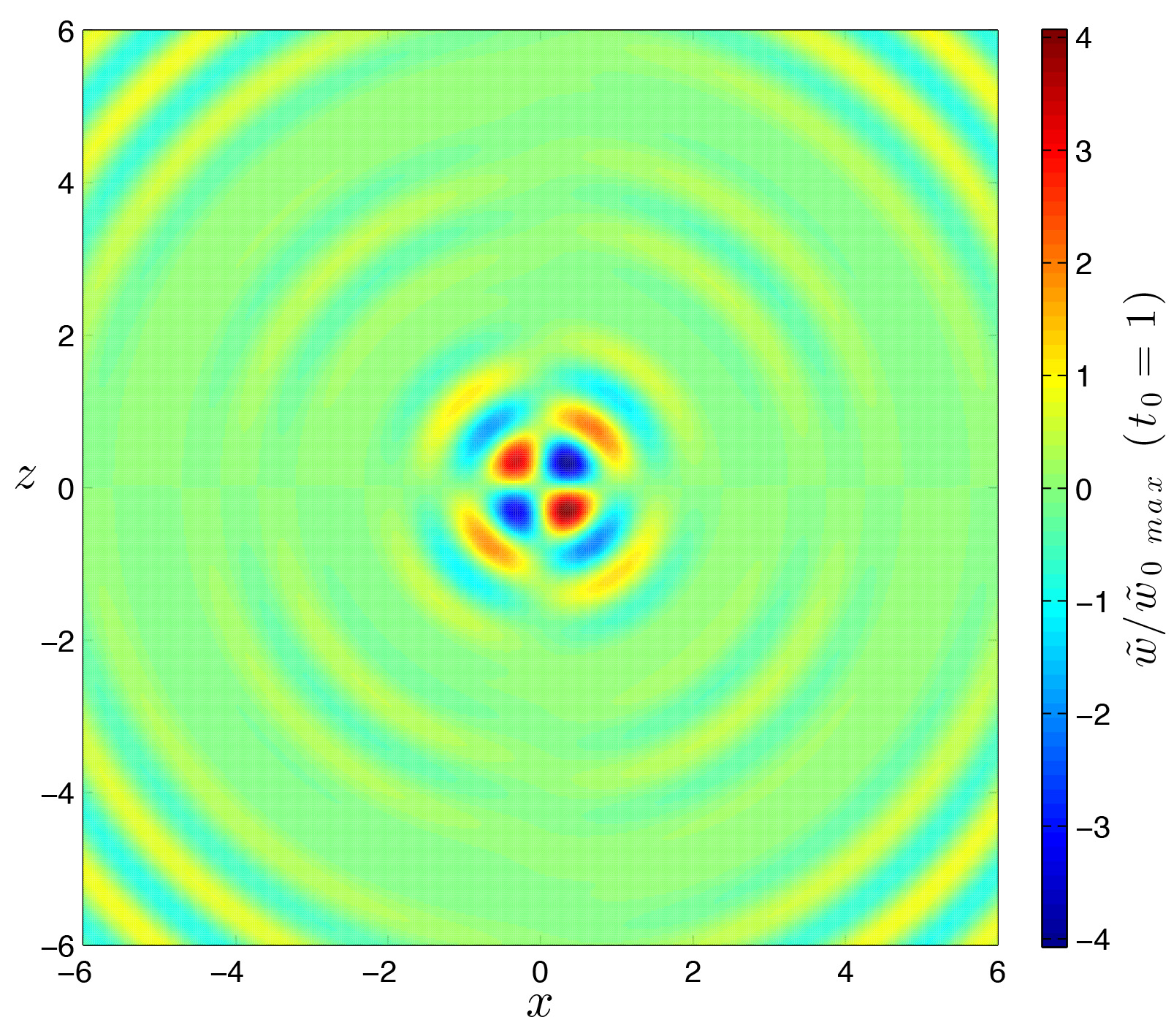}
	\vspace{0cm}
	 \end{subfigure}
        \begin{subfigure}{0.6\textwidth}
        \centering
\includegraphics[width=9.0cm]{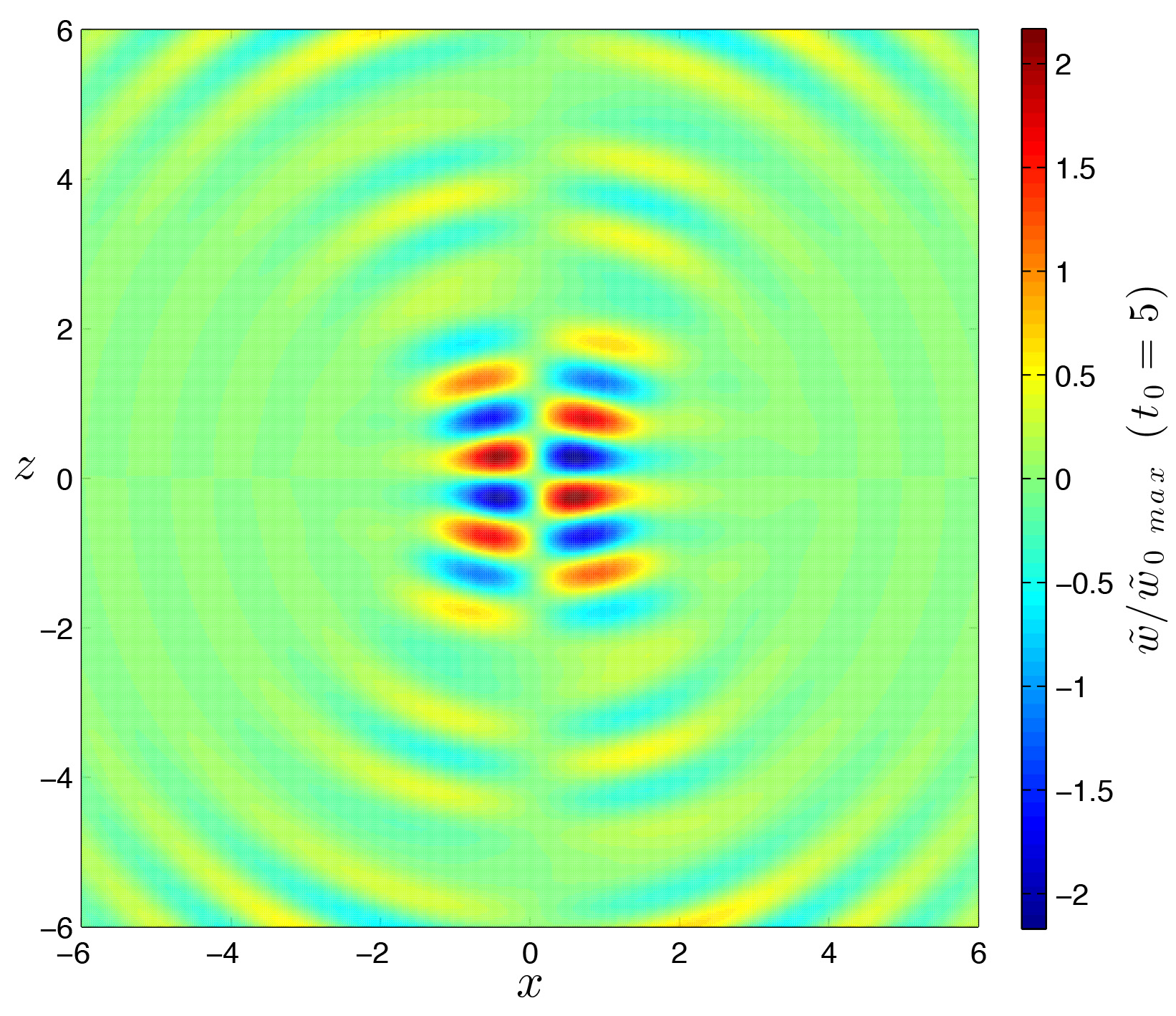}
	\vspace{0.5pt}
	 \end{subfigure}
        \begin{subfigure}{0.6\textwidth}
        \centering 
\includegraphics[width=9.0cm]{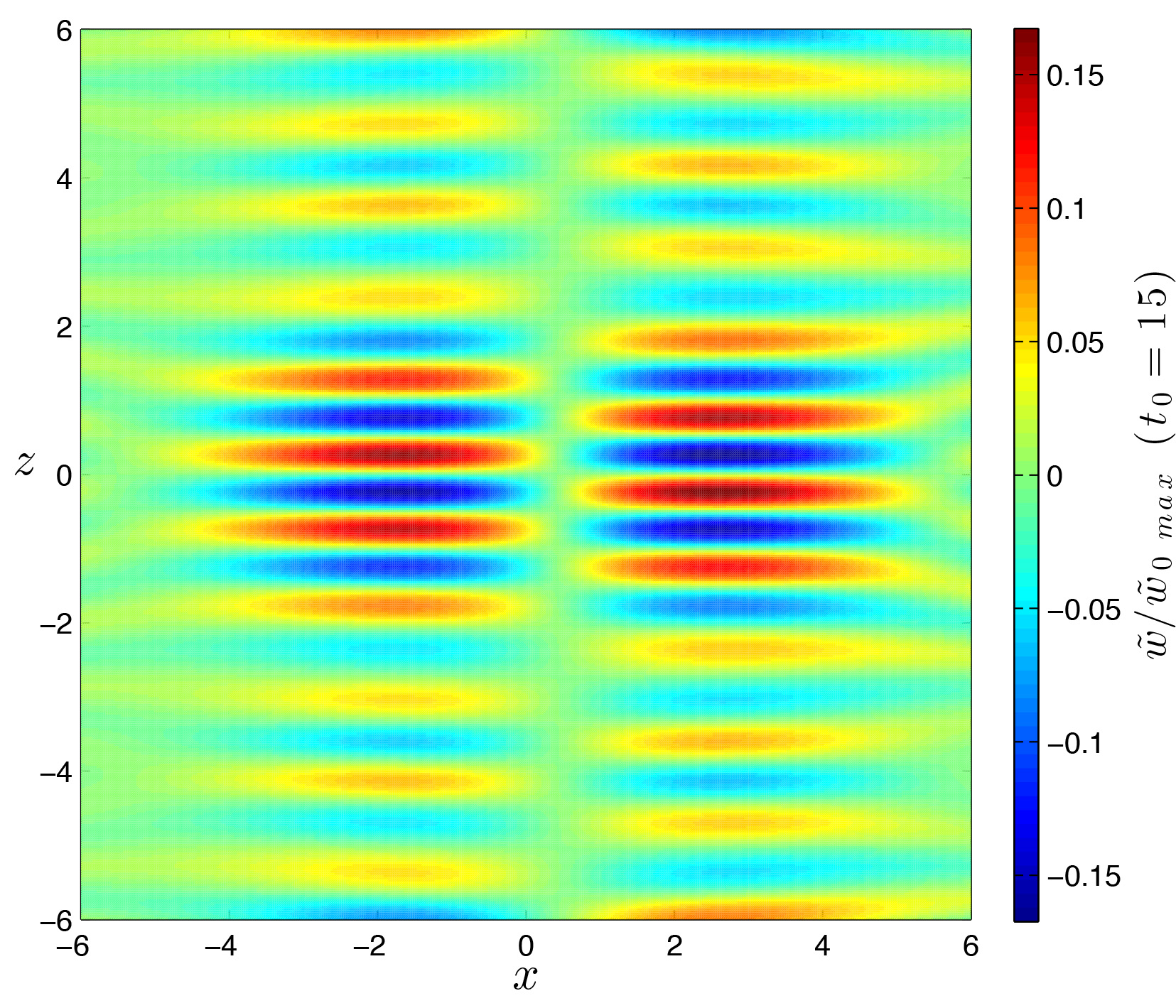}
	\vspace{0.5pt}
	 \end{subfigure}	
	\caption{Visualizations of the spanwise velocity $\tilde w$ for PCf with $Re=500$.  Views of $xz$ plane at the
channel symmetry plane at different times. The evolution of a localized perturbation is obtained by superposition of 220
waves with polar wavenumber $k=\{5.7,\ 6.5,\ 7.3\}$, obliquity angle spanning the  full circle, $\phi\in\{-90^\circ,\
+90^\circ\}$,
with both \textit{sym} and \textit{asym} initial conditions. The values are normalized with respect to the maximum at
time $t=0$.}
\label{fig:CO_Re500_localized_W}
\end{figure}

\begin{figure}[h!]
        \centering
         \advance\leftskip-2.5cm
         \advance\rightskip-2cm
        \begin{subfigure}{0.6\textwidth}
        \centering
\includegraphics[width=9.0cm]{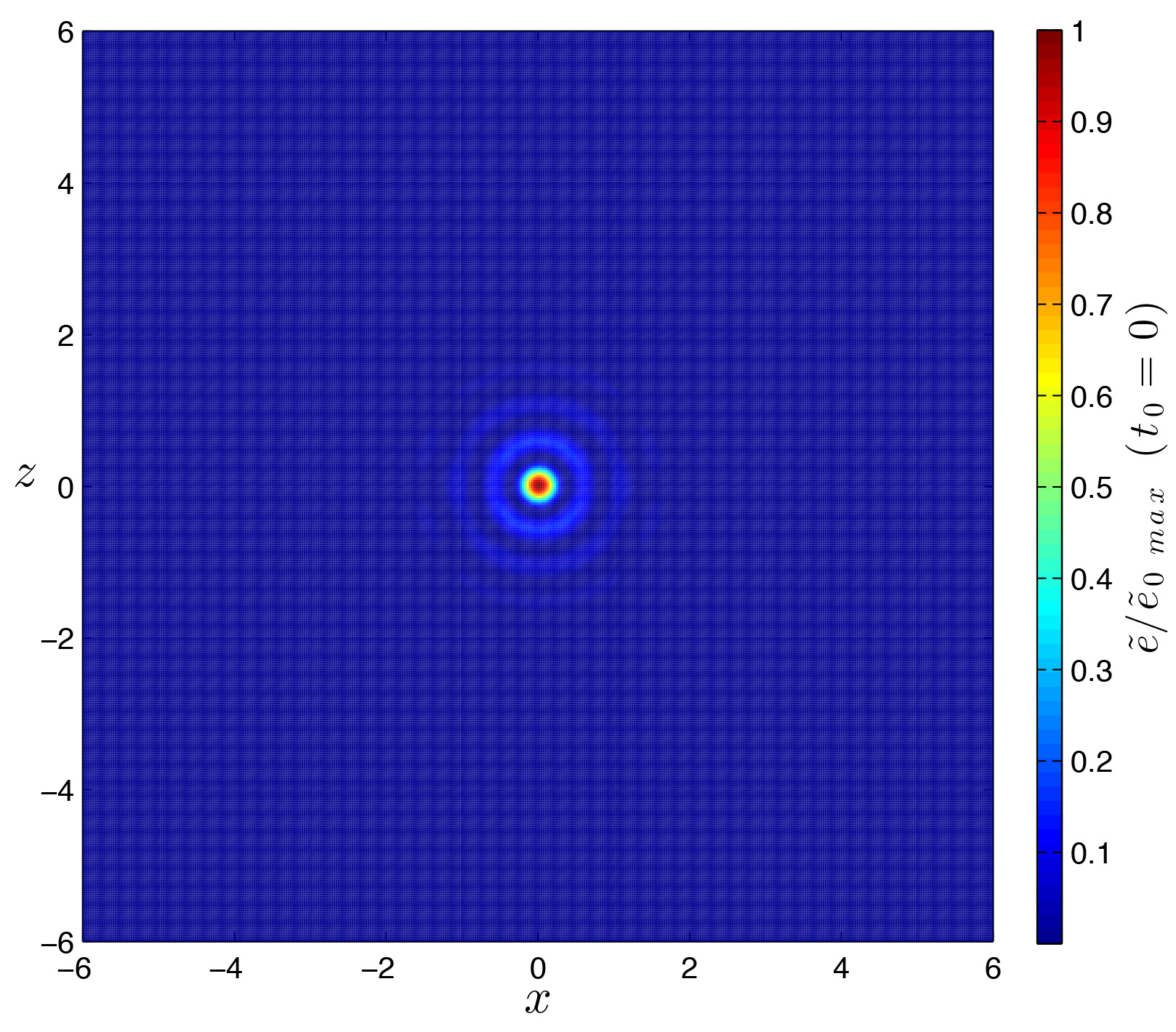}
	\vspace{0cm}
	 \end{subfigure}
        \begin{subfigure}{0.6\textwidth}
        \centering 
\includegraphics[width=9.0cm]{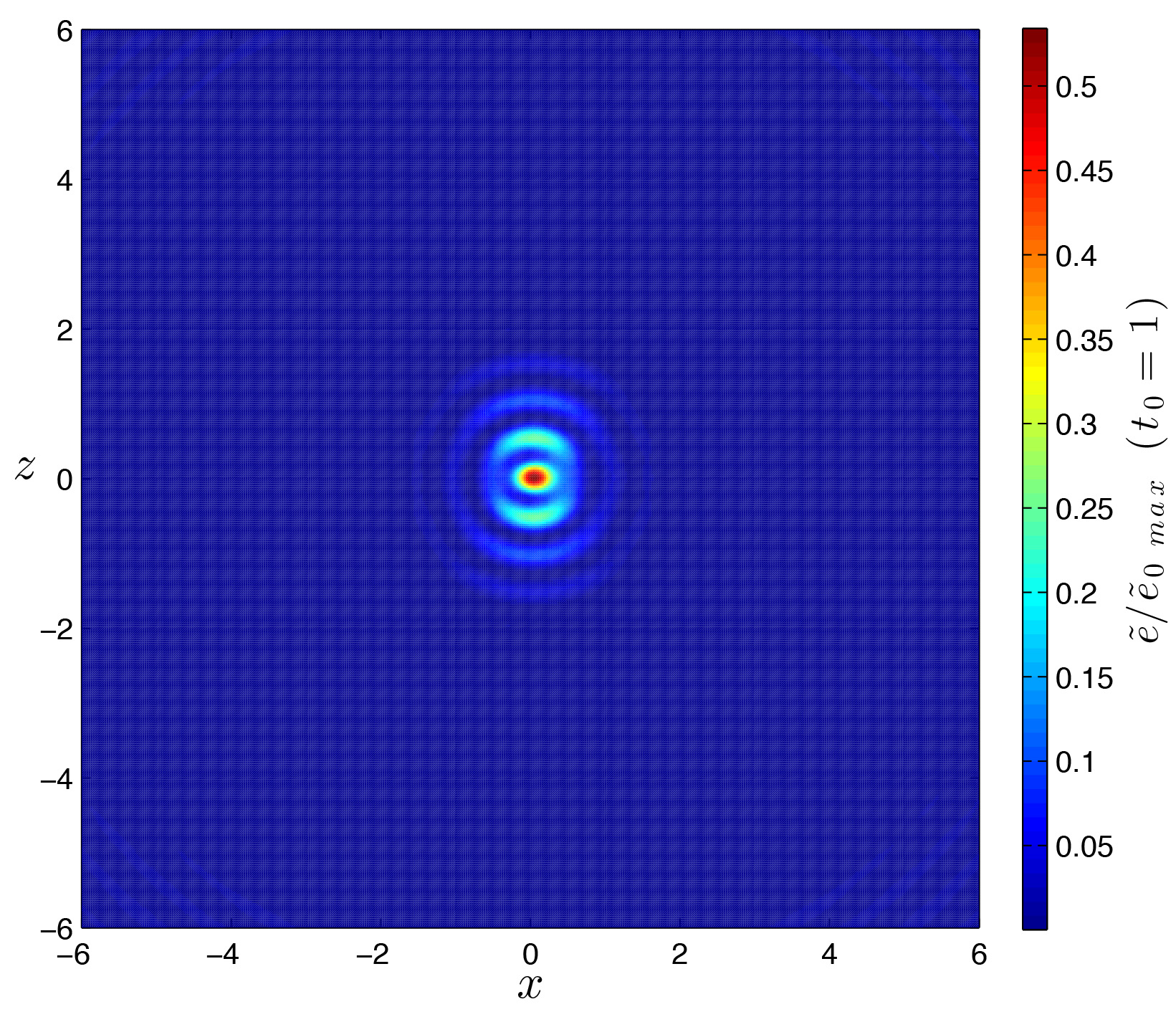}
	\vspace{0cm}
	 \end{subfigure}
        \begin{subfigure}{0.6\textwidth}
        \centering
\includegraphics[width=9.0cm]{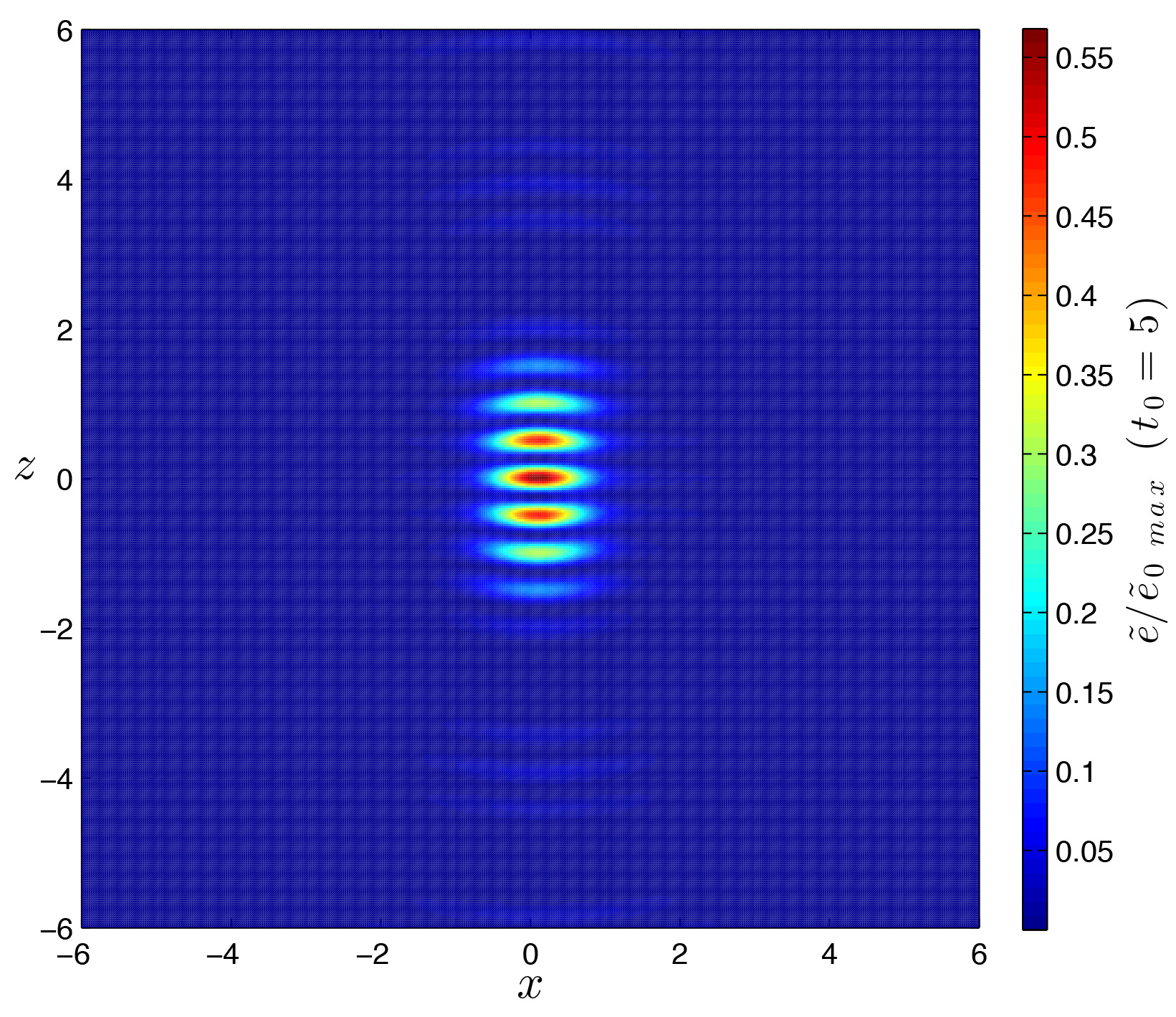}
	\vspace{0.5pt}
	 \end{subfigure}
        \begin{subfigure}{0.6\textwidth}
        \centering 
\includegraphics[width=9.0cm]{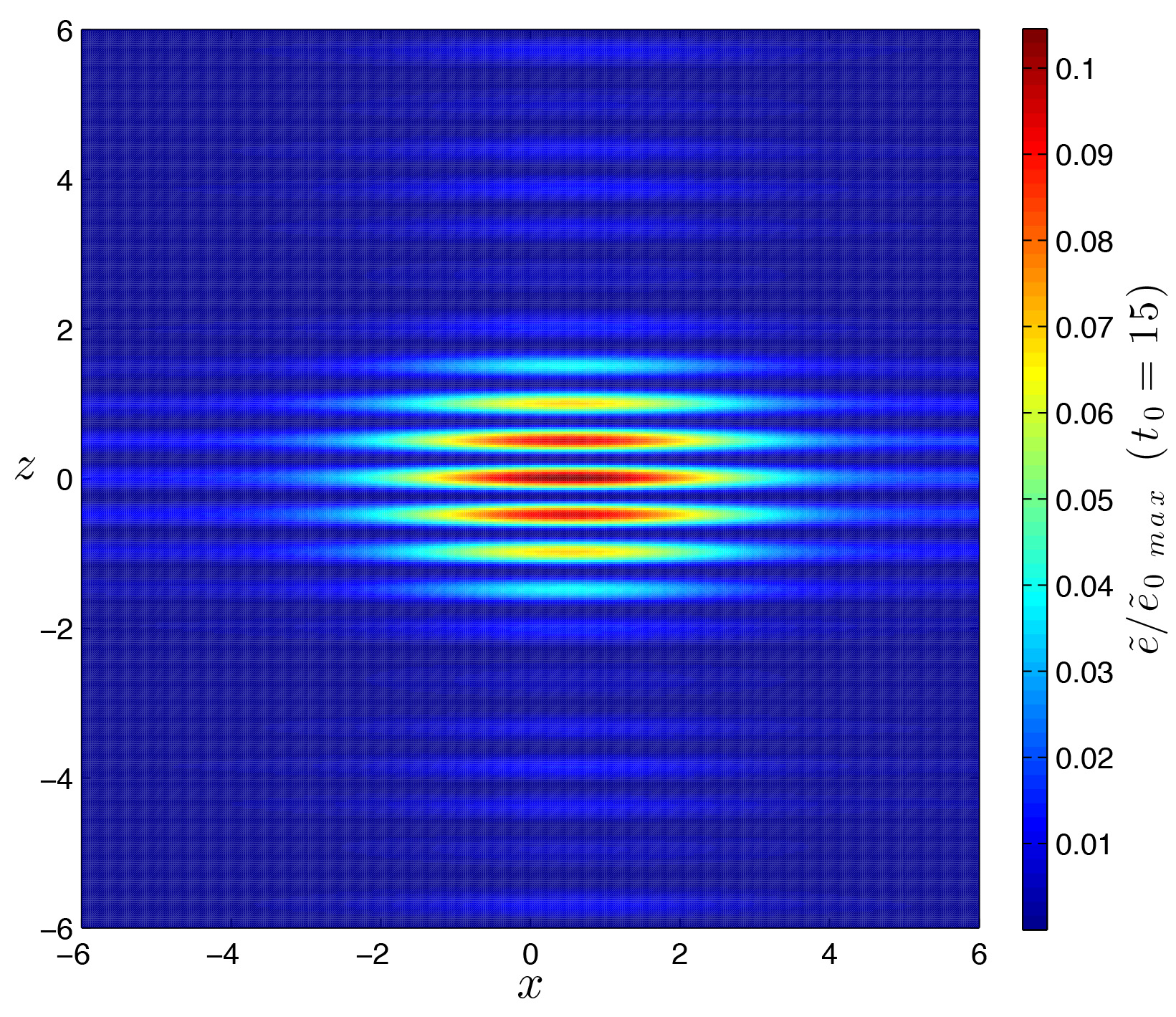}
	\vspace{0.5pt}
	 \end{subfigure}	
	\caption{Visualization of the kinetic energy $\tilde e$ for PCf with $Re=500$. Views of $xz$ plane at the
channel symmetry plane at different times. The evolution of a localized
perturbation is obtained by superposition of 220 waves with
polar wavenumber $k=\{5.7,\ 6.5,\ 7.3\}$, obliquity angle spanning the  full circle, $\phi\in\{-90^\circ,\
+90^\circ\}$,
with both \textit{sym} and \textit{asym} initial conditions. The values are normalized with respect to the maximum at
time $t=0$.}
\label{fig:CO_Re500_localized_E}
\end{figure}

The results of the linear superposition of a large number of waves are shown in 
\figref{fig:CO_Re500_localized_U}-\ref{fig:CO_Re500_localized_E}. The purpose of these visualizations is to confirm the
role of the linear transient dynamics in the complex transitional scenario, showing that in the evolution of a wave
packet some of the typical features of a transitional flow may be encountered. Differently from the works by the cited
authors, here a localized disturbance is simply obtained by a superposition, with zero phase-shift, of a large number
of waves with obliquity angle $\phi$ spanning the full circle. The polar wavenumber $k$ is
restricted to a few values
chosen accordingly to the experimental evidences found in literature. For Plane Couette flow the chosen values are
$k=\{5.7,\ 6.5,\ 7.3\}$, while the Reynolds number is $500$, with reference to the experimental work by
\citet{Hegseth1996}. Both odd and even initial condition (the same introduced and used in the previous chapters) are
considered.
The solutions to the initial value problem ~\eqref{Orr-Somm}-\eqref{Squire} are inverse-transformed, according to the
relations
\eqref{eq:invtrans_1} and \eqref{eq:invtrans_2}, and then superimposed, so the complete flow field in an arbitrary
domain in $x$ and $z$ directions can be easily obtained. If an in-phase superposition of a large number of waves is
considered, a bump-like initial condition in the physical space is obtained, as shown in
\figref{fig:CO_Re500_localized_V}. Clearly,
if the considered domain is wide
enough, a repetitive periodic scheme can be observed. The equivalence with a two-dimensional Fourier transform is
straightforward. Hence, this procedure is a simple way to qualitatively represent a localized perturbation in the
physical space, containing the contributions of all obliquity angles. Moreover, correlations with the transient
evolution of  single waves, shown in Chapter \ref{chap:wave_transient}, are possible. \par
In the following, the results are reported in terms of the three components of velocity $\tilde u$,  $\tilde v$, 
$\tilde w$, and the pointwise kinetic energy $\tilde e(x,y,z,t)=0.5*({\tilde u}^2+{\tilde w}^2+{\tilde w}^2)$. Remind
that the amplitude of the initial perturbation does not have any influence on the results, since the analysis is linear.
That is the reason why all the reported fields are normalized to the maximum value gained at $t=0$. For
Plane Couette flow is convenient to visualize the quantities at the channel symmetry plane $y=0$, since the
mean flow is zero. The first stages of the evolution of the linear spot seem to be characterized by a dominant spanwise
rate of spreading, due to the faster waves with $\beta\to0$. When the orthogonal wave ($\alpha=0$) becomes dominant, a
\textit{streaky} flow structure is found: note that a negative band of the component $\tilde u$ corresponds
to a positive one
of $\tilde v$. \par
\begin{figure}[h!]
        \centering
\includegraphics[width=14.0cm]{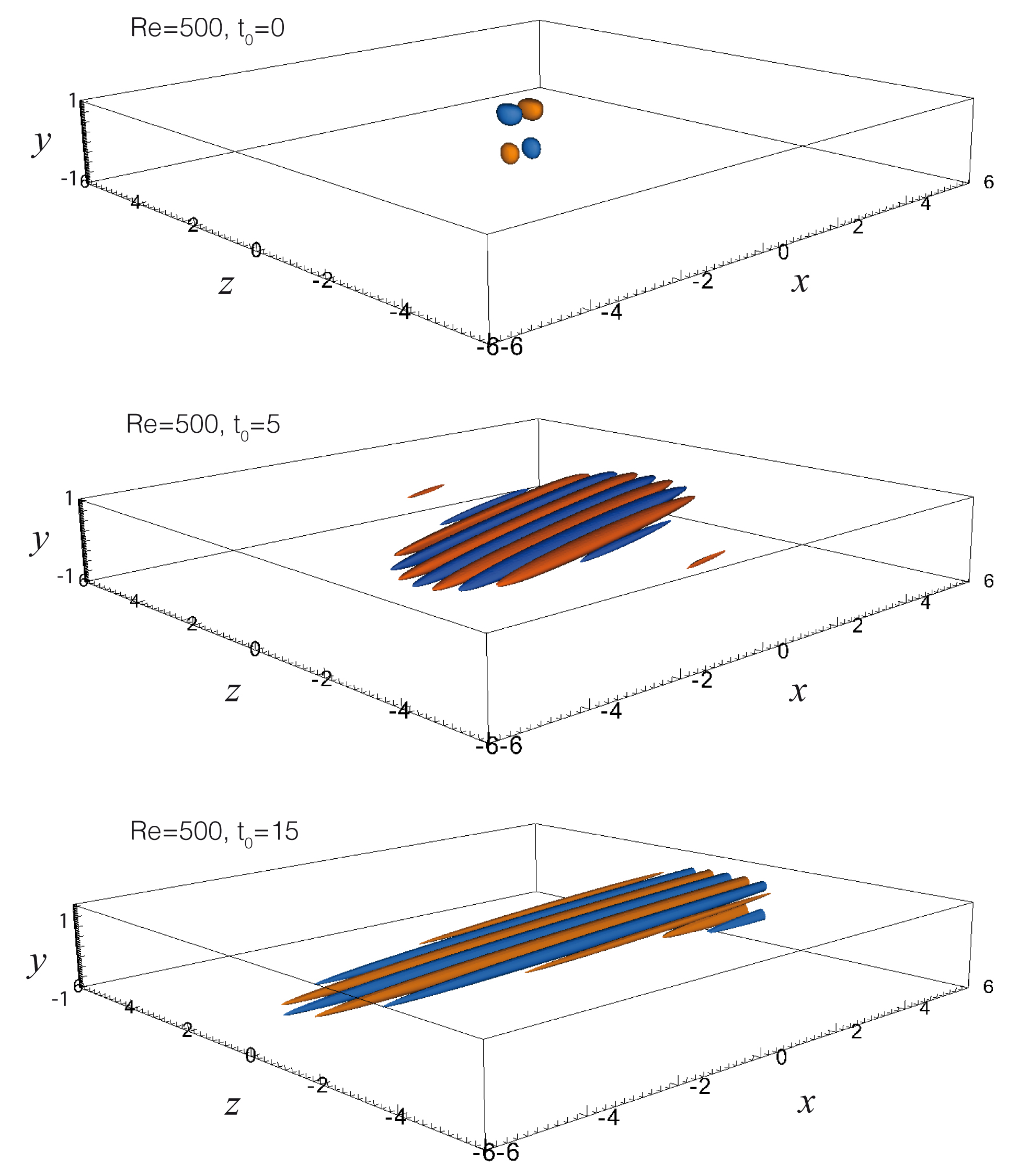}
	\caption{3D visualization of streamwise velocity for PCf with $Re=500$. (a) Initial condition, $t_0=0$;
\textit{orange surface}: $\tilde u / {\tilde u}_{0\ max}=0.5$; \textit{blue surface}: $\tilde u / {\tilde u}_{0\
max}=-0.5$; (b) $t_0=5$; \textit{orange surface}: $\tilde u / {\tilde u}_{0\ max}=1$; \textit{blue surface}: $\tilde u
/ {\tilde u}_{0\
max}=-1$;  (b) $t_0=15$; \textit{orange surface}: $\tilde u / {\tilde u}_{0\ max}=1$; \textit{blue surface}: $\tilde u
/ {\tilde u}_{0\
max}=-1$. Here ${\tilde u}_{0\ max}$ is the maximum value in the whole 3D domain at the initial time (${\tilde u}_{0\
max}=20.8$). Remind that the upper wall moves in the $x$ direction, and the lower wall moves in the opposite direction.
As can be noticed, the initial perturbation is stretched, resulting in a streaky structure which is inclined with
respect to the symmetry plane of the channel.}
\label{fig:CO_3D}
\end{figure}

In order to show the evolution of the initial perturbation in the three-dimensional domain, the open-source
\textit{VisIt} tool has been used. A \matlab script has been written to create the VTK file needed by \textit{VisIt},
and the \textit{points structured grid} format has been used. In \figref{fig:CO_3D}, isosurfaces of
streamwise velocity are shown. 
\newpage
\FloatBarrier
\subsubsection{Streamwise velocity - Plane Couette flow, $Re=500$}
\FloatBarrier
\FloatBarrier
\ \newpage
\subsubsection{Wall-normal velocity - Plane Couette flow with $Re=500$}
\FloatBarrier
\FloatBarrier
\ \newpage
\subsubsection{Spanwise velocity - Plane Couette flow with $Re=500$}
\FloatBarrier
\FloatBarrier
\ \newpage
\subsubsection{Kinetic energy - Plane Couette flow with $Re=500$}
\FloatBarrier
\ \newpage
\subsubsection{3D visualization of streamwise velocity, PCf with $Re=500$}
\vspace{1cm}
\FloatBarrier 

\FloatBarrier
\ \newpage
\section{Linear spot in Blasius boundary-layer flow}
Concluding the present work, the evolution of a localized perturbation in boundary-layer flow is
shown. The base flow here considered is the one corresponding to a flat plate with zero incidence \citep[Blasius
boundary-layer, see e.g.][]{Schlichting_book, Rosenhead}. The chosen value for Reynolds number, defined with the
\textit{displacement thickness} is 1000, while five values of the polar wavenumber are considered, $k=\{1.26,\ 1.57,\
2.09,\ 3.14,\ 6.28\}$. Remind also that the spatial coordinates are here normalized with the displacement thickness,
while the reference velocity is the free stream velocity $U_\infty$. Simulations have been performed with two different
initial conditions in order to get a wide
database with a variety of transient behaviors, whose expressions are the following
\begin{gather}
 \hv_0^{(1)}=y^2e^{-y^2}\ \ \ \ \he_0^{(1)}=0\\ \nonumber
 \hv_0^{(2)}=y^2e^{-y^2}sin(\pi y)\ \ \ \ \he_0^{(2)}=0\nonumber
\end{gather}

\begin{figure}[h!]
        \centering
        \advance\leftskip-1.2cm
        \begin{subfigure}{1\textwidth}
        \centering
\includegraphics[width=16.0cm]{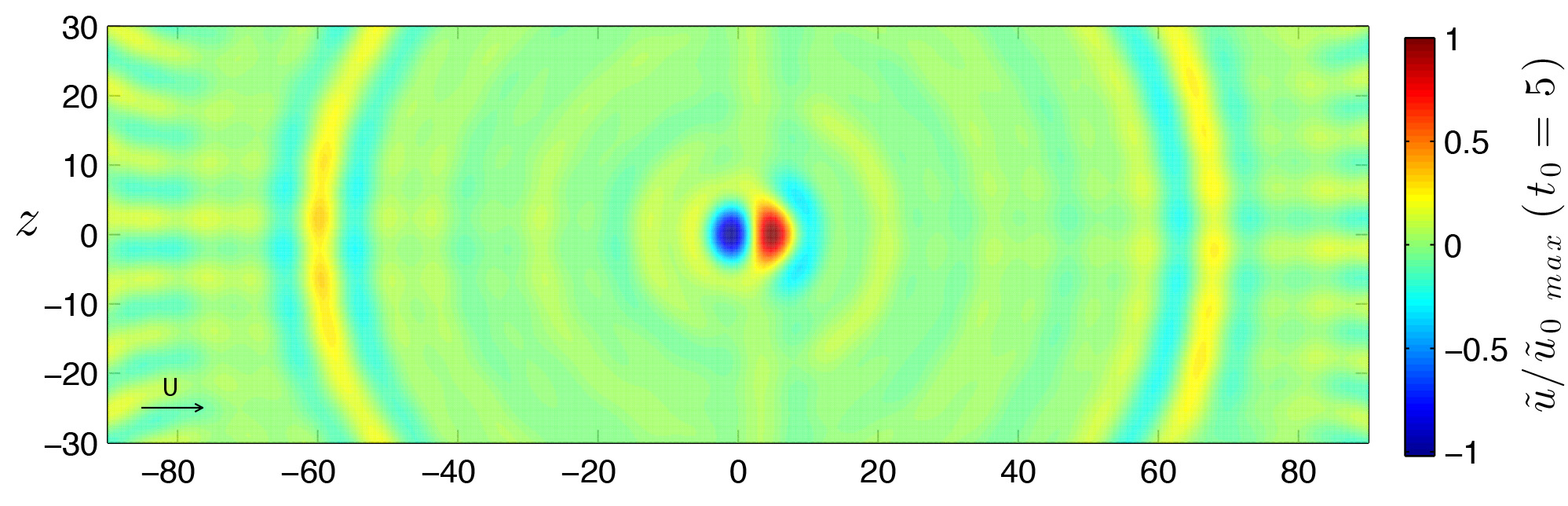}
	\vspace{-0.6cm}
	 \end{subfigure}
        \begin{subfigure}{1\textwidth}
        \centering 
\includegraphics[width=16cm]{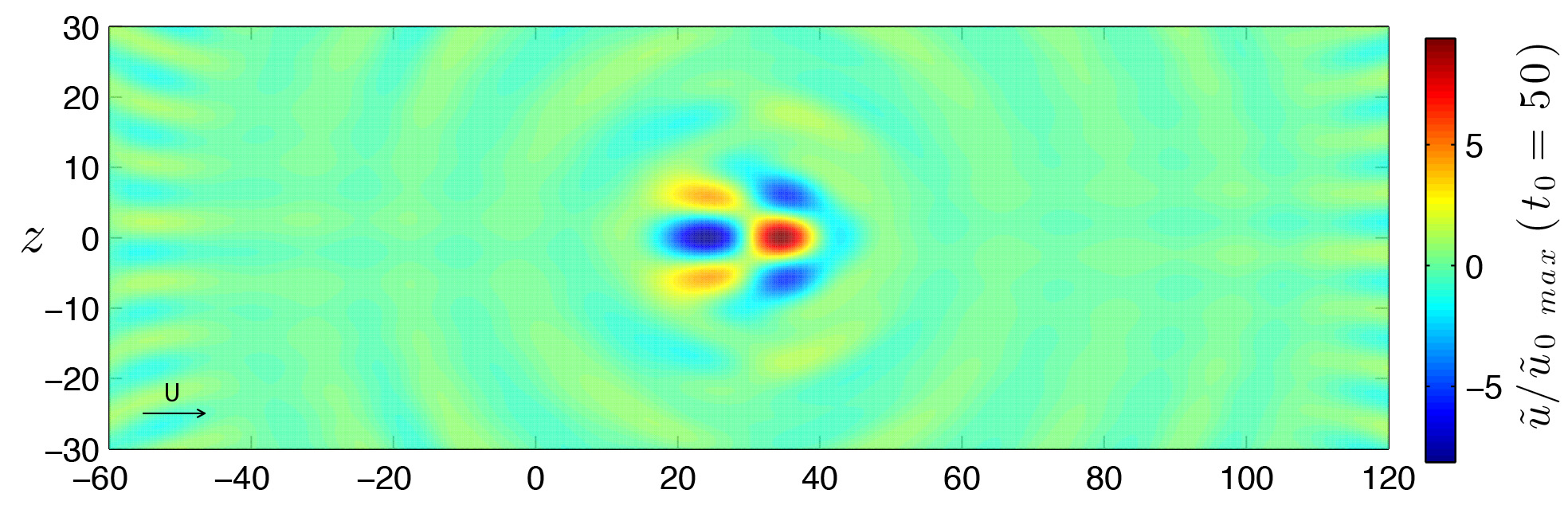}
	\vspace{-0.6cm}
	 \end{subfigure}
	\begin{subfigure}{1\textwidth}
        \centering
\includegraphics[width=16.0cm]{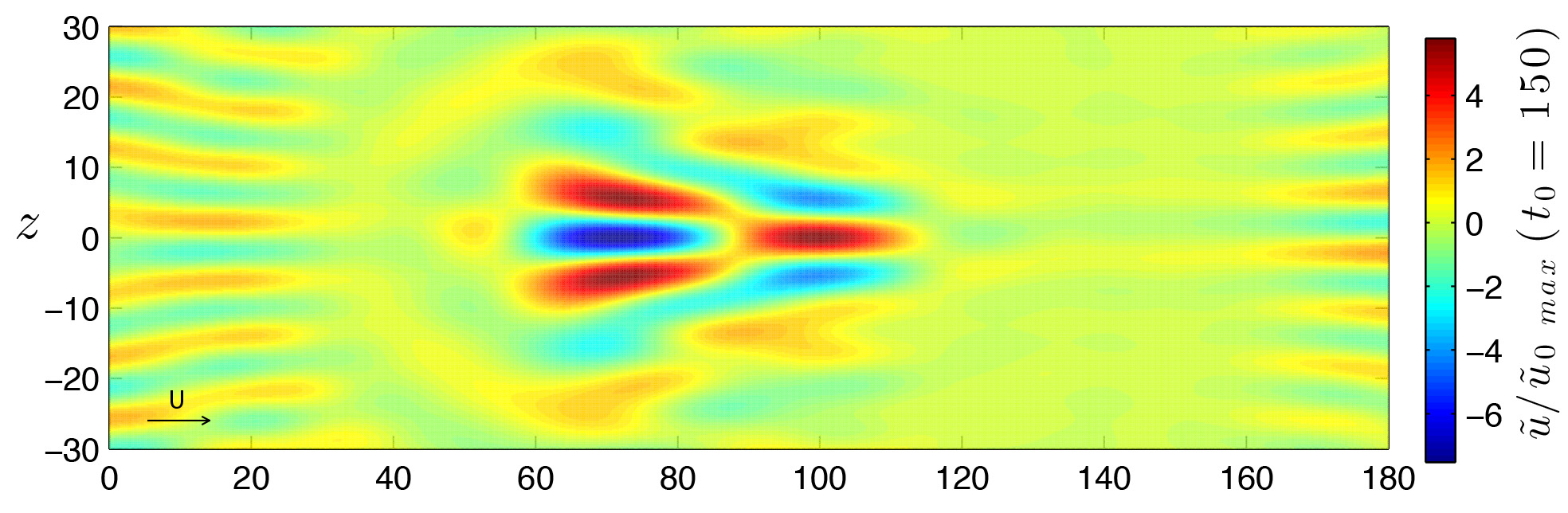}
	\vspace{-0.6cm}
	 \end{subfigure}
	 \begin{subfigure}{1\textwidth}
        \centering
\includegraphics[width=16.0cm]{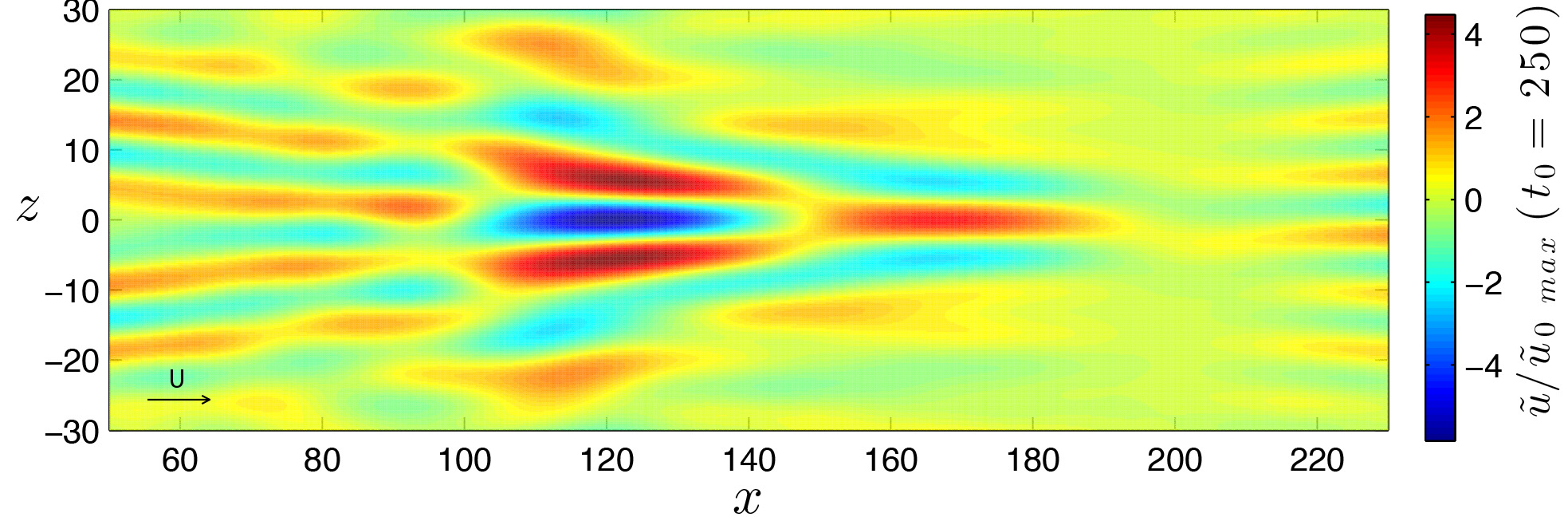}
	\vspace{-0.6cm}
	 \end{subfigure}
	\caption{Visualizations of the longitudinal velocity $\tilde u$ for Bbl with $Re=1000$. Views of $xz$ plane at
$y_0=1.5$. The evolution of a localized perturbation is obtained by superposition of 365
waves with polar wavenumber $k=\{1.26,\ 1.57,\ 2.09,\ 3.14,\ 6.28\}$, obliquity angle spanning the  full circle,
$\phi\in\{-90^\circ,\
+90^\circ\}$, and two different initial conditions. Remind that the mean flow moves in the right direction.}
\label{fig:BL_Re1000_localized_U}
\end{figure}

\begin{figure}[h!]
        \centering
\includegraphics[width=14.0cm]{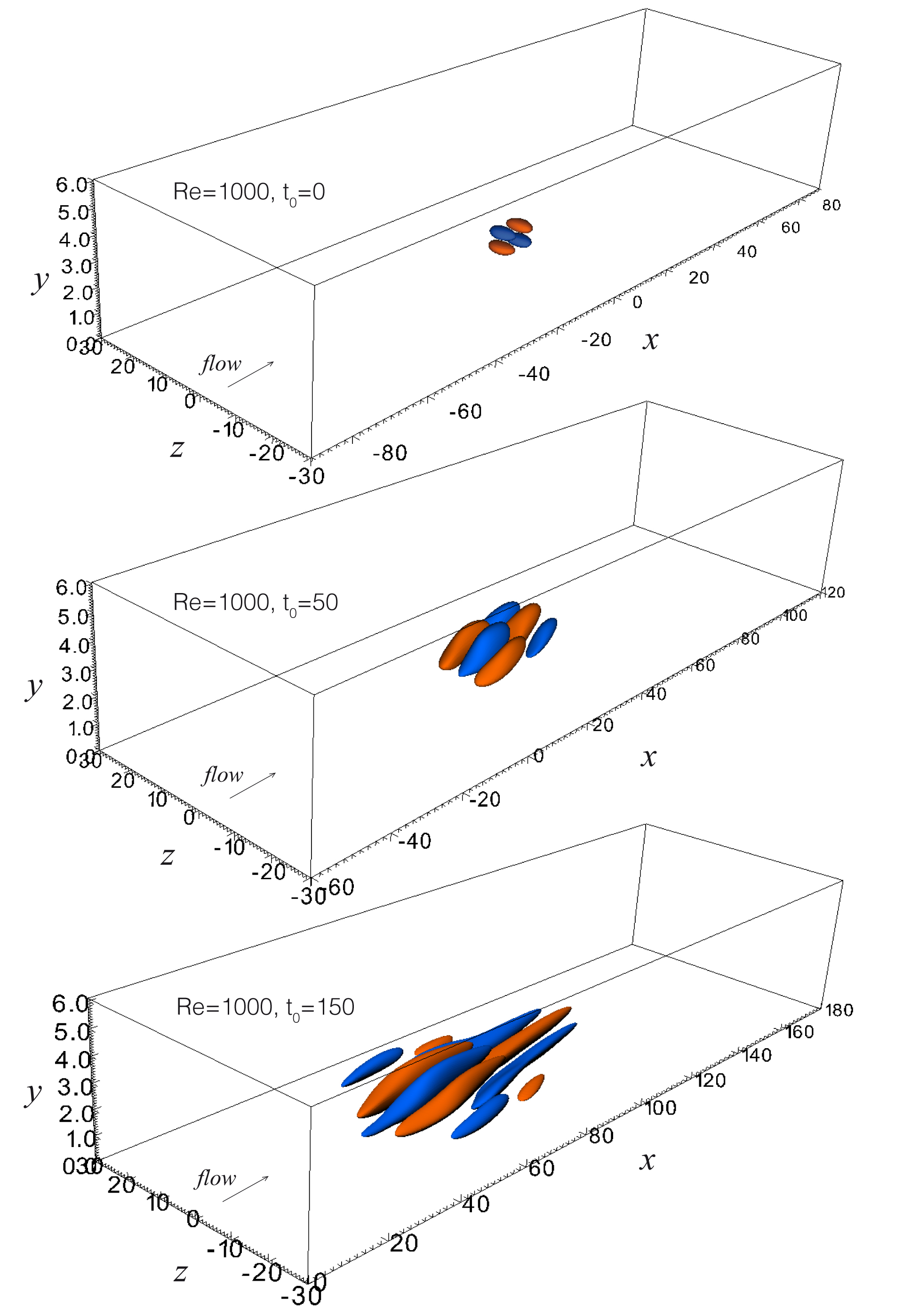}
	\caption{3D visualization of streamwise velocity for Bbl with $Re=1000$. (a) Initial condition, $t_0=0$;
\textit{orange surface}: $\tilde u / {\tilde u}_{0\ max}=0.5$; \textit{blue surface}: $\tilde u / {\tilde u}_{0\
max}=-0.5$; (b) $t_0=50$; \textit{orange surface}: $\tilde u / {\tilde u}_{0\ max}=1$; \textit{blue surface}: $\tilde u
/ {\tilde u}_{0\
max}=-1$;  (b) $t_0=150$; \textit{orange surface}: $\tilde u / {\tilde u}_{0\ max}=1$; \textit{blue surface}: $\tilde u
/ {\tilde u}_{0\
max}=-1$. Here ${\tilde u}_{0\ max}$ is the maximum value in the whole 3D domain at the initial time (${\tilde u}_{0\
max}=214$).}
\label{fig:BL_3D}
\end{figure}

The former is always positive while the latter is oscillating, they both satisfy the boundary conditions and have their
maximum near the wall, inside the boundary layer. 
Also for this case, the evolution of a localized disturbance obtained by in-phase
superposition is shown. The total number of considered waves is 365. However, some trials have been made with a smaller
number of waves, randomly chosen, leading to the same general conclusions. In these cases, a more irregular shape of
the spot is observed. The affinity of the shape acquired by the wave packet with the one of a
turbulent spots, is noticeable (see e.g. \figref{fig:BL_Re1000_localized_U}): the initial disturbance evolves
elongating mainly in the streamwise direction, and a  $\Lambda$-structure can be clearly observed. Even from
noisy or dynamic initial condition cases, carried out by random waves superposition or random  inputs in time (not
presented in this work), it is possible to observe a flow field dominated by not exactly rectilinear \textit{streaks},
and often a $\Lambda$ - pattern can be recognised. \\
Also in the case of Bbl, the origin of the $x$ and $z$ axis for non-dimensional coordinates is considered to be the
location of the initial disturbance (see e.g. \figref{fig:BL_Re1000_localized_U}a). The three-dimensional evolution of
the linear spot can be observed from \figref{fig:BL_3D}, where isosurfaces for the streamwise velocity are shown. The
qualitative behaviour is in agreement with the one recently shown by \citet{Cherubini2010}. \\
Finally, in \tabref{tab:dimensional} we report the results for a dimensional case with $U_\infty=15 m/s$ and
$\nu=1.45\cdot 10^{-5}$ (air flow). This is helpful for understanding the true order of magnitude of the quantities
involved. 
\ \newpage
\FloatBarrier
\subsubsection{Streamwise velocity - Blasius boundary layer flow with $Re=1000$}
\FloatBarrier
\vspace{-0.8cm}
\FloatBarrier
\subsubsection{Wall-normal velocity - Blasius boundary layer flow with $Re=1000$}
\vspace{-0.6cm}
\FloatBarrier
\begin{figure}[h!]
        \centering
        \advance\leftskip-1.2cm
        \begin{subfigure}{1\textwidth}
        \centering
\includegraphics[width=16.0cm]{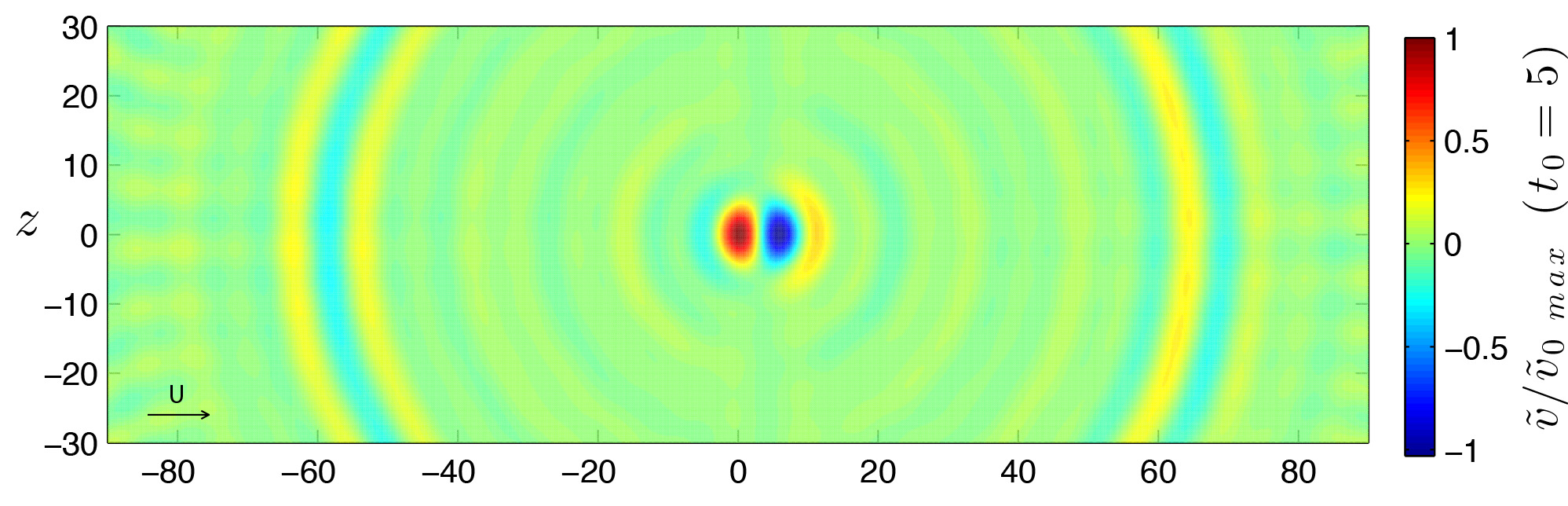}
	\vspace{-0.6cm}
	 \end{subfigure}
        \begin{subfigure}{1\textwidth}
        \centering 
\includegraphics[width=16cm]{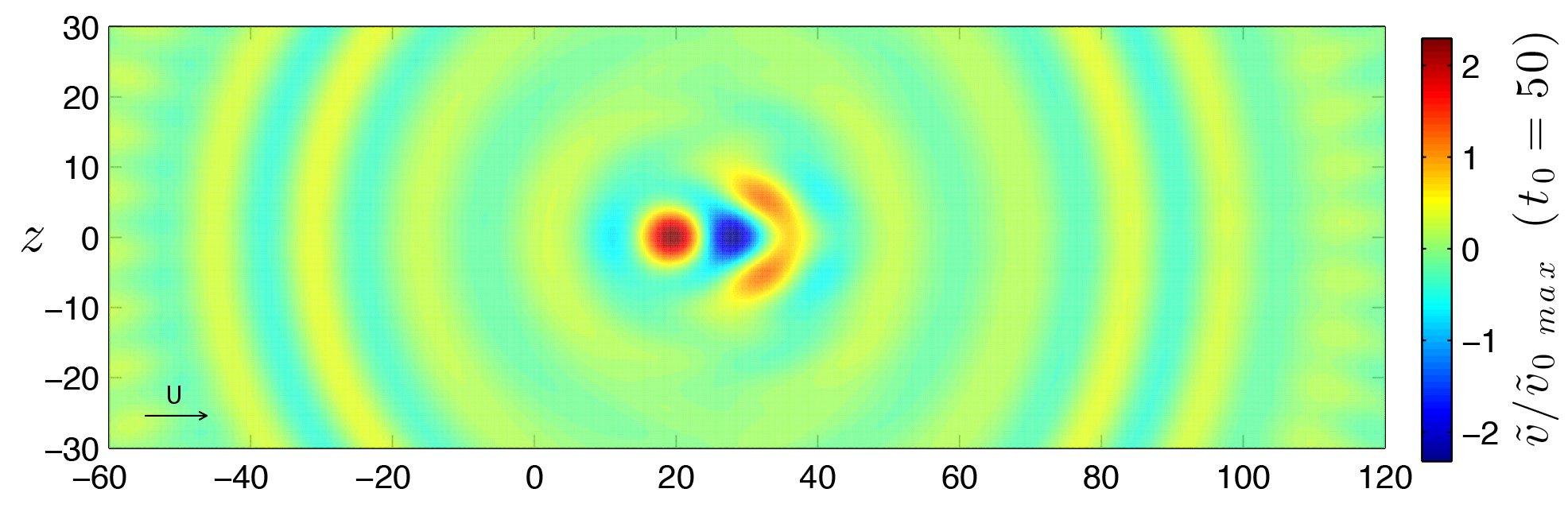}
	\vspace{-0.6cm}
	 \end{subfigure}
	\begin{subfigure}{1\textwidth}
        \centering
\includegraphics[width=16.0cm]{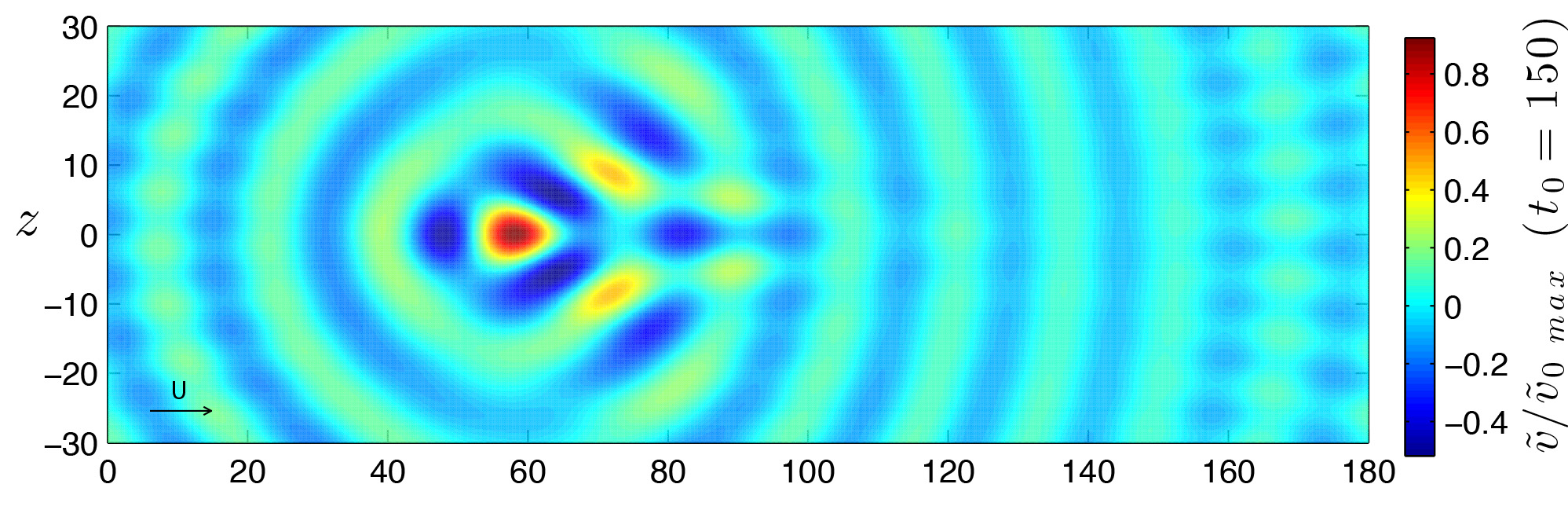}
	\vspace{-0.6cm}
	 \end{subfigure}
	 \begin{subfigure}{1\textwidth}
        \centering
\includegraphics[width=16.0cm]{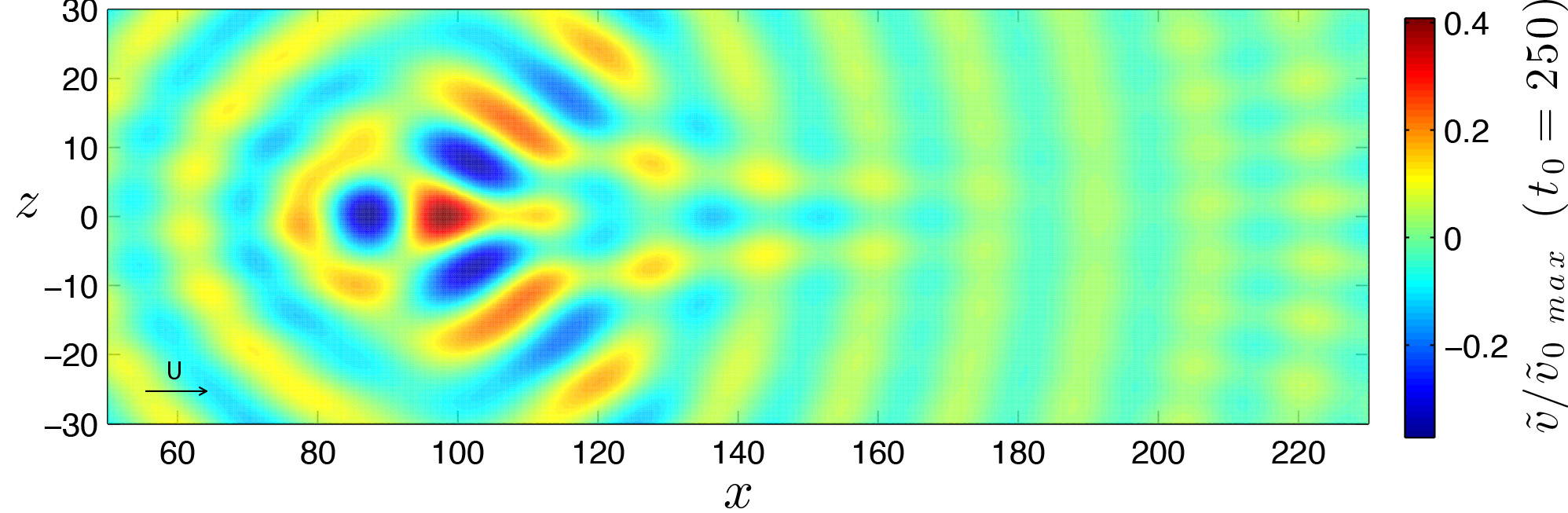}
	\vspace{-0.6cm}
	 \end{subfigure}
	\caption{Visualizations of the wall-normal velocity $\tilde v$ for Bbl with $Re=1000$. Views of $xz$ plane at
$y_0=1.5$. The evolution of a localized perturbation is obtained by superposition of 365
waves with polar wavenumber $k=\{1.26,\ 1.57,\ 2.09,\ 3.14,\ 6.28\}$, obliquity angle spanning the  full circle,
$\phi\in\{-90^\circ,\
+90^\circ\}$, and two different initial conditions.}
\label{fig:BL_Re1000_localized_V}
\end{figure}
\FloatBarrier
\subsubsection{Spanwise velocity - Blasius boundary layer flow with $Re=1000$}
\vspace{-0.6cm}
\FloatBarrier
\begin{figure}[h!]
        \centering
        \advance\leftskip-1.2cm
        \begin{subfigure}{1\textwidth}
        \centering
\includegraphics[width=16.0cm]{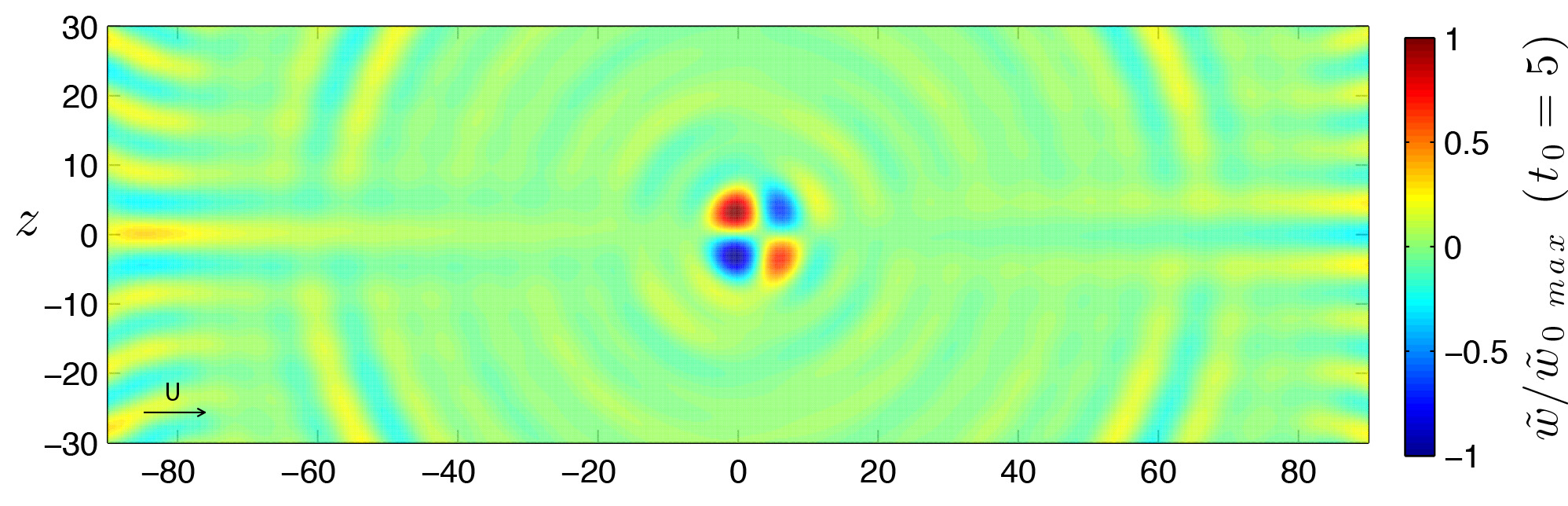}
	\vspace{-0.6cm}
	 \end{subfigure}
        \begin{subfigure}{1\textwidth}
        \centering 
\includegraphics[width=16cm]{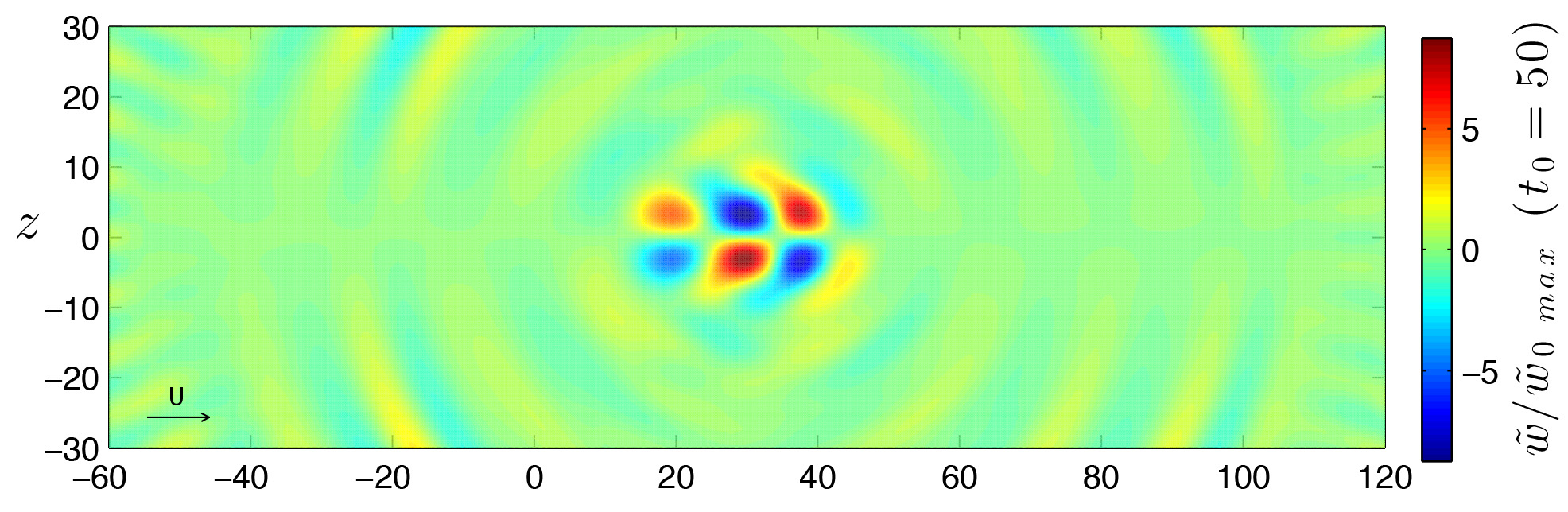}
	\vspace{-0.6cm}
	 \end{subfigure}
	\begin{subfigure}{1\textwidth}
        \centering
\includegraphics[width=16.0cm]{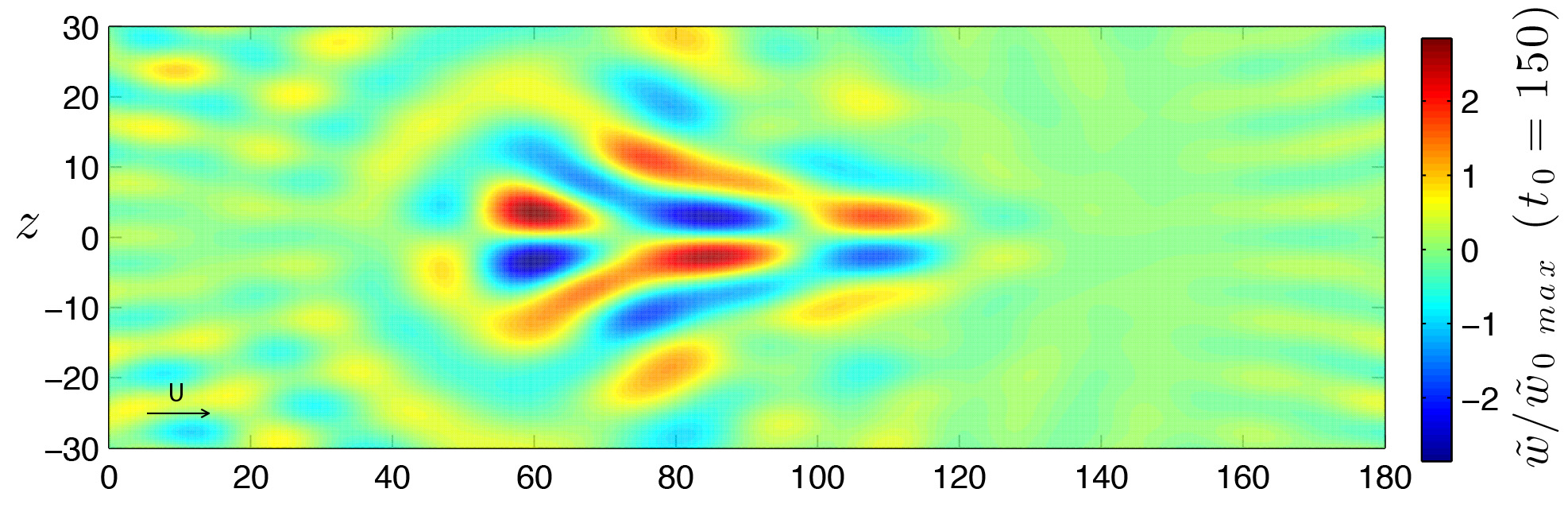}
	\vspace{-0.6cm}
	 \end{subfigure}
	 \begin{subfigure}{1\textwidth}
        \centering
\includegraphics[width=16.0cm]{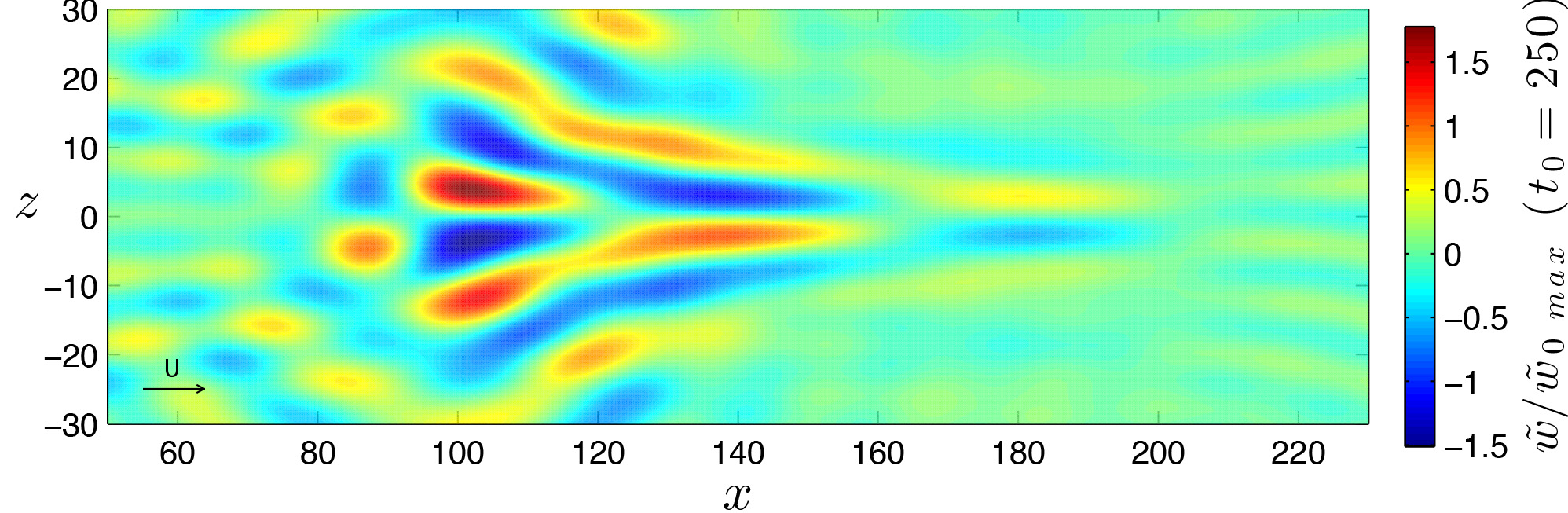}
	\vspace{-0.6cm}
	 \end{subfigure}
	\caption{Visualizations of the spanwise velocity $\tilde w$ for Bbl with $Re=1000$. Views of $xz$ plane at
$y_0=1.5$. The evolution of a localized perturbation is obtained by superposition of 365
waves with polar wavenumber $k=\{1.26,\ 1.57,\ 2.09,\ 3.14,\ 6.28\}$, obliquity angle spanning the  full circle,
$\phi\in\{-90^\circ,\
+90^\circ\}$, and two different initial conditions.}
\label{fig:BL_Re1000_localized_W}
\end{figure}
\FloatBarrier
\subsubsection{Kinetic energy - Blasius boundary layer flow with $Re=1000$}
\vspace{-0.6cm}
\FloatBarrier
\begin{figure}[h!]
        \centering
        \advance\leftskip-1.2cm
        \begin{subfigure}{1\textwidth}
        \centering
\includegraphics[width=16.0cm]{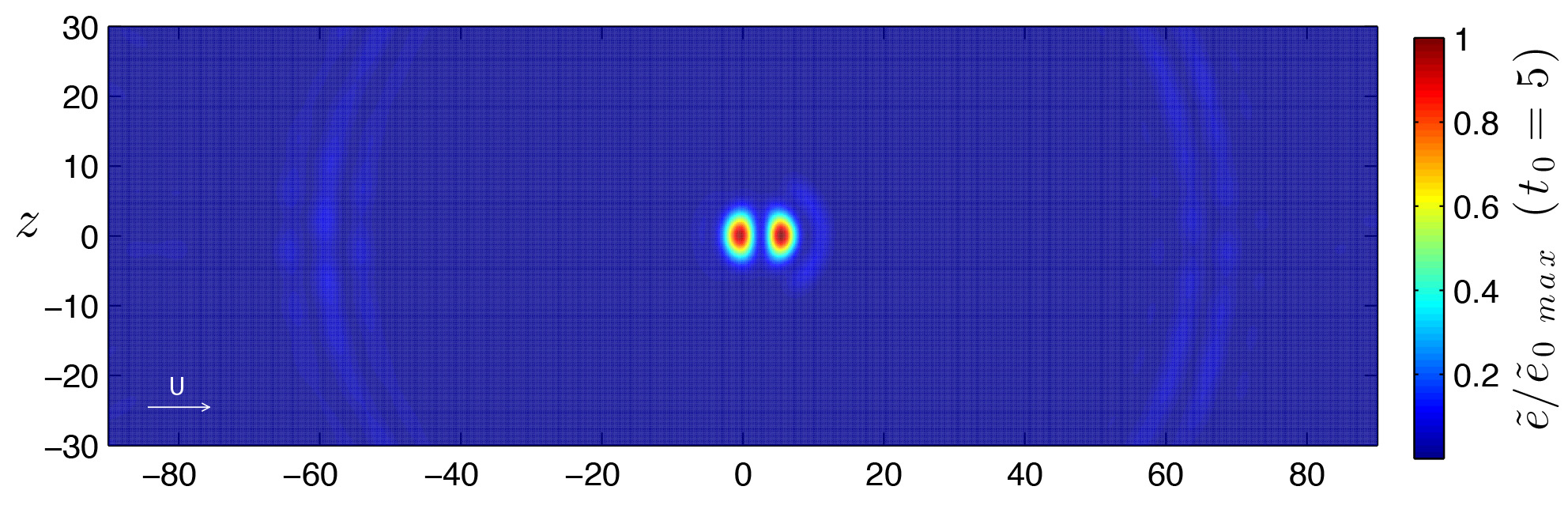}
	\vspace{-0.6cm}
	 \end{subfigure}
        \begin{subfigure}{1\textwidth}
        \centering 
\includegraphics[width=16cm]{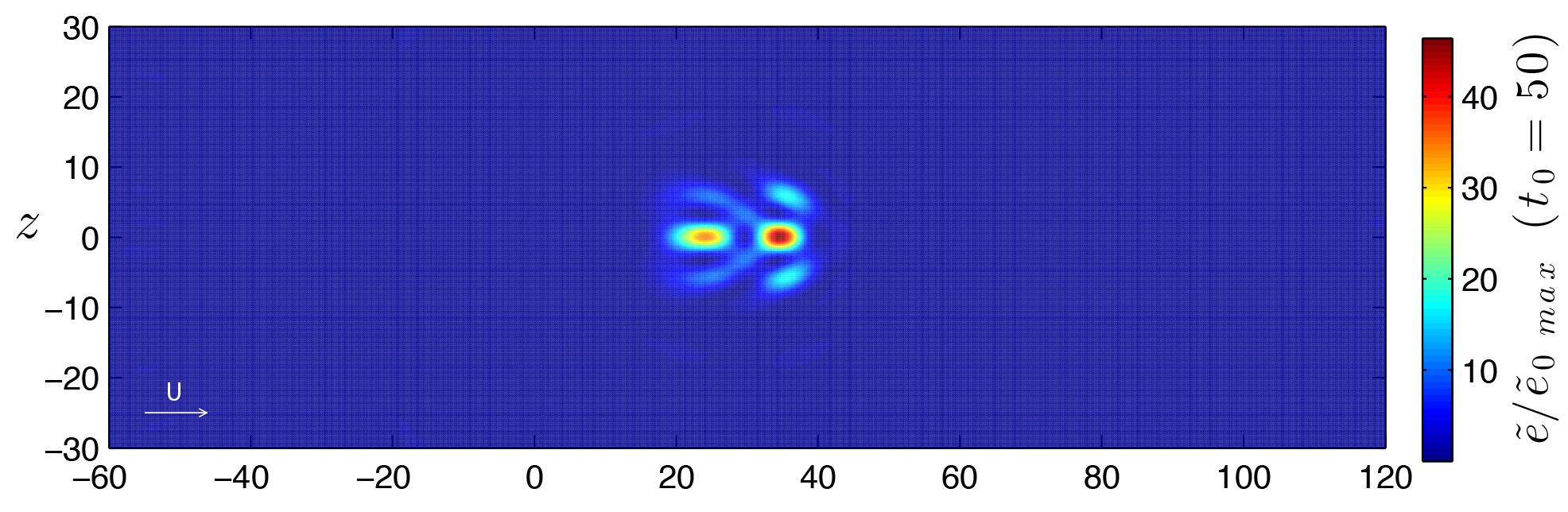}
	\vspace{-0.6cm}
	 \end{subfigure}
	\begin{subfigure}{1\textwidth}
        \centering
\includegraphics[width=16.0cm]{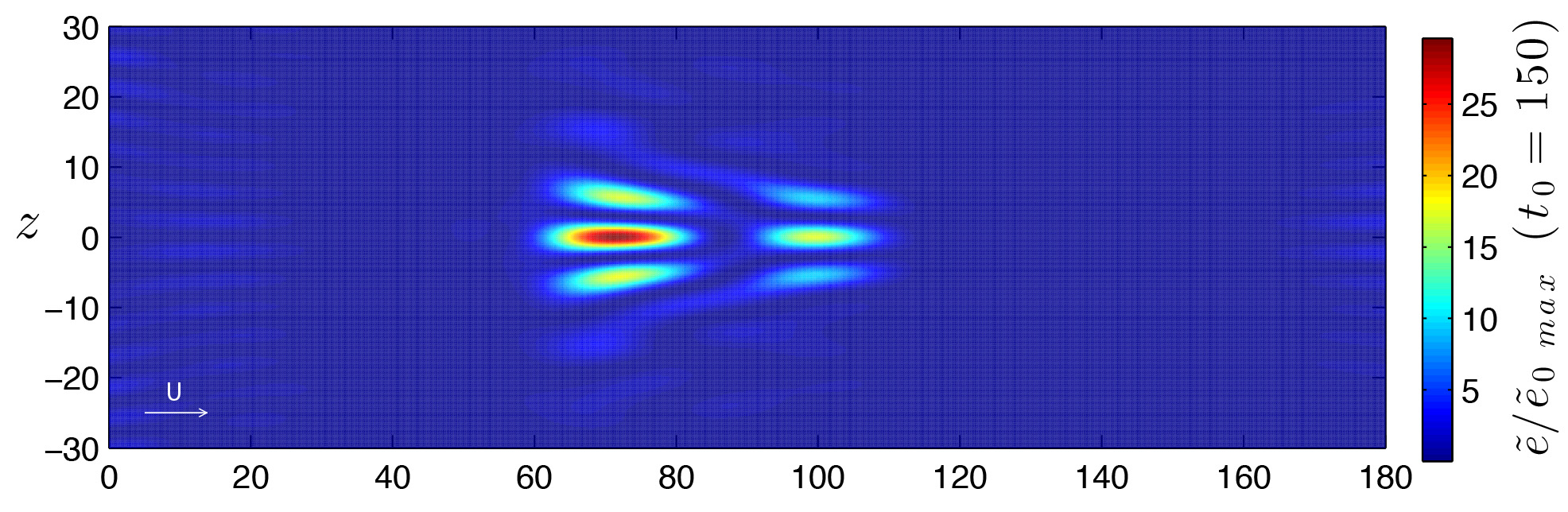}
	\vspace{-0.6cm}
	 \end{subfigure}
	 \begin{subfigure}{1\textwidth}
        \centering
\includegraphics[width=16.0cm]{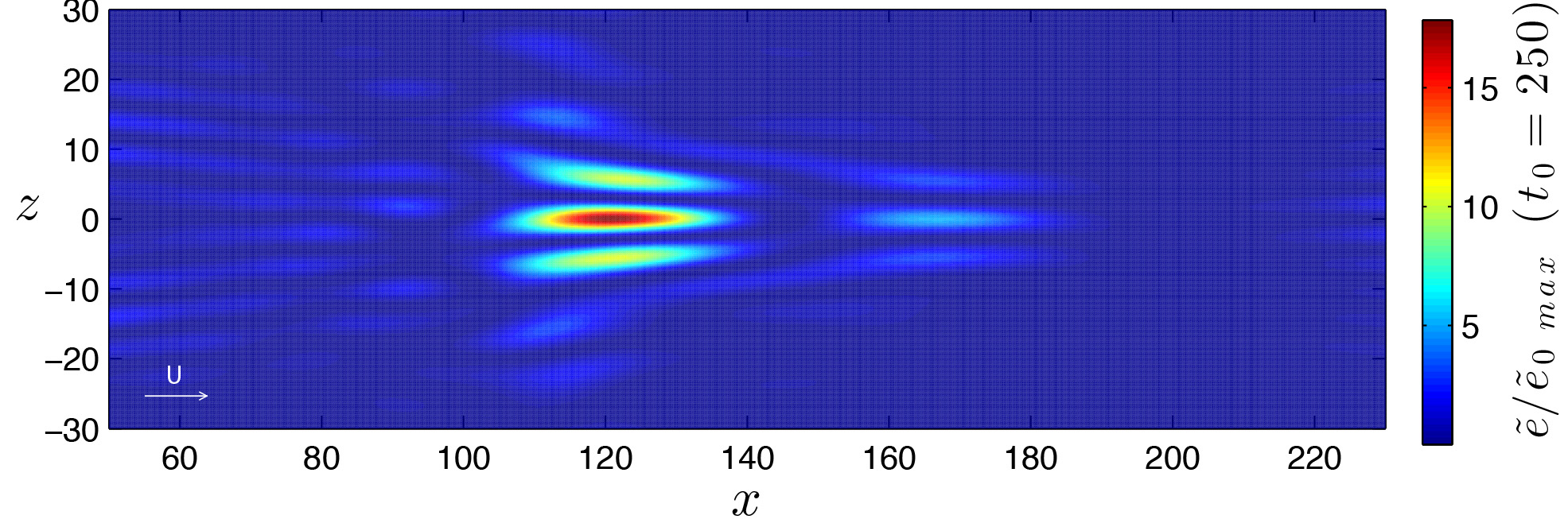}
	\vspace{-0.6cm}
	 \end{subfigure}
	\caption{Kinetic energy visualizations for Bbl with $Re=1000$. Views of $xz$ plane at
$y_0=1.5$. The evolution of a localized perturbation is obtained by superposition of 365
waves with polar wavenumber $k=\{1.26,\ 1.57,\ 2.09,\ 3.14,\ 6.28\}$, obliquity angle spanning the  full circle,
$\phi\in\{-90^\circ,\
+90^\circ\}$, and two different initial conditions.}
\label{fig:BL_Re1000_localized_E}
\end{figure}

\subsubsection{3D visualization of streamwise velocity, Bbl with $Re=1000$}
\vspace{-0.6cm}
\FloatBarrier 

In the following table, a summary of the linear spot evolution for the analyzed case is shown, and also the dimensional
quantities are reported for a specific case with $U_\infty=15 m/s$ and air flow ($\nu=1.45\cdot10^{-5}$). In this
conditions, a Reynolds number of 1000 is found at a distance $x-x_{le}=328$ mm from the leading edge of the flat plate,
where $\delta^*=0.97$ mm. This is considered the origin of the non-dimensional $x$ axis, where the linear spot is
triggered. Actually, in the evolution of the perturbation the boundary-layer thickness and the Reynolds number change,
but in this analysis the approximation of \textit{nearly-parallel} flow applies. In \tabref{tab:dimensional} $x_c$ is
the location of the center of the spot, $U_c$ is the longitudinal velocity of the spot center, $L_x$ and $L_z$ are the
longitudinal and spanwise sizes for $y_0=1.5$, respectively. All these quantities are esteemed from visualizations of
the streamwise perturbation velocity.

\begin{table}[h!]
\centering
  \begin{tabular}{cccccc}
 \hline\hline
  \rule[-0.3cm]{0mm}{0.8cm}
  \multirow{2}{*}{} & \boldmath $t=0$  & \boldmath
$t=5$ & \boldmath $t=50$ & \boldmath $t=150$ & \boldmath
$t=250$\\ 
  \boldmath $ $  & \boldmath \textcolor{blue}{$0$ ms}  & \boldmath
\textcolor{blue}{$0.323$ ms} & \boldmath \textcolor{blue}{$32.3$ ms} & \boldmath \textcolor{blue}{$9.70$ ms} & \boldmath
\textcolor{blue}{$16.2$ ms}\\ 
  \hline\hline \rule[0 cm]{0mm}{0.5cm}   
  \boldmath $\delta^*$  &\textcolor{blue}{$0.967$ mm}  &\textcolor{blue}{$0.972$ mm} &\textcolor{blue}{$1.01$ mm} &
\textcolor{blue}{$1.08$ mm} & \textcolor{blue}{$1.16$ mm} \\ \hline
  \boldmath $\delta_{0.99}$  &\textcolor{blue}{$2.770$ mm}  &\textcolor{blue}{$2.773$ mm} &\textcolor{blue}{$2.88$ mm}
& \textcolor{blue}{$3.08$ mm} & \textcolor{blue}{$3.31$ mm} \\ \hline
  \multirow{2}{*}{ \boldmath $x_c$}  & 0 & 2 & 29.6 & 88.0 & 148  \\
            & \textcolor{blue}{$328$ mm}  & \textcolor{blue}{$330$ mm}  & \textcolor{blue}{$357$ mm}  &
\textcolor{blue}{$413$ mm}  & \textcolor{blue}{$471$ mm}  \\ \hline
 \multirow{2}{*}{ \boldmath $U_c$}  & 0 & 0.8 & 0.67 & 0.64 & 0.64 \\
            &\textcolor{blue}{$0$ m/s} & \textcolor{blue}{$12$ m/s}  & \textcolor{blue}{$10$ m/s}  &
\textcolor{blue}{$9.6$ m/s}  & \textcolor{blue}{$9.6$ m/s} \\ \hline
 \multirow{2}{*}{ \boldmath $L_x$}  & 6 & 12 & 25 & 60 & 94 \\
            &\textcolor{blue}{$5.82$ mm} & \textcolor{blue}{$11.6$ mm}  & \textcolor{blue}{$24.3$ mm}  &
\textcolor{blue}{$58.2$ mm}  & \textcolor{blue}{$91.2$ mm} \\ \hline
 \multirow{2}{*}{ \boldmath $L_z$}  & 6 & 12 & 18 & 24 & 24 \\
            &\textcolor{blue}{$5.82$ mm} & \textcolor{blue}{$11.6$ mm}  & \textcolor{blue}{$17.5$ mm}  &
\textcolor{blue}{$23.3$ mm}  & \textcolor{blue}{$23.3$ mm} \\  \hline\hline
 \end{tabular}
\caption{Dimensional quantities for the evolution of a localized perturbation in a Blasius boundary-layer flow with
$Re=1000$. The approximation of nearly-parallel flow is applied, in fact the Reynolds number for every simulation is
fixed. It can be noticed that the true spatial and temporal scales are quite small.}
\label{tab:dimensional}
\end{table}

\chapter{Conclusions}
The present work deals with the hydrodynamic stability non-modal analysis. In the first part, a close form
solution to the three-dimensional Orr-Sommerfeld and Squire IVP in the form of orthogonal functions
expansion was researched. The Galerkin variational method was then successfully implemented in the \matlab environment
to
numerically compute approximate solutions to the coupled equations, for bounded flows. The Chandrasekhar functions
revealed to ensure a convergence rate scaling as $N^5$ to the correct solution even for the non-modal analysis. The
advantages of this method can be summarized in the independence of the accuracy on both the temporal and
spatial grids, which can be considered arbitrary, and the very low time computational cost. Moreover, since
there is no ``marching'' in time, no stiffness problems are encountered and accurate solutions can be obtained up to
very high times. The
spectra of the Orr-Sommerfeld and Squire operators are computed with high precision, as well.\par
The code has been intensively used to focus on the temporal evolution of the wave frequency and phase velocity,
poorly investigated in the past. The results confirmed recent observations about the frequency jump in the $\tilde v$
component of flow velocity, considered as the end of the \textit{Early transient}. After this first jump, the frequency
of $\tilde v$ for Plane Couette flow experiences a
periodic modulation about the asymptotic value, which has been motivated and investigated in detail.
\par A new result is the presence of a second jump in the phase velocity of the vorticity component $\tilde \eta$ and
consequently of
the other components of
velocity, typically for high times. The presence of a second jump and the possibility for different values
of asymptotic
frequency of the signals were motivated and no contradictions with the results of the modal theory subsist. This is the
proof of the existence of an \textit{Intermediate transient}, in fact, only after the last jump the solution reaches
its asymptotic state. 
Moreover, a connection between the frequency jumps and the establishing of a self-similarity condition in time for both
the
velocity and vorticity profiles was found and investigated for both Plane Couette flow and Plane Poiseuille flow. The
behavior in the physical space was also shown.\par
The last result deals with the linear evolution of wave packets. Through superposition of waves with limited wavenumber
range a wave packet is
reconstructed for Plane Couette flow and  Blasius boundary-layer flow. The structure of the linear spot revealed to
have many common features with the early stages of a \textit{turbulent spot}, particularly the \textit{streaky}
structure and
the shape. This is in agreement with recent ideas and observations and supports the thesis of the underrated importance
of the linear mechanisms such as the transient growth, in the transitional scenario.
\chapter*{Acknowledgements}
\addcontentsline{toc}{chapter}{Acknowledgements}
I wish to express my sincere gratitude to my supervisor, Prof. Daniela Tordella, for introducing me to the Hydrodynamic
Stability, for her guidance and for giving me the opportunity to join her collaboration with Prof. Gigliola
Staffilani.\par
I would like to thank Prof. Gigliola Staffilani for the time she dedicated to me, for her precious guidance and
teachings.\par
I wish also to acknowledge the personnel of the DIMEAS, especially Daniela Foravalle for her precious help during this
year.\par
I am deeply grateful to my parents, my sister and my grandparents, which supports me constantly every day,
and a special thank is for my uncle Claudio, who helped me with the organization of my stage.\par
I wish to thank Ted and all my friends from Boston for the good time spent together during my stay.\par
Thanks to Marco, Vito, Stefano e Andrea for their friendship, and for the  amazing
discussions of these last years. Let me also express my deep gratitude to two special persons, named Luca, for their
friendship. \par
I'd like to dedicate this work to my dear girlfriend, Cecilia.
\newpage \ \thispagestyle{plain}

 \begin{appendices}
 \titleformat{\chapter}[display]
 {\normalfont\sc\Large\centering}
 {{Appendix\ }{\thechapter}}{30pt}{\rm\huge}
 \titlespacing{\chapter}{0pt}{120pt}{54.5pt}
 
\chapter{}
\section{The basis eigenfunctions} \label{sec:Appendix_A1}
\begin{figure}
        \centering
	\includegraphics[width=12cm]{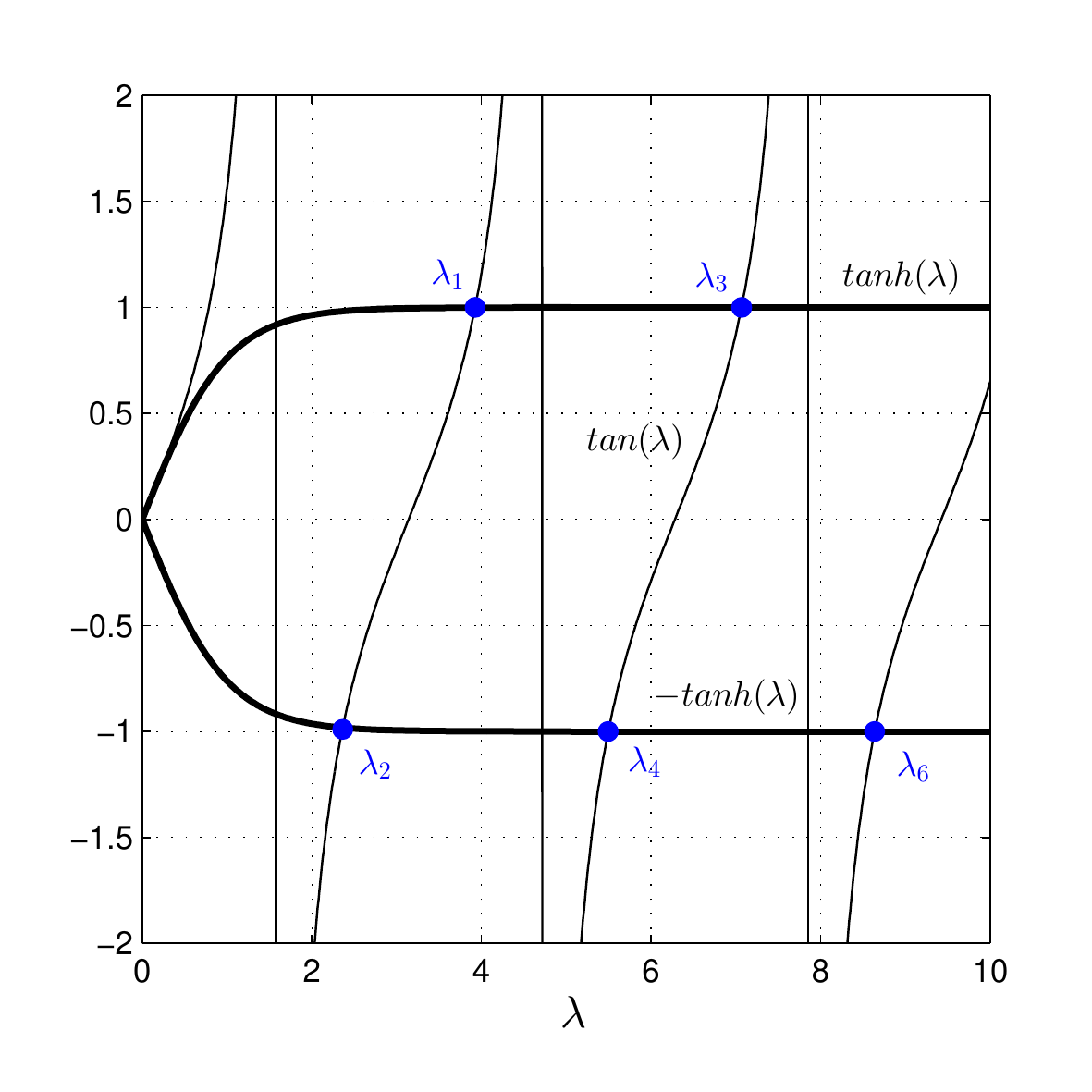}
	\caption{Graphic solution of the eigenvalue problem \eqref{system1}.
Odd indices indicate the eigenvalues corresponding to odd eigenfunctions and
even indices correspond to even eigenfunctions. Since both sets are
antisymmetric with resprect to $\lambda=0$ axis, only positive eigenvalues
are considered.}
\label{fig:graphic_sol}
\end{figure}

The solutions of the problem \eqref{problem4} with homogeneous
boundary conditions \eqref{bc_problem4} can be found by seeking a solution of 
the following form
\begin{gather}
X(y)=C_1e^{\lambda y}+C_2e^{-\lambda y}+C_3e^{i\lambda y}+C_4 e^{-i\lambda y}
\end{gather}
where $C_i$ are constants. Since we are interested in real eigenfunctions, the above
expression becomes
\begin{gather}
X(y)=Asinh(\lambda y)+Bcosh(\lambda y)+Csin(\lambda y)+Dcos(\lambda y)
\end{gather}
where $A$, $B$, $C$ and $D$ are constants, to be determined. Substituting the
solution in the expressions of the boundary conditions, a system of four
algebraic equations where the constant are the unknown is found. The following
relations are obtained
\begin{gather}
A=-C\frac{sin(\lambda)}{sinh(\lambda)}\hspace{1cm}
B=-D\frac{cos(\lambda)}{cosh(\lambda)}
\end{gather}
\begin{gather}
\label{system1}
\begin{bmatrix}
 -\tfrac{sin(\lambda)}{tanh(\lambda)}+cos(\lambda) & -cos(\lambda)
tanh(\lambda)-sin(\lambda)\\
 -\tfrac{sin(\lambda)}{tanh(\lambda)}+cos(\lambda)  & cos(\lambda)
tanh(\lambda)+sin(\lambda)
\end{bmatrix} \begin{pmatrix}C\\D\end{pmatrix}=\begin{pmatrix}0\\0\end{pmatrix}
\end{gather}
Nontrivial solution to the system \eqref{system1} are possible if the
determinant of the coefficients matrix vanishes, this leads to the equation
\begin{gather}
tan^2(\lambda)-tanh^2(\lambda)=0
\end{gather}
from which the couple of trascendental equations \eqref{trasc_odd} and
\eqref{trasc_even} are derived. The solution to these equations consists in two
sets of infinite and discrete eigenvalues $\lambda_n$, that can be
computed numerically by the bisection or the Newton-Raphson method
(\figref{fig:graphic_sol}). In \tabref{Eig} the first twenty
eigenvalues are shown.

\begin{table}[h!]
\centering
\begin{subtable}{0.3\textwidth}
\centering
  \begin{tabular}{cc}
 \hline
  \rule[-0.3cm]{0mm}{0.8cm}
      $n$  & $\lambda_{n\ (odd)}$ \\ 
   \hline \rule[0 cm]{0mm}{0.5cm} 1   & 3.926602\\  
       3   & 7.068582 \\
       5   & 10.21017 \\
       7   & 13.35176 \\
       9   & 16.49336 \\ 
       11  & 19.63495 \\
       13  & 22.77654 \\
       15  & 25.91813 \\
       17  & 29.05973 \\
       19  & 32.20132 \\ \hline
  \end{tabular}
   \vspace{15pt}
  \subcaption{Odd eigenvalues}
\end{subtable}
\begin{subtable}{0.3\textwidth}
\centering
  \begin{tabular}{cc}
\hline
  \rule[-0.3cm]{0mm}{0.8cm}
     $n$  & $\lambda_{n\ (even)}$ \\ 
   \hline \rule[0 cm]{0mm}{0.5cm} 2   & 2.365020\\  
       4   & 5.497803 \\
       6   &  8.639379 \\
       8   & 11.78097 \\
       10  & 14.92256 \\
       12  & 18.06415 \\ 
       14  & 21.20575 \\
       16  & 24.34734 \\
       18  & 27.48893 \\
       20  & 30.63052 \\ \hline
         \end{tabular}
         \vspace{15pt}
           \subcaption{Even eigenvalues}
\end{subtable}
\caption{First 20 eigenvalues, numerically computed using the Newton-Raphson
method.}
\label{Eig}
\end{table}
 \chapter{}
\section{\matlab scripts for channel flows Galerkin method}
\subsection{Main program: ``main\_ivp\_galerkin.m''}\label{sec:Appendix_B11}
\begin{lstlisting}[language=matlab]
%% IVP SOLUTION BY GALERKIN METHOD for Channel flows
% The stability IVP is reduced to an ODE system and solved by
% Galerkin variational method.
clear all,close all,clc
tic
%% Simulation parameters
global tol N U DU D2U type Re k_polar alfa beta phi ss y y_ic t
type='Couette';
Re=500;        % Reynolds number
k_polar=6.5;   % Polar wavenumber
phi=5;         % Obliquity perturbation angle
alfa=k_polar*cosd(phi);
beta=k_polar*sind(phi);
ss='sym';      % Type of initial condition
h=0.01;  
y=-1:h:1;       % Grid for solution
y_ic=-1:1e-4:1; % Fine grid for initial condition computaiton
N=200;          % Num. of eigenmodes chosen for calculation (must be even).
tol=1e-10;      % Tolerance for error checking.
col=colormap(jet(15));
y0=0.5;
temp=find(y>=y0);
idxy=temp(1);
dt=0.5;
t=0:dt:100;     % Time grid

disp '%%%%%%%%%%%%%%%%%%%%%%%%%%%%%%%%%%%%%%%%%%%%%%%%%%%%%%%%%%%%%%%%%%%%'
disp(['Simulation of Plane ',num2str(type),' flow'])
disp(['Re= ',num2str(Re),', k_polar= ',num2str(k_polar),', phi= ',...
     num2str(phi),' deg, ',num2str(ss),' Initial Condition.'])
disp(['N= ',num2str(N),' eigenfunctions'])
disp '%%%%%%%%%%%%%%%%%%%%%%%%%%%%%%%%%%%%%%%%%%%%%%%%%%%%%%%%%%%%%%%%%%%%'

%% Base flow definition
if strcmp(type,'Couette')==1
    U=y;
    DU=1;
    D2U=0;
elseif strcmp(type,'Poiseuille')==1
    U=(1-y.^2);
    DU=-2*y;
    D2U=-2;
end

%-------------------------------------------------------------------------%
%% NORMAL-VELOCITY SOLUTION v(y,t)
%-------------------------------------------------------------------------%
%% Eigenmodes computation
% The chosen base of functions is composed by two set of eigenfuncions of
% the problem X_yyyy=lam^4*X with b.c. X(-1)=X(1)=X'(-1)=X'(1)=0.
% One set is made of odd modes, the other contains even modes. Both sets are
% necessary to obtain a correct solution of the 3D problem, independently 
% on the parity of the initial condition.
% Here we put togheter the two set in following order:[1o 1e 2o 2e 3o 3e...]
%(e=even, o=odd), and the corresponding eigenvalues are ordered in this
%sequence too. See "compute_eigenvalues.m" for their computation.

load 'eigenvalues.mat';% "eigenvalues.mat", contains the eigenvalues
                       %corresponding to form functions.
g=gtot(1:N);           % g are the eigenvalues.

eigenmodes=zeros(length(g),length(y));
D1eigenmodes=zeros(length(g),length(y)); % first derivatives of eigenmodes
D2eigenmodes=zeros(length(g),length(y)); % second derivatives of eigenmodes
for i=1:length(g)
 if mod(i,2)==1 % odd eigenfuncion
  eigenmodes(i,:)=1/sqrt(2)*(sinh(g(i)*y)/sinh(g(i))-sin(g(i)*y)/sin(g(i)));
  eigenmodes_ic(i,:)=1/sqrt(2)*(sinh(g(i)*y_ic)/sinh(g(i))-sin(g(i)*y_ic)/sin(g(i)));
  D1eigenmodes(i,:)=g(i)/sqrt(2)*(cosh(g(i)*y)/sinh(g(i))-cos(g(i)*y)/sin(g(i)));
  D2eigenmodes(i,:)=g(i)^2/sqrt(2)*(sinh(g(i)*y)/sinh(g(i))+sin(g(i)*y)/sin(g(i)));
 elseif mod(i,2)==0 % even eigenfuncion
  eigenmodes(i,:)=1/sqrt(2)*(cosh(g(i)*y)/cosh(g(i))-cos(g(i)*y)/cos(g(i)));
  eigenmodes_ic(i,:)=1/sqrt(2)*(cosh(g(i)*y_ic)/cosh(g(i))-cos(g(i)*y_ic)/cos(g(i)));
  D1eigenmodes(i,:)=g(i)/sqrt(2)*(sinh(g(i)*y)/cosh(g(i))+sin(g(i)*y)/cos(g(i)));
  D2eigenmodes(i,:)=g(i)^2/sqrt(2)*(cosh(g(i)*y)/cosh(g(i))+cos(g(i)*y)/cos(g(i)));
 end   
end

%% Computation of coefficients Cn0, from prescribed initial condition
if strcmp(ss,'sym')==1
    ic=(1-y.^2).^2;
    ic_a=(1-y_ic.^2).^2;    % symmetric initial condition
elseif strcmp(ss,'asym')==1
    ic=(1-y.^2).^2;
    ic_a=y.*(1-y_ic.^2).^2; % asymmetric initial condition
end

A=zeros(1,N);
approx_ic=zeros(1,length(y));
Cn0=zeros(1,N);
for i=1:N;
    num=trapz(y_ic,ic_a.*eigenmodes_ic(i,:));
    den=1;
    Cn0(i)=num/den;
    approx_ic(1,:)= approx_ic(1,:)+Cn0(i)*eigenmodes(i,:);
end

%% Computation of matrices D S F U1 U2 U3 H G
% ODE system is [H]{dc/dt}-[G]{c}=0
D=zeros(N,N);
S=zeros(N,N);
H=zeros(N,N);
F=zeros(N,N);
U1=zeros(N,N);
U2=zeros(N,N);
G=zeros(N,N);
H=zeros(N,N);

%% D and F
Dnm=zeros(1,N);
Fnm=zeros(1,N);
for i=1:N
    Dnm(i)=1;
    Fnm(i)=Dnm(i)*g(i)^4;
end
D=diag(Dnm,0);
F=diag(Fnm,0);

bbb=(cosh(2*g)-cos(2*g))./(sinh(2*g)-sin(2*g));
idNaN=find(isnan(bbb)==1);
bbb(idNaN)=1;
S=zeros(N,N);

for m=1:N
    for n=1:N
        %% S
        if mod((m+n),2)==0 && n~=m
            S(m,n)=+4*g(n)^2*g(m)^2/(g(n)^4-g(m)^4)*(g(n)*bbb(n)-g(m)*bbb(m));
        elseif mod((m+n),2)==1 && n~=m
            S(m,n)=0;
        elseif n==m
            S(m,n)=-(g(n)^2*bbb(n)^2-g(n)*bbb(n));
        end
        
        %% U1 and U2
        if strcmp(type,'Couette')==1  % analytical expressions for Couette
            
            if  mod((m+n),2)==1 && n~=m
                U1(m,n)=4*g(n)^2*g(m)^2/(g(n)^4-g(m)^4)*(-1+g(n)*bbb(n)...
             -g(m)*bbb(m))-8*(g(n)^4+g(m)^4)/(g(n)^4-g(m)^4)^2*g(n)^2*g(m)^2;
            end
            if mod((m+n),2)==1 && n~=m
                U2(m,n)=16*g(n)^3*g(m)^3*bbb(n)*bbb(m)/(g(n)^4-g(m)^4)^2;
            elseif mod((m+n),2)==0 || n==m
                U2(m,n)=0;
            end
        elseif strcmp(type,'Poiseuille')==1 % numerical integration
            
            prod=U.*D2eigenmodes(n,:).*eigenmodes(m,:);
            U1(m,n)=trapz(y,prod);
            clear prod
            prod=U.*eigenmodes(n,:).*eigenmodes(m,:);
            U2(m,n)=trapz(y,prod);
        end
        
    end
end
%% U3
U3=D2U*D;

%% H and G

H=S-k_polar^2*D;
G=-1i*alfa*U1...
    +1i*alfa*k_polar^2*U2...
    +1i*alfa*U3...
    +1/Re*F...
    -2*k_polar^2/Re*S...
    +k_polar^4/Re*D;

%% ODE system solution
% Compute the series. The solution V has length(y) rows and length(t)
% columns.
%  V(y,t)=sum( Cn(t)*Xn(y))

A=H\G;
%----------------- Check error in H matrix inversion ---------------------%
relerrA_norm2=norm((H*A-G),2)/norm(G,2); 
relerrA_norminf=norm((H*A-G),inf)/norm(G,inf);
disp('-------------------------------------------------------------')
disp(['Absolute error of A computation using \ in norm 2: ',...
    num2str(relerrA_norm2*norm(G,2))])
disp(['Relative error of A computation using \ in norm 2: ',...
    num2str(relerrA_norm2)])
disp(['Relative error of A computation using \ in norm Inf: ',...
    num2str(relerrA_norminf)])
disp('-------------------------------------------------------------')
%-------------------------------------------------------------------------%

% Diagonalizing matrix A
[L,LAMBDA]=eig(A);
 
Cn=zeros(N,length(t));
hh=zeros(N,length(t));
h0=L\(Cn0');
    
%----------------- Check error in L matrix inversion -------------------
relerrL_norm2=norm((L*h0-Cn0'),2)/norm(Cn0,2);
relerrL_norminf=norm((L*h0-Cn0'),inf)/norm(Cn0,inf);
disp('-------------------------------------------------------------')
disp(['Absolute error of h0 computation using \ in norm 2: ',...
    num2str(relerrL_norm2*norm(Cn0,2))])
disp(['Relative error of h0 computation using \ in norm 2: ',...
    num2str(relerrL_norm2)])
disp(['Relative error of h0 computation using \ in norm Inf: ',...
    num2str(relerrL_norminf)])
    
if relerrL_norm2>tol
        
[h02,flag_h0,relerrL2_norm2] = gmres(L,Cn0',[],tol,N);
  if (relerrL2_norm2)>(relerrL_norm2)
     disp(['Won''t use GMRS.'])
  elseif flag_h0==0 && (relerrL2_norm2)<(relerrL_norm2)
     disp(['GMRS converged within tolerance. '])
     disp(['Absolute error of h0 computation using GMRS in norm 2: ',...
         num2str(norm((L*h02-Cn0'),2))])
     disp(['Relative error of h0 computation using GMRS in norm 2: ',...
         num2str(relerrL2_norm2)])
     disp(['Relative error of h0 computation using \ in norm Inf: ',...
         num2str(relerrL_norminf)])
     h0=h02;
  end
end
disp('-------------------------------------------------------------------')


% Compute coefficients Cn(t)
hh=zeros(N,length(t));
for i=1:N
    hh(i,:)=h0(i)*exp(LAMBDA(i,i)*t);
end

Cn=L*(hh); % Coefficients of the series Cn(t) (N rows,length(t) columns)

% Compute v(y,t), v_y(y,t), v_yy(y,t)
v=zeros(length(y),length(t));
Dv=zeros(length(y),length(t));
D2v=zeros(length(y),length(t));
for j=1:length(y)
    for i=1:N
        v(j,:)=v(j,:)+ Cn(i,:)*eigenmodes(i,j);
        Dv(j,:)=Dv(j,:)+Cn(i,:)*D1eigenmodes(i,j);
        D2v(j,:)=D2v(j,:)+Cn(i,:)*D2eigenmodes(i,j);
    end
end

%------------ Check error in initial condition computation----------------%
abserr_v_norm2=norm(v(:,1)-ic',2);
abserr_v_norminf=norm(v(:,1)-ic',inf);
disp('--------------------------------------------------------------------')
disp(['Absolute error on v(t=0) in norm 2: ',num2str(abserr_v_norm2)])
disp(['Absolute error on v(t=0) in norm inf: ',num2str(abserr_v_norminf)])
disp('--------------------------------------------------------------------')

%-------------------------------------------------------------------------%
%% NORMAL-VORTICITY SOLUTION
%-------------------------------------------------------------------------%
% calls function solve_squire.m
% oy(y,t) is omega_y (eta)
oy=solve_squire(L,LAMBDA,h0,eigenmodes,g);

%-------------------------------------------------------------------------%
%% POSTPROCESSING
% v and eta modulus at all y and times
mod_v=abs(v);
mod_oy=abs(oy);
%% Streamwise velocity u(y,t)
u=1/(1i*k_polar^2)*(beta*oy-alfa*Dv);
mod_u=abs(u); % modulus at all y and all times
wrphase_u=phase(u(idxy,:));
phase_u=unwrap(wrphase_u);

%% Spanwise velocity w(y,t)
w=1/(1i*k_polar^2)*(-beta*Dv-alfa*oy);
mod_w=abs(w);
wrphase_w=phase(w(idxy,:));
phase_w=unwrap(wrphase_w);
%% Frequency and phase velocity
% Calculated from v (4th order finite differences centered scheme)
wrphase_v=phase(v(idxy,:));
phase_v=unwrap(wrphase_v);
jj=3:length(t)-2;
omega_v=zeros(1,length(t));
omega_v(jj)=(+phase_v(jj-2)-8*phase_v(jj-1)+8*phase_v(jj+1)...
    -phase_v(jj+2))/(12*dt);
omega_v(1)=(-25*phase_v(1)+48*phase_v(2)-36*phase_v(3)+16*phase_v(4)...
    -3*phase_v(5))/(12*dt);
omega_v(2)=(-3*phase_v(1)-10*phase_v(2)+18*phase_v(3)-6*phase_v(4)+...
    phase_v(5))/(12*dt);
omega_v(end-1)=-(-3*phase_v(end)-10*phase_v(end-1)+18*phase_v(end-2)...
    -6*phase_v(end-3)+phase_v(end-4))/(12*dt);
omega_v(end)=-(-25*phase_v(end)+48*phase_v(end-1)-36*phase_v(end-2)...
    +16*phase_v(end-3)-3*phase_v(end-4))/(12*dt);

c_v=abs(omega_v)/k_polar;  % Phase velocity


%% Energy Growth Factor G = integral(u2+v2+w2) in [-1 1]
ke=mod_u.^2+mod_v.^2+mod_w.^2; % kinetic energy in wavespace
e0=trapz(y,ke(:,1));
G=trapz(y,ke,1)/e0;

%% PLOTS
toc


\end{lstlisting}
\newpage
\subsection{Function: ``solve\_squire.m''}
\begin{lstlisting}[language=matlab]
function [E]=squire_particular_solution(L,LAMBDA,h0,eigenmodes,g)

global tol type  U DU N Re k_polar alfa beta  y t
%% Basic eigenfunctions computation
j=1:N/2;
g_sq_odd=j*pi;
g_sq_even=(2*j-1)*pi/2;

g_sq=zeros(1,N);
g_sq(1:2:N)=g_sq_odd;
g_sq(2:2:N)=g_sq_even;

eigenmodes_squire=zeros(length(g_sq),length(y));
D2eigenmodes_squire=zeros(length(g_sq),length(y));
for i=1:2:length(g_sq)
    eigenmodes_squire(i,:)=sin(g_sq(i)*y);
    D2eigenmodes_squire(i,:)=-g_sq(i)^2*sin(g_sq(i)*y);
end
for i=2:2:length(g_sq)
    eigenmodes_squire(i,:)=cos(g_sq(i)*y);
    D2eigenmodes_squire(i,:)=-g_sq(i)^2*cos(g_sq(i)*y);
end

%% Compute matrices

DD=zeros(N,N);
SS=zeros(N,N);
UU=zeros(N,N);
GG=zeros(N,N);
BB=zeros(N,N);
FORZ=zeros(N,N);

%% D
DD=eye(N);
%% SS
SS=zeros(N,N);
for n=1:N
    SS(n,n)=-g_sq(n)^2  ;
end
%% UU and FORZ
for m=1:N
    for n=1:N
        if strcmp(type,'Couette')==1   % Analytical expression
            if mod((n+m),2)==1
              UU(m,n)=(-1)^((m+n+1)/2)*4*g_sq(m)*g_sq(n)...
                    /(g_sq(n)^2-g_sq(m)^2)^2;
            end
            if mod(m,2)==1 && mod(n,2)==1
              FORZ(m,n)=sqrt(2)*2*g_sq(m)*g(n)^2*(-1)^((m+1)/2)/(g_sq(m)^4-g(n)^4);
            elseif mod(n,2)==0 && mod(m,2)==0
              FORZ(m,n)=sqrt(2)*2*g_sq(m)*g(n)^2*(-1)^(m/2)/(g_sq(m)^4-g(n)^4);
            end
        elseif strcmp(type,'Poiseuille')==1 % Numerical integration 
            prod=U.*eigenmodes_squire(n,:).*eigenmodes_squire(m,:);
            UU(m,n)=trapz(y,prod);
            clear prod
            prod=DU.*eigenmodes(n,:).*eigenmodes_squire(m,:);
            FORZ2(m,n)=trapz(y,prod);
        end
    end
end

%% Find particular solution E_p(y,t)

GG=-1i*alfa*UU+1/Re*SS-k_polar^2/Re*DD;
BB=-1i*beta*FORZ*L;

% Compute coefficient matrix aa 
aa=zeros(N,N);
idx_gmrs=0;

for i=1:N
    TT=LAMBDA(i,i)*eye(N)-GG;
    rhs=BB(:,i)*h0(i);
    aa(:,i)=TT\rhs;
    
    %----------------- Check error in TT matrix inversion ----------------%
    relerrTT_norm2(i)=norm((TT*aa(:,i)-rhs),2)/norm(rhs,2);
    relerrTT_norminf(i)=norm((TT*aa(:,i)-rhs),inf)/norm(rhs,inf);
    abserrTT_norm2(i)= relerrTT_norm2(i)*norm(rhs,2);
    if relerrTT_norm2(i)>tol      
        [aa2(:,i),flag_TT,relerrTT2_norm2(i)] = gmres(TT,rhs,[],tol,N);  
        if (relerrTT2_norm2(i))<(relerrTT_norm2(i))
            aa(:,i)=aa2(:,i);
            relerrTT_norm2(i)=relerrTT2_norm2(i)
            relerrTT_norminf(i)=norm((TT*aa(:,i)-rhs),inf)/norm(rhs,inf);
            abserrTT_norm2(i)= relerrTT_norm2(i)*norm(rhs,2);
            idx_gmrs=[idx_gmrs i];     
        end   
    end
    %---------------------------------------------------------------------%
end
if idx_gmrs ~=0
    disp(['Used GMRS for aa, at columns of indices: ',num2str(idx_svd)])
end
disp(['Max. absolute error for aa in norm 2: ', num2str(max(abserrTT_norm2))])
disp(['Max. relative error for aa in norm 2: ', num2str(max(relerrTT_norm2))])
disp(['Max. relative error for aa in norm Inf: ', num2str(max(relerrTT_norminf))])

% Compute coefficients for particular solution
expon=zeros(N,length(t));
for i=1:N
    expon(i,:)=exp(LAMBDA(i,i)*t);
end
Jp=aa*expon;
Jp0=KK*ones(N,1);% for zero initial vorticity
% Compute solution
E_p=zeros(length(y),length(t));
for j=1:length(y)
    for i=1:N
        E_p(j,:)=E_p(j,:)+ Jp(i,:)*eigenmodes_squire(i,j);
    end
end

%% Compute homogeneous solution E_h(y,t)

[L_eta,LAMBDA_eta]=eig(GG);
hh0_eta=-L_eta\Jp0;  % Here we use initial condition eta(t=0,y)=0

%----------------- Check error in L_eta matrix inversion --------------------
relerrLeta_norm2=norm((L_eta*hh0_eta+Jp0),2)/norm(-Jp0,2);
relerrLeta_norminf=norm((L_eta*hh0_eta+Jp0),inf)/norm(-Jp0,inf);
abserrLeta_norm2= relerrLeta_norm2*norm(-Jp0,2);
disp('-------------------------------------------------------------')
disp(['Absolute error of hh0_eta computation using \ in norm 2: ',...
    num2str(relerrLeta_norm2*norm(-Jp0,2))])
disp(['Relative error of hh0_eta computation using \ in norm 2: ',...
    num2str(relerrLeta_norm2)])
disp(['Relative error of hh0_eta computation using \ in norm Inf: ',...
    num2str(abserrLeta_norm2)])

if relerrLeta_norm2<tol
    [hh0_eta2,flag_hh0_eta,relerrLeta2_norm2] = gmres(L_eta,-Jp0,[],tol,N);
    if (relerrLeta_norm2)>(relerrLeta2_norm2)
        disp(['Won''t use GMRS.'])
    elseif flag_hh0_eta==0 &&  (relerrLeta2_norm2)<(relerrLeta_norm2)
        disp(['GMRS converged within tolerance. '])
        disp(['Absolute error of hh0_eta computation using GMRS in norm 2: ',...
            num2str(relerrLeta_norm2*norm(-Jp0,2))])
        disp(['Relative error of hh0_eta computation using GMRS in norm 2: ',...
            num2str(relerrLeta_norm2)])
        disp(['Relative error of hh0_eta computation using \ in norm Inf: ',...
            num2str(relerrLeta_norminf)])
        hh0_eta=hh0_eta2;
    end
end
%-------------------------------------------------------------------------%

hh_eta=zeros(N,length(t));
for i=1:N
    hh_eta(i,:)=hh0_eta(i)*exp(LAMBDA_eta(i,i)*t);
end
% Compute coefficients for homogeneous solution
Jh=L_eta*(hh_eta);   % Size of Jh is ( N x length(t) )
% Compute homogeneous solution
E_h=zeros(length(y),length(t));
for j=1:length(y)
    for i=1:N
        E_h(j,:)=E_h(j,:)+ Jh(i,:)*eigenmodes_squire(i,j);
    end
end

%% Compute complete solution E(y,t)
E=E_p+E_h;

return
\end{lstlisting}
\ \thispagestyle{plain}
 \end{appendices}

\vspace{-20pt}
\bibliographystyle{jfm}
\addcontentsline{toc}{chapter}{References}
\bibliography{Federico_Fraternale_Master_Thesis.bib}

\end{document}